\documentclass[3p]{elsarticle}
\journal{Physics Reports}

\usepackage{amsmath,amssymb,array,booktabs,dsfont,equationarray,graphicx,mathtools,multirow,rotating,stmaryrd,tensor,verbatim}
\usepackage[bookmarksnumbered,linktocpage,pdfstartview=FitH]{hyperref}
\usepackage[all]{hypcap}

\newcolumntype{M}[1]{>{$}{#1}<{$}}
\newcolumntype{B}[1]{>{\mathbf\bgroup}{#1}<{\egroup}}
\newcolumntype{K}{>{\lvert}{c}<{\rangle}}

\DeclarePairedDelimiter{\bra}{\langle}{\rvert}
\DeclarePairedDelimiter{\ket}{\lvert}{\rangle}

\numberwithin{equation}{section}

\DeclareMathOperator{\Aut}{Aut}
\DeclareMathOperator{\Str}{Str}
\DeclareMathOperator{\tr}{tr}
\DeclareMathOperator{\Tr}{Tr} 
\DeclareMathOperator{\Det}{Det}
\DeclareMathOperator{\Pf}{Pf}
\DeclareMathOperator{\diag}{diag}

\newcommand{\be}{\begin{equation}}
\newcommand{\ee}{\end{equation}}
\newcommand{\bea}{\begin{eqnarray}}
\newcommand{\eea}{\end{eqnarray}}
\newcommand{\sfx}{\textsf{x}}
\newcommand{\sfo}{\textsf{o}}
\newcommand{\half}{\tfrac{1}{2}}
\newcommand{\braket}[2]{\langle#1|#2\rangle}

\begin{document}

\begin{frontmatter}

\title{Black Holes, Qubits and Octonions}

\author[Imperial]{L.~Borsten}
\ead{leron.borsten@imperial.ac.uk}
\author[Imperial]{D.~Dahanayake}
\ead{duminda.dahanayake@imperial.ac.uk}
\author[Imperial]{M.~J.~Duff}
\ead{m.duff@imperial.ac.uk}
\author[Imperial,IPM,Brandeis]{H.~Ebrahim}
\ead{hebrahim@brandeis.edu}
\author[Imperial]{W.~Rubens}
\ead{william.rubens06@imperial.ac.uk}

\address[Imperial]{The Blackett Laboratory, Imperial College London,\\Prince Consort Road, London SW7 2BZ, U.K.}
\address[IPM]{Institute for Research in Fundamental Sciences (IPM),\\P.O. Box 19395-5531, Tehran, Iran}
\address[Brandeis]{Theory Group, Martin Fisher School of Physics, Brandeis University,\\MS057, 415 South St., Waltham, MA 02454, U.S.A.}

\begin{abstract}

We review the recently established relationships between black hole entropy in string theory and the quantum entanglement of qubits and qutrits in quantum information theory.  The first example is provided by the measure of the tripartite entanglement of three qubits (Alice, Bob and Charlie), known as the 3-tangle, and the entropy of the 8-charge $STU$ black hole of $\mathcal{N}=2$ supergravity, both of which are given by  the $[SL(2)]^3$ invariant \emph{hyperdeterminant}, a quantity first introduced by Cayley in 1845. Moreover the classification of three-qubit entanglements is related to the classification of $\mathcal{N}=2$ supersymmetric $STU$ black holes.  There are further relationships between the attractor mechanism and local distillation protocols and between supersymmetry and the suppression of bit flip errors. At the microscopic level, the black holes are described by intersecting $D3$-branes whose wrapping around the six compact dimensions $T^6$ provides the string-theoretic interpretation of the charges and we associate the three-qubit basis vectors, $\ket{ABC}$ $(A,B,C=0$ or $1$), with the corresponding 8 wrapping cycles. The black hole/qubit correspondence extends to the 56 charge $\mathcal{N}=8$ black holes and the tripartite entanglement of seven qubits where the measure is provided by Cartan's $E_7 \supset [SL(2)]^7$ invariant. The qubits are naturally described by the seven vertices $ABCDEFG$ of the Fano plane, which provides the multiplication table of the seven imaginary octonions, reflecting the fact that $E_7$ has a natural structure of an $\mathds{O}$-graded algebra. This in turn provides a novel imaginary octonionic interpretation of the $56=7 \times 8$ charges of $\mathcal{N}=8$: the $24=3 \times 8$ NS-NS charges correspond to the  three imaginary quaternions and the $32=4 \times 8$ R-R to the four complementary imaginary octonions. We contrast this approach with that based on Jordan algebras and the Freudenthal triple system. $\mathcal{N}=8$ black holes (or black strings) in five dimensions are also related to the bipartite entanglement of three qutrits (3-state systems), where the analogous measure is Cartan's $E_6 \supset [SL(3)]^3$ invariant. Similar analogies exist for \emph{magic} $\mathcal{N}=2$ supergravity black holes in both four and five dimensions. Despite the ubiquity of octonions, our analogy between black holes and quantum information theory is based on conventional quantum mechanics but for completeness we also provide a more exotic one based on octonionic quantum mechanics. Finally, we note some intriguing, but still mysterious, assignments of entanglements to cosets, such as the 4-way entanglement of eight qubits to $E_8/[SL(2)]^8$.

\end{abstract}


\end{frontmatter}

\newpage

\tableofcontents
\newpage
\phantomsection
\addcontentsline{toc}{section}{List of Tables}
\listoftables
\newpage
\phantomsection
\addcontentsline{toc}{section}{List of Figures}
\listoffigures

\newpage

\section{\texorpdfstring{INTRODUCTION}{Introduction}}
\label{sec:Introduction}

\subsection{Overview}

It sometimes happens that two very different areas of theoretical physics share the same mathematics. This may eventually lead to the realisation that they are, in fact, dual descriptions of the same physical phenomena, or it may not.  Either way, it frequently leads to new insights in both areas. In this Review, the two areas in question are black hole entropy in string theory and qubit entanglement in quantum information theory. Quantum entanglement lies at the heart of quantum information theory, with applications to quantum computing, teleportation, cryptography and communication. In the apparently separate world of quantum gravity, the Bekenstein-Hawking entropy \cite{Bekenstein:1973ur,Hawking:1974sw} of black holes has also occupied centre stage. Despite their apparent differences, recent work \cite{Duff:2006uz,Kallosh:2006zs,Levay:2006kf,Duff:2006ue,Levay:2006pt,Duff:2007wa,Levay:2007nm,Bellucci:2007zi,Borsten:2008ur,Ferrara:2008hw,Borsten:2008,levay-2008,Bellucci:2008sv,Levay:2008mi,Borsten:2008fts} has demonstrated a correspondence between the two. Although we still do not know whether there are any physical reasons underlying these mathematical coincidences, there is now a growing dictionary, which translates a variety of phenomena in one language to those in the other. For example:
\begin{description}
\item[Qubits and $D=4$ black holes]\ \\
\begin{enumerate}
\item The measure of tripartite entanglement of three qubits (Alice, Bob and Charlie), known as the (unnormalised) 3-tangle $\tau_{ABC}$ \cite{Coffman:1999jd}, and the entropy $S$ of the 8-charge $STU$ black hole of supergravity \cite{Duff:1995sm,Behrndt:1996hu,Bellucci:2008sv} are related by \cite{Duff:2006uz}
\begin{equation}\label{eq:entropy}
S=\tfrac{\pi}{2}\sqrt{\tau_{ABC}}
\end{equation}
where $\tau_{ABC}$ is given by the \emph{hyperdeterminant} \cite{Miyake:2002}, a quantity first introduced by Cayley in 1845
\cite{Cayley:1845}.
\item The classification of three-qubit entanglements is related to the classification of $\mathcal{N}=2$ supersymmetric $STU$ black holes \cite{Kallosh:2006zs} shown in \autoref{tab:3QubitEntangClassif} and explained in more detail in \autoref{sec:blackholesandthree-qubit states}.
\begin{table}[ht]
\begin{tabular*}{\textwidth}{@{\extracolsep{\fill}}ccrrrrccc}
\toprule
& Class         & $S_A$ & $S_B$ & $S_C$ & $\Det a$ & Black hole & SUSY & \\
\midrule
& $A$-$B$-$C$   & 0     & 0     & 0     & 0        & small      & 1/2  & \\
& $A$-$BC$      & 0     & $>0$  & $>0$  & 0        & small      & 1/2  & \\
& $B$-$CA$      & $>0$  & 0     & $>0$  & 0        & small      & 1/2  & \\
& $C$-$AB$      & $>0$  & $>0$  & 0     & 0        & small      & 1/2  & \\
& W             & $>0$  & $>0$  & $>0$  & 0        & small      & 1/2  & \\
& GHZ           & $>0$  & $>0$  & $>0$  & $<0$     & large      & 1/2  & \\
& GHZ           & $>0$  & $>0$  & $>0$  & $>0$     & large      & 0    & \\
\bottomrule
\end{tabular*}
\caption[$D=4$, three-qubit entanglement classification]{The values of the local entropies $S_A, S_B$, and $S_C$ and the hyperdeterminant $\Det a$ (defined in \autoref{sec:Tangles} and \autoref{sec:Hyperdet}) are used to partition three-qubit states into entanglement classes. The entropy/entanglement correspondence relates these to $D=4,\mathcal{N}=2, STU$ model black holes (see \autoref{sec:STU}). Specifically, the to absence/presence of a horizon (small/large) and the extent of supersymmetry.}\label{tab:3QubitEntangClassif}
\end{table}
\begin{table}[ht]
\begin{tabular*}{\textwidth}{@{\extracolsep{\fill}}ccrrrrccc}
\toprule
& Class         & $S_A$ & $S_B$ & $S_C$ & $\Det a$ & Black hole & SUSY & \\
\midrule
& $A$-$B$-$C$   & 0     & 0     & 0     & 0        & small      & 1/2  & \\
& $A$-$BC$      & 0     & $>0$  & $>0$  & 0        & small      & 1/4  & \\
& $B$-$CA$      & $>0$  & 0     & $>0$  & 0        & small      & 1/4  & \\
& $C$-$AB$      & $>0$  & $>0$  & 0     & 0        & small      & 1/4  & \\
& W             & $>0$  & $>0$  & $>0$  & 0        & small      & 1/8  & \\
& GHZ           & $>0$  & $>0$  & $>0$  & $<0$     & large      & 1/8  & \\
& GHZ           & $>0$  & $>0$  & $>0$  & $>0$     & large      & 0    & \\
\bottomrule
\end{tabular*}
\caption[$D=4$, seven-qubit entanglement classification]{As in \autoref{tab:3QubitEntangClassif} entanglement measures are used to classify states, but this time concerning the tripartite entanglement of seven qubit states. The correspondence relates these to the $D=4,\mathcal{N}=8$ black holes discussed in \autoref{sec:N=8 generalisation}.}\label{tab:7QubitEntangClassif}
\end{table}
\item The attractor mechanism on the black hole side is related to optimal local distillation protocols on the QI side. Moreover, supersymmetric and non-supersymmetric black holes correspond to the suppression or non-suppression of bit flip errors \cite{Levay:2007nm}.
\item There is also a quantum information theoretic interpretation of the 56 charge $\mathcal{N}=8$ black hole in terms of a Hilbert space consisting of seven copies of the three-qubit Hilbert space \cite{Duff:2006ue,Levay:2006pt}.  It relies on the decomposition $E_{7(7)} \supset [SL(2)]^7$ and admits the interpretation, via the Fano plane \cite{Pegg}, of a tripartite entanglement of seven qubits\footnote{As explained in \autoref{sec:seven}, our terminology \emph{tripartite entanglement of seven qubits} differs from that used in some of the QI literature.} with the entanglement measure given by Cartan's  quartic $E_{7(7)}$ invariant.
\item The classification of tripartite entanglements of seven qubits is related to the classification of $(\mathcal{N}=8,D=4)$ supersymmetric black holes \cite{Borsten:2008ur} shown in \autoref{tab:7QubitEntangClassif} and explained in more detail in \autoref{sec:blackholesand7-qubitstates}.
\item Since the Fano plane provides the multiplication table of the octonions, this means that the octonions (often written off as a lost cause in physics \cite{Penrose:2005,Streater:2007}) may actually be testable in the laboratory.
\item There are similar correspondences with the black holes of the $\mathcal{N}=2$ \textit{magic} supergravities  in $D=4$  \cite{Duff:2006rf}.
\item Turning to the  \textit{microscopic} interpretation of black hole entropy in Type IIB string theory, one can consider configurations of intersecting D3-branes, whose wrapping around the six compact dimensions $T^6$ provides the microscopic string-theoretic interpretation of the charges. The three-qubit basis vectors $\ket{ABC}$, ($A,B,C=0$ or 1) are associated with the corresponding 8 wrapping cycles, where  $\ket{0}$ corresponds to \textsf{xo} and $\ket{1}$ to \textsf{ox} in \autoref{tab:3QubitIntersect}. Performing a T-duality transformation, one obtains a Type IIA interpretation with $N_0$ D0-branes, $N_1$, $N_2$, $N_3$ D4-branes plus effective D2-brane charges, where  $\ket{0}$ now corresponds to \textsf{xx} and $\ket{1}$ to \textsf{oo}.
\item In particular, one can relate a well-known fact of quantum information theory, that the most general real three-qubit state can be parameterised by four real numbers and an angle, to a well-known fact of string theory, that the most general $STU$ black hole can be described by four D3-branes intersecting at an angle.
\end{enumerate}
\item[Qutrits and $D=5$ black holes]\ \\
\begin{enumerate}
\item The measure of the bipartite entanglement of two qutrits \cite{Duff:2006rf}, known as the 2-tangle $\tau_{AB}$, is also related to the entropy of the 9-charge black hole of $D=5$ supergravity.
\begin{gather}
S=2\pi\sqrt{|\det a_{AB}|},
\shortintertext{where}
\tau_{AB}=27 |\det a_{AB}|^2.
\end{gather}
This corresponds to the $\mathcal{N}=8$ black hole with just 9 of the 27 charges switched on.
\item There is also a quantum information theoretic interpretation of the full 27 charge $\mathcal{N}=8,D=5$ black hole in terms of a Hilbert space consisting of three copies of the two-qutrit Hilbert space \cite{Duff:2007wa}.  It relies on the decomposition $E_{6(6)} \supset [SL(3)]^3$ and admits the interpretation of a bipartite entanglement of three qutrits\footnote{As explained in \autoref{sec:three}, our terminology {\it bipartite entanglement of three qutrits} differs from that used in some of the QI literature.}, with the entanglement measure given by Cartan's  cubic $E_{6(6)}$ invariant.
\item The classification of the bipartite entanglements of three qutrits is related to the classification of $\mathcal{N}=8,D=5 $ supersymmetric black holes \cite{Borsten:2008ur} shown in \autoref{tab:2QutritEntangClassif} and explained in more detail in \autoref{sec:blackholesandthree-qutritstates}.
\begin{table}[ht]
\begin{tabular*}{\textwidth}{@{\extracolsep{\fill}}ccrrccc}
\toprule
& Class       & $C_2$ & $\tau_{AB}$ & Black hole & SUSY & \\
\midrule
& $A$-$B$     & 0     & 0           & small      & 1/2  & \\
& Rank 2 Bell & $>0$  & 0           & small      & 1/4  & \\
& Rank 3 Bell & $>0$  & $>0$        & large      & 1/8  & \\
\bottomrule
\end{tabular*}
\caption[$D=5$, two-qutrit entanglement classification]{The $D=5$ analogue of \autoref{tab:3QubitEntangClassif} and \autoref{tab:2QutritEntangClassif} relates two-qutrit entanglements and their corresponding $D=5, \mathcal{N}=8$ black holes.}\label{tab:2QutritEntangClassif}
\end{table}
\item There are similar correspondences with the black holes of the $\mathcal{N}=2$ \emph{magic} supergravities  in $D=5$ \cite{Duff:2006rf}.
\item Turning to the \emph{microscopic} interpretation of black hole entropy in M- theory, one can consider configurations of intersecting M2-branes, whose wrapping around the six compact dimensions $T^6$ provides the microscopic M-theoretic interpretation of the charges. The two-qutrit basis vectors $\ket{AB}$, ($A,B=0$ or 1 or 2) are associated with the corresponding 9 wrapping cycles, where  $\ket{0}$ corresponds to \textsf{xoo}, $\ket{1}$ to \textsf{oxo} and $\ket{2}$ to \textsf{oox} as in \autoref{tab:2QutritIntersect}.
\item In particular, one can relate a well-known fact of quantum information theory, that the most general real two-qutrit state can be parameterised by three real numbers, to a well-known fact of M- theory, that the $9$-charge black hole can be described by three intersecting M2-branes.
\end{enumerate}
\item[M-theory and octonions]\ \\ \\
This is a two-way process and the qubit interpretation can also teach us new things about M-theory:
\begin{enumerate}
\item The role played by the theory of hyperdeterminants \cite{Gelfand:1994} in constructing U-duality invariants.
\item The remarkable fact that $E_7$ has  a natural structure of an $\mathds{O}$-graded algebra \cite{Manivel:2005,Elduque:2005}, compatible with its action on the minimal 56-dimensional representation, provides a new octonionic view of $M$-theory compactified on $T^7$, and hence a new quaternionic view of string theory compactified on $T^6$.  Explicitly, as discussed in \autoref{sec:graded}
    \begin{gather}
    \begin{gathered}
    \textstyle\mathfrak{e}_7=\times_l~\mathfrak{sl}(A_l)e_0\oplus\bigoplus_{1\leq i \leq 7}\left[\bigotimes_{i\notin l} A_l\right]e_i, \\
    \textstyle\mathbf{56} = \bigoplus_{1\leq i \leq 7}\left[\bigotimes_{i\in l} A_l\right]e_i, \label{eq:man561}
    \end{gathered}
    \shortintertext{where}
    i\in\{1,\dotsc,7\}
    \shortintertext{are the seven points of the Fano plane,}
    l\in\{124,235,346,457,561,672,713\}
    \end{gather}
    are the seven lines, and we have attached to each line a two-dimensional vector space $A_l$.  $e_0$ is the real octonion and $e_i$ are the imaginary ones. The same formulae hold good if we go to the dual Fano plane by swapping the roles of points and lines. There is a quaternionic analogue of this construction where we consider just three of the seven lines. This describes the $\mathcal{N}=4$ subsector.

    Strings can carry two kinds of charge: NS-NS coming from right and left moving bosonic modes and R-R coming from right and left moving fermionic modes (NS = Neveu-Schwarz and R = Ramond). When the U-duality group $E_{7(7)}$ is decomposed under the $SL(2)$ S-duality and $SO(6,6)$ T-duality
    \begin{equation}
    \begin{array}{c*{4}{@{\ }c}}
    E_{7(7)}    & \supset &  SL(2)                   & \times & SO(6,6), \\
    \textbf{56} & \to     & (\textbf{2},\textbf{12}) & +      & (\textbf{1},\textbf{32}),
    \end{array}
    \end{equation}
    the first term describes the $\mathcal{N}=4$ subsector with 24 NS-NS charges and the second term describes the 32 R-R charges.  In terms of the above seven lines of the Fano plane of \autoref{fig:FanoPlane} (which describe the seven imaginary octonions), the NS-NS charges correspond to the three lines $124,561,713$ (which describe the three imaginary quaternions) and the R-R to the four lines $235,346,457,672$ (which describe the four complementary imaginary octonions).
\item Noting that $24= 8 \times 3$, we show in \autoref{sec:Cayleyimquat}, that these NS-NS charges may be interpreted as the 8 charges  of the $STU$ model defined over the 3 imaginary quaternions. Accordingly, the $\mathcal{N}=4$ Cartan invariant with $SL(2)  \times SO(6,6)$ symmetry, may be written as Cayley's hyperdeterminant defined over the imaginary quaternions, provided we adopt a suitable operator ordering\footnote{All this suggests that $D=10$ string theory compactified on $T^6$ is dual to $D=6$ string theory compactified on $T^2$ defined over the imaginary quaternions, at least for the NS-NS sector. We hope to return to this elsewhere.}.
\item Noting that $56= 8 \times 7$, it is tempting to employ a similar construction replacing imaginary quaternions by imaginary octonions to describe the full 56 charges, including the 32 R-R. Here, however, the relationship is more subtle. Since the $\mathcal{N}=8$ Cartan invariant is just the singlet in $\mathbf{56}\times\mathbf{56}\times\mathbf{56}\times\mathbf{56}$, it follows from \eqref{eq:man561} that it may indeed be expressed as a quartic combination of octonions.  However, as explained in \autoref{sec:Cayley} it is not simply given by Cayley over the octonions, at least if Cayley is defined in the usual way. Nevertheless, the two are closely related\footnote{More speculatively, all this might suggest a similarity between $D=11$ M-theory compactified on $T^7$ and $D=6$ string theory compactified on $T^2$ defined over the imaginary octonions.}.
\item We emphasise that this way of incorporating the octonions, inspired by the Fano plane qubit interpretation, is completely different from the role of octonions in the description of $\mathcal{N}=8$ supergravity using Jordan algebras and Freudenthal triple systems \cite{Ferrara:1997uz,Kallosh:2006zs,Pioline:2006ni} , described in  \autoref{sec:Freudenthal}. For example, in the former we are dealing with the 7 imaginary octonions, while in the latter we are dealing with the 8 \emph{split} octonions.  There are other important differences. These are described in \autoref{sec:Freudenthal-Fano}, where we provide a dictionary to go from one language to the other.
\end{enumerate}
\end{description}

\subsection{Summary of Report}

The purpose of this Report is twofold. First we summarise the progress made so far in this interplay between black hole entropy and quantum information theory and secondly we describe some new so-far unpublished developments.  Both black holes and entanglement are subjects that can be quite technical, but our aim is to make our exposition understandable to both communities.

We begin in \autoref{sec:black} with some background material on supergravity, string theory and M-theory, focussing on the issue of U-duality and black hole entropy required in subsequent sections.  In particular we identify the U-duality group of the $STU$ model as $[SL(2)]^3$.

\hyperref[sec:quantum]{Section~\ref*{sec:quantum}} serves a similar purpose for quantum information theory, introducing some elementary concepts such as entanglement and SLOCC (Stochastic Local Operations and Classical Communication). In particular we identify the group of invertible SLOCC transformations for three qubits as $[SL(2)]^3$, which in fact first suggested the link to black holes. We continue in \autoref{sec:qubits} with more on qubits, focussing especially on three qubits and the all-important Cayley's hyperdeterminant which provides the measure of tripartite entanglement known as the 3-tangle. The hyperdeterminant will also determine the $D=4$ black hole entropy. \hyperref[sec:qutrits]{Section~\ref*{sec:qutrits}} performs the analogous role for the two qutrit system and the corresponding 2-tangle that will be  related to the $D=5$  black hole entropy.

The black holes of the $D=4,\mathcal{N}=2$ $STU$ model are introduced in \autoref{sec:STU} and in \autoref{sec:correspondence}  we make the connection with quantum information theory, showing that the black hole entropy as a function of the 8 charges is also given by Cayley's hyperdeterminant. One may then go further and match the classification of $\mathcal{N}=2$ black holes to the classification of three-qubit entanglements. The higher-order corrections to the black hole entropy formula also admit a QI interpretation.   This section also provides yet more entries in the dictionary by relating the attractor mechanism on the black hole side to SLOCC on the QI side and by relating supersymmetric and non-supersymmetric solutions to suppression or non-suppression of bit flip errors.

The generalisation from $\mathcal{N}=2$ to $\mathcal{N}=8$ black holes is the subject of \autoref{sec:N=8}. Here we encounter Cartan's quartic invariant which provides the entropy of the 56-charge black hole and is invariant under the $E_{7(7)}$ U-duality. The QI interpretation is that of the tripartite entanglement of seven qubits (Alice, Bob, Charlie, Daisy, Emma, Fred and George). The entanglement is encoded in the Fano plane which also provides the multiplication table of the seven imaginary octonions. The seven vertices of the Fano plane $ABCDEFG$ describe the seven qubits, while the seven lines $abcdefg$, each passing through three vertices, describe the intricate tripartite entanglement.

While the Fano plane basis is tailored to describe the seven qubits, the $\mathcal{N}=8$ black hole is most elegantly discussed within the framework of Jordan algebras and the Freudenthal triple system (FTS), which is the subject of \autoref{sec:Freudenthal}. In fact FTS provides a unified description of both the $\mathcal{N}=8$ black holes and the magic $\mathcal{N}=2$ black holes of \autoref{sec:magic}. The magic supergravities correspond to Jordan algebras over the reals $\mathds{R}$, complex $\mathds{C}$, quaternions $\mathds{H}$ and octonions $\mathds{O}$ and the $\mathcal{N}=8$ supergravity to the split octonions $\mathds{O}^s$. The $\mathcal{N}=2$ $STU$ model also fits within this scheme with the Jordan algebra being trivial.

In \autoref{sec:Cartan} we provide the three descriptions of the group $E_7$ and its quartic invariant that we will find useful and the dictionaries that link them: Cartan-Fano, Fano-Freudenthal, and Freudenthal-Cartan.

\hyperref[sec:Cayley]{Section~\ref*{sec:Cayley}} shows how the $\mathcal{N}=4$ $SL(2) \times SO(6,6)$ invariant may, with a suitable choice of operator ordering, be expressed as Cayley's quartic hyperdeterminant over the imaginary quaternions. The $\mathcal{N}=8$ invariant is more subtle. Although it can be expressed as a quartic product of imaginary octonions, it  not the same as Cayley's hyperdeterminant over the imaginary octonions.

In \autoref{sec:five}, we turn our attention to five-dimensional black holes, which also exhibit non-trivial entropy. In particular, we find a relation between $\mathcal{N}=8$ black holes and the bipartite entanglement of three qutrits, with Cartan's cubic $E_6$ invariant playing the dual roles of black hole entropy and qutrit entanglement measure.

Magic supergravities in both four and five dimensions can also be incorporated into the black hole/qubit and black hole/qutrit correspondence in a way similar to the $STU$ model and the $\mathcal{N}=8$ models, as described in \autoref{sec:magic}. However, involving as they do the Jordan algebras over $\mathds{R,C,H,O}$, they also offer an alternative interpretation in terms of unconventional quantum mechanics defined over $\mathds{R,C,H,O}$. Indeed, it was just such variations of standard quantum mechanics which led Jordan to propose his algebras in the first place. We compare this new interpretation with the standard one, listing the pros and cons. The main objection, in our view, is that while it seems to work well in five dimensions, we have been unable to get it to work in four.

One might ask why black holes should display any kind of two-valuedness at all. We answer this question in \autoref{sec:wrap}  by turning to the \emph{microscopic} interpretation of black hole entropy in Type IIB string theory. One can associate configurations of intersecting D3-branes, whose wrapping around the six compact dimensions $T^6$ provides the microscopic string-theoretic interpretation of the charges. The three-qubit basis vectors $\ket{ABC}$, ($A,B,C=0$ or 1) are associated with the corresponding 8 wrapping cycles, where $\ket{0}$ corresponds to \textsf{xo} and $\ket{1}$ to \textsf{ox} as in \autoref{tab:3QubitIntersect}. Performing a T-duality transformation, one obtains a Type IIA interpretation with $N_0$ D0-branes, $N_1$, $N_2$, $N_3$ D4-branes plus effective D2-brane charges, where  $\ket{0}$ now corresponds to \textsf{xx} and $\ket{1}$ to \textsf{oo}. To wrap or not to wrap; that is the qubit. Similarly, in M-theory, one can associate configurations of intersecting M2-branes, whose wrapping around the six compact dimensions $T^6$ provides the microscopic M-theoretic interpretation of the charges. The two-qutrit basis vectors $\ket{AB}$, ($A,B=0$ or 1 or 2) are associated with the corresponding 9 wrapping cycles, where  $\ket{0}$ corresponds to \textsf{xoo}, $\ket{1}$ to \textsf{oxo} and $\ket{2}$ to \textsf{oox} as in \autoref{tab:2QutritIntersect}.

Finally, in \autoref{sec:conclusions} we list some unsolved problems and directions for future research.

\autoref{sec:Transvectants} gives a historical overview of Cayley's original 1845 treatment of the hyperdeterminant.

In \autoref{sec:compact} we look at superalgebras \cite{Townsend:1995gp} for M-theory in $D=11$, Type IIA theory on $D=10$ and Type IIB theory in $D=10$ in order to see how the central charges give rise to the black hole charges in $D=4$ after compactification on either $T^7$ or $T^6$.

The SLOCC group for an $n$-qudit state is given by $G=[SL(d,\mathds{C})]^n$, but subspaces of the $d^n$-dimensional Hilbert space may display hidden symmetries not contained in $G$. As we shall see,  these include an $E_7(\mathds{C})$ symmetry of a 56-dimensional subspace of seven qutrits and $E_6(\mathds{C})$ symmetry of a 27-dimensional subspace of three 7-dits. In \autoref{sec:hidden} we explore some more possibilities, involving in particular  the observation that a 224-dimensional subspace of eight qutrits, describing the 4-way entanglement of 8 qubits, may be assigned to the coset-space $E_8/[SL(2)]^8$.  We have confined these coset constructions to the \hyperref[sec:hidden]{Appendix} because we have as yet no good application of them within quantum information theory.

In \autoref{sec:DiscreteSymmFano} we discuss the $PSL(2,\mathds{F}_7)$, the discrete subgroup of $E_7$ which is the symmetry of the Fano plane, and its 56-dimensional representation.

\section{\texorpdfstring{BLACK HOLES}{Black Holes}}
\label{sec:black}

\subsection{Supergravity, string theory and M-theory}
\label{sec:string}

One of the central dilemmas of XXI century physics is that the two main pillars of XX century physics, quantum mechanics and Einstein's general theory of relativity, seem to be mutually incompatible. General relativity fails to comply with the quantum rules that govern the behaviour of elementary particles, while black holes are challenging the very foundations of quantum mechanics. Something big has to give.

Until recently, the best hope for a theory that would unite gravity with quantum mechanics and describe all physical phenomena was based on strings: one-dimensional objects whose modes of vibration represent the elementary particles. In 1995, however, strings were subsumed by M-theory. In the words of Edward Witten \cite{Horava:1995qa,Witten:1995em}, M stands for magic, mystery or membrane, according to taste. New evidence in favour of this theory continues to appear.

M-theory, like string theory, relies crucially on the idea of supersymmetry \cite{Duff:1996aw}. Supersymmetry requires that for each known boson there is a fermion of equal mass. However, no such superpartners have yet been found. The symmetry, if it exists at all, must be broken, so that the postulated particles do not have the same mass as known ones but instead are too heavy to be seen in current accelerators. The Large Hadron Collider at CERN Geneva, will be looking for just these particles.  Theorists persist with supersymmetry because it provides a framework within which the weak, electromagnetic and strong forces may be united with gravity.

Conventional gravity does not place any limits on the possible dimensions of spacetime: its equations can, in principle, be formulated in any dimension. Not so with supergravity \cite{Salam:1989fm,deWit:2002vz,Tanii:1998px}, which places an upper limit of 11 on the dimensions of space-time (10 of space and one of time). In 1978 Cremmer, Julia and Scherk \cite{Cremmer:1978km} realised that supergravity not only permits up to seven extra dimensions but is most elegant when written in 11-dimensional form. The kind of real, four-dimensional world the theory ultimately predicts depends on how the extra dimensions are curled up, in the way suggested by Kaluza and Klein in the 1920s in their unified theory of gravity and electromagnetism. Seven curled dimensions could conceivably allow the appearance of the strong and weak nuclear forces, in addition to electromagnetism, For these reasons, many physicists began to look to supergravity in 11 dimensions for the unified theory.

In 1984, however, 11-dimensional supergravity was knocked off its pedestal by 10-dimensional superstring theory.  There were five consistent anomaly-free theories: the $E_8 \times E_8$ heterotic, the $SO(32)$ heterotic, the $SO(32)$ Type I, and the Type IIA and Type IIB strings. The Type I is an \emph{open} string consisting of just a segment; the others are \emph{closed} strings that form loops. The $E_8 \times E_8$ seemed, at least in principle, capable of explaining the elementary particles and forces, including their handedness. Furthermore, strings seemed to provide a theory of gravity consistent with quantum effects without the ultraviolet divergences that plagued general relativity.

After the initial euphoria over strings, however, doubts began to creep in. First, important questions, especially how to confront the theory with experiment, seemed incapable of being answered by perturbative methods.  Second, why were there five different string theories? If one is looking for a unique Theory of Everything, this seems like an embarrassment of riches. Third, if supersymmetry permits 11 dimensions, why do superstrings stop at 10? Finally, if we are going to conceive of pointlike particles as strings, why not as membranes or more generally as p-dimensional objects, inevitably dubbed $p$-branes?

Supersymmetry severely restricts the possible dimensions of a $p$-brane. A spacetime of 11 dimensions permits a supermembrane and a super 5-brane, called the M2-brane and M5-brane respectively. In 1987 it was shown \cite{Duff:1987bx} that if one of the 11 dimensions is a circle, one can wrap the membrane around it once, pasting the edges together to form a tube. If the radius becomes sufficiently small, the rolled-up membrane ends up looking like a string in 10 dimensions; it yields precisely the Type IIA superstring.

In a landmark talk at the University of Southern California in 1995, Witten \cite{Witten:1995ex} drew together all this work on strings, branes and 11 dimensions under the umbrella of M-theory in 11 dimensions, which has 11-dimensional supergravity as its low-energy limit.   In particular, he pointed out that the strength of the string coupling constant $g_{\text{s}}$ grows with the radius of the 11th dimension: M-theory is intrinsically non-perturbative! Since then, thousands of papers have appeared confirming that whatever M-theory may be, it certainly involves branes in an important way.

In 1995 Polchinski \cite{Polchinski:1995mt} realised that certain $p$-branes, appearing in Type IIA and Type IIB and carrying R-R charge, admit the dual interpretation as Dirichlet branes or \emph{D-branes}; surfaces on which open strings can end. Such breakthroughs have led to a new interpretation of black holes as intersecting black-branes wrapped around the six curled dimensions of string theory or seven of M-theory. As a result, there are strong hints that M-theory may even clear up the paradoxes of black holes raised by Hawking, who in 1974 showed that black holes are not entirely black but may radiate energy. In that case, black holes must possess entropy, which measures the disorder by counting for the number of quantum states available. Yet the microscopic origin of these states stayed a mystery. The D-brane technology  has enabled Strominger and Vafa \cite{Strominger:1996sh} to count the number of quantum states in black-branes. They find an entropy that agrees with Hawking's prediction, placing another feather in the cap of M-theory. Thus, branes are no longer the ugly ducklings of string theory. They have taken centre stage as the microscopic constituents of M-theory, as the higher-dimensional progenitors of black holes and as entire universes in their own right.

Despite all these successes, theorists are glimpsing only small corners of M-theory; the big picture is still lacking. In trying to discover what M-theory really is, the understanding black holes will be an essential prerequisite.

\subsection{U-duality}
\label{sec:U}

String theory may be described as a worldsheet sigma-model with the background spacetime as its target space \cite{Green:1987sp,Green:1987mn}. Different backgrounds generally correspond to different quantum string theories. But it might happen that some of the backgrounds produce physically equivalent theories. In such a case the different backgrounds are mapped into each other by discrete transformations coming from the symmetry groups of string dualities. These dualities which transform one theory into the other are classified into T, S and U-dualities where the latter one is a combination of the other two. In order to explain these transformations we start by a brief overview on T and S-dualities.

The simplest example of T-duality is provided by superstring theory compactified on a circle. The worldsheet theory with a circle of radius R is dual to that on a circle of radius ${\alpha'}/{R}$ where $\alpha'$ is the tension of the string. This can be generalised to toroidal compactification of string theories where the compact space is a k-dimensional torus $T^k$. In this case the dimensionally reduced theory is invariant under the discrete symmetry group $SO(k,k,\mathds{Z})$ where its discreteness is due to the quantisation of the charges. This is true not only in the low energy effective theory but also in the full interacting theory. In fact, T-duality is true perturbatively order by order in the string coupling constant $g_{\text{s}}$. The fields  of the dimensionally reduced theory  transform as representations of this T-duality symmetry group. The space of scalar fields, or \emph{moduli space}, of the theory is given by the coset space
\begin{equation}
\frac{SO(k,k,\mathds{R})}{SO(k,\mathds{R})\times SO(k,\mathds{R})}
\end{equation}
factored by the T-duality group $SO(k,k,\mathds{Z})$. See \cite{Giveon:1994fu} for a review of T-duality. From the open string or D-brane point of view, T-duality transforms Dirichlet boundary conditions into Neumann boundary conditions and vice versa \cite{Polchinski:1995mt}.

The other important string theory duality is the S-duality \cite{Font:1990gx,Rey:1989xj,Schwarz:1993mg,Duff:1993ij,Schwarz:1993vs,Sen:1992ch,Sen:1992fr} which generalises the electric/magnetic duality of supersymmetric Yang-Mills theories \cite{Montonen:1977sn}. In contrast to T-duality, S-duality acts non-perturbatively and is, strictly speaking, still a conjecture, although the evidence in its favour is now overwhelming. It relates the theory at strong coupling to the same theory at weak coupling. One example of this duality is provided by Type IIB string theory \cite{Hull:1994ys}. The S-duality group transformation is similar to electric-magnetic duality in super Yang-Mills theory, in which the coupling $g^2_{\text{YM}}$ is mapped into ${1}/{g^2_{\text{YM}}}$. In order to explain this duality more explicitly we consider the massless spectrum of the Type IIB theory. This spectrum contains two scalar fields $\phi$ and $C$ which belong to NS-NS and R-R sectors, respectively. The former one is called dilaton and the latter one axion. One can write the low energy effective action of the Type IIB theory in terms of a complex scalar ($\tau$) defined by
\begin{equation}
\tau=C+ie^{-\phi}.
\end{equation}
It can be shown that the equations of motion are invariant under the symmetry group $SL(2,\mathds{R})$ under which $\tau$ transforms as
\begin{equation}
\tau\to\frac{a\tau+b}{c\tau+d},
\end{equation}
where $a$, $b$, $c$ and $d$ specify the $SL(2,\mathds{R})$ transformation and satisfy the condition $ad-bc=1$. Therefore there exists a transformation which maps $\tau$ into ${-1}/{\tau}$. In the case of zero axion ($C=0$) it maps weak coupling  into strong coupling, where $e^{\phi}=g_{\text{s}}$ is the string coupling constant. The other massless fields which transform as a doublet under this symmetry group are the fundamental string which carries the charge of NS-NS 2-form $B_2$ and the D-string which has the charge of the R-R 2-form $C_2$. Noting that the F- and D-string bound states can be formed and their charges quantised, the correct symmetry group is the integer subgroup  $SL(2,\mathds{Z})$. This is called the S-duality symmetry group of Type IIB.

Another context in which S-duality arises is in the compactification of string theory from ten to four dimensions where an $SL(2,\mathds{Z})$ emerges  as an electric-magnetic duality that transforms field equations into Bianchi identities.  It may be shown to be a consequence of six-dimensional string/string duality \cite{Duff:1994zt} which in turn is a consequence of membrane/fivebare duality of eleven-dimensional M-theory \cite{Duff:1996rs}.
For $\mathcal{N}=2$ compactifications, the combined S- and T -dualities are then given by
\begin{equation}\label{eq:d=4n=2}
SL(2,\mathds{Z}) \times SO(l,2,\mathds{Z}),
\end{equation}
where the resulting low-energy limit is $(D=4,\mathcal{N}=2)$ supergravity coupled to $l+1$ vector multiplets \cite{Cremmer:1984hc}. We shall encounter the $l=2$ case again in \autoref{sec:stu}. For $\mathcal{N}=4$ compactifications, the combined S- and T-dualities are given by
\begin{equation}\label{eq:d=4n=4}
SL(2,\mathds{Z}) \times SO(6,m,\mathds{Z}),
\end{equation}
where the resulting low-energy limit is $(D=4,\mathcal{N}=4)$ supergravity coupled to $m$ vector multiplets \cite{Bergshoeff:1985ms}. We shall encounter the $m=6$ case again in \autoref{sec:N=8 generalisation}.

A interesting phenomenon happens when one considers the strong coupling limit of Type IIA theory. In that case D0-branes are the lightest objects in the spectrum since their mass is $\tau_0={1}/{g_{\text{s}} \alpha'^{1/2}}$. Therefore the state of any number $n$ of D0-branes becomes light at strong coupling and reaches a continuum. Such a continuum limit acts as an extra dimension where the theory becomes noncompact. Therefore one can conclude that the strong coupling limit of Type IIA theory is the 11-dimensional theory M-theory \cite{Witten:1995ex}. The low energy effective action of M-theory is 11-dimensional supergravity and the non-perturbative objects in this theory are M2-branes and their magnetic duals, the M5-branes.

The S and T-duality groups can be unified in a larger group called U-duality  \cite{Hull:1994ys}.  See also \cite{Duff:1990hn} for similar conjectures. U-duality which is a non-perturbative symmetry group mixes the sigma model and string coupling constants, $\alpha'$ and $g_{\text{s}}$ respectively. For M-theory  on $\mathds{R}^d\times T^{k}$ or string theory on $\mathds{R}^d\times T^{k-1}$, where $d+k=11$, the reduced theories are invariant under a global symmetry group called the U-duality group. The reduced d-dimensional low energy effective action, which is the d-dimensional supergravity theory, is invariant under a continuous symmetry group and the discrete subgroup of it, the U-duality group, is the symmetry of the full theory.

A classification of symmetry groups of the supergravities with 32 supercharges in different dimensions has been given in \autoref{tab:Uduality} \cite{Polchinski:1998rr}. A general prescription is given for $3\leq k \leq 8$ where the global symmetry group can be identified as the exceptional group $E_{k(k)}$ which is the maximally noncompact form of $E_k$ \cite{Cremmer:1979up}. The maximal discrete subgroups of the groups $G$ are in fact the U-duality groups of the compactified theories in different dimensions. Regarding the 10-dimensional Type IIA and IIB theories, $SL(2,\mathds{Z})$ is the S-duality group of Type IIB while $SO(1,1,\mathds{Z})$ is the symmetry group of Type IIA. Specifically for $d=3$, 4 and 5 dimensions the exceptional groups $E_{k(k)}(\mathds{Z})$ contain the subgroup $SL(2,\mathds{Z})\times SO(k-1,k-1,\mathds{Z})$. This can be identified as S-duality group $SL(2,\mathds{Z})$ and T-duality group $SO(k-1,k-1,\mathds{Z})$ coming from the compactification of string theory on $T^{k-1}$.

Using  the maximal compact subgroup $H$ of $G$ in \autoref{tab:Uduality}, one can form the coset space $G/H$ which defines the homogeneous space to which the moduli belong.  The number of the scalar fields of the compactified theory is equal to the dimension of the coset space, $\dim G-\dim H$.

Of special interest is the compactification to four dimensions where the U-duality group is
\begin{equation}\label{eq:d=4n=8}
E_{7(7)},
\end{equation}
the non-compact form of $E_7$ with 63 compact and 70 non-compact generators. The resulting low-energy limit is $(D=4,\mathcal{N}=8)$ supergravity with 28 abelian vector fields. The 28 electric and 28 magnetic black hole charges transform as an irreducible 56 of $E_{7(7)}$ as shown in \autoref{tab:PeterWest} which we have taken from \cite{Riccioni:2007au,Cook:2008bi}. We shall encounter these again in \autoref{sec:N=8 generalisation}.

For compactification to five dimensions, on the other hand, the analogue of  \eqref{eq:d=4n=2} is
\begin{equation}\label{eq:d=5n=2}
SO(1,1,\mathds{Z}) \times SO(l-1,1,\mathds{Z}),
\end{equation}
for which the resulting low-energy limit is $(D=5,\mathcal{N}=2)$ supergravity coupled to $l$ vector multiplets. The analogue of  \eqref{eq:d=4n=4} is
\begin{equation}\label{eq:d=5n=4}
SO(1,1,\mathds{Z}) \times SO(m-1,5,\mathds{Z}),
\end{equation}
for which the resulting low-energy limit is $(D=5,\mathcal{N}=4)$ supergravity coupled to $m-1$ vector multiplets. The analogue of \eqref{eq:d=4n=8} is
\begin{equation}\label{eq:d=5n=8}
E_{6(6)},
\end{equation}
the non-compact form of $E_6$ with 36 compact and 42 non-compact generators. The resulting low-energy limit is $(D=5,\mathcal{N}=8)$ supergravity which has 27 abelian vector fields. This gives rise to 27 electric black hole charges and 27 magnetic black string charges as shown in \autoref{tab:PeterWest}.

We shall encounter all three analogues again in \autoref{sec:fivedimensionalsupergravity}. Note that upon dimensional reduction from $D=5$ to $D=4$ we recover the corresponding three four-dimensional cases which each have one more abelian vector (the Kaluza-Klein photon).
\begin{table}[ht]
\begin{tabular*}{\textwidth}{@{\extracolsep{\fill}}ccccM{c}M{c}c}
\toprule
& $D$   & scalars & vectors & G                                         & H                                        & \\
\midrule
& 10A   & 1       & 1       & SO(1,1,\mathds{R})                        & -                                        & \\
& 10B   & 2       & 0       & SL(2,\mathds{R})                          & SO(2,\mathds{R})                         & \\
& 9     & 3       & 3       & SL(2,\mathds{R})\times SO(1,1,\mathds{R}) & SO(2,\mathds{R})                         & \\
& 8     & 7       & 6       & SL(2,\mathds{R})\times SL(3,\mathds{R})   & SO(2,\mathds{R})\times SO(3,\mathds{R})  & \\
& 7     & 14      & 10      & SL(5,\mathds{R})                          & SO(5,\mathds{R})                         & \\
& 6     & 25      & 16      & SO(5,5,\mathds{R})                        & SO(5,\mathds{R})\times SO(5,\mathds{R})  & \\
& 5     & 42      & 27      & E_{6(6)}(\mathds{R})                      & USP(8)                                   & \\
& 4     & 70      & 28      & E_{7(7)}(\mathds{R})                      & SU(8)                                    & \\
& 3     & 128     & -       & E_{8(8)}(\mathds{R})                      & SO(16,\mathds{R})                        & \\
\bottomrule
\end{tabular*}
\caption[U-duality groups]{The symmetry groups ($G$) of the low energy supergravity theories with 32 supercharges in different dimensions ($D$) and their maximal compact subgroups ($H$).}
\label{tab:Uduality}
\end{table}
\begin{sidewaystable}
\begin{tabular*}{\textwidth}{@{\extracolsep{\fill}}*{14}{c}}
\toprule
& \multirow{2}{*}{$D$} & \multirow{2}{*}{$G$} & \multicolumn{10}{c}{Form Valence}      & \\
\cmidrule(r){4-14}
&                      &                      & 1 & 2 & 3 & 4 & 5 & 6 & 7 & 8 & 9 & 10 & \\
\midrule
& \multirow{2}{*}{10A} & \multirow{2}{*}{$\mathds{R}^{+}$} & \multirow{2}{*}{\textbf{1}} & \multirow{2}{*}{\textbf{1}} & \multirow{2}{*}{\textbf{1}} & & \multirow{2}{*}{\textbf{1}} & \multirow{2}{*}{\textbf{1}} & \multirow{2}{*}{\textbf{1}} & \multirow{2}{*}{\textbf{1}} & \multirow{2}{*}{\textbf{1}} & \textbf{1} & \\
& & & & & & & & & & & & \textbf{1} & \\
\midrule
& \multirow{2}{*}{10B} & \multirow{2}{*}{$SL(2,\mathds{R})$} & & \multirow{2}{*}{\textbf{2}} & & \multirow{2}{*}{\textbf{1}} & & \multirow{2}{*}{\textbf{2}} & & \multirow{2}{*}{\textbf{3}} & & \textbf{4} & \\
& & & & & & & & & & & & \textbf{2} & \\
\midrule
& \multirow{3}{*}{9} & \multirow{3}{*}{$SL(2,\mathds{R})\times\mathds{R}^{+}$} & \textbf{2} & \multirow{3}{*}{\textbf{2}} & \multirow{3}{*}{\textbf{1}} & \multirow{3}{*}{\textbf{1}} & \multirow{3}{*}{\textbf{2}} & \textbf{2} & \textbf{3} & \textbf{3} & \textbf{4} & & \\
& & & & & & & & & & & \textbf{2} & & \\
& & & \textbf{1} & & & & & \textbf{1} & \textbf{1} & \textbf{2} & \textbf{2} & & \\
\midrule
& \multirow{4}{*}{8} & \multirow{4}{*}{$SL(3,\mathds{R})\times SL(2,\mathds{R})$} & \multirow{4}{*}{$(\overline{\textbf{3}},\textbf{2})$} & \multirow{4}{*}{$(\textbf{3},\textbf{1})$} & \multirow{4}{*}{$(\textbf{1},\textbf{2})$} & \multirow{4}{*}{$(\overline{\textbf{3}},\textbf{1})$} & \multirow{4}{*}{$(\textbf{3},\textbf{2})$} & \multirow{2}{*}{$(\textbf{8},\textbf{1})$} & \multirow{2}{*}{$(\textbf{6},\textbf{2})$} & $(\textbf{15},\textbf{1})$ & & & \\
& & & & & & & & & & $(\textbf{3},\textbf{3})$ & & & \\
& & & & & & & & \multirow{2}{*}{$(\textbf{1},\textbf{3})$} & \multirow{2}{*}{$(\overline{\textbf{3}},\textbf{2})$} & $(\textbf{3},\textbf{1})$ & & & \\
& & & & & & & & & & $(\textbf{3},\textbf{1})$ & & & \\
\midrule
& \multirow{3}{*}{7} & \multirow{3}{*}{$SL(5,\mathds{R})$} & \multirow{3}{*}{$\overline{\textbf{10}}$} & \multirow{3}{*}{\textbf{5}} & \multirow{3}{*}{$\overline{\textbf{5}}$} & \multirow{3}{*}{\textbf{10}} & \multirow{3}{*}{\textbf{24}} & $\overline{\textbf{40}}$ & \textbf{70} & & & & \\
& & & & & & & & & \textbf{45} & & & & \\
& & & & & & & & $\overline{\textbf{15}}$ & \textbf{5} & & & & \\
\midrule
& \multirow{3}{*}{6} & \multirow{3}{*}{$SO(5,5)$} & \multirow{3}{*}{\textbf{16}} & \multirow{3}{*}{\textbf{10}} & \multirow{3}{*}{$\overline{\textbf{16}}$} & \multirow{3}{*}{\textbf{45}} & \multirow{3}{*}{\textbf{144}} & \textbf{320} & & & & & \\
& & & & & & & & $\overline{\textbf{126}}$ & & & & & \\
& & & & & & & & \textbf{10} & & & & & \\
\midrule
& \multirow{2}{*}{5} & \multirow{2}{*}{$E_{6(+6)}$} & \multirow{2}{*}{\textbf{27}} & \multirow{2}{*}{$\overline{\textbf{27}}$} & \multirow{2}{*}{\textbf{78}} & \multirow{2}{*}{\textbf{351}} & $\overline{\textbf{1,728}}$ & & & & & & \\
& & & & & & & $\overline{\textbf{27}}$ & & & & & & \\
\midrule
& \multirow{2}{*}{4} & \multirow{2}{*}{$E_{7(+7)}$} & \multirow{2}{*}{\textbf{56}} & \multirow{2}{*}{\textbf{133}} & \multirow{2}{*}{\textbf{912}} & \textbf{8,645} & & & & & & & \\
& & & & & & \textbf{133} & & & & & & & \\
\midrule
& \multirow{3}{*}{3} & \multirow{3}{*}{$E_{8(+8)}$} & \multirow{3}{*}{\textbf{248}} & \textbf{3,875} & \textbf{147,250} & & & & & & & & \\
& & & & & \textbf{3,875} & & & & & & & & \\
& & & & \textbf{1} & \textbf{248} & & & & & & & & \\
\bottomrule
\end{tabular*}
\caption[U-duality representations]{The representations of the U-duality group $G$ of all the forms of maximal supergravities in any dimension. The $(D-2)$-forms dual to the scalars always belong to the adjoint representation. The scalars, parameterising the coset $G/H$, are not included in the table.}\label{tab:PeterWest}
\end{sidewaystable}
The importance of these U-duality symmetries for this Report is that the black hole entropies must be  U-duality invariants \cite{Andrianopoli:1996ve,Gunaydin:2005gd,Hull:1997kt,Sen:2008sp}, and it is these invariants that will also play the role of the new entanglement measures on the QI side. A discussion of more conventional entanglement measures may be found in \cite{Wootters:1997id,Grassl:1997sq,Kempe:1999vk,Bennett:1999,Brylinski:2000x,Brylinski:2000y,Albeverio:2001ey,Carteret:2000-1,Carteret:2000-2,Verstraete:2003,Levay:2003pb,Toumazet:2006,Abascal:2007,Plenio:2007,
Amico:2007ag,Horodecki:2007}.

\subsection{Black hole entropy}
\label{sec:entropy}

As we have seen, the low energy limit of string theory gives rise to gravity coupled to matter fields which include vectors and scalars. Therefore these theories typically admit black hole solutions.

Indeed, string theory, D-branes and M-theory  have provided a useful theoretical framework in which to study the classical and quantum properties of black holes. Since the work of Bekenstein and Hawking in the 1970s we know that  black holes behave like  thermodynamic systems characterised by their entropy and other thermodynamic quantities.  Black hole physics provides a nice relationship between geometrical properties of space-time and thermodynamic properties of a statistical system. This relation has been explicitly shown by a set of laws, called the laws of black hole mechanics, in analogy with the laws of thermodynamics \cite{Bardeen:1973gs}. See \autoref{tab:BHThermAnalogy}.
\begin{table}[ht]
\begin{tabular*}{\textwidth}{@{\extracolsep{\fill}}ccp{5cm}p{5cm}c}
\toprule
& Law    & \centering{Thermodynamics} & \centering{Black Holes} & \\
\midrule
& \multirow{3}{*}{Zeroth} &	If two thermodynamic systems are each in thermal equilibrium with a third, then they are in thermal equilibrium with each other. & The horizon has constant surface gravity for a stationary black hole. & \\
& First  & \centering{$dU = TdS - pdV + \mu dN$} & \centering{$dM = \frac{\kappa}{8\pi}dA + \Omega dJ = \Phi dQ $} & \\
& Second & \centering{$dS \geq 0$}&  \centering{$dA \geq 0$} & \\
& \multirow{2}{*}{Third}  & It is impossible to reach absolute zero temperature in a physical process. & It is impossible to form a black hole with vanishing surface gravity. & \\
\bottomrule
\end{tabular*}
\caption[Laws of black hole mechanics and classical thermodynamics]{The Laws of black hole mechanics and classical thermodynamics.}\label{tab:BHThermAnalogy}
\end{table}
The zeroth law states that the surface gravity $\kappa_S$ of a stationary black hole is constant over the event horizon. Surface gravity is a quantity which measures the strength of the gravitational field on the event horizon. Compared to the first law of thermodynamics one can conclude that the surface gravity acts like the temperature. In fact if one analyses black holes in quantum field theory in a curved background, where gravity is considered classically and matter fields are treated quantum mechanically, one can see that they emit radiation, called Hawking radiation. In comparison with the black body spectrum one finds that the temperature of the radiation is
\begin{equation}\label{eq:tem}
T_{\text{H}}=\frac{\hbar\kappa_S}{2\pi}\;.
\end{equation}
For a stationary black hole, the first law of black hole mechanics relates the variation of the energy to the other conserved quantities of the black hole such as angular momentum $J$ and charge $Q$. It states that
\begin{equation}\label{eq:flaw}
\delta M=\frac{\kappa_S}{8\pi}\delta A+\mu\delta Q+\Omega\delta J\;,
\end{equation}
where $\mu$ is the electric potential, $\Omega$ is the angular velocity of rotation and $A$ is the area of the event horizon. This is an interesting statement because it relates the quantities measured at infinity such as mass, charge and angular momentum to the quantities measured on the event horizon like surface gravity and the area of the horizon. Having the temperature given by \eqref{eq:tem} and comparing the relation \eqref{eq:flaw} with the first law of thermodynamics suggests taking the area as the entropy of the black hole:
\begin{equation}
S_{\text{BH}}=\frac{A}{4\hbar G_4}\;,
\end{equation}
where $G_4$ is the four dimensional Newton constant. This is called the Bekenstein-Hawking entropy.  Note that this formula is reliable only in classical gravity and changes as one considers higher order terms in the action. The analogy of the area and the entropy is confirmed by the second law of black hole mechanics which states the total area of all event horizons is non-decreasing. This is a statement about non-stationary processes in space-time such as black hole collisions.

The laws of black hole mechanics raise two important questions. The first concerns the statistical interpretation of Bekenstein-Hawking entropy as the logarithm of the number of quantum states associated with the black hole. Although we do not have a complete answer to this question yet, string theory has provided an answer to a special class of black holes, extremal black holes. Since they have zero Hawking temperature they do not radiate and are stable. Some of these black holes are stable by virtue of preserving some supersymmetry and one has control over the dynamics of microscopic configurations. In the string theory context these microscopic configurations which represent the black hole, involve D-branes, fundamental (F-) strings and other solitonic objects \cite{Duff:1991sz,Duff:1994an,Polchinski:1995mt}. In the next subsection we give an overview on extremal black holes.

The second question concerns the information loss problem: the thermal nature of the radiation from the evaporating black hole appears to involve a loss of information, in contradiction to standard quantum mechanics.  String theory provides some clues to this problem but it is still an open question which we will not discuss here. Reviews on black holes in string theory can be found in
\cite{Peet:2000hn, Pioline:2006ni,Ferrara:2008hw}.

\subsection{Extremal black holes}
\label{sec:extremal}

Let us consider a static, spherically symmetric four-dimensional black hole whose line element can be written
\begin{equation}\label{eq:bhsolution}
ds^2 = - e^{2 h(r)} dt^2 + e^{2 k(r)} dr^2 + r^2 (d\theta^2 + sin^2 \theta d\phi^2).
\end{equation}
The most general static black hole solution of Einstein-Maxwell theory is given by Reissner-Nordstr\"om solution which has the metric of the above form where
\begin{equation}
e^{2 h(r)} = e^{- 2 k(r)} = 1-\frac{2M}{r}+\frac{Q^2}{r^2},
\end{equation}
with $Q$ and $M$ the charge and the mass of the black hole. The mass of an asymptotically flat space-time is determined by the non-relativistic motion of a test particle in the asymptotically flat region. Such a particle sees a Newtonian gravitational potential $V=- \frac{M}{r}$ where M is related to $\frac{1}{r}$ deviation of the metric from flat space-time given by $g_{tt}=-(1-\frac{2M}{r}+\dotsb)$. Like the mass, the charge of the black hole is measured at asymptotic region by
\begin{equation}
Q=\frac{1}{4\pi} \oint_{S^2_{\infty}}\star F,
\end{equation}
where $S^2_{\infty}$ is a spacelike sphere at infinity and$\;\star F$ is the field strength dual.  (The black hole can also carry magnetic charge
\begin{equation}
P=\frac{1}{4\pi} \oint_{S^2_{\infty}}F,
\end{equation}
and we just need to replace $Q^2$ with $Q^2+P^2$ in all the formulae). The electric field strength at infinity is
\begin{equation}
F_{rt}=\frac{Q}{r^2}.
\end{equation}
Besides the mass and the charge which are measured at infinity there are two other quantities, surface gravity and the area, measured on the event horizon that are given by
\begin{equation}\label{eq:rn}
\kappa_S=\frac{\sqrt{M^2-Q^2}}{2M(M+\sqrt{M^2-Q^2})-Q^2},\quad A=4\pi(M+\sqrt{M^2-Q^2})^2.
\end{equation}
The solution \eqref{eq:bhsolution} has two horizons determined by $g_{rr}=0$ which gives
\begin{equation}
r_\pm=M\pm\sqrt{M^2-Q^2}.
\end{equation}
Using the above formulae one can see that the condition to have the singularity hidden behind the horizon for the charged black holes is to have $M\geq |Q|$. These black holes have smooth geometries at the horizon and  free-falling observers would not feel anything as they fall through the horizon.

For the case where we have $M<|Q|$, the event horizon vanishes and the solution has a naked singularity which is considered to be unphysical according to the cosmic censorship hypothesis.

An interesting case is where one considers $M=|Q|$. For these kinds of black hole, which are called \emph{extremal} black holes, the two horizons, $r_+$ and $r_-$, coincide and the radius of the horizon gets fixed in terms of the charge of the black hole, $r_0=|Q|$. Similarly the area of the horizon gets simplified and reduces to $A=4\pi Q^2$, as can be seen from \eqref{eq:rn}. Therefore the entropy which is proportional to the area is completely determined in terms of the charges of the extremal black hole. This is an important observation for extremal black holes  resulting from  the attractor mechanism. We will elaborate more on this phenomenon in \autoref{sec:attractor}. It is related to the fact that, for static solutions in four dimensions, the near horizon geometry of the extremal black holes reduces to $\text{AdS}_2\times S^2$ where AdS is anti-de Sitter space.

Another noteworthy observation is that the surface gravity for extremal black holes is zero, using the relation \eqref{eq:rn}. Therefore the temperature which is  for the static solutions in four dimensions given by
\begin{equation}
T_{\text{H}}=\frac{\hbar \kappa_S}{2\pi}
\end{equation}
is also zero. Since the Hawking radiation is proportional to the temperature, one can conclude that the extremal black holes are stable.

The supergravity theories we shall consider in this Report have more than the one photon of Einstein-Maxwell theory. The $(D=4,\mathcal{N}=2)$ $STU$ model has 4; the $(D=4,\mathcal{N}=8)$ model has 28, so the black holes will carry 8 or 56 electric and magnetic charges, respectively. Similarly, the $(D=5,\mathcal{N}=8)$ black hole and black string have 27 electric and 27 magnetic charges, respectively. They also involve scalar fields \cite{Ferrara:1997tw}. Consequently the extremal black holes obey generalised $\text{mass}=\text{charge}$ conditions described in subsequent sections. A black hole that preserves some unbroken supersymmetry (admitting  one or more Killing spinors) is said to be \emph{BPS} (after Bogomol'nyi-Prasad-Sommerfield) and non-BPS otherwise. All BPS black holes are extremal but extremal black holes can be BPS or non-BPS.

\newpage
\section{\texorpdfstring{QUANTUM INFORMATION THEORY}{Quantum Information Theory}}
\label{sec:quantum}

\subsection{Qubits and quantum information theory}
\label{sec:bits}

In completely separate developments, exciting things were happening in the world of quantum information theory. Quantum entanglement is a phenomenon in which the quantum states of two or more objects must be described with reference to each other, even though the individual objects may be spatially separated \cite{Bais:2007pk,Bell:1964kc,Bohm:1951,Einstein:1935rr,Einstein:1970,Mermin:1990,Mermin:2006}. This leads to correlations between observable physical properties of the systems that are classically forbidden. For example, as described in \autoref{sec:Bell}, it is possible to prepare two particles in a single quantum state such that when one is observed to be spin-up, the other one will always be observed to be spin-down and vice versa, this despite the fact that it is impossible to predict, according to quantum mechanics, which set of measurements will be observed. As a result, measurements performed on one system seem to be instantaneously influencing other systems entangled with it. Note, however, that quantum entanglement does not enable the transmission of classical information faster than the speed of light.

Quantum entanglement has applications in the emerging technologies of quantum computing and quantum cryptography, and has been used to realise quantum teleportation experimentally. At the same time, it prompts interesting discussions concerning the interpretation of quantum theory. The correlations predicted by quantum mechanics, and observed in experiment, reject the principle of local realism, which is that information about the state of a system should only be mediated by interactions in its immediate surroundings and that the state of a system exists and is well-defined before any measurement.

\subsection{LOCC and SLOCC}
\label{sec:LOCC}

The concept of entanglement is the single most important feature distinguishing classical information theory from quantum information theory. It has become clear to the quantum information community that, in order to exploit quantum states for communication, computation and other such purposes, it is necessary  to develop a theory that naturally describes how entanglement works and how to harness its uses. The principle of Local Operations and Classical Communication, or LOCC for short, has become the paradigm by which we may achieve this goal. Heuristically, entanglement can be understood as correlations between two or more quantum systems which cannot be of a classical origin. Thus, in order properly to understand  entanglement we must distinguish correlations of a classical nature from those of a quantum nature. The LOCC paradigm is used to precisely characterise all possible classical correlations. Any classical correlation may be experimentally established using LOCC. Conversely, all correlations not achievable via LOCC are attributed to genuine quantum correlations. That is, LOCC cannot create entanglement \cite{Plenio:2007}. What is more, it has been shown, using the LOCC paradigm, that all non-separable states can perform some task which is not possible with any separable state. This underlies the interchangeability of the terms entanglement and non-separability \cite{Plenio:2007}.

LOCC essentially describes a multi-step process for transforming any input state to a different output state while obeying certain rules. Given any multipartite state, we may split it up into its relevant parts and send each of them to different labs around the world. We allow the respective scientists to perform any experiment they see fit; they may then communicate these results to each other classically (using email or phone or carrier pigeon). Furthermore, for the most general LOCC, we allow them to do this as many times as they like.

Using the aforementioned protocol one may  create only separable states and hence it must be impossible for any entanglement to be generated since entangled states are non-separable. Therefore, we may demand that any would-be function for measuring entanglement \emph{must} be a monotonically decreasing function of \emph{any} LOCC transformation.  Any function fulfilling this and a couple other criteria will be deemed an \emph{entanglement monotone}. We will shortly arrive at a mathematical description of what this means, but first we must discuss LOCC.

In order to understand the mathematical formulation of LOCC transformations, we should briefly cover Positive Operator Valued Measures (POVMs). The field of POVMs is rather large and technical see for example \cite{Nielsen:2000,Jaeger:2006,Li:2007}, for now we will content ourselves with a basic functional introduction. The following introduction leans heavily on Nielsen and Chuang \cite{Nielsen:2000}.

POVMs are a generalisation of what is called projective measurement. Projective measurements are the usual quantum measuring operators that square to one and form an orthonormal basis. They have the following form
\begin{align}
P_i = \ket{\Psi_i}\bra{\Psi_i}
\end{align}
with the following familiar properties
\begin{subequations}
\begin{equationarray}{cccccl}
P_i^\dag                & \ =\  & P_i             &          &   & \qquad\text{Hermiticity,} \\
P_i P_j                 & \ =\  & \delta_{ij} P_i &          &   & \qquad\text{Orthogonality,} \label{eq:proj_orthogonality} \\
\bra{\Phi}P_i\ket{\Phi} & \ =\  & p_i             & \ \geq\  & 0 & \qquad\text{Positivity,} \\
\textstyle\sum_i P_i    & \ =\  & \mathds{1}      &          &   & \qquad\text{Completeness.}
\end{equationarray}
\end{subequations}
According to the Copenhagen Interpretation, any state $\ket{\Phi}$ can expanded in the measurement basis
\begin{equation}
\ket{\Phi} = \sum_i a_i \ket{\Psi_i},\qquad a_i \in\mathds{C},
\end{equation}
and a measurement will result in the collapse of the state $\ket{\Phi}$ to one of the eigenvectors $\ket{\Psi_i}$, with a probability given by $a_i^*a_i$. This much is known already. The problem is that this language is not ideally suited for real life quantum experiments nor does it adequately describe the space of possible quantum operators. For this we use density matrices and POVMs.

One way to obtain POVMs is to relax the orthogonality condition \eqref{eq:proj_orthogonality} to allow a set of operators $E_i$ that are positive and complete (the Hermiticity condition is implied as this is always true for positive operators). In this formalism, we may take any set of matrices that are positive and complete, $\{E_i\}$; these will then form set of POVM elements $E_i$, for which the measurement operators, $M_i$, will be defined as
\begin{equation}
M_i = \sqrt{E_i},
\end{equation}
where the square root of a matrix is defined as the positive square root of the eigenvalues in the basis in which the matrix is diagonal. Hence we have $\sum_i M_i^\dag M_i = \sum_i E_i = I$ and the probability that the state $\ket{\Psi}$ will go to the state $M_i\ket{\Psi}$ is given by $p_i =  \bra{\Psi}M_i^\dag M_i\ket{\Psi}$

We will shortly further generalise POVMs to quantum operators, which will form the basis of our discussion of LOCC, but before we do that, we must briefly discuss the density matrix.  The density matrix is essentially a reformulation of the usual state vector formulation but it has the added bonus that it deals with any quantum state that contains mixtures of classical and quantum correlations in a completely consistent way. For example, if we know that the output from a certain source is likely to be a certain ensemble of states $\ket{\Psi_i}$ with probabilities $p_i$, then the density matrix $\rho$ is defined to be
\begin{equation}\label{eq:mixed_state_defn}
\rho = \sum_i p_i \ket{\Psi_i}\bra{\Psi_i}.
\end{equation}
Instead of dealing with a vector $\ket{\Psi_i}$ and keeping track of various $p_i$ associated with which state the system might be in, we now deal with a matrix $\rho$ that takes care of all of that.

When we know the exact state of the system, then all but one of the $p_i$'s will be zero and the density matrix reduces to $\rho = \ket{\Psi}\bra{\Psi}$. This allows us to distinguish two types of density matrices, those which are \emph{pure} like the one in the previous sentence and can be expressed in a form such that $\rho = \ket{\Psi}\bra{\Psi}$ and those that are \emph{mixed} which can only be expressed in the form \eqref{eq:mixed_state_defn}, as an ensemble average of density matrices.

Let us now consider how the density matrix transforms under a general POVM. The probability of getting the result associated with measurement operator $M_m$ from initial state $\ket{\psi_i}$ is given by
\begin{equation}
\begin{split}
p\left(m|i\right) &= \bra{\psi_i}M_m^\dag M_m \ket{\psi_i} \\
&= \tr \left(M_m^\dag M_m \ket{\psi_i} \bra{\psi_i} \right).
\end{split}
\end{equation}
But $p(m) = \sum_i p(m|i)p_i$ so
\begin{equation}
\begin{split}
p(m) &= \sum_i p\left(m|i \right) p_i \\
&= \sum_i \tr \left(M_m^\dag M_m \ket{\psi_i}\bra{\psi_i}\right) p_i \\
&= \tr\left(M_m^\dag M_m \rho \right).
\end{split}
\end{equation}
Now, under the action of POVM elements $M_m$, on the state $\ket{\psi_i}$, we have
\begin{equation}\label{eq:psi_povm}
\ket{\psi_i}\to\ket{\psi_i^m} = \frac{M_m\ket{\psi_i}}{\sqrt{\bra{\psi_i}M^\dag_m M_m\ket{\psi_i}}}.
\end{equation}
To see how the density matrix transforms, let us consider $\rho_m$ associated with result $m$ from an ensemble of states $\ket{\psi^m_i}$, which will then have probabilities $p(i|m)$
\begin{equation}
\begin{split}\label{eq:rho_m_povm_transform}
\rho_m &= \sum_i p(i|m) \ket{\psi^m_i}\bra{\psi^m_i} \\
&= \sum_i p(i|m)\frac{M_m\ket{\psi_i}\bra{\psi_i}M_m^\dag}{\bra{\psi_i} M^\dag_m M_m\ket{\psi_i}},
\end{split}
\end{equation}
where we substituted $ \ket{\psi_i^m}$ from \eqref{eq:psi_povm}. Now, from probability theory, we know $p(i|m) = p(m,i)/p(m) = p(m|i)p_i/p(m)$, which on substituting in \eqref{eq:rho_m_povm_transform}, gives us
\begin{equation}
\begin{split}
\rho_m &= \sum_i p(i) \frac{M_m \ket{\psi_i}\bra{\psi_i} M_m^\dag}{\bra{\psi_i} M^\dag_m M_m \ket{\psi_i}} \\
&= \frac{M_m \rho M_m^\dag}{\tr\left( M^\dag_m M_m \rho \right)}.
\end{split}
\end{equation}

We now prove that the necessary and sufficient conditions for a matrix to be a density operator are that its trace is 1
and that it is positive. Let us suppose that $\rho = \sum_i p_i \ket{\psi_i}\bra{\psi_i}$, then
\begin{equation}
\tr\rho = \sum_i p_i \tr \left( \ket{\psi_i}\bra{\psi_i} \right) = \sum_i p_i = 1,
\end{equation}
and for an arbitrary state $\ket{\phi}$ we have
\begin{equation}
\begin{split}
\bra{\phi} \rho \ket{\phi} &= \sum_i p_i \braket{\phi}{\psi_i}\braket{\psi_i}{\phi} \\
&= \sum_i p_i | \braket{\phi}{\psi_i} |^2 \\
& \geq 0.
\end{split}
\end{equation}

Conversely, suppose that $\rho$ is positive and has trace 1, then it must have a spectral decomposition
\begin{equation}
\rho = \sum_i \lambda_i \ket{i}\bra{i},
\end{equation}
where the $\lambda_i$ are real, positive eigenvalues of $\rho$, and from the trace condition we have $\sum_i \lambda_i= 1$ which implies $\rho$ describes the ensemble $\{ \lambda_i, \ket{i}\}$.

We may now extend our earlier analysis of POVMs to quantum operators. Considering quantum states as positive matrices with trace 1, we may now define any valid quantum process as a map
\begin{equation}
\rho \to \mathcal{E}(\rho)
\end{equation}
from the domain to the codomain of valid density operators. We include those functions that enlarge or reduce the Hilbert space, either by locally appending ancilla, the new larger joint system being allowed to evolve unitarily, or by tracing out local subsystems. These superoperators must obey three axioms:
\begin{enumerate}
\item $\tr\mathcal{E}(\rho)$ is the probability that the transformation $\rho \to\mathcal{E}(\rho)$ occurs.
\item $\mathcal{E}(\rho)$ is a linear convex map on the set of density operators, that is,
    \begin{equation}
    \textstyle\mathcal{E}\left(\sum_i p_i \rho_i \right) = \sum_i p_i \mathcal{E}\left(\rho_i\right).
    \end{equation}
    This is necessary so that $\mathcal{E}(\rho)$ makes sense as a density matrix in its own right.
\item $\mathcal{E}(\rho)$ is a completely positive map. Requiring only positivity allows one to construct maps that are positive on a subsystem but not on the complete Hilbert space; \emph{complete} positivity cures that problem.
\end{enumerate}
Using these three axioms, it is not too hard to show that
\begin{equation}
\mathcal{E}\left(\rho\right) = \sum_i E_i \rho E_i^\dag,
\end{equation}
for some set ${E_i}$ of operators that map the input Hilbert space to the output Hilbert space and obey $\sum_i E_i^\dag E_i \leq \mathds{1}$. These $\mathcal{E}$ are called Completely Positive maps (CP maps), and if they satisfy the stronger condition $\sum_i E_i^\dag E_i = \mathds{1}$ then they are called Completely Positive Trace Preserving maps (CPTP maps) and are like the POVMs. We are now in a position to describe LOCC.

As explained earlier, LOCC is the name of a class of protocols that act on any multi-party quantum state that is split between a number of parties (say Alice in France, Bob in the States, and Charlie on the moon, etc...). Alice, Bob and Charlie are allowed to perform any local quantum operation they want (LO) and communicate the results back to the others (CC) for further experimentation. It is deemed self evident that one cannot introduce non-local quantum correlations using this process, and hence that entanglement can never be created by this process.

For now we will assume that any LOCC process is deterministic but not necessarily reversible (i.e. we may consider maps that enlarge/reduce the Hilbert space, including measurements, but only if they do so with certainty). This requires us to consider only CPTP maps $\mathcal{E}$. The classification of LOCC protocols may be found in \cite{Horodecki:2007,Lamata:2006}.

Equipped with a mathematically precise prescription for LOCC (it ought to be noted that this is still, to some extent, an open question) we may begin to address the issues of using and quantifying entanglement. Generally, in quantum information theory, we are interested in how we can usefully convert $n$ copies of a state $\ket{\psi}$ into $m$ maximally entangled states, and the protocol for doing so will be a LOCC protocol. In the limit in which $n \to \infty$ it is possible to use these protocols to get bounds on  asymptotic entanglement measures such as the entanglement of formation and the entanglement of dissolution \cite{Bennett:1999,Plenio:2007}.

For these protocols one would generally use irreversible LOCC, meaning that the protocol will always succeed, but it is impossible return the final states to the initial states. This is fine if we have $n$ copies of a state, but if we only  have one copy, we do not have this `asymptotic freedom'. If we only have one copy of the state we are going to be  interested in classifying it under reversible LOCC, so that we may equate any state with any other state that can be  reached by a reversible LOCC protocol.

\subsection{LOCC and SLOCC equivalence}\label{sec:LOCCequivalence}

As emphasised LOCC cannot create entanglement. Consequently, from a quantum information theoretic perspective, any two states which may be interrelated using LOCC ought to be physically equivalent with respect to their entanglement properties. This motivates the concept of LOCC equivalence, introduced in \cite{Bennett:1999}. Two states lie in the same LOCC equivalence class if and only if they may be transformed into one another using LOCC operations. Since LOCC cannot create entanglement any two LOCC equivalent states necessarily have the same entanglement. The set of LOCC transformations relating equivalent states forms a group. It may be thought of as a gauge group with respect to entanglement in the sense that it mods out the physically redundant information. It was shown in \cite{Bennett:1999} that two states of a composite system are LOCC equivalent if and only if they may be transformed into one another using the group of \emph{local unitaries} (LU), unitary transformations which factorise into separate transformations on the component parts.  In the case of $n$ qudits, the LU group (up to a phase) is given by $\left[SU(d)\right]^n$. The LU orbits partition the Hilbert space into equivalence classes. For a $n$-qudit system the space of orbits is given by \cite{Linden:1997qd,Carteret:2000-1}:
\begin{equation}
\frac{[{\mathds C}^d]^n}{U(1) \times [SU(d)]^n}.
\end{equation}

However, for single copies of pure states this classification is both mathematically and physically too restrictive. Under LU two states of even the simplest bipartite systems will not, in general, be related \cite{Dur:2000}. Continuous parameters are required to describe the space of entanglement classes \cite{Linden:1997qd,Carteret:2000-1,Sudbery:2001,Carteret:2000-2}. In this sense the LU classification is too severe \cite{Dur:2000}, obscuring some of the more qualitative features of entanglement. An alternative classification scheme was proposed in \cite{Bennett:1999, Dur:2000}. Rather than declare equivalence when states are deterministically related to each other by LOCC, we require only that they may be transformed into one another with  some non-zero probability of success.

This coarse graining goes by the name of Stochastic LOCC or SLOCC for short. Stochastic LOCC includes, in addition to LOCC, those quantum operations that are not trace-preserving, i.e. we have $\tr(\mathcal{E}(\rho)) \leq 1$, so that we no longer require that the protocol always succeeds with certainty. It is proved in \cite{Dur:2000} that for $n$ qudits, the SLOCC equivalence group is (up to an overall complex factor) $\left[SL(d, \mathds{C})\right]^n$. Essentially, we may identify two states if there is a non-zero probability that one can be converted into the other and vice-versa, which means we get $[SL(d, \mathds{C})]^n$ orbits rather than the $[SU(d)]^n$ kind of LOCC. This generalisation may be physically motivated by the fact that any set of SLOCC equivalent entangled states may be used to perform the very same non-classical, entanglement dependent, operations, only with varying likelihoods of success.  For a $n$-qudit state the space of SLOCC equivalence classes  is given by \cite{Dur:2000}:
\begin{equation}
\frac{[{\mathds C}^d]^n}{[SL(d, \mathds{C})]^n}.
\end{equation}
For $n$-qubit systems we have
\begin{equation}
\frac{[{\mathds C}^2]^n}{[SL(2, \mathds{C})]^n}
\end{equation}
and, in this case, the lower bound on the number of continuous variables needed to parameterise the space of orbits is $2(2^n-1)-6n$. Note, for three qubits the space of orbits is finite and discrete (see \autoref{tab:3QubitEntangClassif}) giving the concise classification of entanglement classes of \cite{Dur:2000}.

\subsection{Entanglement monotones}

In general, for some function to be an entanglement monotone, we require three axioms:
\begin{enumerate}
\item $E(\rho) = 0$ if $\rho$ is separable. This fixes the bottom end of the scale and ensures that states that are separable have no entanglement
\item $E(\rho_{AB}) = E\left((U_A \otimes U_B) \rho_{AB}(U_A \otimes U_B)^\dag\right)$ which means that states that are related under LU transformations have the same entanglement.
\item $E(\rho_{AB}) \geq E\left(\mathcal{E}(\rho_{AB})\right)$ where $\mathcal{E}$ is a CPTP map. This is perhaps the most important axiom and ensures that our intuitive definition of entanglement, as something non-local that \emph{cannot} be created by LOCC, is correct.
\end{enumerate}

Depending on which source you read \cite{Jaeger:2006, Plenio:2007, Horodecki:2007}, it is possible to add a few more axioms, whose essential point is to guarantee that all pure bipartite measure of entanglement reduce to the Von Neumann entropy function, i.e.
\begin{equation}
E\left(\ket{\psi}_{AB}\right)\equiv S(\rho) = -\tr(\rho \log_2 \rho).
\end{equation}
For our purposes, we need not concern ourselves with these technicalities. See also \cite{Gingrich:2001,Osterloh:2006,Vidal:1998re,Levay:2005b}.

Since SLOCC is a coarse graining of LOCC, proving that something is a monotone under SLOCC operators, is a stronger condition than axiom 3 above (and in most cases simplifies the task of establishing monotonicity). In fact, this is equivalent to saying that
\begin{equation}\label{eq:entanglement_average}
E(\rho) \geq  p_i E(\rho_i)
\end{equation}
under any CP map $\mathcal{E}$ such that $\rho \to \sum_i p_i \rho_i$ (which, historically, was the original monotone axiom 3 in the quantum information literature \cite{Horodecki:2007}). This is the approach taken by \cite{Dur:2000} when verifying that Cayley's hyperdeterminant is a genuine entanglement monotone (as we will show in \autoref{sec:monotone}).

\newpage
\section{\texorpdfstring{QUBITS}{Qubits}}
\label{sec:qubits}

\subsection{\texorpdfstring{One qubit: $SL(2)_A$}{One qubit: SL(2)A}}

 Entanglement may be thought of as a quantum information \emph{resource} in the same sense as entropy or energy are classical resources \cite{Nielsen:2000}. However, its properties profoundly differ from the properties of those familiar concepts. We have, at best, an incomplete description. In order fully to understand entanglement we would like to be able to describe precisely its creation and transformation, to classify the distinct types of entanglement, to quantitatively measure it, to utilise it as a resource and to illustrate, precisely, how it differs from classical resources at a fundamental level.

We will touch upon each of these topics at some stage of our qubit-black hole discussion suggesting, already, that it is both a substantial and a profitable line of thought. However, before we do so we must introduce the basic quantum information theoretic concepts.

A quantum bit, or \emph{qubit}, is the smallest unit of quantum information. It refers to the state of a 2-level quantum system, such as the spin-up/spin-down  states of an electron. The two basis states are labelled $\ket{0}$ and $\ket{1}$. The one qubit system (Alice) is described by the state
\begin{equation}
\ket{\Psi} = a_{A}\ket{A},
\end{equation}
where $A=0,1$, so
\begin{equation}
\ket{\Psi} = a_{0}\ket{0} +a_{1}\ket{1},
\end{equation}
and the Hilbert space has dimension 2. As described in \autoref{sec:LOCC}, the SLOCC group for a single qubit is $SL(2)_{A}$, under which $a_{A}$ transforms as a \textbf{2}:
\begin{equation}
\begin{pmatrix} a_{0} \\ a_{1} \end{pmatrix} \to
\begin{pmatrix} a & b \\ c & d \end{pmatrix}
\begin{pmatrix} a_{0} \\ a_{1} \end{pmatrix},
\end{equation}
where $ad-bc=1$.

The density matrix $\rho$, defined by
\begin{gather}
\rho=\ket{\Psi}\bra{\Psi},
\shortintertext{or}
\rho_{A_1A_2}=a_{A_1}a^*_{A_2},
\shortintertext{obeys}
\tr\rho= \braket{\Psi}{\Psi}.
\end{gather}

\subsection{\texorpdfstring{Two qubits: $SL(2)_{A} \times SL(2)_{B}$}{Two qubits: SL(2)A x SL(2)B}}\label{sec:Twoqubits}

\subsubsection{States}

The two qubit system (Alice and Bob) is described by the state
\begin{gather}
\ket{\Psi} = a_{AB}\ket{AB},
\shortintertext{where $A=0,1$, so}
\ket{\Psi} = a_{00}\ket{00} +a_{01}\ket{01}+
             a_{10}\ket{10} +a_{11}\ket{11}
\end{gather}
and the Hilbert space has dimension $2^2=4$.  $a_{AB}$ transforms as a $\mathbf{(2,2)}$ under the SLOCC group $SL(2)_{A} \times SL(2)_{B}$.

The determinant of the $2 \times 2$ matrix $a_{AB}$ is given by
\begin{equation}\label{eq:det}
\begin{split}
\det a_{AB} &=\half\varepsilon^{A_{1}A_{2}}\varepsilon^{B_{1}B_{2}}a_{A_{1}B_{1}}a_{A_{2}B_{2}} \\
          &=a_{00}a_{11}-a_{01}a_{10} \\
          &=a_{0}a_{3}-a_{1}a_{2},
\end{split}
\end{equation}
where we have made the binary conversion $0, 1, 2, 3$ for $00, 01, 10, 11$.
Note that
\begin{equation}
|\det a|^{2}=\tfrac{1}{4}\varepsilon^{A_{1}A_{2}}\varepsilon^{B_{1}B_{2}}a_{A_{1}B_{1}}a_{A_{2}B_{2}}
                         \varepsilon^{A_{3}A_{4}}\varepsilon^{B_{3}B_{4}}a^{*}_{A_{3}B_{3}}a^{*}_{A_{4}B_{4}}.
\end{equation}
Using the identity
\begin{gather}
\varepsilon^{A_{1}A_{2}}\varepsilon^{A_{3}A_{4}}=\delta^{A_{1}A_{3}}\delta^{A_{2}A_{4}}-\delta^{A_{1}A_{4}}\delta^{A_{2}A_{3}},
\shortintertext{we have}
\begin{split}
|\det a|^{2}&=\tfrac{1}{4}(\delta^{A_{1}A_{3}}\delta^{A_{2}A_{4}}-\delta^{A_{1}A_{4}}\delta^{A_{2}A_{3}}) \varepsilon^{B_{1}B_{2}}a_{A_{1}B_{1}}a_{A_{2}B_{2}}\varepsilon^{B_{3}B_{4}}a^{*}_{A_{3}B_{3}}a^{*}_{A_{4}B_{4}}\\
&=\tfrac{1}{4}\varepsilon^{B_{1}B_{2}}\varepsilon^{B_{3}B_{4}}(\rho_{B_{1}B_{3}}\rho_{B_{2}B_{4}}-\rho_{B_{1}B_{4}}\rho_{B_{2}B_{3}})\\
&=\half\varepsilon^{B_{1}B_{2}}\varepsilon^{B_{3}B_{4}}\rho_{B_{1}B_{3}}\rho_{B_{2}B_{4}} \\
&=\det\rho_{B}=\det\rho_{A},
\end{split}
\end{gather}
where we have defined the reduced density matrices
\begin{equation}
\begin{split}
\rho_{A}=\Tr_B \ket{\Psi}\bra{\Psi},\\
\rho_{B}=\Tr_A \ket{\Psi}\bra{\Psi},\\
\end{split}
\end{equation}
or
\begin{equation}
\begin{split}
(\rho_{A})_{A_{1}A_{2}}&=\delta^{B_{1}B_{2}}a_{A_{1}B_{1}}a^*_{A_{2}B_{2}}=(\rho_{A})_{A_{2}A_{1}}, \\
(\rho_{B})_{B_{1}B_{2}}&=\delta^{A_{1}A_{2}}a_{A_{1}B_{1}}a^*_{A_{2}B_{2}}=(\rho_{B})_{B_{2}B_{1}}.
\end{split}
\end{equation}

Explicitly,
\begin{gather}\label{eq:reduceddensitymatrix}
\begin{split}
\rho_{A}=
\begin{pmatrix}
|a_{0}|^{2}+|a_{1}|^{2}       & a_{0}a^{*}_{2}+a_{1}a^{*}_{3}\\
a_{2}a^{*}_{0}+a_{3}a^{*}_{1} & |a_{2}|^{2}+|a_{3}|^{2}
\end{pmatrix}, \\
\rho_{B}=
\begin{pmatrix}
|a_{0}|^{2}+|a_{2}|^{2}       & a_{0}a^{*}_{1}+a_{2}a^{*}_{3}\\
a_{1}a^{*}_{0}+a_{3}a^{*}_{2} & |a_{1}|^{2}+|a_{3}|^{2}
\end{pmatrix},
\end{split}
\shortintertext{and}
\det\rho_{A}=\det\rho_{B}=|a_{0}|^{2}|a_{3}|^{2}+|a_{1}|^{2}|a_{2}|^{2}-(a_{0}a^{*}_{2}a_{3}a^{*}_{1}+a_{1}a^{*}_{3}a_{2}a^{*}_{0}).
\end{gather}
Alternatively
\begin{gather}
\begin{split}
|\det a|^{2}&=\tfrac{1}{4}(\delta^{A_{1}A_{3}}\delta^{A_{2}A_{4}}-\delta^{A_{1}A_{4}}\delta^{A_{2}A_{3}})
\varepsilon^{B_{1}B_{2}}a_{A_{1}B_{1}}a_{A_{2}B_{2}}\varepsilon^{B_{3}B_{4}}a^{*}_{A_{3}B_{3}}a^{*}_{A_{4}B_{4}} \\
&=\tfrac{1}{4}(\delta^{A_{1}A_{3}}\delta^{A_{2}A_{4}} -\delta^{A_{1}A_{4}}\delta^{A_{2}A_{3}}) (\sigma_{A})_{A_{1}A_{2}}(\sigma_{A})^*_{A_{3}A_{4}},
\end{split}
\shortintertext{where}
\begin{split}
(\sigma_{A})_{A_{1}A_{2}}&=\varepsilon^{B_{1}B_{2}}a_{A_{1}B_{1}}a_{A_{2}B_{2}}=-(\sigma_{A})_{A_{2}A_{1}}, \\
(\sigma_{B})_{B_{1}B_{2}}&=\varepsilon^{A_{1}A_{2}}a_{A_{1}B_{1}}a_{A_{2}B_{2}}=-(\sigma_{B})_{B_{2}B_{1}}.
\end{split}
\end{gather}

\subsubsection{Entanglement classification}

For bipartite pure states, $\ket{\psi}=a_{AB}\ket{AB}$, where $A=0,\ldots m$ and $B=0,\ldots n$\footnote{Without loss of generality we take $n\leq m$.}, one can always answer the question of whether a state is entangled or not. Such a state is separable if and only if $a_{AB}$ is rank one.

The SLOCC classification is particularly simple in this case. The set of local unitaries is contained in SLOCC  and consequently, using the Schmidt decomposition, any state $\ket{\psi}$ can be written as
\begin{equation}
\ket{\psi}=\sum_{i=1}^{n_\psi} \sqrt{\alpha_i} \ket{ii},\quad \alpha_i > 0,
\end{equation}
$n_\psi \leq n$, where $n$ is the dimension of the smaller of the two  sub-systems. The Schmidt number $n_\psi$ is given by the rank of either one of the reduced density matrices \eqref{eq:reduceddensitymatrix}. Since their rank cannot be changed using $SL(m, \mathds{C})\times SL(n, \mathds{C})$ there are $n$ entanglement classes under SLOCC \cite{Dur:2000}. A state of a given rank may be transformed into any state of a lower rank with some non-zero probability using non-invertible SLOCC operations. No SLOCC operation can increase the rank of a reduced density matrix. Hence, the SLOCC classification is stratified: the higher the rank the stronger the entanglement \cite{Dur:2000}.

For a two-qubit system there are then only two SLOCC classes: entangled and separable corresponding to rank 2 or  rank 1 reduced density matrices, respectively. The bipartite entanglement is measured by the concurrence \cite{Wootters:1997id}
\begin{gather}
 C_{AB}=2\sqrt{\det\rho_{A}}=2\sqrt{\det\rho_{B}}=2~|\det a|,\label{eq:conc2}
\shortintertext{or the 2-tangle}
\tau_{AB}=C_{AB}^{2}=4(|a_{0}|^{2}|a_{3}|^{2}+|a_{1}|^{2}|a_{2}|^{2}-(a_{0}a^{*}_{2}a_{3}a^{*}_{1}+a_{1}a^{*}_{3}a_{2}a^{*}_{0})).
\end{gather}

Recall that the eigenvalues of a $2\times 2$ matrix obey the characteristic equation
\begin{equation}
\det\rho-\tr ~\rho \lambda +\lambda^{2}=0.
\end{equation}
Hence the eigenvalues of $\rho_{A}$ are
\begin{subequations}
\begin{gather}
\lambda_{0}= \half[\tr\rho+\sqrt{(\tr\rho)^{2}-4\det\rho}],
\shortintertext{and}
\lambda_{1}= \half[\tr\rho-\sqrt{(\tr\rho)^{2}-4\det\rho}],
\shortintertext{obeying}
\begin{array}{c@{\ =\ }c}
\lambda_{0}+\lambda_{1} & \tr\rho, \\
\lambda_{0} \lambda_{1} & \det\rho.
\end{array}
\end{gather}
\end{subequations}
so for an entangled state, with non-zero concurrence, $\rho_{A}$ has rank 2 but only rank 1 for a product state as required by the SLOCC classification. Note that the 2-tangle may also be written
\begin{equation}\label{eq:conc3}
\tau_{AB}=2[(\tr\rho)^{2}-\tr\rho^{2}].
\end{equation}

\subsubsection{Bell states}
\label{sec:Bell}

An example of a separable state is
\begin{equation}
\ket{\Psi} = \tfrac{1}{\sqrt{2}}(\ket{|00}+ \ket{01}),
\end{equation}
since when Alice measures $0$, Bob can measure either $0$ or $1$ with equal probability. Here
\begin{equation}
\tau_{AB}=0.
\end{equation}

An example of an entangled state is the Bell state:
\begin{equation}
\ket{\Psi} = \tfrac{1}{\sqrt{2}}(\ket{00}+ \ket{11}), \\
\end{equation}
for which
\begin{equation}
\tau_{AB}=1,
\end{equation}
since if Alice measures 0, Bob must also measure 0 and if Alice measures 1, Bob must also measure 1. This is the origin of the famous EPR ``paradox'', since Alice may be in South Kensington and Bob on Alpha Centauri, yet their measurements are correlated.

For normalised states,
\begin{subequations}
\begin{gather}
\braket{\Psi}{\Psi}=1
\shortintertext{implies}
|a_{0}|^{2}+|a_{1}|^{2}+|a_{2}|^{2}+|a_{3}|^{2}=1,
\shortintertext{or}
|a_{0}|=\pm (1-a^{2})^{1/2},
\shortintertext{where}
a^{2} \equiv |a_{1}|^{2}+|a_{2}|^{2}+|a_{3}|^{2}.
\end{gather}
\end{subequations}

One can verify that the 2-tangle is maximised (and equal to unity) when
\begin{subequations}
\begin{gather}
\begin{split}
a_{0}&=a_{3}=\tfrac{1}{\sqrt{2}}, \\
a_{1}&=a_{2}=0,
\end{split}
\shortintertext{and when}
\begin{split}
a_{1}&=a_{2}=\tfrac{1}{\sqrt{2}}, \\
a_{0}&=a_{3}=0.
\end{split}
\end{gather}
\end{subequations}
These states are obviously entangled since if Alice measures 0 or 1 then Bob must measure 0 or 1, respectively, in the first case and 1 or 0, respectively, in the second.

\subsection{\texorpdfstring{Three qubits: $SL(2)_{A} \times SL(2)_{B} \times SL(2)_{C}$}{Three qubits: SL(2)A x SL(2)B x SL(2)C}}

\subsubsection{States}

The three qubit system (Alice, Bob, Charlie) is described by the state
\begin{gather}
\ket{\Psi} = a_{ABC}\ket{ABC},\label{eq:hypermatrix}
\shortintertext{where $A=0,1$, so}
\begin{split}
\ket{\Psi} =&\phantom{+\:\:} a_{000}\ket{000}+a_{001}\ket{001}+a_{010}\ket{010}+a_{011}\ket{011} \\
&+a_{100}\ket{100}+a_{101}\ket{101}+a_{110}\ket{110}+a_{111}\ket{111},
\end{split}
\end{gather}
and the Hilbert space has dimension $2^3=8$. $a_{ABC}$ transforms as a $\mathbf{(2,2,2)}$ under the SLOCC group $SL(2)_{A} \times SL(2)_{B} \times SL(2)_{C}$.

\subsubsection{Cayley's hyperdeterminant}
\label{sec:Hyperdet}

\begin{figure}[ht]
 \centering
\includegraphics[width=7cm]{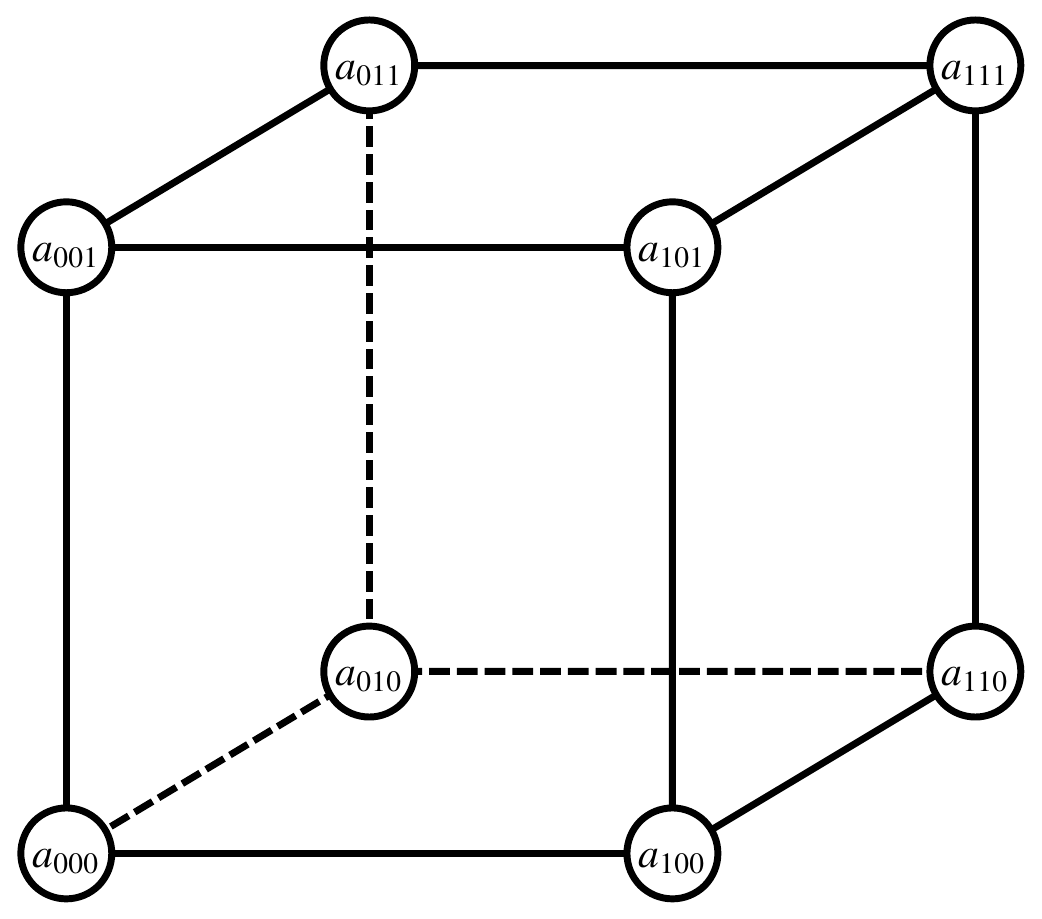}
 \caption[Hypermatrix cube]{The 3-index quantity $a_{ABC}$ is an example of a hypermatrix, here depicted as a cube. In 1845 Cayley generalised the determinant of a $2\times 2$ matrix to the hyperdeterminant of a $2 \times 2 \times 2$ hypermatrix. The hyperdeterminant may also be assembled from the $2\times 2$ matrices formed by partitioning the hypermatrix cube across its planes of symmetry as shown in \autoref{sec:Freudenthal-Fano}.}
 \label{fig:BinaryFanoBasisCube}
\end{figure}

\begin{figure}[ht]
 \centering
 \includegraphics[width=\textwidth]{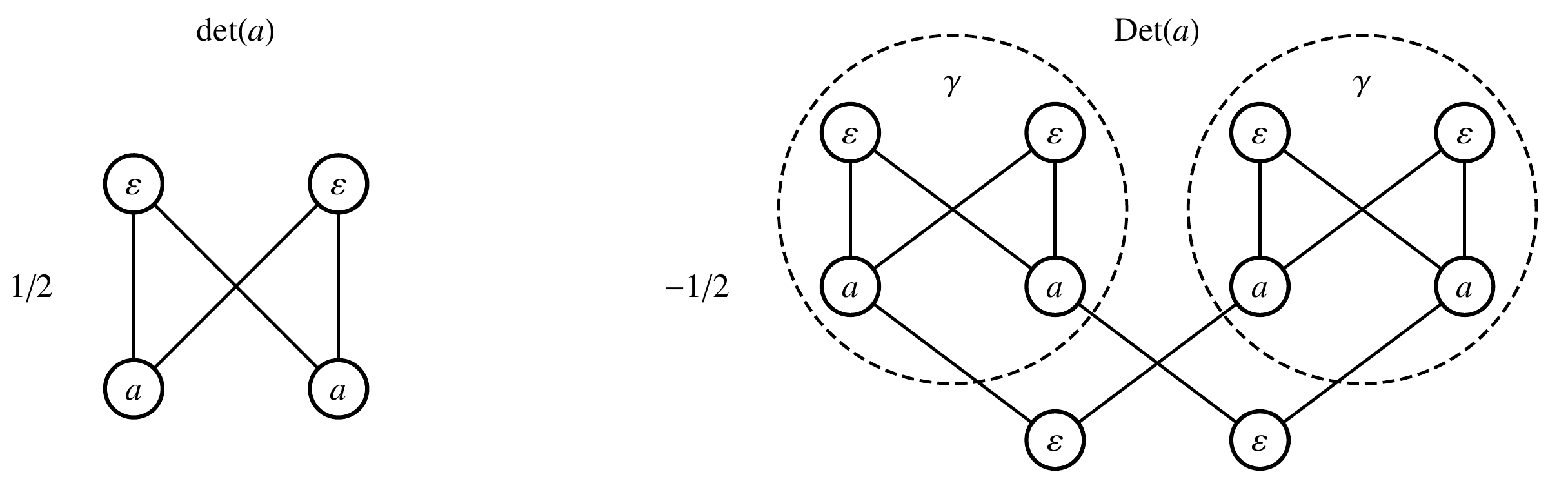}
 \caption[Hyperdeterminant vs. determinant]{A diagrammatic comparison of the hyperdeterminant against the ordinary determinant. The encircled letters denote tensors and the lines are index contractions. The similarities are manifest as the building blocks are virtually identical, the principle difference being the extra indices on the $a$'s. The hyperdeterminant is simply the determinant of the determinant-like object $\gamma$. (The third indices on the $a$'s are the free indices of the $\gamma$'s).}
 \label{fig:Cayleyvsdet}
\end{figure}

The 3-index quantity $a_{ABC}$ is an example of what Cayley termed a \emph{hypermatrix}. Its elements may be represented by the cube shown in \autoref{fig:BinaryFanoBasisCube}.
In 1845 he generalised the determinant of a $2 \times 2$ matrix to the \emph{hyperdeterminant} of a $2 \times 2 \times 2$ hypermatrix $a_{ABC}$ \cite{Cayley:1845}
\begin{subequations}
\begin{gather}
\begin{split}\label{eq:CayleyHyperdeterminant}
\Det a &:= -\half~\varepsilon^{A_1A_2}\varepsilon^{B_1B_2}\varepsilon^{A_3A_4}\varepsilon^{B_3B_4}\varepsilon^{C_1C_4}\varepsilon^{C_2C_3}\\ &\phantom{:=\quad}\times a_{A_1B_1C_1}a_{A_2B_2C_2}a_{A_3B_3C_3}a_{A_4B_4C_4}
\end{split}\\
\begin{split}
=~a_{000}^2 a_{111}^2 + a_{001}^2 a_{110}^2 & +a_{010}^2 a_{101}^2 + a_{100}^2 a_{011}^2 \\
-~2\,(a_{000}a_{001}a_{110}a_{111}          & +a_{000}a_{010}a_{101}a_{111}              \\
+~a_{000}a_{100}a_{011}a_{111}              & +a_{001}a_{010}a_{101}a_{110}              \\
+~a_{001}a_{100}a_{011}a_{110}              & +a_{010}a_{100}a_{011}a_{101})             \\
+~4\,(a_{000}a_{011}a_{101}a_{110}          & +a_{001}a_{010}a_{100}a_{111})
\end{split}\\
\begin{split}
=a_{0}^2 a_{7}^2 + a_{1}^2 a_{6}^2~+   & ~a_{2}^2 a_{5}^2 + a_{3}^2 a_{4}^2 \\
-~2\,(a_{0}a_{1}a_{6}a_{7} +a_{0}a_{2} & a_{5}a_{7} +a_{0}a_{4}a_{3}a_{7} \\
+~a_{1}a_{2}a_{5}a_{6} +a_{1}a_{3}     & a_{4}a_{6} +a_{2}a_{3}a_{4}a_{5}) \\
+~4\,(a_{0}a_{3}a_{5}a_{6}~+           & ~a_{1}a_{2}a_{4}a_{7}),
\end{split}
\end{gather}
\end{subequations}
where we have made the binary conversion 0, 1, 2, 3, 4, 5, 6, 7 for 000, 001, 010, 011, 100, 101, 110, 111. (The cyclic permutations are 0, 4, 1, 5, 2, 6, 3, 7 and 0, 2, 4, 6, 1, 3, 5, 7. See \autoref{tab:BinaryPerm}.)
\begin{table}[ht]
\begin{tabular*}{\textwidth}{@{\extracolsep{\fill}}*{8}{c}}
\toprule
& \multicolumn{3}{c}{Binary} & \multicolumn{3}{c}{Decimal} & \\
& $ABC$ & $CAB$ & $BCA$            & $ABC$ & $CAB$ & $BCA$             & \\
\midrule
& 000   & 000   & 000              & 0     & 0     & 0                 & \\
& 001   & 100   & 010              & 1     & 4     & 2                 & \\
& 010   & 001   & 100              & 2     & 1     & 4                 & \\
& 011   & 101   & 110              & 3     & 5     & 6                 & \\
& 100   & 010   & 001              & 4     & 2     & 1                 & \\
& 101   & 110   & 011              & 5     & 6     & 3                 & \\
& 110   & 011   & 101              & 6     & 3     & 5                 & \\
& 111   & 111   & 111              & 7     & 7     & 7                 & \\
\bottomrule
\end{tabular*}
\caption[Three cyclic permutations of binary notation]{Three cyclic permutations of the binary notation. The hyperdeterminant, \eqref{eq:CayleyHyperdeterminant}, is invariant under this triality.}\label{tab:BinaryPerm}
\end{table}
The hyperdeterminant, which we denote by $\Det$ with a capital D, is also expressed diagrammatically in \autoref{fig:Cayleyvsdet}.
The hyperdeterminant vanishes iff the following system of equations in six unknowns $u^{A},v^{B},w^{C}$ has a nontrivial solution, not allowing any of the pairs to be both zero:
\begin{equation}\label{eq:VanishingHyperdetConditions}
\begin{split}
a_{ABC}u^{A}v^{B}&=0, \\
a_{ABC}u^{A}w^{C}&=0, \\
a_{ABC}v^{B}w^{C}&=0.
\end{split}
\end{equation}
For our purposes, the important properties of the hyperdeterminant are that it is a quartic invariant under $[SL(2)]^{3}$ and under a triality that interchanges $A$, $B$ and $C$. These properties are valid whether the $a_{ABC}$ are complex, real or integer. For more on three qubit entanglement see \cite{Gelfand:1994,Gibbs:2001,Fernando:2006,Brody:2007,Lee:2005,Lee:2007}.

One way to understand this triality is to think of having three different metrics (Alice, Bob and Charlie)
\begin{gather}
\begin{split}\label{eq:ABCgammas}
\gamma^{1}(a)_{A_{1}A_{2}}&=\varepsilon^{B_{1}B_{2}}\varepsilon^{C_{1}C_{2}}a_{A_{1}B_{1}C_{1}}a_{A_{2}B_{2}C_{2}}, \\
\gamma^{2}(a)_{B_{1}B_{2}}&=\varepsilon^{C_{1}C_{2}}\varepsilon^{A_{1}A_{2}}a_{A_{1}B_{1}C_{1}}a_{A_{2}B_{2}C_{2}}, \\
\gamma^{3}(a)_{C_{1}C_{2}}&=\varepsilon^{A_{1}A_{2}}\varepsilon^{B_{1}B_{2}}a_{A_{1}B_{1}C_{1}}a_{A_{2}B_{2}C_{2}}.
\end{split}
\shortintertext{Explicitly,}
\begin{split}
\gamma^{1}(a)&=
\begin{pmatrix}
2(a_{0}a_{3}-a_{1}a_{2}) &  a_{0}a_{7}-a_{1}a_{6}+a_{4}a_{3}-a_{5}a_{2}\\
a_{0}a_{7}-a_{1}a_{6}+a_{4}a_{3}-a_{5}a_{2}  & 2(a_{4}a_{7}-a_{5}a_{6})
\end{pmatrix}, \\
\gamma^{2}(a)&=
\begin{pmatrix}2(a_{0}a_{5}-a_{4}a_{1}) & a_{0}a_{7}-a_{4}a_{3}+a_{2}a_{5}-a_{6}a_{1}\\
a_{0}a_{7}-a_{4}a_{3}+a_{2}a_{5}-a_{6}a_{1} & 2(a_{2}a_{7}-a_{6}a_{3})
\end{pmatrix}, \\
\gamma^{3}(a)&=
\begin{pmatrix}
2(a_{0}a_{6}-a_{2}a_{4}) &  a_{0}a_{7}-a_{2}a_{5}+a_{1}a_{6}-a_{3}a_{4}\\
a_{0}a_{7}-a_{2}a_{5}+a_{1}a_{6}-a_{3}a_{4} & 2(a_{1}a_{7}-a_{3}a_{5})
\end{pmatrix}.
\end{split}
\shortintertext{All are equivalent, however, since}
\det\gamma^{1}(a)=\det\gamma^{2}(a)=\det\gamma^{3}(a)=-\Det a.\label{eq:hdet_gamma}
\shortintertext{If we make the identifications}
\begin{array}{l@{\qquad\qquad}l}\label{eq:MiniDict}
a_{0} = \phantom{-}\tfrac{1}{\sqrt{2}}(P^{0}-P^{2}) & a_{1} = -\tfrac{1}{\sqrt{2}}(Q_{0}+Q_{2})\\
a_{2} = \phantom{-}\tfrac{1}{\sqrt{2}}(P^{1}-P^{3}) & a_{3} = -\tfrac{1}{\sqrt{2}}(Q_{3}+Q_{1})\\
a_{4} = \phantom{-}\tfrac{1}{\sqrt{2}}(P^{1}+P^{3}) & a_{5} = \phantom{-}\tfrac{1}{\sqrt{2}}(Q_{3}-Q_{1})\\
a_{6} = -\tfrac{1}{\sqrt{2}}(P^{0}+P^{2}) & a_{7} = \phantom{-}\tfrac{1}{\sqrt{2}}(Q_{0}-Q_{2}),
\end{array}
\shortintertext{or inversely,}
\begin{array}{l@{\qquad\qquad}l}\label{eq:dictionary2}
P^{0} = \phantom{-}\tfrac{1}{\sqrt{2}}(a_{0}-a_{6}) & Q_{0} = \phantom{-}\tfrac{1}{\sqrt{2}}(a_{7}-a_{1}) \\
P^{1} = \phantom{-}\tfrac{1}{\sqrt{2}}(a_{4}+a_{2}) & Q_{1} = -\tfrac{1}{\sqrt{2}}(a_{5}+a_{3}) \\
P^{2} = -\tfrac{1}{\sqrt{2}}(a_{0}+a_{6}) & Q_{2} = -\tfrac{1}{\sqrt{2}}(a_{7}+a_{1}) \\
P^{3} = \phantom{-}\tfrac{1}{\sqrt{2}}(a_{4}-a_{2}) & Q^{3} = \phantom{-}\tfrac{1}{\sqrt{2}}(a_{5}-a_{3}),
\end{array}
\shortintertext{then}
\begin{array}{r@{~=~}c@{~=~}c*{6}{@{\ }c}}
2(-a_{0}a_{6}+a_{2}a_{4})                    & P^{2}      & {P^{0}}^{2} & + & {P^{1}}^{2} & - & {P^{2}}^{2} & - & {P^{3}}^{2}, \\
2(-a_{1}a_{7}+a_{3}a_{5})                    & Q^{2}      & {Q_{0}}^{2} & + & {Q_{1}}^{2} & - & {Q_{2}}^{2} & - & {Q_{3}}^{2}, \\
a_{0}a_{7}-a_{2}a_{5}+a_{1}a_{6}-a_{3}a_{4}  & P\cdot Q   & P^{0}Q_{0}  & + & P^{1}Q_{1}  & + & P^{2}Q_{2}  & + & P^{3}Q_{3},
\end{array}
\shortintertext{and}
\gamma^{3}=\begin{pmatrix}
-P^{2}& P\cdot Q \\
P\cdot Q  & -Q^{2}
\end{pmatrix}.
\shortintertext{Hence}
\Det a=-P^{2}Q^{2}+(P\cdot Q)^{2}.
\end{gather}

In two component spinor notation
\begin{gather}
\begin{split}
P^{AB}&=\tfrac{1}{\sqrt{2}}\begin{pmatrix*}[r]P^{0}-P^{2} & P^1-P^3 \\P^1+P^3 & -P^0-P^2\end{pmatrix*},\\
Q^{AB}&=\tfrac{1}{\sqrt{2}}\begin{pmatrix*}[r]-Q_0-Q_2 & -Q_1-Q_3 \\-Q_1+Q_3 &Q_0-Q_2\end{pmatrix*},
\end{split}
\shortintertext{we have}
a_{ABC}=\begin{pmatrix}P^{AB} \\ Q^{AB}\end{pmatrix},
\shortintertext{and}
\gamma^{3}=\varepsilon^{A_{1}A_{2}}\varepsilon^{B_{1}B_{2}}
\begin{pmatrix}
P_{A_{1}B_{1}}P_{A_{2}B_{2}} &  P_{A_{1}B_{1}}Q_{A_{2}B_{2}} \\
Q_{A_{1}B_{1}}P_{A_{2}B_{2}} & Q_{A_{1}B_{1}}Q_{A_{2}B_{2}}
\end{pmatrix}.
\end{gather}

This is manifestly invariant under  $SO(2,2)_{AB}$ and transforms as a $3$ under $SL(2)_{C}$.

\subsubsection{2-tangles and 3-tangles}
\label{sec:Tangles}

\begin{figure}
 \centering
 \includegraphics[width=10cm]{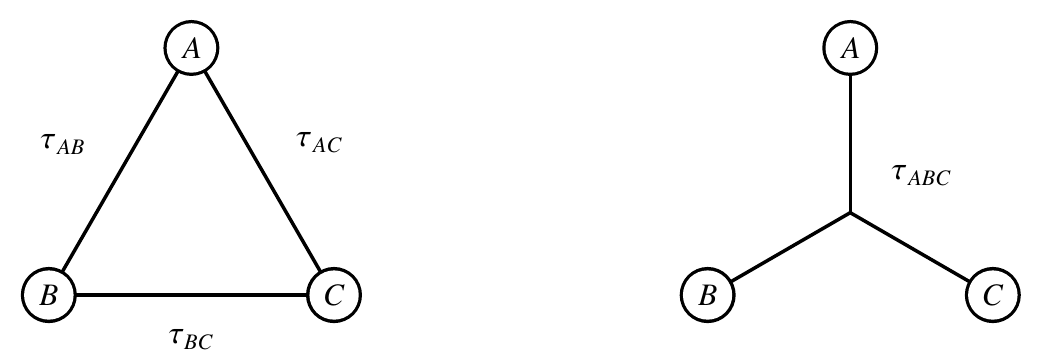}
 \caption[2-tangles vs. 3-tangle]{A schematic of four of the entanglement measures of a three-qubit system: the 2-tangles, $\tau_{AB}, \tau_{AC}$, and $\tau_{BC}$ and the 3-tangle, $\tau_{ABC}$. The 2-tangles give the bipartite entanglements between pairs, while the 3-tangle is a measure of the genuine three-way entanglement.}
 \label{fig:Tangles}
\end{figure}

It is useful to define the reduced density matrices
\begin{gather}
\begin{split}
\rho_A&=\Tr_{BC}\ket{\Psi}\bra{\Psi}, \\
\rho_B&=\Tr_{CA}\ket{\Psi}\bra{\Psi}, \\
\rho_C&=\Tr_{AB}\ket{\Psi}\bra{\Psi},
\end{split}
\shortintertext{so}
\begin{split}
(\rho_A)_{A_1A_2}&=\delta^{B_1B_2}\delta^{C_1C_2}a_{A_1B_1C_1}a^{*}_{A_2B_2C_2}=(\rho_A)^{*}_{A_2A_1}, \\
(\rho_B)_{B_1B_2}&=\delta^{C_1C_2}\delta^{A_1A_2}a_{A_1B_1C_1}a^{*}_{A_2B_2C_2}=(\rho_B)^{*}_{B_2B_1}, \\
(\rho_C)_{C_1C_2}&=\delta^{A_1A_2}\delta^{B_1B_2}a_{A_1B_1C_1}a^{*}_{A_2B_2C_2}=(\rho_C)^{*}_{C_2C_1}.
\end{split}
\shortintertext{Explicitly,}
\begin{split}
\rho_A&=
\begin{pmatrix}
|a_{0}|^{2}+|a_{1}|^{2}+|a_{2}|^{2}+|a_{3}|^{2} & a_{0}a_{4}^{*}+a_{1}a_{5}^{*}+a_{2}a_{6}^{*}+a_{3}a_{7}^{*}\\
         a_{4}a_{0}^{*}+a_{5}a_{1}^{*}+a_{6}a_{2}^{*}+a_{7}a_{3}^{*}& |a_{4}|^{2}+|a_{5}|^{2}+|a_{6}|^{2}+|a_{7}|^{2}\end{pmatrix},\\
\rho_B&=
\begin{pmatrix}
|a_{0}|^{2}+|a_{4}|^{2}+|a_{1}|^{2}+|a_{5}|^{2} & a_{0}a_{2}^{*}+a_{4}a_{6}^{*}+a_{1}a_{3}^{*}+a_{5}a_{7}^{*}\\
         a_{2}a_{0}^{*}+a_{6}a_{4}^{*}+a_{3}a_{1}^{*}+a_{7}a_{5}^{*}& |a_{2}|^{2}+|a_{6}|^{2}+|a_{3}|^{2}+|a_{7}|^{2}\end{pmatrix},\\
\rho_C&=
\begin{pmatrix}
|a_{0}|^{2}+|a_{2}|^{2}+|a_{4}|^{2}+|a_{6}|^{2} &a_{0}a_{1}^{*}+a_{2}a_{3}^{*}+a_{4}a_{5}^{*}+a_{6}a_{7}^{*}\\
         a_{1}a_{0}^{*}+a_{3}a_{2}^{*}+a_{5}a_{4}^{*}+a_{7}a_{6}^{*}& |a_{1}|^{2}+|a_{3}|^{2}+|a_{5}|^{2}+|a_{7}|^{2}\end{pmatrix}.
\end{split}
\end{gather}
The 2-tangles between Alice and the Bob-Charlie system, Bob and the Charlie-Alice system and Charlie and the Alice-Bob system are given by \cite{Coffman:1999jd}
\begin{equation}
\begin{split}
\tau_{A(BC)}&=4\det\rho_A, \\
\tau_{B(CA)}&=4\det\rho_B, \\
\tau_{C(AB)}&=4\det\rho_C,
\end{split}
\end{equation}
and are also known as local entropies, denoted $S_A, S_B$, and $S_C$. Explicitly,
\begin{equation}
\begin{array}{c@{}r@{}c*{5}{@{\ +\ }c}@{}c}
\tau_{A(BC)}= & 4[     & |a_{0}|^{2}|a_{7}|^{2} & |a_{1}|^{2}|a_{6}|^{2} & |a_{2}|^{2}|a_{5}|^{2} & |a_{3}|^{2}|a_{4}|^{2} & |a_{0}|^{2}|a_{5}|^{2} & |a_{0}|^{2}|a_{6}|^{2} & \\
              & +\ \   & |a_{1}|^{2}|a_{4}|^{2} & |a_{1}|^{2}|a_{7}|^{2} & |a_{2}|^{2}|a_{4}|^{2} & |a_{2}|^{2}|a_{7}|^{2} & |a_{3}|^{2}|a_{5}|^{2} & |a_{3}|^{2}|a_{6}|^{2} & \\
              & -2\Re( & a_{0}a_{4}^{*}a_{5}a_{1}^{*} & a_{0}a_{4}^{*}a_{6}a_{2}^{*} & a_{0}a_{4}^{*}a_{7}a_{3}^{*} & a_{1}a_{5}^{*}a_{6}a_{2}^{*} & a_{1}a_{5}^{*}a_{7}a_{3}^{*} & a_{2}a_{6}^{*}a_{7}a_{3}^{*} & )],\\[5pt]
\tau_{B(CA)}= & 4[     & |a_{0}|^{2}|a_{7}|^{2} & |a_{4}|^{2}|a_{3}|^{2} & |a_{1}|^{2}|a_{6}|^{2} & |a_{5}|^{2}|a_{2}|^{2} & |a_{0}|^{2}|a_{6}|^{2} & |a_{0}|^{2}|a_{3}|^{2}\\
              & +\ \   & |a_{4}|^{2}|a_{2}|^{2} & |a_{4}|^{2}|a_{7}|^{2} & |a_{1}|^{2}|a_{2}|^{2} & |a_{1}|^{2}|a_{7}|^{2} & |a_{5}|^{2}|a_{6}|^{2} & |a_{5}|^{2}|a_{3}|^{2}\\
              & -2\Re( & a_{0}a_{2}^{*}a_{6}a_{4}^{*} & a_{0}a_{2}^{*}a_{3}a_{1}^{*} & a_{0}a_{2}^{*}a_{7}a_{5}^{*} & a_{1}a_{5}^{*}a_{6}a_{2}^{*} & a_{1}a_{5}^{*}a_{7}a_{3}^{*} & a_{2}a_{6}^{*}a_{7}a_{3}^{*} & )],\\[5pt]
\tau_{C(AB)}= & 4[     & |a_{0}|^{2}|a_{7}|^{2} & |a_{2}|^{2}|a_{5}|^{2} & |a_{4}|^{2}|a_{3}|^{2} & |a_{6}|^{2}|a_{1}|^{2} & |a_{0}|^{2}|a_{3}|^{2} & |a_{0}|^{2}|a_{5}|^{2}\\
              & +\ \   & |a_{2}|^{2}|a_{1}|^{2} & |a_{2}|^{2}|a_{7}|^{2} & |a_{4}|^{2}|a_{1}|^{2} & |a_{4}|^{2}|a_{7}|^{2} & |a_{6}|^{2}|a_{3}|^{2} & |a_{6}|^{2}|a_{5}|^{2}\\
              & -2\Re( & a_{0}a_{1}^{*}a_{3}a_{2}^{*} & a_{0}a_{1}^{*}a_{5}a_{4}^{*} & a_{0}a_{1}^{*}a_{7}a_{6}^{*} & a_{2}a_{3}^{*}a_{5}a_{4}^{*} & a_{2}a_{3}^{*}a_{7}a_{6}^{*} & a_{4}a_{5}^{*}a_{7}a_{6}^{*} & )].
\end{array}
\end{equation}
The 2-tangles between Alice and Bob, Bob and Charlie and Charlie and Alice within the $ABC$ system are given by \cite{Coffman:1999jd}
\begin{equation}\label{eq:concur}
\begin{split}
\tau_{AB}=C_{AB}^2&=\half\left(-\tau_{C(AB)}+\tau_{A(BC)}+\tau_{B(CA)}-\tau_{ABC}\right), \\
\tau_{BC}=C_{BC}^2&=\half\left(-\tau_{A(BC)}+\tau_{B(CA)}+\tau_{C(AB)}-\tau_{ABC}\right), \\
\tau_{CA}=C_{CA}^2&=\half\left(-\tau_{B(CA)}+\tau_{C(AB)}+\tau_{A(BC)}-\tau_{ABC}\right),
\end{split}
\end{equation}
where $C_{AB}$,$C_{BC}$ and $C_{CA}$ are the corresponding concurrences, see \autoref{fig:Tangles}.
Thus
\begin{gather}
\tau_{AB}+\tau_{BC}+\tau_{CA}=\half\left(\tau_{A(BC)}+\tau_{B(CA)}+\tau_{C(AB)}-3\tau_{ABC}\right),
\shortintertext{and}
\begin{split}
\tau_{ABC}&=\tau_{A(BC)}-\tau_{AB}-\tau_{CA}, \\
\tau_{ABC}&=\tau_{B(CA)}-\tau_{BC}-\tau_{AB}, \\
\tau_{ABC}&=\tau_{C(AB)}-\tau_{CA}-\tau_{BC},
\end{split}
\end{gather}
where $\tau_{ABC}$ is the 3-tangle \cite{Miyake:2002,Miyake:2003}
\begin{equation}\label{eq:3-tangle}
\tau_{ABC}=4|\Det a|.
\end{equation}
Explicitly, for $\Det a>0$,
\begin{equation}
\begin{split}
\tau_{AB}=4(a_0a_6-a_4a_2+a_1a_7-a_5a_3)^2, \\
\tau_{BC}=4(a_0a_3-a_2a_1+a_4a_7-a_6a_5)^2, \\
\tau_{CA}=4(a_0a_5-a_1a_4+a_2a_7-a_3a_6)^2.
\end{split}
\end{equation}

It is also useful to define
\begin{equation}
(\sigma_A)_{A_1A_2}=\delta^{B_1B_2}\varepsilon^{C_1C_2}a_{A_1B_1C_1}a_{A_2B_2C_2}=-(\sigma_{A})_{A_2A_1},
\end{equation}
similarly $(\sigma_B)_{B_1B_2}$ and $(\sigma_C)_{C_1C_2}$.  Explicitly,
\begin{gather}
\begin{split}
\sigma_A=
\begin{pmatrix}
0 &  -a_2a_1+a_0a_3-a_5a_6+a_4a_7 \\
a_2a_1-a_0a_3+a_5a_6-a_4a_7& 0
\end{pmatrix}, \nonumber \\
\sigma_B=
\begin{pmatrix}
0 & -a_1a_4+a_0a_5-a_6a_3+a_2a_7 \\
a_1a_4-a_0a_5+a_6a_3-a_2a_7& 0
\end{pmatrix}, \\
\sigma_C=
\begin{pmatrix}
0 & -a_4a_2+a_0a_6-a_3a_5+a_1a_7 \\
a_4a_2-a_0a_6+a_3a_5-a_1a_7& 0
\end{pmatrix}.
\end{split}
\shortintertext{Then using the identity}
\varepsilon^{A_1A_2}\varepsilon^{A_3A_4}= \delta^{A_1A_3}\delta^{A_2A_4}-\delta^{A_1A_4}\delta^{A_2A_3},
\shortintertext{we have, restricting to real $a_{ABC}$,}
\begin{split}
\Det a&=-\half(\delta^{A_1A_3}\delta^{A_2A_4}-\delta^{A_1A_4}\delta^{A_2A_3})
(\delta^{B_1B_3}\delta^{B_2B_4}-\delta^{B_1B_4}\delta^{B_2B_3}) \\
&\phantom{~=-\half(}\times\varepsilon^{C_1C_4}\varepsilon^{C_2C_3}a_{A_1B_1C_1}a_{A_2B_2C_2}a_{A_3B_3C_3}a_{A_4B_4C_4} \\
&=\half\big((\rho_C)_{C_1C_3}(\rho_C)_{C_2C_4}\varepsilon^{C_1C_4}\varepsilon^{C_2C_3} -(\sigma_A)_{A_1A_4}(\sigma_A)_{A_2A_3}\delta^{A_1A_3}\delta^{A_2A_4} \\
&\phantom{\;=\half\big((\rho_C)_{C_1C_3}(\rho_C)_{C_2C_4}\varepsilon^{C_1C_4}\varepsilon^{C_2C_3}}
-(\sigma_B)_{B_1B_4}(\sigma_B)_{B_2B_3}\delta^{B_1B_3}\delta^{B_2B_4}\big) \\
&=\det(\rho_C)+\half\tr(\sigma_A)^2+\half\tr(\sigma_B)^2,
\end{split}
\shortintertext{or equivalently}
\Det a=\det(\rho_C)-\det(\sigma_B)-\det(\sigma_A).
\shortintertext{Hence}
\begin{split}
\tau_{AB}&=4\det(\sigma_C), \\
\tau_{BC}&=4\det(\sigma_A), \\
\tau_{CA}&=4\det(\sigma_B).
\end{split}
\end{gather}

\subsubsection{Entanglement classification}

Recall, two states of a composite quantum system are regarded as equivalent if they are related by a unitary transformation which factorises into separate transformations on the component parts, so-called \emph{local unitaries}. See \autoref{sec:LOCCequivalence} and \cite{Albeverio:2001ey,Bennett:1999,Carteret:2000-2,Kempe:1999vk,Luque:2005,Verstraete:2003} for more details. The Hilbert space decomposes into equivalence classes, or \emph{orbits} under the action of the group of local unitaries.  For unnormalised three-qubit states, the number of parameters \cite{Linden:1997qd} needed to describe inequivalent states or, what amounts to the same thing, the number of algebraically independent invariants \cite{Sudbery:2001} is given by the dimension of the space of orbits
\begin{equation}
\frac{{\mathds C}^2 \times {\mathds C}^2 \times {\mathds C}^2}{U(1) \times SU(2) \times SU(2) \times SU(2)},
\end{equation}
namely $16-10=6$.  For subsequent comparison with the $STU$ black hole, however, we restrict our attention to states with \emph{real} coefficients $a_{ABC}$. In this case, one can show that there are five algebraically independent invariants: $\Det a$, $S_A$, $S_B$, $S_C$ and the norm $\braket{\Psi}{\Psi}$, corresponding to the dimension of
\begin{equation}
\frac{{\mathds R}^2 \times {\mathds R}^2 \times {\mathds R}^2}{SO(2) \times SO(2) \times SO(2)},
\end{equation}
namely $8-3=5$.  Hence, the most general real three-qubit state can be described by just five parameters \cite{Acin:2001}, conveniently taken as four real numbers $N_0,N_1,N_2,N_3$ and an angle $\theta$:
\begin{equation}\label{eq:five}
\begin{split}
\ket{\Psi} &= -N_3\cos^2\theta\ket{001}-N_2\ket{010}+N_3\sin\theta\cos\theta\ket{011}\\
&\phantom{=}-N_1\ket{100}-N_3\sin\theta\cos\theta\ket{101}+(N_0+N_3\sin^2\theta)\ket{111}.
\end{split}
\end{equation}
Further, under the coarser SLOCC classification, D\"ur et al. \cite{Dur:2000} used simple arguments concerning the conservation of ranks of reduced density matrices to show that there are only six three-qubit equivalence classes; only two of which show \emph{genuine} tripartite entanglement, see \autoref{tab:3QubitEntangClassif}. These are:
\begin{description}
\item[Class $A$-$B$-$C$, product states] represent completely separable states like
    \begin{equation}
    \ket{\psi_{\text{$A$-$B$-$C$}}}= N_0\ket{111},
    \end{equation}
    for which all 2-tangles and 3-tangles vanish.
\item[Classes $A$-$BC$, $AB$-$C$ and $C$-$AB$, bipartite entanglement.] These states have only bipartite entanglement between two of the qubits and zero entanglement with the third, for the class $A$-$BC$ a representative is
    \begin{equation}
    \ket{\psi_{\text{$A$-$BC$}}} = N_0\ket{111}-N_1\ket{100},
    \end{equation}
    and they have their respective $\tau_{BC} \neq 0$ but $\tau_{ABC} = 0$
\item[Class W, tripartite entanglement.] The canonical state is
    \begin{equation}
    \ket{\text{W}} =-N_1\ket{100}- N_2\ket{010}-N_3\ket{001},
    \end{equation}
    for which all 2-tangles do not vanish but the 3-tangle is still zero. These do not violate Bell's inequalities but are resilient to information loss: tracing out one qubit leaves a maximally entangled bipartite state.
\item[Class GHZ (Greenberger-Horne-Zeilinger \cite{Greenberger:1989,Greenberger:1990})] the other genuine tripartite entangled state, for which the canonical example is
    \begin{equation}
    \ket{\text{GHZ}} = N_0\ket{111}-N_1\ket{100}- N_2\ket{010}-N_3\ket{001},
    \end{equation}
    which has non-zero 2-tangles and non-zero $\tau_{ABC}$. These states are the ones that maximally violate Bell's inequalities; they are, however, fragile, as tracing out one qubit leaves you with a mixed bipartite state.
\end{description}
Using parametrisation \eqref{eq:five}, representatives from each SLOCC class can be found in \autoref{tab:SLOCCRepresentatives}.

The state \eqref{eq:five} is obtained from the canonical form for real states \cite{Acin:2000},
\begin{equation}\label{eq:ThreeQubitGen}
\ket{\Psi} =a_0\ket{000}+a_4\ket{100}+a_5\ket{101}+a_6\ket{110}+a_7\ket{111}
\end{equation}
by applying two different $SO(2)$ transformations on the second and third bits, $\mathds{1}_A\otimes T_B \otimes T_C$ where
\begin{equation}
T_B = \begin{pmatrix*}[r]\sin\theta_1 & \cos\theta_1 \\ -\cos\theta_1 & \sin\theta_1 \end{pmatrix*},
\quad
T_C = \begin{pmatrix*}[r]\sin\theta_2 & \cos\theta_2 \\ -\cos\theta_2 & \sin\theta_2 \end{pmatrix*}.
\end{equation}

Applying this to \eqref{eq:ThreeQubitGen} and setting it equal to \eqref{eq:five}, we obtain the following two solutions
\begin{equation}
\begin{split}
N_0 &= \frac{\csc\theta_1}{4 a_4}\Big(-4 a_4 a_6 \cos\theta_2-4 a_5  a_7 \cos\theta_2+a_5^2 \sin(2\theta_1-\theta_2) \\
    &\phantom{=\frac{\csc\theta_1}{4 a_4}\Big(}+ a_7^2 \sin(2\theta_1-\theta_2)+ a_5^2\sin(2\theta_1+ \theta_2)+ a_7^2 \sin(2\theta_1+ \theta_2)\Big), \\
N_1 &=  - a_7 \cos\theta_1\cos\theta_2- a_5 \cos\theta_2\sin\theta_1 \\
    &\phantom{=}+\frac{a_6}{a_0 a_4} \Big{(}a_0 a_5 \cos\theta_1 \cos\theta_2-a_0  a_7 \cos\theta_2\sin\theta_1\Big{)},\\
N_2 &= \frac{1}{a_4}\Big{(}a_0 a_7 \cos\theta_2\sin\theta_1 - a_0 a_5 \cos\theta_1\cos\theta_2\Big{)},\\
N_3 &= \frac{1}{a_4}\Big{(}\cos\theta_2 \csc\theta_1\sec\theta_1\left(a_4 a_6 \cos\theta_1+ a_5  a_7 \cos\theta_1\right) \\
    &\phantom{=\frac{1}{a_4}\Big{(}}+ a_4^2 \sin\theta_1- a_7^2 \sin\theta_1\Big{)} \\
\theta &= \theta_1\pm\tfrac{\pi}{2}.
\end{split}
\end{equation}
\begin{center}
\begin{table}[ht]
\begin{tabular*}{\textwidth}{@{\extracolsep{\fill}}*{5}{c}}
\toprule
& Class                        & Entanglement                & Representative                                          & \\
\midrule
& $A$-$B$-$C$                  & Separable                   & $N_0\ket{111}$                                          & \\
& $A$-$BC$, $AB$-$C$, $AC$-$B$ & Bi-separable                & $N_0\ket{111}-N_1\ket{100}$                             & \\
& W                            & Full bipartite entanglement & $-N_1\ket{100}- N_2\ket{010}-N_3\ket{001}$              & \\
& GHZ                          & Tripartite entanglement     & $N_0\ket{111}-N_1\ket{100}- N_2\ket{010}-N_3\ket{001} $ & \\
\bottomrule
\end{tabular*}
\caption[Representatives from each SLOCC class]{The values of entanglement measures permit the partitioning of state space into SLOCC entanglement classes. Special cases of \eqref{eq:five} are selected as representatives of these classes.}
 \label{tab:SLOCCRepresentatives}
\end{table}
\end{center}

\subsubsection{Monotonicity of the 3-tangle}
\label{sec:monotone}

In this section we prove (following \cite{Dur:2000}) that $E = \sqrt{\tau_{ABC}}$ is a monotone. In general we would want to prove that $\sqrt{\tau_{ABC}}$ is a monotone under any transformations on Alice, Bob or Charlie. However, the triality invariance of Cayley's hyperdeterminant under interchange of parties affords us a simplification. Since any SLOCC protocol can always be broken down into transformations on one party followed by some classical communication, and since ${\tau_{ABC}}$ is invariant under interchange, we need only consider transformations on Alice. We further note that any local POVM can be implemented by a sequence of two outcome POVMs \cite{Dur:2000}, so we only have to consider POVMs $A_1$ and $A_2$, with the following condition
\begin{equation}
A_1^\dag A_1 + A_2^\dag A_2 = \mathds{1}.
\end{equation}
Under $A_i$, state $\ket{\psi}$ goes to $\ket{\tilde{\phi_i}} = A_i \ket{\psi}$ with $p_i = \braket{\tilde{\phi_i}}{\tilde{\phi_i}}$ representing the probability that operation $A_i$ succeeds. After normalising, we have $\ket{\phi_i} = \tilde{\phi_i}/\sqrt{p_i}$. So, we require
\begin{gather}
E(\psi) \geq p_1 E(\phi_1) + p_2 E(\phi_2),\label{eq:tau_under_povm}
\shortintertext{where}
E(\psi) = \sqrt{\tau_{ABC}} = \sqrt{4|\Det(\psi)|}.
\shortintertext{For Cayley we have \eqref{eq:hdet_gamma}}
\Det\left(\psi\right) = -\det\gamma_{A_1 A_2},
\shortintertext{where}
\gamma_{A_1 A_2} = \varepsilon^{B_1 B_2} \varepsilon^{C_1 C_2}a_{A_1 B_1 C_1} a_{A_2 B_2 C_2}.
\end{gather}
Under a transformation on Alice ($A_i \otimes \mathds{1}_B \otimes \mathds{1}_C$), we have
\begin{gather}
\begin{split}
\gamma_{A_1 A_2} &\to \varepsilon^{B_1 B_2} \varepsilon^{C_1 C_2}a_{A_1 B_1 C_1} a_{A_2 B_2 C_2} \\
                 & \quad \times \left(A_{A_1 \bar{A}_1} \otimes \mathds{1}_B \otimes \mathds{1}_C \right)\left(A_{A_2 \bar{A}_2}\otimes \mathds{1}_B \otimes \mathds{1}_C \right) \\
                 & = A^{\textsf{T}}_{\bar{A}_1A_1} \gamma_{A_1 A_2} A_{A_2 \bar{A}_2},
\end{split}
\shortintertext{which gives}
\begin{split}\label{eq:cayley_under_povm}
\det \gamma_{A_1 A_2} &\to \det \left[ A^{\textsf{T}}_{\bar{A}_1A_1} \gamma_{A_1 A_2} A_{A_2 \bar{A}_2} \right] \\
                      & = \det A^{\textsf{T}} \det \gamma \det A.
\end{split}
\end{gather}

Next, we use the singular value decomposition, which allows us to express any positive matrix as a product of a unitary matrix, a diagonal matrix and a different unitary matrix \cite{Nielsen:2000}. We have
\begin{equation}
A_i = U_i D_i V,
\end{equation}
where $U_i \in SU(2)$, $V \in U(2)$ and the $D_i$ are diagonal such that
\begin{equation}
D_1 = \begin{pmatrix} a & \\ & b \end{pmatrix},
\qquad
D_2 = \begin{pmatrix} +\sqrt{(1 - a^2)} & \\ & +\sqrt{(1 - b^2)} \end{pmatrix},
\end{equation}
where $a,b \in [0,1]$; hence, we have $\det U_i = 1$, $\det V = e^{i\theta}$, $\det D_1 = ab$ and $\det D_2 = \sqrt{(1 - a^2)(1-b^2)}$. This gives $\det A_1 = ab e^{i\theta}$. For $A_1$, if we put all this in \eqref{eq:cayley_under_povm} and recalling that $\phi_i = \tilde{\phi_i}/p_i$, we get
\begin{gather}
\begin{split}
\det \gamma_{A_1 A_2}(\phi_1) &= \det A_1^{\textsf{T}} \det \gamma(\psi / \sqrt{p_1}) \det A_1 \\
                              &= \frac{a^2b^2 }{ p_1^2} \gamma(\psi) e^{2i\theta},
\end{split}
\shortintertext{and for $A_2$, we get}
\begin{split}
\det \gamma_{A_1 A_2}(\phi_2) &= \det A_2^{\textsf{T}} \det \gamma(\psi / \sqrt{p_2}) \det A_2 \\
                              &= \frac{(1 - a^2)(1-b^2)}{ p_2^2} \gamma(\psi)  e^{2i\theta}.
\end{split}
\shortintertext{Substituting into $E$, we have}
\begin{split}\label{eq:tau_after_povm}
E(\psi) &\to p_1 \sqrt{\tau}(\phi_1)+ p_2 \sqrt{\tau}(\phi_2) \\
        &= a b \sqrt{\tau}(\psi) +  \sqrt{(1 - a^2)(1-b^2)}\sqrt{\tau}(\psi )\\
        &= \left(a b + \sqrt{(1 - a^2)(1-b^2)}\right)E(\psi).
\end{split}
\end{gather}
Finally we note that \eqref{eq:tau_after_povm} is maximised for $a=b$ and hence the condition in \eqref{eq:entanglement_average} is satisfied and $\sqrt{\tau_{ABC}}$ is always decreasing, and hence a monotone.

\newpage
\section{\texorpdfstring{QUTRITS}{Qutrits}}
\label{sec:qutrits}

\subsection{\texorpdfstring{One qutrit: $SL(3)_A$}{One qutrit: SL(3)A}}

The one qutrit system (Alice) is described by the state
\begin{gather}
\ket{\Psi} = {a}_{{A}}\ket{{A}},
\shortintertext{where ${A}=0,1,2$, so}
\ket{\Psi} ={a}_{0}\ket{0} +{a}_{1}\ket{1}+{a}_{2}\ket{2},
\end{gather}
and the Hilbert space has dimension 3. $a_A$ transforms as a \textbf{3} under SLOCC group $SL(3)_{A}$.

\subsection{\texorpdfstring{Two qutrits: $SL(3)_A \times SL(3)_A$}{Two qutrits: SL(3)A x SL(3)A}}

\subsubsection{States}

The two qutrit system (Alice and Bob) is described by the state
\begin{gather}
\ket{\Psi} ={a}_{{A}{B}}\ket{{A}{B}},
\shortintertext{where ${A}=0,1,2$, so}
\begin{split}
\ket{\Psi}=&\phantom{+\:\:}{a}_{00}\ket{00} + {a}_{01}\ket{01} +{a}_{02}\ket{02}\\
           &+ {a}_{10}\ket{10} +{a}_{11}\ket{11} + {a}_{12}\ket{12} \\
           &+ {a}_{20}\ket{20} + {a}_{21}\ket{21} + {a}_{22}\ket{22},
\end{split}
\end{gather}
and the Hilbert space has dimension $3^2=9$. $\widehat{a}_{\widehat{A}\widehat{B}}$ transforms as a $\mathbf{(3,3)}$ under $SL(3)_{A} \times SL(3)_{B}$.

We  define the reduced density matrices
\begin{equation}
\begin{split}
\rho_{A}=\Tr_B \ket{\Psi}\bra{\Psi},\\
\rho_{B}=\Tr_A \ket{\Psi}\bra{\Psi},\\
\end{split}
\end{equation}
or
\begin{gather}
\begin{split}
(\rho_A)_{A_1A_2}&=\delta^{B_1B_2}a_{A_1B_1}a^{*}_{A_2B_2}=(\rho_A)^{*}_{A_2A_1}, \\
(\rho_B)_{B_1B_2}&=\delta^{A_1A_2}a_{A_1B_1}a^{*}_{A_2B_2}=(\rho_B)^{*}_{B_2B_1}.
\end{split}
\shortintertext{Explicitly,}
\begin{split}
\rho_A&=
\begin{pmatrix}
|a_{00}|^{2}+|a_{01}|^{2}+|a_{02}|^{2} & a_{00}a^{*}_{10}+a_{01}a^{*}_{11}+a_{02}a^{*}_{12}&a_{00}a^{*}_{20}+a_{01}a^{*}_{21}+a_{02}a^{*}_{22}\\
         a_{10}a^{*}_{00}+a_{11}a^{*}_{01}+a_{12}a^{*}_{02} & |a_{10}|^{2}+|a_{11}|^{2}+|a_{12}|^{2}&a_{10}a^{*}_{20}+a_{11}a^{*}_{21}+a_{12}a^{*}_{22}\\
         a_{20}a^{*}_{00}+a_{21}a^{*}_{01}+a_{22}a^{*}_{02} & a_{20}a^{*}_{10}+a_{21}a^{*}_{11}+a_{22}a^{*}_{12}& |a_{20}|^{2}+|a_{21}|^{2}+|a_{22}|^{2}\end{pmatrix}, \\
\rho_B&=
\begin{pmatrix}
|a_{00}|^{2}+|a_{10}|^{2}+|a_{02}|^{2} & a_{00}a^{*}_{01}+a_{10}a^{*}_{11}+a_{20}a^{*}_{21}& a_{00}a^{*}_{02}+a_{10}a^{*}_{12}+a_{20}a^{*}_{22}\\
         a_{01}a^{*}_{00}+a_{11}a^{*}_{10}+a_{21}a^{*}_{20} & |a_{01}|^{2}+|a_{11}|^{2}+|a_{21}|^{2}& a_{01}a^{*}_{02}+a_{11}a^{*}_{12}+a_{21}a^{*}_{22}\\
         a_{02}a^{*}_{00}+a_{12}a^{*}_{10}+a_{22}a^{*}_{20} & a_{02}a^{*}_{01}+a_{12}a^{*}_{11}+a_{22}a^{*}_{21} & |a_{02}|^{2}+|a_{12}|^{2}+|a_{22}|^{2}\end{pmatrix}.
\end{split}
\end{gather}
Recall that the eigenvalues of the $3 \times 3$ density matrices obey the characteristic equation
\begin{equation}
\det\rho-C_2\lambda +\tr\rho~\lambda^2-\lambda^3=0.
\end{equation}
The determinant of a $3 \times 3$ matrix $a_{AB}$ is given by
\begin{gather}
\begin{split}
\det a=&\tfrac{1}{3!}\varepsilon^{A_1A_2A_3}\varepsilon^{B_1B_2B_3}a_{A_1B_1}a_{A_2B_2}a_{A_3B_3} \\
=&\phantom{-~}a_{00}(a_{11}a_{22}-a_{12}a_{21}) \\
&-a_{01}(a_{10}a_{22}-a_{12}a_{20}) \\
&+a_{02}(a_{01}a_{21}-a_{11}a_{20}),
\end{split}
\shortintertext{and}
\begin{gathered}
\det\rho_{A}=\det\rho_{B} =\\
\begin{array}{c*{4}{@{\ }c@{\ }}c@{}c}
  & |a_{0 2}|^2 |a_{1 1}|^2 |a_{2 0}|^2 & + & |a_{0 1}|^2 |a_{1 2}|^2 |a_{2 0}|^2 & + & |a_{0 2}|^2 |a_{1 0}|^2 |a_{2 1}|^2 \\
+ & |a_{0 0}|^2 |a_{1 2}|^2 |a_{2 1}|^2 & + & |a_{0 1}|^2 |a_{1 0}|^2 |a_{2 2}|^2 & + & |a_{0 0}|^2 |a_{2 1}|^2 |a_{2 2}|^2 \\
- & a_{00}a_{11}|a_{22}|^2a_{01}^{*}a_{10}^{*} & - & a_{01}a_{10}|a_{22}|^2a_{00}^{*}a_{11}^{*} & - & a_{00}a_{12}|a_{21}|^2a_{02}^{*}a_{10}^{*}\\
- & a_{02}a_{10}|a_{21}|^2a_{00}^{*}a_{12}^{*} & - & a_{01}a_{12}|a_{20}|^2a_{02}^{*}a_{11}^{*} & - & a_{02}a_{11}|a_{20}|^2a_{01}^{*}a_{12}^{*}\\
- & |a_{02}|^2a_{10}a_{21}a_{11}^{*}a_{20}^{*} & - & |a_{02}|^2a_{11}a_{20}a_{10}^{*}a_{21}^{*} & - & a_{01}|a_{12}|^2a_{20}a_{00}^{*}a_{21}^{*}\\
- & a_{00}|a_{12}|^2a_{21}a_{01}^{*}a_{20}^{*} & - & a_{02}|a_{11}|^2a_{20}a_{00}^{*}a_{22}^{*} & - & a_{00}|a_{11}|^2a_{22}a_{02}^{*}a_{20}^{*}\\
- & a_{02}|a_{10}|^2a_{21}a_{01}^{*}a_{22}^{*} & - & a_{01}|a_{10}|^2a_{22}a_{02}^{*}a_{21}^{*} & - & |a_{01}|^2a_{12}a_{20}a_{10}^{*}a_{22}^{*}\\
- & |a_{01}|^2a_{10}a_{22}a_{12}^{*}a_{20}^{*} & - & |a_{00}|^2a_{12}a_{21}a_{11}^{*}a_{22}^{*} & - & |a_{00}|^2a_{11}a_{22}a_{12}^{*}a_{21}^{*}\\
+ & a_{00}a_{12}a_{21}a_{02}^{*}a_{11}^{*}a_{20}^{*} & + & a_{02}a_{11}a_{20}a_{00}^{*}a_{12}^{*}a_{21}^{*} & + & a_{01}a_{10}a_{22}a_{02}^{*}a_{11}^{*}a_{20}^{*}\\
+ & a_{02}a_{11}a_{20}a_{01}^{*}a_{10}^{*}a_{22}^{*} & + & a_{02}a_{10}a_{21}a_{01}^{*}a_{12}^{*}a_{20}^{*} & + & a_{01}a_{12}a_{20}a_{02}^{*}a_{10}^{*}a_{21}^{*}\\
+ & a_{00}a_{11}a_{22}a_{01}^{*}a_{12}^{*}a_{20}^{*} & + & a_{02}a_{10}a_{21}a_{00}^{*}a_{11}^{*}a_{22}^{*} & + & a_{00}a_{11}a_{22}a_{02}^{*}a_{10}^{*}a_{21}^{*}\\
+ & a_{01}a_{12}a_{20}a_{00}^{*}a_{11}^{*}a_{22}^{*} & + & a_{01}a_{10}a_{22}a_{00}^{*}a_{12}^{*}a_{21}^{*} & + & a_{00}a_{12}a_{21}a_{01}^{*}a_{10}^{*}a_{22}^{*}&.
\end{array}
\end{gathered}
\shortintertext{Here $C_2$ is sum of the principal minors of the density matrix:}
\begin{split}\label{eq:2QutritConcurrence}
C_2&=\phantom{+\!}|a_{00}a_{11}-a_{01}a_{10}|^2+|a_{02}a_{10}-a_{00}a_{12}|^2+|a_{01}a_{12}-a_{02}a_{11}|^2 \\
&\phantom{=}+|a_{01}a_{20}-a_{00}a_{21}|^2+|a_{00}a_{22}-a_{02}a_{20}|^2+|a_{10}a_{21}-a_{11}a_{20}|^2 \\
&\phantom{=}+|a_{12}a_{20}-a_{10}a_{22}|^2+|a_{02}a_{21}-a_{01}a_{22}|^2+|a_{11}a_{22}-a_{12}a_{21}|^2.
\end{split}
\end{gather}

The eigenvalues of the density matrix $\lambda_0,\lambda_1,\lambda_2$ satisfy
\begin{equation}
\begin{array}{c@{\ =\ }c}
\lambda_0+\lambda_1+\lambda_2                            & 1,   \\
\lambda_0\lambda_1+\lambda_1\lambda_2+\lambda_2\lambda_0 & C_2, \\
\lambda_0 \lambda_1 \lambda_2                            & |\det a|^2.
\end{array}
\end{equation}

\subsubsection{2-tangle}

 The bipartite entanglement of Alice and Bob is given by the 2-tangle \cite{Fan:2003,Cereceda:2003,HerrenoFierro:2005,Pan:2006,Rai:2005}
\begin{gather}
\tau_{AB}=27\det\rho_{A}=27\,|\det a_{AB}|^2,\label{eq:qutrit2-tangle}
\shortintertext{where $\rho_A$ is the reduced density matrix}
\rho_A=\Tr_B\ket{\Psi}\bra{\Psi}.
\end{gather}
The determinant is invariant under $SL(3)_A \times SL(3)_B$, with $a_{AB}$ transforming as a $\mathbf{(3,3)}$, and under a discrete duality that interchanges $A$ and $B$.

\subsubsection{Entanglement classification}

Once again, for subsequent comparison with the $D=5$ black hole, we restrict our attention to unnormalised states with real coefficients $a_{AB}$. In this case, one can show \cite{Dur:2000} that there are three algebraically independent invariants:  $\tau_{AB}, C_2$ and the norm $\braket{\Psi}{\Psi}$, corresponding to the dimension of the space of orbits
\begin{equation}
\frac{\mathds{R}^3 \times \mathds{R}^3}{SO(3) \times SO(3)},
\end{equation}
namely $9-6=3$. Hence, the most general two-qutrit state can be described by just three parameters, which may conveniently taken to be three real numbers $N_0$, $N_1$, $N_2$:
\begin{equation}\label{eq:three}
\ket{\Psi} = N_0\ket{00}+N_1\ket{11}+N_2\ket{22}.
\end{equation}
An SLOCC classification of two-qutrit entanglements, depending on the rank of the density matrix, is given in \autoref{tab:2QutritEntangClassif}.

\newpage
\section{\texorpdfstring{$STU$ BLACK HOLES}{STU Black Holes}}
\label{sec:STU}

\subsection{\texorpdfstring{The $STU$ model}{The STU model}}
\label{sec:stu}

\subsubsection{The Lagrangian}

An interesting subsector of string compactification to four dimensions is provided by the $STU$ model whose low energy limit is described by $\mathcal{N}=2$ supergravity coupled to three vector multiplets \cite{Sen:1995ff,Duff:1995sm,Gregori:1999ns,Bellucci:2008sv}.  One may regard it as a truncation of an $\mathcal{N}=2$ theory obtained by compactifying the Type IIA string on Calabi-Yau. Alternatively,  one may regard it as a truncation of an $\mathcal{N}=4$ theory obtained by compactifying the heterotic string on $T^{6}$, where $S,T,U$ correspond to the dilaton/axion, complex K\"{a}hler form and complex structure fields respectively. It exhibits an $SL(2, \mathds{Z})_S$ strong/weak coupling duality and an $SL(2, \mathds{Z})_T \times SL(2, \mathds{Z})_U$ target space duality\footnote{This $SL(2, \mathds{Z})_U$ should not be confused with U-duality.}. By string/string duality, this is equivalent to a Type IIA string on $K3 \times T^{2}$ with $S$ and $T$ exchanging roles \cite{Duff:1993ij,Hull:1994ys,Duff:1994zt}. Moreover, by mirror symmetry this is in turn equivalent to a Type IIB string on the mirror manifold with $T$ and $U$ exchanging roles. Hence the truncated theory has a combined $[SL(2, \mathds{Z})]^{3}$ duality and complete $S$-$T$-$U$ triality symmetry \cite{Duff:1995sm}. Alternatively, one may simply start with this $\mathcal{N}=2$ theory directly as an interesting four-dimensional supergravity in its own right, as described below.

The model admits extremal black holes solutions carrying four electric and four magnetic charges \cite{Duff:1995sm,Cvetic:1995uj}. Below we organise these 8 charges into the  $2 \times 2 \times 2$ \emph{hypermatrix} and display the $S$-$T$-$U$ symmetric Bogomol'nyi mass spectrum \cite{Duff:1995sm}. Associated with this hypermatrix is a \emph{hyperdeterminant}, discussed in \autoref{sec:Hyperdet}, first introduced by Cayley in 1845 \cite{Cayley:1845}. The black hole entropy, first calculated in \cite{Behrndt:1996hu}, is quartic in the charges and must be invariant under $[SL(2, \mathds{Z})]^{3}$ and under triality. As we shall see, the important observation for its link to entanglement is that it is given by the square root of Cayley's hyperdeterminant.

Consider the three complex scalars axion/dilaton field $S$, the complex K\"{a}hler form field $T$ and the complex structure field $U$
\begin{equation}\label{eq:stu}
\begin{split}
S&=S_1+i S_2, \\
T&=T_1+i T_2, \\
U&=U_1+i U_2.
\end{split}
\end{equation}
This complex parametrisation allows for a natural transformation under the various $SL(2, \mathds{Z})$ symmetries. The action of $SL(2, \mathds{Z})_S$ is given by
\begin{equation}\label{eq:SL2ZAction}
S \to \frac{aS+b}{cS+d},
\end{equation}
where $a,b,c,d$ are integers satisfying $ad-bc=1$, with similar expressions for $SL(2, \mathds{Z})_T$ and $SL(2, \mathds{Z})_U$.  Defining the matrices $\mathcal{M}_S,\mathcal{M}_T$ and $\mathcal{M}_U$ via
\begin{equation}\label{eq:SL2Matrix}
\mathcal{M}_S=\frac{1}{S_2}
\begin{pmatrix}1 & S_1  \\S_1 & |S|^2\end{pmatrix},
\end{equation}
the action of $SL(2, \mathds{Z})_S$ now takes the form
\begin{gather}
\mathcal{M}_S \to {\omega_S}^{\textsf{T}}\mathcal{M}_S\omega_S,
\shortintertext{where}
\omega_S=\begin{pmatrix}d & b \\c & a\end{pmatrix},
\end{gather}
with similar expressions for $\mathcal{M}_T$ and $\mathcal{M}_U$. We also define the $SL(2, \mathds{Z})$ invariant tensors
\begin{equation}
\varepsilon_S=\varepsilon_T=\varepsilon_U=\begin{pmatrix*}[r] 0 & 1 \\-1 & 0\end{pmatrix*}.
\end{equation}
Starting from the heterotic string,the bosonic action for the graviton $g_{\mu\nu}$, dilaton $\eta$, two-form $B_{\mu\nu}$ four $U(1)$ gauge fields $A_S^a$ and two complex scalars $T$ and $U$ is \cite{Duff:1995sm}
\begin{equation}
\begin{split}\label{eq:STUAction}
I_{STU}=\frac{1}{16\pi G}\int d^4x\sqrt{-g}e^{-\eta}\Bigl[&R_g + g^{\mu\nu}\partial_{\mu}\eta\partial_{\nu}\eta - \tfrac{1}{12}g^{\mu\lambda}g^{\nu\tau}g^{\rho\sigma}H_{\mu\nu\rho}H_{\lambda\tau\sigma} \\
&+\tfrac{1}{4}\tr(\partial {\mathcal{M}_T}^{-1}\partial\mathcal{M}_T) + \tfrac{1}{4}\tr(\partial {\mathcal{M}_U}^{-1}\partial\mathcal{M}_U) \\
&-\tfrac{1}{4}{{F_S}_{\mu\nu}}^{\textsf{T}}(\mathcal{M}_T \times \mathcal{M}_U){F_S}^{\mu\nu}\Bigr].
\end{split}
\end{equation}
where the metric $g_{\mu\nu}$ is related to the four-dimensional canonical Einstein metric $g^c_{\mu\nu}$ by $g_{\mu\nu}=e^{\eta}{g^c}_{\mu\nu}$ and where
\begin{equation}
H_{\mu\nu\rho}=3\left(\partial_{[\mu}B_{\nu\rho]} - \half A_{S[\mu}^{\textsf{T}} (\varepsilon_T\times\varepsilon_U){F_S}_{\nu\rho]}\right).
\end{equation}
This action is manifestly invariant under $T$-duality and $U$-duality, with
\begin{align}
{F_S}_{\mu\nu}&\to({\omega_T}^{-1}\times {\omega_U}^{-1}){F_S}_{\mu\nu},
&\mathcal{M}_{T/U}&\to {\omega_{T/U}}^{\textsf{T}} \, \mathcal{M}_{T/U} \, \omega_{T/U},
\end{align}
and with $\eta$, $g_{\mu\nu}$ and $B_{\mu\nu}$ inert.  Its equations of motion and Bianchi identities (but not the action itself) are also invariant under $S$-duality \eqref{eq:SL2ZAction}, with $T$ and ${g^c}_{\mu\nu}$ inert and with
\begin{gather}
\begin{pmatrix}{F_S}\indices{_{\mu\nu}^a} \\\widetilde{F}\indices{_S _{\mu\nu}^a}\end{pmatrix}
\to
{\omega_S}^{-1}\begin{pmatrix}{F_S}\indices{_{\mu\nu}^a} \\\widetilde{F}\indices{_S _{\mu\nu}^a}\end{pmatrix},
\shortintertext{where}
\widetilde{F}\indices{_S _{\mu\nu}^a}=-S_2[({\mathcal{M}_T}^{-1} \times {\mathcal{M}_U}^{-1})(\varepsilon_T \times \varepsilon_U)]\indices{^a_b}{\star{F_S}}\indices{_{\mu\nu}^b}-S_1 {F_S}\indices{_{\mu\nu}^a},
\shortintertext{where the axion field $a$ is defined by}
\varepsilon^{\mu\nu\rho\sigma}\partial_{\sigma}a=
\sqrt{-g}e^{-\eta}g^{\mu\sigma}g^{\nu\lambda}g^{\rho\tau}
H_{\sigma\lambda\tau},
\end{gather}
and where $S=S_1+i S_2=a+i e^{-\eta}$.

Thus $T$-duality transforms Kaluza-Klein electric charges $({F_S}^3,{F_S}^4)$ into winding electric charges $({F_S}^1,{F_S}^2)$ (and Kaluza-Klein magnetic charges into winding magnetic charges), $U$-duality transforms the Kaluza-Klein and winding electric charge of one circle $({F_S}^3,{F_S}^2)$ into those of the other $({F_S}^4,{F_S}^1)$ (and similarly for the magnetic charges) but $S$-duality transforms Kaluza-Klein electric charge $({F_S}^3,{F_S}^4)$ into winding magnetic charge $({\widetilde{F}_S}^3,{\widetilde{F}_S}^4)$ (and winding electric charge into Kaluza-Klein magnetic charge). In summary we have $SL(2, \mathds{Z})_T \times SL(2, \mathds{Z})_U$ and $T \leftrightarrow U$ off-shell but $SL(2, \mathds{Z})_S \times SL(2, \mathds{Z})_T \times SL(2, \mathds{Z})_U$ and an $S$-$T$-$U$ interchange on-shell.

One may also consider the Type IIA action  $I_{TUS}$ and the Type IIB action $I_{UST}$ obtained by cyclic permutation of the fields $S,T,U$. Finally, one may consider an action \cite{Behrndt:1996hu} where the $S$, $T$ and $U$ fields enter democratically with a prepotential
\begin{equation}\label{eq:HAxion}
F=STU,
\end{equation}
which off-shell has the full $STU$ interchange but none of the $SL(2, \mathds{Z})$. All four versions are on-shell equivalent.

\subsubsection{The Bogomol'nyi spectrum}

Following \cite{Duff:1995sm}, it is now straightforward to write down an $S$-$T$-$U$ symmetric Bogomol'nyi mass formula. Let us define electric and magnetic charge vectors $\alpha_S^a$ and $\beta_S^a$ associated with the field strengths ${F_S}^a$ and $\widetilde{F_S}^a$ in the standard way. The electric and magnetic charges $Q_S^a$ and $P_S^a$ are given by
\begin{gather}
\begin{align}
{F_S}\indices{_{0r}^a}&\sim\frac{Q_S^a}{r^2},&{\star F_S}\indices{_{0r}^a}&\sim\frac{P_S^a}{r^2},
\end{align}
\shortintertext{giving rise to the charge vectors}
\begin{pmatrix}\alpha_S^a \\ \beta_S^a\end{pmatrix}=
\begin{pmatrix}
S_2^{(0)} {\mathcal{M}_T}^{-1}\times {\mathcal{M}_U}^{-1} & S_1^{(0)} \varepsilon_T \times \varepsilon_U  \\
0 & -\varepsilon_T \times \varepsilon_U
\end{pmatrix}^{ab}
\begin{pmatrix} Q_S^b \\ P_S^b\end{pmatrix}.
\end{gather}
Interestingly enough, as was shown in \cite{Duff:1995sm,Behrndt:1996hu}, the 8 charges may usefully be represented as a $2 \times 2 \times 2$ hypermatrix $a_{ABC}$
\begin{gather}
\begin{array}{c@{\Big(}*{7}{c@{,\ }}c@{\Big)}c}
   & a_{000}    & a_{001}    & a_{010}    & a_{011}    & a_{100}    & a_{101}    & a_{110}    & a_{111}    & \\[5pt]
=~ & -\beta_S^1 & -\beta_S^2 & -\beta_S^3 & -\beta_S^4 & \alpha_S^1 & \alpha_S^2 & \alpha_S^3 & \alpha_S^4 & ,
\end{array}
\shortintertext{transforming as}
a^{ABC}\to{\omega_S}\indices{^A_E}{\omega_T}\indices{^B_F}{\omega_U}\indices{^C_G}a^{EFG}.
\end{gather}
Then the mass formula is
\begin{equation}\label{eq:mass1}
\begin{split}
M^2_{\text{BPS}}=\tfrac{1}{8}{a}^{\textsf{T}}\big(&{\mathcal M}_U^{-1}\otimes{\mathcal M}_T^{-1}\otimes{\mathcal M}_S^{-1}-{\varepsilon}_U\otimes{\varepsilon}_T\otimes{\mathcal M}_S^{-1}\\
&-{\varepsilon}_U\otimes{\mathcal M}_T^{-1}\otimes{\varepsilon}_S-{\mathcal M}_U^{-1}\otimes{\varepsilon}_T\otimes{\varepsilon}_S\big){a},
\end{split}
\end{equation}

This is consistent with the general $\mathcal{N}=2$ Bogomol'nyi formula \cite{Ceresole:1994cx}. Although all theories have the same mass spectrum, there is clearly a difference of interpretation with electrically charged elementary states in one picture being solitonic monopole or dyon states in the other.

\subsubsection{Black hole entropy}

\begin{figure}[ht]
 \centering
 \includegraphics[width=7cm]{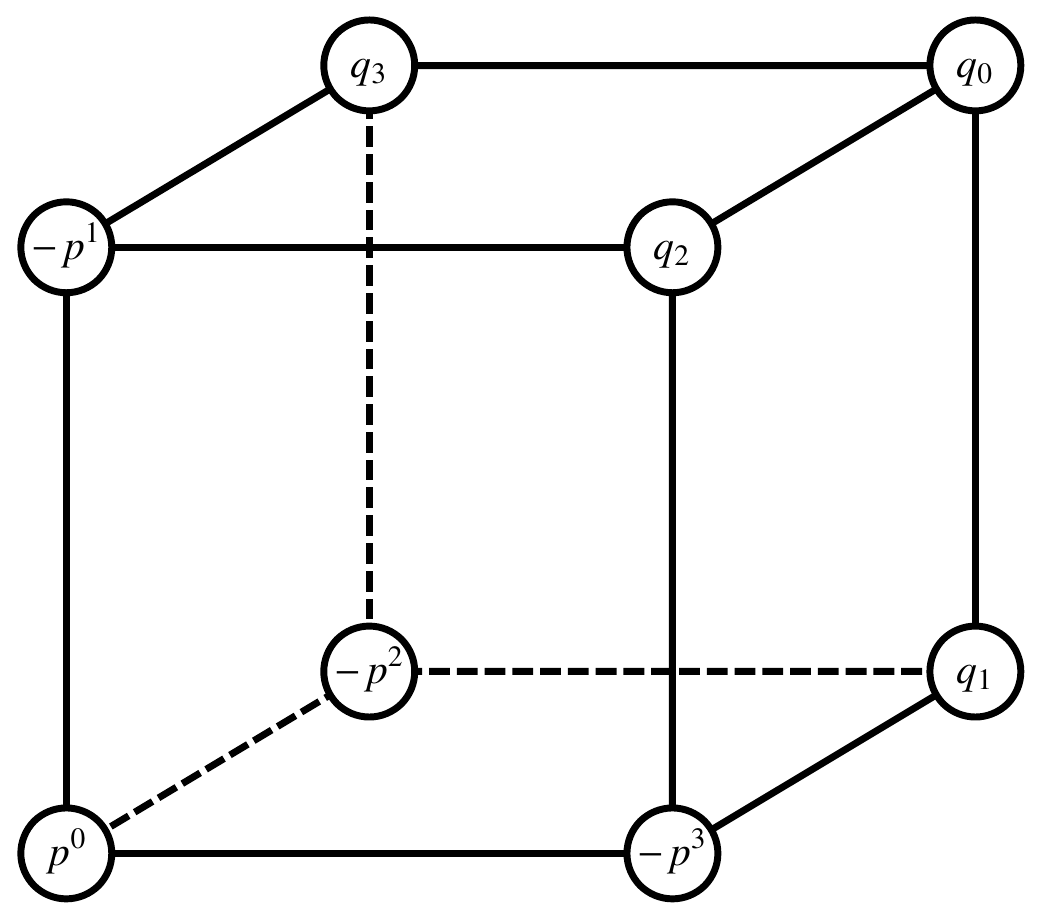}
 \caption[$STU$ charge hypermatrix cube]{The vertices of the hypermatrix cube from \autoref{fig:BinaryFanoBasisCube} are transformed under the dictionary \eqref{eq:charges7} to electric and magnetic charges of the $STU$ black hole. The black hole entropy is related to the hyperdeterminant of this hypermatrix, and can also be computed from the three pairs of slicings of the cube along its symmetry planes (see \autoref{sec:Freudenthal-Fano}).}
 \label{fig:Freudenthalbasiscube}
\end{figure}

The $STU$ model admits extremal black hole solutions satisfying the Bogomol'nyi mass formula. As usual, their entropy is given by one quarter the area of the event horizon. However, to calculate this area requires evaluating the mass not with the asymptotic values of the moduli, but with their frozen values on the horizon which are fixed in terms of the charges \cite{Ferrara:1995ih}. This ensures that the entropy is moduli-independent, as it should be. The relevant calculation was carried out in \cite{Behrndt:1996hu} for the model with the $STU$ prepotential. The electric and magnetic charges of that paper are denoted $(p^0, q_0),\, (p^1, q_1),\, (p^2, q_2),\, (p^3,q_3)$ with $O(2,2)$ scalar products
\begin{equation}
\begin{array}{c@{\ =\ }c@{\ +\ }c*{4}{@{\ }c}}
p^2      & (p^0)^2  & (p^1)^2  & - & (p^2)^2  & - & (p^3)^2,\\
q^2      & (q_0)^2  & (q_1)^2  & - & (q_2)^2  & - & (q_3)^2,\\
p\cdot q & (p^0q_0) & (p^1q_1) & + & (p^2q_2) & + & (p^3q_3).
\end{array}
\end{equation}
These eight charges may be represented by the cube shown in \autoref{fig:Freudenthalbasiscube}.
This charge basis is related to the one above by\footnote{Note that this is the convention of \cite{Kallosh:2006zs}; in  \cite{Duff:2006uz}   the sign of $a_{0},a_{3},a_{4},a_{7}$ is flipped, which gives the same answer.}

\begin{gather}
\begin{array}{c@{\ \big(}*{7}{c@{,\ }}c@{\big)}c@{}c}\label{eq:charges7}
  & p^0   & p^1    & p^2    & p^3    & q_0   & q_1   & q_2   & q_3   & \\[3pt]
= & a_{0} & -a_{1} & -a_{2} & -a_{4} & a_{7} & a_{6} & a_{5} & a_{3} & ,
\end{array}
\shortintertext{In these variables, the entropy is given by}
S=\pi \sqrt{|D(p^\Lambda,q_\Lambda)|},
\shortintertext{where}
\begin{split}\label{eq:STUBHEntropy}
D(p^\Lambda ,q_\Lambda) &= -{(p\cdot q)}^2+4\bigl((p^1q_1)(p^2q_2)+(p^1q_1)(p^3q_3)+(p^3q_3)(p^2q_2)\bigr) \\
                        &\phantom{=}-4 p^0 q_1 q_2 q_3 + 4q_0 p^1 p^2 p^3.
\end{split}
\end{gather}
The function $D(p^\Lambda ,q_\Lambda)$ is symmetric under transformations: $p^1\leftrightarrow p^2 \leftrightarrow p^3$ and  $q_1\leftrightarrow q_2 \leftrightarrow q_3$.

\section{\texorpdfstring{$STU$ BLACK HOLE/QUBIT CORRESPONDENCE}{STU Black Hole/Qubit Correspondence}}
\label{sec:correspondence}

\subsection{Entropy/entanglement correspondence}

The black hole/qubit correspondence now comes about by identifying \cite{Duff:2006uz} the black hole charge hypermatrix \eqref{eq:charges7} with the three qubit hypermatrix \eqref{eq:hypermatrix}. Then we recognise from \eqref{eq:CayleyHyperdeterminant} that
\begin{equation}
D(p^\Lambda ,q_\Lambda)=-\Det a,
\end{equation}
and hence, as advertised in the Introduction, the $STU$ black hole entropy and Alice-Bob-Charlie 3-tangle are related by
\begin{equation}
S=\tfrac{\pi}{2}\sqrt{\tau_{ABC}}.
\end{equation}
Thus Cayley's hyperdeterminant provides an interesting connection, at least at the level of mathematics, between string theory and quantum entanglement. Other mathematical similarities are provided by the division algebras \cite{Bernevig:2003} and by twistors \cite{Levay:2004}. What about physics? The near horizon geometry of the black holes is $\text{AdS}_2 \times S^{2}$ and one might expect a relation between the black hole entropy and the entanglement entropy of the conformal quantum mechanics that lives on the boundary \cite{Hawking:2000da}, although the nature of this particular AdS/CFT duality is not well-understood \cite{BrittoPacumio:1999ax}. In any event, the 3-tangle is not the same as the entropy of entanglement \cite{Sudbery:2001}. So the appearance of the Cayley hyperdeterminant in these two different contexts of stringy black hole entropy and three-qubit quantum entanglement remains, for the moment, a purely mathematical coincidence.

To illustrate more these coincidences we compare some examples of $\mathcal{N}=2$ black holes with examples of three-qubit states, following \cite{Kallosh:2006zs}.

\subsection{Rebits}

In the $STU$ stringy black hole context \cite{Duff:2006uz,Duff:1995sm,Behrndt:1996hu,Kallosh:2006zs} the $a_{ABC}$ are integers (corresponding to quantised charges) and hence the symmetry group is $[SL(2, \mathds{Z})]^{3}$ rather than $[SL(2,\mathds{C})]^{3}$. However, as discussed by Levay \cite{Levay:2006kf}, it is possible within quantum information theory to focus on real qubits, called \emph{rebits}, for which the $a_{ABC}$ are real \cite{Batle:2003,Caves:2000}. (One difference remains, however: one may normalise the wavefunction, whereas for black holes there is no such restriction on the charges $a_{ABC}$). It turns out that there are three reality classes which can be characterised by the hyperdeterminant
\begin{subequations}
\begin{align}
  \Det a &< 0, \label{eq:RebitCase1}\\
  \Det a &= 0, \label{eq:RebitCase2}\\
  \Det a &> 0. \label{eq:RebitCase3}
\end{align}
\end{subequations}
Case \eqref{eq:RebitCase1} corresponds to the non-separable or GHZ class \cite{Greenberger:1989}, for example,
\begin{equation}\label{eq:RebitGHZ}
\ket{\Psi}= \half(-\ket{000} +\ket{011}+\ket{101} +\ket{110}).
\end{equation}
Case \eqref{eq:RebitCase2} corresponds to the separable ($A$-$B$-$C$, $A$-$BC$, $B$-$CA$, $C$-$AB$) and W classes, for example
\begin{equation}
\ket{\Psi}=\tfrac{1}{\sqrt{3}}(\ket{100}+\ket{010}+\ket{001}).
\end{equation}

In the string/supergravity interpretation \cite{Duff:2006uz}, cases \eqref{eq:RebitCase1} and \eqref{eq:RebitCase2} were shown to correspond to BPS black holes, for which half of the supersymmetry is preserved.  Case \eqref{eq:RebitCase1} has non-zero horizon area and entropy (``large'' black holes), and case \eqref{eq:RebitCase2} has    vanishing horizon area and entropy (``small'' black holes), at least at the semi-classical level. However, small black holes may acquire a non-zero entropy through higher order quantum effects. As discussed in \autoref{sec:higher}, this entropy also has a quantum information interpretation involving bipartite entanglement of the three qubits \cite{Kallosh:2006zs}.

Case \eqref{eq:RebitCase3} is also GHZ, for example the above GHZ state \eqref{eq:RebitGHZ} with a sign flip
\begin{gather}
\ket{\Psi}= \half(\ket{000} +\ket{011}+\ket{101} +\ket{110}).
\shortintertext{The canonical GHZ state}
\ket{\Psi}= \tfrac{1}{\sqrt{2}}(\ket{111} +\ket{000})
\end{gather}
also belongs to case \eqref{eq:RebitCase3}.

\subsection{\texorpdfstring{Classification of $\mathcal{N}=2$ black holes and three-qubit states}{Classification of N=2 black holes and three-qubit states}}
\label{sec:blackholesandthree-qubit states}

\begin{description}
\item[$A$-$B$-$C$ states and singly charged black holes:]
A black hole with just one charge, say $q_0$ as in  \autoref{fig:q_0Cube}, has vanishing entropy and corresponds to a completely separable $A$-$B$-$C$ state
    \begin{gather}
    \ket{\Psi}=q_0\ket{111}\label{eq:q_0State},
    \shortintertext{with}
    \begin{gathered}
    S_A=S_B=S_C=0,\\
    \tau_{ABC}=0.\label{eq:A-B-C}
    \end{gathered}
    \end{gather}
    \begin{figure}[ht]
    \centering
    \includegraphics[width=5cm]{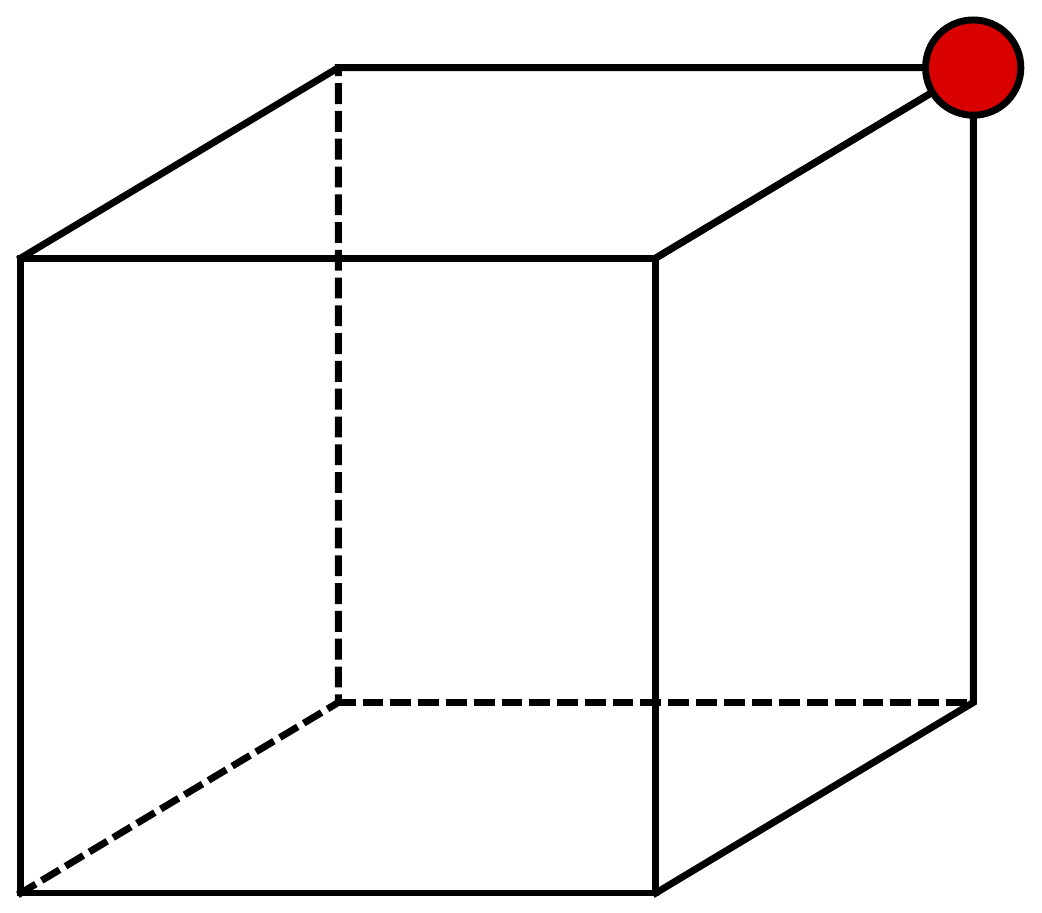}
    \caption[Completely separable, single-charge hypermatrix cube]{The hypermatrix cube of \autoref{fig:Freudenthalbasiscube} is restricted to correspond to the state \eqref{eq:q_0State} by retaining only a single nonzero entry at the $q_0$ vertex, denoted by the red disc. The state is completely separable and accordingly, the entropy vanishes for this cube.}
    \label{fig:q_0Cube}
    \end{figure}

    Examples of singly charged supersymmetric black hole solutions \cite{Duff:1994jr} are provided by the electric Kaluza-Klein black hole with $\alpha=(1,0,0,0)$ and $\beta=(0,0,0,0)$; the electric winding black hole with $\alpha=(0,0,0,-1)$ and $\beta=(0,0,0,0)$; the magnetic Kaluza-Klein black hole with $\alpha=(0,0,0,0)$ and $\beta=(0,-1,0,0)$; the magnetic winding black hole with $\alpha=(0,0,0,0)$ and $\beta=(0,0,-1,0)$. These are characterised by a scalar-Maxwell coupling parameter $a=\sqrt{3}$.

    By combining these 1-particle states, we may build up 2-, 3- and 4-particle bound states at threshold \cite{Duff:1994jr,Duff:1995sm,Duff:1996qp}, characterised by scalar-Maxwell coupling parameter $a=1$,$1/\sqrt{3}$ and $0$ respectively. The 1-, 2- and 3-particle states all yield vanishing contributions to $\Det a$.  A non-zero value is obtained for the 4-particle example, however, which is just the Reissner-Nordstr\"om black hole, as explained below.

    One could also consider a black hole with two charges, say $q_0$ and $q_1$ as in \autoref{fig:q_0q_1Cube}, having vanishing entropy and corresponding to another completely separable state
    \begin{equation}\label{eq:q_0q_1State}
    \ket{\Psi}=q_0\ket{111}+q_1\ket{110}
    \end{equation}
    also satisfying \eqref{eq:A-B-C}.
    \begin{figure}[ht]
    \centering
    \includegraphics[width=5cm]{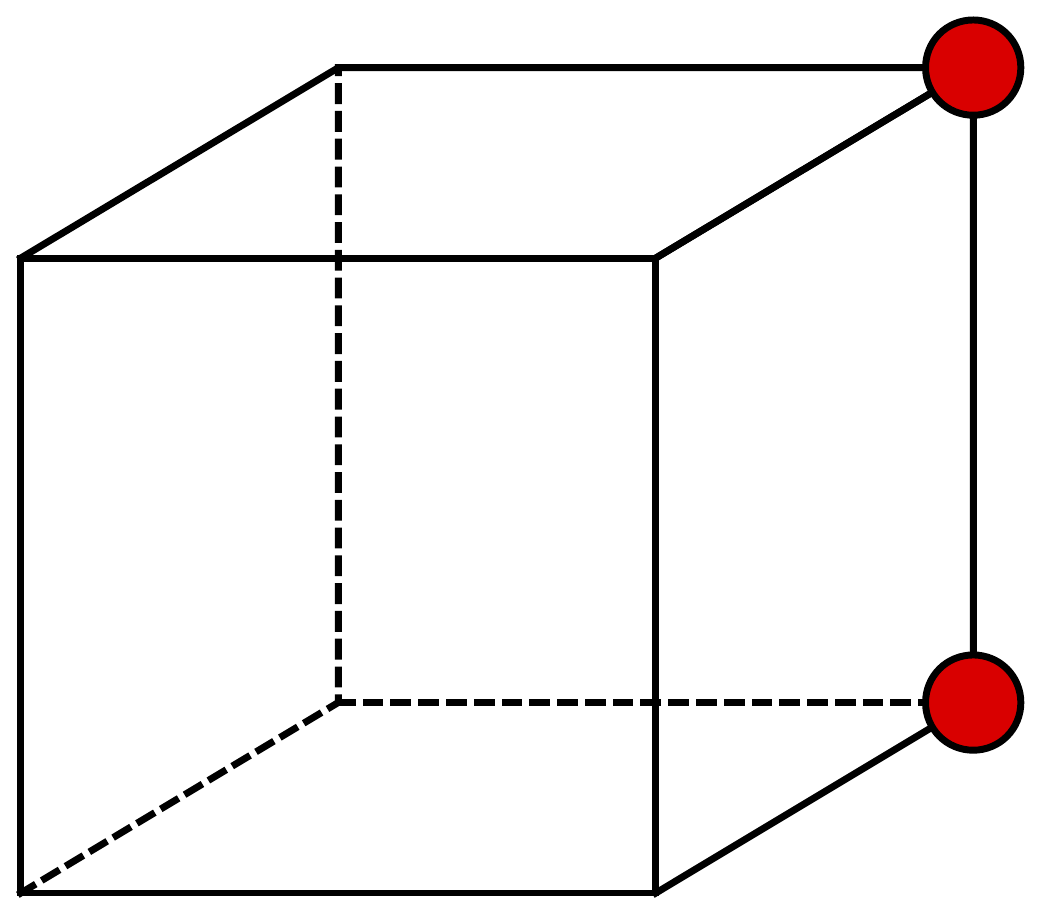}
    \caption[Completely separable, two-charge hypermatrix cube]{The hypermatrix cube of \autoref{fig:Freudenthalbasiscube} is restricted to correspond to the state \eqref{eq:q_0q_1State} by retaining nonzero entries at the $q_0$ and $q_1$ vertices, denoted by the red discs. Despite having two nonzero vertices the cube's entropy vanishes since the state is again completely separable.}
    \label{fig:q_0q_1Cube}
    \end{figure}
\item[$A$-$BC$ states and doubly charged black holes:]
    A black hole with just two charges, say $q_0$ and $p^1$ as in \autoref{fig:q_0p^1Cube}, has vanishing entropy and corresponds to a bipartite entangled state
    \begin{gather}
    \ket{\Psi}=q_0\ket{111}-p^1\ket{001}\label{eq:q_0p^1State},
    \shortintertext{with}
    \begin{gathered}
    S_A=S_B=4(q_0p^1)^2,\\
    S_C=0,\\
    \Det a=0.
    \end{gathered}
    \end{gather}
    \begin{figure}[ht]
    \centering
    \includegraphics[width=5cm]{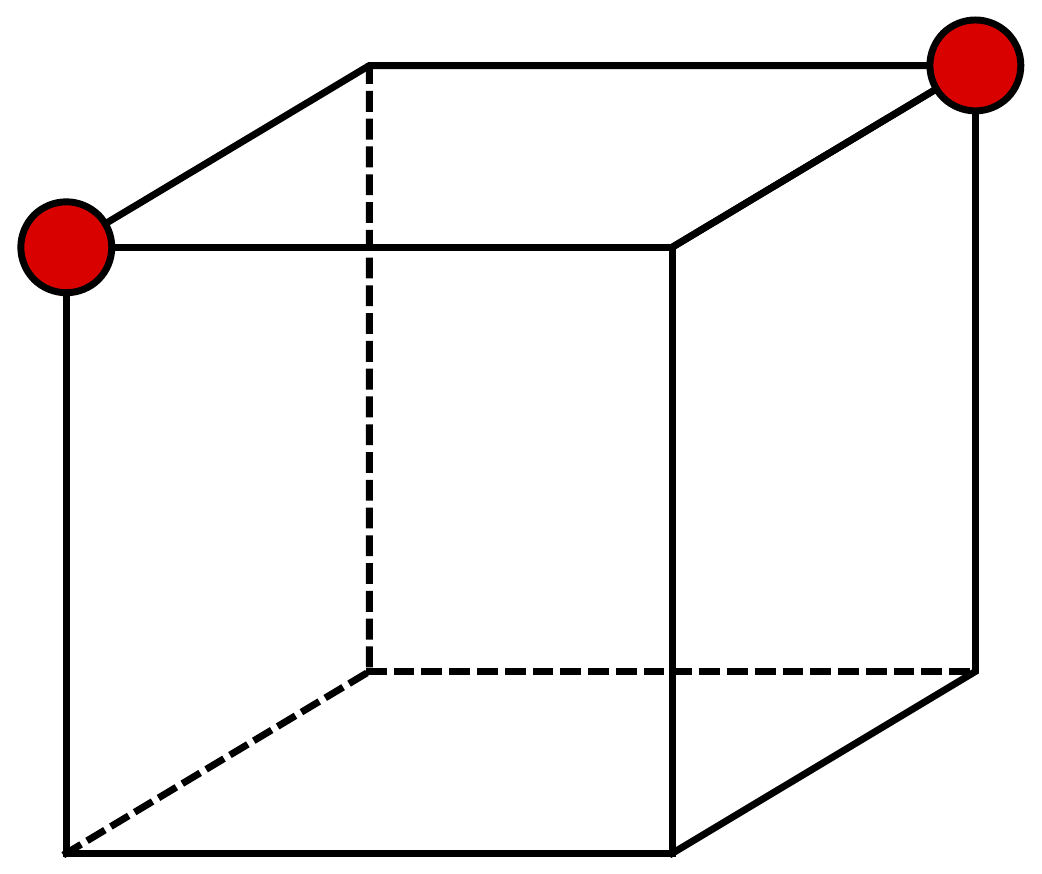}
    \caption[Bi-separable, two-charge hypermatrix cube]{The hypermatrix cube of \autoref{fig:Freudenthalbasiscube} is restricted to correspond to the state \eqref{eq:q_0p^1State} by retaining nonzero entries at the $q_0$ and $p^1$ vertices, denoted by the red discs. In this case the state is bi-separable, so the entropy vanishes once more.}
    \label{fig:q_0p^1Cube}
    \end{figure}
\item[W states and triply charged black holes:]
    A black hole with just three charges, say $q_0$, $p^1$ and $p^2$ as in \autoref{fig:q_0p^1p^2Cube}, has vanishing entropy and corresponds to a W state
    \begin{gather}
    \ket{\Psi}=q_0\ket{111}-p^1\ket{001}-p^2\ket{010},\label{eq:q_0p^1p^2State}
    \shortintertext{with}
    \begin{gathered}
    \begin{aligned}
    S_A&=4(q_0)^2((p^1)^2+(p^2)^2), \\
    S_B&=4(p^1)^2((q_0)^2+(p^2)^2), \\
    S_C&=4(p^2)^2((q_0)^2+(p^1)^2),
    \end{aligned}\\
    \Det a=0.
    \end{gathered}
    \end{gather}
    \begin{figure}[ht]
    \centering
    \includegraphics[width=5cm]{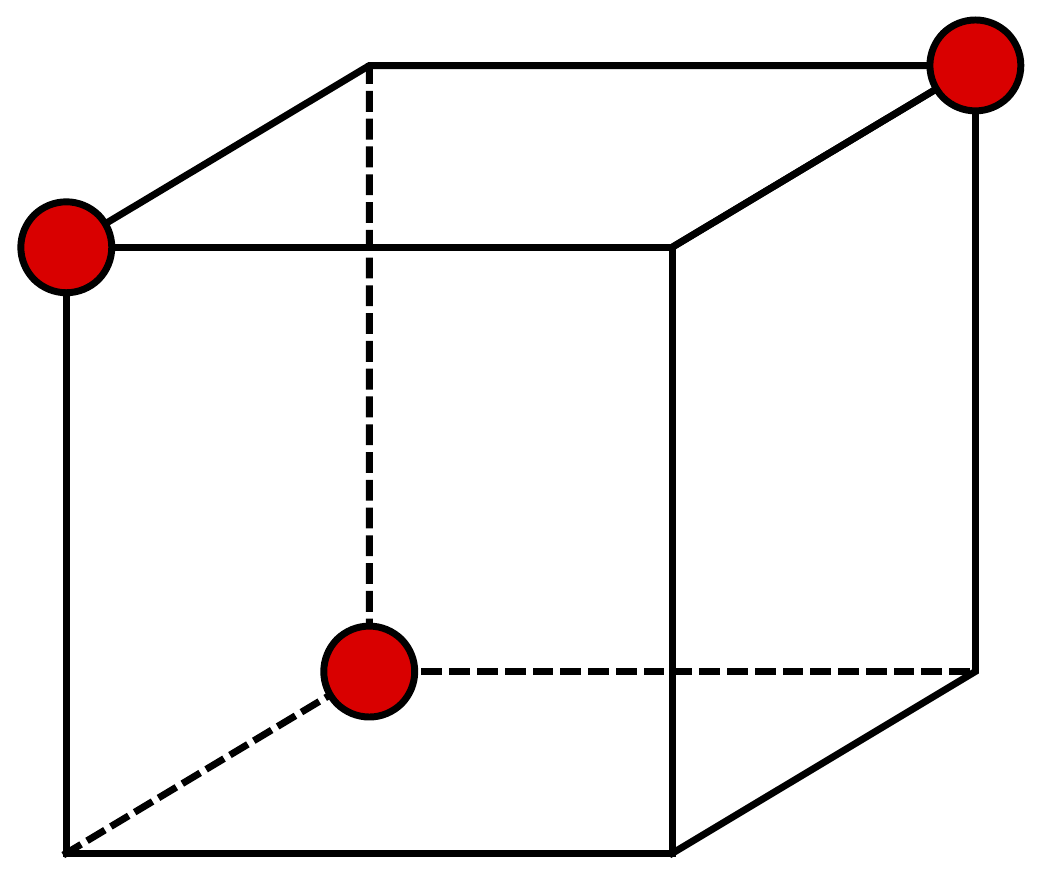}
    \caption[Three charge, W state hypermatrix cube]{The hypermatrix cube of \autoref{fig:Freudenthalbasiscube} is restricted to correspond to the state \eqref{eq:q_0p^1p^2State} by retaining nonzero entries at the $q_0, p^1$ and $p^2$ vertices, denoted by the red discs. The entropy vanishes this time since the cube corresponds to a W state.}
    \label{fig:q_0p^1p^2Cube}
    \end{figure}
\item[GHZ and 4-charge BPS and non-BPS black holes:]
    A black hole with just four charges, say $q_0$, $p^1$,$p^2$ and $p^3$ as in \autoref{fig:q_0p^1p^2p^3Cube}, has non-vanishing entropy and corresponds to a GHZ state
    \begin{gather}
    \ket{\Psi}=q_0\ket{111}-p^1\ket{001}-p^2\ket{010}-p^3\ket{100},\label{eq:q_0p^1p^2p^3State}
    \shortintertext{with}
    \begin{gathered}
    \begin{split}
    S_A&=4((p^3)^2+(q_0)^2)((p^1)^2+(p^2)^2),\\
    S_B&=4((p^1)^2+(p^3)^2)((q_0)^2+(p^2)^2),\\
    S_C&=4((p^2)^2+(p^3)^2)((q_0)^2+(p^1)^2),
    \end{split}\\
    \Det a=-4q_0p^1p^2p^3.
    \end{gathered}
    \end{gather}
    This is a large BPS black hole if $\Det a<0$ and a large non-BPS black hole if $\Det a>0$.
    \begin{figure}[ht]
    \centering
    \includegraphics[width=5cm]{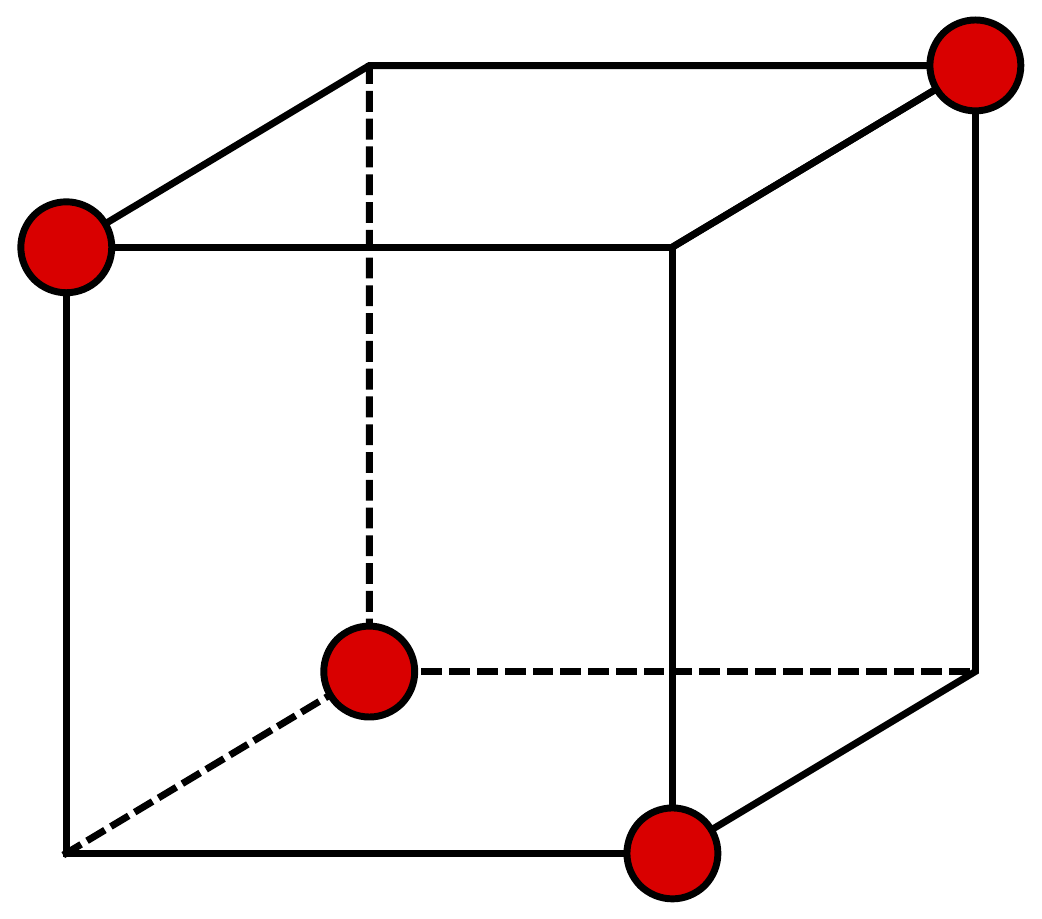}
    \caption[Four charge, GHZ hypermatrix cube]{The hypermatrix cube of \autoref{fig:Freudenthalbasiscube} is restricted to correspond to the state \eqref{eq:q_0p^1p^2p^3State} by retaining nonzero entries at the $q_0, p^1, p^2$ and $p^3$ vertices, denoted by the red discs. This is a GHZ state exhibiting genuine tripartite entanglement and accordingly the cube has nonzero entropy. While the previous cubes corresponded to small BPS black holes, this black hole is large and BPS/non-BPS for $\Det a < 0$/$\Det a > 0$.}
    \label{fig:q_0p^1p^2p^3Cube}
    \end{figure}
\item[GHZ and 2-charge non-BPS black holes:]
    A black hole with just two charges \autoref{fig:q_0p^0Cube}, say $q_0$ and $p^0$ as in \autoref{fig:q_0p^0Cube}, has non-vanishing entropy and corresponds to a GHZ state
    \begin{gather}
    \ket{\Psi}=q_0\ket{111}+p^0\ket{000},\label{eq:q_0p^0State}
    \shortintertext{with}
    S_A=S_B=S_C=4(p^{0})^{2}(q_{0})^{2},
    \shortintertext{and}
    \Det a=(p^{0})^{2}(q_{0})^{2}.
    \end{gather}
    Since $\Det a >0$, this describes a non-BPS large black hole \cite{Kallosh:2006bx,Gimon:2007mh}.
    \begin{figure}[ht]
    \centering
    \includegraphics[width=5cm]{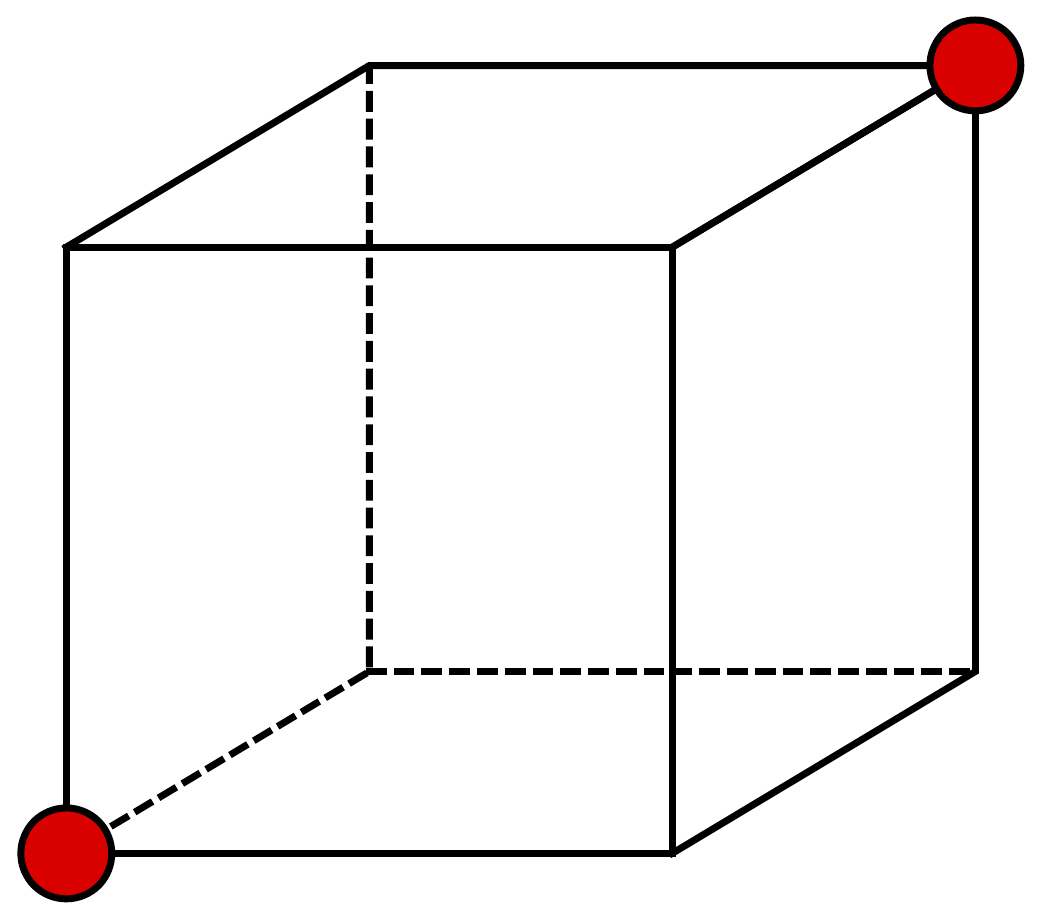}
    \caption[Two charge, GHZ hypermatrix cube]{The hypermatrix cube of \autoref{fig:Freudenthalbasiscube} is restricted to correspond to the state \eqref{eq:q_0p^0State} by retaining nonzero entries at the $q_0$ and $p^0$ vertices, denoted by the red discs. Despite having only two nonzero charges the cube has nonvanishing entropy since the state is GHZ. The corresponding black hole is large and always non-BPS.}
    \label{fig:q_0p^0Cube}
    \end{figure}
\end{description}

\subsection{Higher order corrections}
\label{sec:higher}
The small black holes have a singular horizon with vanishing area and entropy at the classical level, but may acquire nonvanishing area and entropy due to quantum corrections, characterised by higher derivatives in the supergravity lagrangian. One can interpret this as consequence of the quantum stretching of the horizon conjectured by Susskind \cite{Susskind:1993aa} and Sen \cite{Sen:1994eb,Sen:1995in}. See also \cite{Dabholkar:2004dq,Sinha:2006yy,Ooguri:2004zv,Dabholkar:2005dt,Cardoso:2006bg,Alishahiha:2006jd}.

Kallosh and Linde \cite{Kallosh:2006zs} have noted that this quantum entropy also admits an interpretation in terms of qubit entanglement measures. They propose a general formula that correctly reduces to the known special cases. It is given by
\begin{equation}
S_{total}=\frac{\pi}{2}\sqrt{\tau_{ABC}+\frac{4c_2}{3}(C_{AB}+C_{BC}+C_{CA})+\frac{8K^2}{3}|\Psi|},
\end{equation}
where $\tau_{ABC}$ is the 3-tangle \eqref{eq:3-tangle}, $C_{AB}$ is the concurrence \eqref{eq:concur}, $|\Psi|$ is the norm and $c_2$ and $K$ are constants that depend upon the compactification.

For completely separable states with only one nonzero charge, this reduces to
\begin{equation}
S=K\sqrt{\tfrac{2}{3}|\Psi|}=\pi K\sqrt{\tfrac{2}{3}|q_0|}.
\end{equation}
For the bipartite and W-states at large values of the charges, the concurrences are much greater than $|\Psi|$ and the formula reduces to
\begin{align}
S_{total}&=\pi\sqrt{\frac{c_2}{3}(C_{AB}+C_{BC}+C_{CA})}\\
&=4\pi\sqrt{|q_0(p^1+p^2)|}.
\end{align}
Finally, for the GHZ states the (unnormalised) 3-tangle is much greater than the concurrences and we regain
\begin{equation}
S=\tfrac{\pi}{2}\sqrt{\tau_{ABC}}.
\end{equation}
As with the lowest order black hole/qubit correspondence, there is as yet no underlying physical explanation for these mathematical coincidences at higher orders.

\subsection{Attractors and SLOCC}
\label{sec:attractor}

The extremal black holes of \autoref{sec:extremal} occupy a special place in string theory. They have zero Hawking temperature and are therefore  stable. In the string theory context the microscopic configurations of these black holes are explained by D-branes, fundamental strings and other solitonic objects of the theory \cite{Duff:1994an}. The supergravity theories in different dimensions are obtained by compactifying the low energy effective action of string theory on compact manifolds such as a torus or Calabi-Yau. The bosonic field content of these theories consists of the graviton coupled to some gauge and scalar fields. The extremal black hole solutions of these supergravity theories are charged under the gauge fields of the theory. These solutions can be BPS or non-BPS according as they preserve some or none of the supersymmetries.  In the following we restrict our discussion to static spherically symmetric and asymptotically flat charged extremal black holes.

An interesting phenomenon for extremal black holes is the attractor behaviour. In general the entropy of the black hole can depend on the values of moduli or scalar fields but in the case of the extremal black holes these values are fixed on the horizon in terms of the charges, independent of their asymptotic values. In fact the scalar fields describe trajectories in radial coordinate which evolve into the fixed points on the horizon. This mechanism was first observed in $\mathcal{N}=2$ theory \cite{Ferrara:1995ih,Strominger:1996kf,Ferrara:1996dd} and then was extended to theories with a larger number of supersymmetries and also to higher dimensions and non-supersymmetric black holes.

In order to explain the attractor mechanism and the idea of moduli fixing we recall some more facts on the $STU$ model of \autoref{sec:stu}. This theory is explained in terms of the complex scalar fields $S$, $T$ and $U$, a prepotential $F$ which is written in terms of the scalar fields as $F=STU$ and four gauge fields.

Any extended supergravity theory (${\mathcal N}>1$) admits an operator called the central charge which commutes with the other symmetry generators of the theory. The central charge $Z$ of the four-dimensional ${\mathcal N}=2$ theory coupled to $n_V$ vector multiplets can be written in terms of the electric and magnetic charges of the black hole from the $(n_V+1)$ $U(1)$ gauge fields of the theory \cite{Ferrara:1995ih,Bellucci:2006xz,Bellucci:2007gb}
\begin{gather}\label{eq:centcharge}
Z(z,\overline{z},p,q)=e^{K(z,\overline{z})/2}(X^I(z)q_I-F_I(z)p^I),\quad I\in\{0,\dotsc,n_V\},
\shortintertext{where}
K=-\ln i(\overline{X}^IF_I-X^I\overline{F}_I),
\end{gather}
is the K\"ahler potential and  $z^i$, $i\in\{1,\dotsc,n_V\}$, are the scalar fields. For these theories one can introduce a symplectic section $(X^I,F_I)$ which transforms under the symplectic symmetry group $Sp(2n_V+2,\mathds{Z})$ while keeping the action invariant. The $X^I$ represent the complex scalar fields and $F_I$ are the derivatives of the prepotential $F$ defined as $F_I={\partial F}/{\partial X^I}$. The prepotential $F$ is subject to two constraints: being holomorphic (independent of the complex conjugate scalars $\overline{X}^I$) and being homogeneous of degree two, $F(\lambda X^I)=\lambda^2 F(X^I)$.  In the case of the $STU$ model we have
\begin{gather}\label{eq:XF}
\begin{align}
z^i&=\frac{X^i}{X^0},&X^0&=1,
\end{align}
\shortintertext{where}
\begin{align}
X^I(z)&=\left(1, S, T, U\right)^{\textsf{T}},&F_I(z)&=\left(-STU, TU, SU, ST\right)^{\textsf{T}}.
\end{align}
\end{gather}
Note that the index $I\in\{1,\dotsc,4\}$ is raised and lowered by the metric $(+,+,-,-)$. The complex scalars $S$,$T$ and $U$ can be parameterised as in \eqref{eq:stu}. Therefore using \eqref{eq:XF} the K\"{a}hler potential simplifies to $e^{-K}= -8 S_2 T_2 U_2$.

In the Lagrangian of ${\cal N}=2$ supergravity the effective black hole potential is also given in terms of the central charge by
\begin{equation}
V_{\text{BH}}=|Z(z,p,q)|^2+|D_i Z(z,p,q)|^2,
\end{equation}
where $D_i$ is the covariant derivative $\partial_i+\partial_i K$ \cite{Ferrara:1996dd}. As can be seen from this formula the effective potential depends on the values of the scalar fields  and varies as one moves along the radial coordinate.

Let us now consider a supersymmetric extremal  black hole solution of the above theory. The near horizon metric of the black hole takes the form
\begin{gather}
ds^2=-\frac{r^2}{{|Z|}^2_{\text{hor}}}dt^2+\frac{{|Z|}^2_{\text{hor}}}{r^2}dr^2+{|Z|}^2_{\text{hor}}d{\Omega}^2,
\shortintertext{and the entropy of the black hole is given by}
{\mathcal S}_{BH}=\frac{A}{4}=\pi{|Z|}^2_{\text{hor}}.
\end{gather}
The attractor mechanism which is responsible for fixing the moduli at the horizon will consequently fix the entropy \cite{Ferrara:1997tw}. The values of the moduli are given by the critical points of $V_{\text{BH}}$ and hence the attractor equations are obtained by extremising the effective potential:

\begin{gather}
\begin{pmatrix}p^I\\q_J\end{pmatrix}=2\Im\begin{pmatrix}Z\overline{L}^I\\Z\overline{M}_J\end{pmatrix},\label{eq:atteq}
\shortintertext{where}
\begin{align}
L^I&=e^{K/2}X^I, & M_J&=e^{K/2}F_J.
\end{align}
\end{gather}
Note that for supersymmetric solution $D_i Z(z,p,q)=0$ and the potential is just proportional to the central charge. Therefore the critical points of the potential coincide with the critical points of the central charge. The attractor points are the minima of the potential.

For the BPS solution we have $M^2_{\text{BPS}}=|Z|^2$, the critical points are hence obtained by extremising the mass formula \eqref{eq:mass1}, which simplifies to:

\begin{equation}\label{eq:mass2}
M^2_{\text{BPS}}=\frac{1}{8S_2T_2U_2}{\lvert q_0+q_1S+q_2T+q_3U+p^0STU-p^1TU-p^2SU-p^3ST\rvert}^2.
\end{equation}

Following \cite{Levay:2006kf}, we now want to demonstrate the similarity between this process of moduli stabilisation and a quantum information process called a \emph{distillation protocol}.  In order to make this explicit we re-write the mass formula \eqref{eq:mass1} using an SLOCC transformation on the familiar three qubit system of Alice, Bob and Charlie:
\begin{gather}
M^2_{\text{BPS}}= \bra{{\Psi}}({\mathcal C}^{\textsf{T}}_U\otimes{\mathcal B}^{\textsf{T}}_T\otimes{\mathcal A}^{\textsf{T}}_S){\rho}^{\prime}({\mathcal C}_U\otimes{\mathcal B}_T\otimes{\mathcal A}_S)\ket{\Psi},
\shortintertext{where}
\ket{\Psi} = a_{ABC} \ket{A} \otimes \ket{B }\otimes \ket{C},\label{eq:Psi}
\shortintertext{and where}
\begin{align}
{\mathcal M}_S^{-1}&={\mathcal A}^{\textsf{T}}_S{\mathcal A}_S, & {\mathcal A}_S&=\frac{1}{\sqrt{S_2}}\begin{pmatrix*}[r]S_2&0\\-S_1&-1\end{pmatrix*},
\end{align}
\end{gather}
and similarly ${\mathcal M}_T^{-1}={\mathcal B}^{\textsf{T}}_T{\mathcal B}_T,\;{\mathcal M}_U^{-1}=\mathcal{C}^{\textsf{T}}_U\mathcal{C}_U$. In this equation $\rho'$ is defined as
\begin{equation}
{\rho}^{\prime}\equiv\tfrac{1}{8}(\mathds{1}\otimes \mathds{1}\otimes \mathds{1}+{\sigma}_2\otimes{\sigma}_2\otimes \mathds{1}+{\sigma}_2\otimes \mathds{1}\otimes{\sigma}_2+\mathds{1}\otimes{\sigma}_2\otimes{\sigma}_2),
\end{equation}
where we use the notation $\sigma_2=-i\varepsilon$. In fact $\rho'$ is an example of three-qubit mixed state. This relation can be simplified further by diagonalising it using the unitary matrix ${\mathcal U}$ on each qubit as follows
\begin{align}
{\sigma}_3&={\mathcal U}{\sigma}_2{\mathcal U}^{\dagger},&
{\mathcal U}&=\frac{1}{\sqrt{2}}\begin{pmatrix*}[r]i&1\\i&-1\end{pmatrix*}.
\end{align}
Therefore the new density matrix which is given by
\begin{gather}
\rho={\mathcal U}_U\otimes\,{\mathcal U}_T\otimes\,{\mathcal U}_S\,\rho'\,{\mathcal U}^{\dagger}_U\otimes\,{\mathcal U}^{\dagger}_T\otimes\,{\mathcal U}^{\dagger}_S=\half\diag(1,0,0,0,0,0,0,1)
\shortintertext{takes the form}
\rho=\half(\ket{000}\bra{000} +\ket{111}\bra{111}).
\end{gather}
Using these unitary transformations we obtain a complex representation for $M^2_{\text{BPS}}$ as follows
\begin{equation}
M^2_{\text{BPS}}=\bra{\Psi}(C^{\dagger}_U\otimes B^{\dagger}_T\otimes A^{\dagger}_S)\rho(C_U\otimes B_T\otimes A_S)\ket{\Psi},
\end{equation}
where $A_S$, $B_T$ and $C_U$ are now scaled SLOCC i.e. $GL(2, \mathds{C})$ transformations of the form
\begin{equation}
A_S=\frac{1}{\sqrt{2S_2}}\begin{pmatrix*}[r] -\overline{S}&-1\\S&1\end{pmatrix*},
\end{equation}
with $B_T$ and $C_U$ defined similarly. If we define
\begin{equation}\label{eq:wavefun}
\ket{\Psi'}\equiv (C_U\otimes B_T\otimes A_S)\ket{\Psi}
\end{equation}
the black hole potential can be written in a particularly concise form
\begin{equation}
V_{\text{BH}}=\half\braket{\Psi'}{\Psi'}
\end{equation}
and the mass formula is simply given by
\begin{equation}
M^2_{\text{BPS}}=\tfrac{1}{2}(|{a}^{\prime}_{000}|^2+|{a}^{\prime}_{111}|^2).
\end{equation}
The SLOCC transformations are in fact the transformations which map a general three qubit state to the canonical GHZ form. Comparing this formula with the result obtained in \eqref{eq:mass2} one can verify that the components of $\Psi$ are those given in \eqref{eq:charges7}. In order to calculate the values of the scalar fields at the near horizon limit one can extremise the mass or central charge formula with respect to the complex scalar fields $S$, $T$ and $U$. This results in the attractor equations \eqref{eq:atteq} and the stabilised scalar values are \cite{Behrndt:1996hu}
\begin{align}
S&=\frac{((p.q)-2p^1q_1)-i\sqrt{|D|}}{2(p^3p^2-p^0q_1)}, & T&=\frac{p^2 \bar{S}-q_3}{p^0 \bar{S}-p^1}, & U&=\frac{p^3 \bar{S}-q_2}{p^0 \bar{S}-p^1},
\end{align}
where $D$ is given by \eqref{eq:STUBHEntropy}. Now one can substitute these frozen or fixed values of the scalar fields in the equation for the transformed wavefunction \eqref{eq:wavefun} to get
\begin{equation}\label{eq:GGHZ1}
\ket{\Psi'}=\sqrt{\frac{2 T_2 U_2}{S_2}}\left((p^1- p^0 \overline{S})\ket{000} -(p^1-p^0 S)\ket{111}\right),
\end{equation}
which is of the form of a generalised GHZ state. The expression \eqref{eq:GGHZ1} can be multiplied by a phase without changing the mass;
\begin{gather}
\ket{\Psi'} =|D|^{1/4}\left(\ket{000} +e^{i\delta}\ket{111}\right),\label{eq:genGHZ}
\shortintertext{where}
\delta=\pi+2\arctan\frac{p^0 S_2}{p^0 S_1-p^1}.
\end{gather}
Interestingly, this result means that the process of finding the fixed values of the scalar fields at the horizon for BPS $STU$ model black holes is equivalent to finding the canonical form of the corresponding three-qubit state using complex SLOCC transformations. In other words the process of finding the frozen values of the moduli is equivalent to the quantum information theoretic one of performing an optimal set of SLOCC transformations on the initial three-qubit state with integer amplitudes to arrive at a state of GHZ form \cite{Levay:2006kf}. It is important to notice that although the transformed states seem to be complex, one can find a basis in which they are real.

\subsection{Bit flip errors and black holes}

We now turn our attention to BPS and non-BPS black holes and their relation to the suppression or non-suppression of bit flip errors, following \cite{Levay:2007nm}.  We begin with some definitions. For a qubit system denoted by $\ket{0}$ and $\ket{1}$ the phase flip operator ${\mathcal Z}$ acts like the Pauli matrix $\sigma_3$ as
\begin{align}
{\mathcal Z}\ket{0}&=\ket{0},&{\mathcal Z}\ket{1}&=-\ket{1}.
\end{align}
This is one of the single qubit gates in quantum information theory which leaves the state $\ket{0}$ unchanged but flips the sign of $\ket{1}$. Another useful operator is $X$, which is the same as Pauli matrix $\sigma_1$ and is used to represent bit flips. The action of this bit flip operator is
\begin{align}\label{eq:bitflip}
X\ket{0}&=\ket{1}, & X\ket{1}&=\ket{0}.
\end{align}
In addition, we have the Hadamard gate which is one of the building blocks of quantum circuit theory and is especially useful in quantum error correction. Quantum error correction is used to protect the information from errors due to decoherence and other quantum noise over transmission lines. The Hadamard gate acts on the qubits as
\begin{align}
\ket{\bar{0}}&= H\ket{0} =\tfrac{1}{\sqrt{2}}(\ket{0}+\ket{1}), & \ket{\bar{1}}&=H\ket{1}=\tfrac{1}{\sqrt{2}}(\ket{0}-\ket{1}),
\end{align}
where $\ket{\bar{0}}$ and $\ket{\bar{1}}$ are the Hadamard transformed basis which are, in fact, the eigenvectors of the bit flip operator $X$ with $+1$ and $-1$ eigenvalues, respectively. The matrix representation of the operator $H$ is given by
\begin{equation}
H=\frac{1}{\sqrt{2}}\begin{pmatrix*}[r] 1 & 1 \\  1 & -1 \end{pmatrix*}.
\end{equation}
For the three-qubit system the Hadamard transformations are given by an $8\times 8$ matrix defined as $H^{\otimes 3}=H\otimes H \otimes H$. These transformations are
\begin{equation}
H^{\otimes 3}=\frac{1}{\sqrt{8}}
\begin{pmatrix*}[r]
1& 1& 1& 1& 1& 1& 1& 1\\
1&-1& 1&-1& 1&-1& 1&-1\\
1& 1&-1&-1& 1& 1&-1&-1\\
1&-1&-1& 1& 1&-1&-1& 1\\
1& 1& 1& 1&-1&-1&-1&-1\\
1&-1& 1&-1&-1& 1&-1& 1\\
1& 1&-1&-1&-1&-1& 1& 1\\
1&-1&-1& 1&-1& 1& 1&-1
\end{pmatrix*}.
\end{equation}
For example the state $\ket{\overline{110}}$ in Hadamard transformed basis reads
\begin{equation}
\ket{\overline{110}} =\tfrac{1}{\sqrt{8}}\left(\ket{000}+\ket{001}-\ket{010} -\ket{011} -\ket{100} -\ket{101} +\ket{110} +\ket{111}\right),
\end{equation}
coming from the sign combination of the sixth row.

Returning now to the BPS and non-BPS black hole solutions, we saw that one can introduce a new wavefunction $\ket{\Psi'}$ given by \eqref{eq:wavefun} that for the fixed values of the scalar fields at the horizon takes the form of a generalised GHZ state \eqref{eq:genGHZ}. In general, the coefficients of the state $\ket{\Psi'}$ before stabilising the value of the scalar fields at the horizon are
\begin{equation}
\begin{aligned}
a'_{000}&=-e^{K/2}W(\overline{U},\overline{T},\overline{S}), & a'_{111}&=e^{K/2}W(U,T,S),\\
a'_{110}&=-e^{K/2}W({U},{T},\overline{S}),                   & a'_{001}&=e^{K/2}W(\overline{U},\overline{T},S),\\
a'_{101}&=-e^{K/2}W({U},\overline{T},{S}),                   & a'_{010}&=e^{K/2}W(\overline{U},T,\overline{S}),\\
a'_{011}&=-e^{K/2}W(\overline{U},{T},{S}),                   & a'_{100}&=e^{K/2}W(U,\overline{T},\overline{S}),
\end{aligned}
\end{equation}
where we have defined the central charge of \eqref{eq:centcharge} as $Z=e^{K/2}W(S,T,U)$ , and used the definition \eqref{eq:wavefun}. Here the function $W$ is the superpotential. Using the above relations one can easily see that the wavefunction coefficients satisfy these properties
\begin{align}
a'_{000}&=-\overline{a'_{111}}, &
a'_{110}&=-\overline{a'_{001}}, &
a'_{101}&=-\overline{a'_{010}}, &
a'_{011}&=-\overline{a'_{100}}.
\end{align}
To see the relation between quantum error correction and black holes we use the definition of the covariant derivative of the superpotential. This is given by $D_a W=\partial_a W + (\partial_a K)W$ where $a$ represents the scalar fields $S$, $T$ and $U$. Using this definition one can see that for example
\begin{equation}
D_SW(U,T,S)=\frac{W(U,T,\overline{S})}{\overline{S}-S}.
\end{equation}
Therefore if we define the vielbeins $e^a_{\hat{a}}$ as
\begin{align}
e_{\hat{S}}^S&=i(S-\overline{S})=-2S_2, &  e_{\hat{T}}^T&=i(T-\overline{T})=-2T_2, &  e_{\hat{U}}^U&=i(U-\overline{U})=-2U_2,
\end{align}
we see that the covariant derivatives of the wavefunction coefficients reduce to
\begin{equation}
\begin{gathered}
\begin{aligned}
D_{\hat{S}} a'_{111} &=i a'_{110}, & D_{\hat{T}} a'_{111} &=i a'_{101}, & D_{\hat{U}} a'_{111} &=i a'_{011},
\end{aligned}\\
D_{\hat{\overline{S}}} a'_{111}=D_{\hat{\overline{T}}} a'_{111}=D_{\hat{\overline{U}}} a'_{111}=0,
\end{gathered}
\end{equation}
where we have used $D_{\hat{a}}=e_{\hat{a}}^a D_a$. Note that one can define the combination $(D_{\hat{a}}-D_{\hat{\bar{a}}})/i$ which acts on the state $\ket{\Psi'}$ as
\begin{equation}\label{eq:IIX}
\tfrac{1}{i}(D_{\hat{S}}-D_{\hat{\overline{S}}})\ket{\Psi'} =(\mathds{1}\otimes\mathds{1}\otimes X)\ket{\Psi'},\quad\text{etc,}
\end{equation}
where the operator $\mathds{1}\otimes\mathds{1}\otimes X$ represents a bit flip error \eqref{eq:bitflip} on the third qubit. Therefore the derivatives of the central charge, which play a key role in the BPS conditions for the black holes, correspond to the bit flip errors in quantum information theory. We shall now  see this more explicitly for BPS and non-BPS black hole solutions.

For the BPS extremal black hole solution, the covariant derivatives of the central charge are zero $(D_a Z =0)$ and this guarantees the result \eqref{eq:genGHZ}, in which all the wavefunction coefficients are zero at the horizon except $\Psi'^{000}$ and $\Psi'^{111}$. Consider the case in which the charges $q_0,\;p^1,\;p^2$ and $p^3$ are zero, for which  \eqref{eq:genGHZ} reduces to
\begin{equation}
\ket{\Psi'}=i\sqrt{2}(-p^0 q_1 q_2 q_3)^{1/4}\left(\ket{000} + \ket{111}\right),
\end{equation}
which is explicitly  the unnormalised canonical GHZ state. Using $D_a Z=0$ and \eqref{eq:IIX}, we  conclude that for the BPS case the GHZ state at the horizon $(\ket{\Psi'})$ is protected from bit flip errors. This means that the operators $\mathds{1}\otimes\mathds{1}\otimes X$, $\mathds{1}\otimes X\otimes\mathds{1}$ and $X\otimes\mathds{1}\otimes\mathds{1}$ cannot change the BPS state $\ket{\Psi'}$ at the horizon. In fact the BPS conditions precisely suppress the bit flip errors for the three qubit state at the horizon.

The state $\ket{\Psi'}$ can also be written in Hadamard transformed basis as
\begin{equation}\label{eq:HadBPS}
\ket{\Psi'}=i(-p^0q_1q_2q_3)^{1/4}(\ket{\overline{000}}+\ket{\overline{011}}+\ket{\overline{101}}+\ket{\overline{110}}).
\end{equation}
Acting on it with the bit flip operator (here on the third qubit) gives
\begin{equation}
(\mathds{1}\otimes\mathds{1}\otimes X)\ket{\Psi'}=i(-p^0q_1q_2q_3)^{1/4}(\ket{\overline{000}}-\ket{\overline{011}}-\ket{\overline{101}}+\ket{\overline{110}}),
\end{equation}
telling us that bit flip errors in the normal basis correspond to phase flip errors in the Hadamard basis \cite{Levay:2007nm}.

The situation changes for non-BPS extremal black holes; the fixed values of the moduli at the horizon are given by \cite{Tripathy:2005qp,Kallosh:2006bt}
\begin{align}
S&=\pm i\sqrt{\frac{q_2q_3}{p^0q_1}}, & T&=\pm i\sqrt{\frac{q_1q_3}{p^0q_2}}, & U&=\pm i\sqrt{\frac{q_1q_2}{p^0q_3}},
\end{align}
where in this case we have $p^0 q_1 q_2 q_3>0$. The sign combinations which do not violate the positivity of the K\"ahler potential are
\begin{equation}\label{eq:signs}
\{(-,-,-), (-,+,+), (+,-,+), (+,+,-)\}.
\end{equation}
For example if we let the $S$, $T$ and $U$ scalars be negative and the charges be all positive (which is consistent with the positivity of their multiplication), the state can be written as
\begin{equation}\label{eq:nonBPShor}
\ket{\Psi'}_{---}=\frac{i}{\sqrt{2}}(p^0q_1q_2q_3)^{1/4}(\ket{000} -\ket{001} -\ket{010} -\ket{011} -\ket{100} -\ket{101} -\ket{110}+\ket{111}).
\end{equation}
This state can be written in the Hadamard transformed basis as
\begin{equation}
\ket{\Psi'}_{---}=-i(p^0q_1q_2q_3)^{1/4}(\ket{\overline{000}} -\ket{\overline{011}} -\ket{\overline{101}} -\ket{\overline{110}}).
\end{equation}
Comparing this equation with \eqref{eq:HadBPS},  the BPS state in Hadamard transformed basis, one can see that (in addition to the sign difference of $p^0 q_1 q_2 q_3$) the BPS and non-BPS states may be distinguished by a nontrivial relative phase between the Hadamard basis vectors. In general, for all sign combinations of the $S$, $T$ and $U$ scalars, the state at the horizon takes the form
\begin{equation}
\ket{\Psi'}_{\gamma\beta\alpha}=-i(p^0q_1q_2q_3)^{1/4}\{\ket{\overline{000}}+\gamma\ket{\overline{011}} +\beta\ket{\overline{101}}+\alpha\ket{\overline{110}}\}\;,
\end{equation}
where $\alpha, \beta$ and $\gamma$ are the combinations in \eqref{eq:signs}. In the non-BPS case the action of the bit flip errors on the states at the horizon is
\begin{align}
(\mathds{1}\otimes\mathds{1}\otimes X){\ket{\Psi'}}_{---}&={\ket{\Psi'}}_{++-}, &
(\mathds{1}\otimes\mathds{1}\otimes X){\ket{\Psi'}}_{-++}&={\ket{\Psi'}}_{+-+}, & &\text{etc}.
\end{align}
The bit flip operators $\mathds{1}\otimes\mathds{1}\otimes X$, $\mathds{1}\otimes X\otimes\mathds{1}$ and $X\otimes\mathds{1}\otimes\mathds{1}$ transform the states with different combinations of sign into each other. The rule for this transformation is that those labels which are in the same slot as the bit flip operator $X$ are not changed while the remaining ones are flipped. In contrast to the BPS case, the bit flip errors transform one non-BPS solution another \cite{Levay:2007nm} and the quantum state corresponding to non-BPS extremal black hole is not protected.

We note that non-BPS state at the horizon \eqref{eq:nonBPShor} is an example of a {\it graph-state}, which is particularly important for quantum computation \cite{Levay:2007nm}.

\newpage
\section{\texorpdfstring{$\mathcal{N}=8$ BLACK HOLE/QUBIT CORRESPONDENCE}{N=8 Black Hole/Qubit Correspondence}}
\label{sec:N=8}

\subsection{\texorpdfstring{The $\mathcal{N}=8$ generalisation}{The N=8 generalisation}}
\label{sec:N=8 generalisation}

The black holes described by Cayley's hyperdeterminant are those of $\mathcal{N}=2$ supergravity coupled to three vector multiplets, where the symmetry is $[SL(2,\mathds{Z})]^{3}$.  One might therefore ask whether the black hole/information theory correspondence could be generalised. There are three generalisations we might consider:
\begin{enumerate}
\item $\mathcal{N}=2$ supergravity coupled to $l+1$ vector multiplets where the symmetry is $SL(2,\mathds{Z}) \times SO(l,2,\mathds{Z})$ and the black holes carry charges belonging to the $(2,l+2)$ representation ($l+2$ electric plus $l+2$ magnetic).

\item $\mathcal{N}=4$ supergravity coupled to $m$ vector multiplets where the symmetry is $SL(2,\mathds{Z}) \times SO(6,m,\mathds{Z})$ where the black holes carry charges belonging to the $(2,6+m)$ representation ($m+6$ electric plus $m+6$ magnetic).

\item $\mathcal{N}=8$ supergravity where the symmetry is the non-compact exceptional group $E_{7(7)}(\mathds{Z})$ and the black holes carry charges belonging to the fundamental $56$-dimensional representation (28 electric plus 28 magnetic). See  \autoref{tab:PeterWest}
\end{enumerate}

In all three cases there exist quartic invariants akin to Cayley's hyperdeterminant whose square root yields the corresponding black hole entropy.

Let us first focus on $\mathcal{N}=8$, where the entropy is given by
\begin{equation}\label{eq:N=8BHEntropy}
S = \pi\sqrt{|I_4|},
\end{equation}
where $I_4$ is Cartan's quartic $E_{7(7)}$ invariant given by
\begin{equation}\label{eq:E7InvariantCartanBasis}
I_{4}=-\tr(xy)^2 + \tfrac{1}{4}(\tr xy)^2 - 4\left(\Pf x + \Pf y\right),
\end{equation}
where $x^{IJ}$ and $y_{IJ}$ are $8 \times 8$ antisymmetric matrices and $\Pf$ is the Pfaffian.  An alternative expression has been provided by Cremmer and Julia \cite{Cremmer:1979up}
\begin{gather}
I_4 = \tr(\bar{Z}Z)^2 - \tfrac{1}{4}(\tr\bar{Z}Z)^2 + 4\left(\Pf Z + \Pf\bar{Z}\right).\label{eq:E7InvariantCremmerJuliaBasis}
\shortintertext{Here}
Z_{AB} = -\tfrac{1}{4\sqrt{2}}(x^{IJ} + i y_{IJ})(\gamma^{IJ})_{AB},\label{eq:CartanToCremmerJuliaTransf}
\shortintertext{and}
x^{IJ} + i y_{IJ} = -\tfrac{\sqrt{2}}{4}Z_{AB}(\gamma^{AB})_{IJ}.\label{eq:CremmerJuliaToCartanTransf}
\end{gather}
The matrices of the $SO(8)$ algebra are $(\gamma^{IJ})_{AB}$ where $(I,J)$ are the 8 vector indices and $(A,B)$ are the 8 spinor indices. The $(\gamma^{IJ})_{AB}$ matrices can be considered also as $(\gamma^{AB})_{IJ}$ matrices due to equivalence of the vector and spinor representations of the $SO(8)$ Lie algebra. The exact relation between the Cartan invariant  in \eqref{eq:E7InvariantCartanBasis}  and Cremmer-Julia  invariant \cite{Cremmer:1979up} in \eqref{eq:E7InvariantCremmerJuliaBasis} was established in \cite{Balasubramanian:1997az,Gunaydin:2000xr}. The quartic invariant $I_4$ of $E_{7(7)}$ is also related to the octonionic Jordan algebra $J_3^{\mathds{O}}$ \cite{Ferrara:1997uz} as described in \autoref{sec:Cartan}.

In the stringy black hole context, $Z_{AB}$ is the central charge matrix and $(x,y)$ are the quantised charges of the black hole (28 electric and 28 magnetic).  The relation between the entropy of stringy black holes and the Cartan-Cremmer-Julia $E_{7(7)}$ invariant was established in \cite{Kallosh:1996uy}. The central charge matrix $Z_{AB}$ can be brought to the canonical basis for the skew-symmetric matrix using an $SU(8)$ transformation:
\begin{equation}
Z_{AB} =
\begin{pmatrix}
z_1 & 0   & 0   & 0 \\
0   & z_2 & 0   & 0 \\
0   & 0   & z_3 & 0 \\
0   & 0   & 0   & z_4
\end{pmatrix} \otimes
\begin{pmatrix*}[r]0 & 1\\-1 & 0\end{pmatrix*},
\end{equation}
where $z_i=\rho_i e^{i\varphi_i}$ are complex. In this way the number of entries is reduced from 56 to 8.  In a systematic treatment in \cite{Ferrara:1997ci}, the meaning of these parameters was clarified. From 4 complex values of  $z_i=\rho_i e^{i\varphi_i}$ one can remove 3 phases  by an $SU(8)$ rotation, but the overall phase cannot be removed; it is related to an extra parameter in the class of black hole solutions \cite{Cvetic:1995kv,Cvetic:1995bj}.  In this basis, the quartic invariant takes the form \cite{Kallosh:1996uy}
\begin{equation}\label{eq:5ParameterGeneratingSoln}
\begin{gathered}
I_4 = \sum_i|z_i|^4 - 2\sum_{i<j}|z_i|^2|z_j|^2 + 4\left(z_1z_2z_3z_4 + \bar{z}_1\bar{z}_2\bar{z}_3\bar{z}_4\right) \\
\begin{aligned}
= &\phantom{\times\;}\left(\rho_1 + \rho_2 + \rho_3 + \rho_4\right) \\
&\times\left(\rho_1 + \rho_2 - \rho_3 - \rho_4\right) \\
&\times\left(\rho_1 - \rho_2 + \rho_3 - \rho_4\right) \\
&\times\left(\rho_1 - \rho_2 - \rho_3 + \rho_4\right) \\
&+~8\rho_1\rho_2\rho_3\rho_4\left(\cos\varphi-1\right).
\end{aligned}
\end{gathered}
\end{equation}
Therefore a 5-parameter solution is called a generating solution for other black holes in $\mathcal{N}=8$ supergravity/M-theory \cite{Cvetic:1996zq,Bertolini:1999je}.

\subsection{\texorpdfstring{$E_7$ and the tripartite entanglement of seven qubits}{E7 and the tripartite entanglement of seven qubits}}

We have seen that in the case of three qubits, the tripartite entanglement is described by $[SL(2,\mathds{C})]^{3}$ and that the entanglement measure is given by Cayley's hyperdeterminant. If there is to be a quantum information theoretic interpretation in the $\mathcal{N}=8$ case, however, it cannot just be random entanglement of more qubits, because the general $n$ qubit entanglement is described by the group $[SL(2,\mathds{C})]^{n}$, which, even after replacing $\mathds{Z}$ by $\mathds{C}$, differs from the $\mathcal{N}=8$ $E_{7}$ (and also from the $\mathcal{N}=4$ and $\mathcal{N}=2$ symmetries mentioned above, except when $n=3$, which correspond to case (1) with $l=2$, the case we already know. ) We note, however, that
\begin{gather}
E_{7(7)}(\mathds{Z}) \supset [SL(2,\mathds{Z})]^{7},
\shortintertext{and}
E_{7}(\mathds{C})\supset [SL(2,\mathds{C})]^{7}.
\end{gather}
We shall now show that the corresponding system in quantum information theory is that of seven qubits (Alice, Bob, Charlie, Daisy, Emma, Fred and George). However, the larger symmetry requires that they undergo at most tripartite entanglement of a very specific kind. The entanglement measure will be given by the quartic Cartan $E_{7}(\mathds{C})$ invariant \cite{Cartan,Cremmer:1979up,Kallosh:1996uy,Ferrara:1997uz}.

The crucial ingredient is the observation that $E_7$ contains 7 copies of the single qubit SLOCC group $SL(2)$ and that the $\mathbf{56}$ decomposes in a very particular way. We begin by considering the maximal subgroup $SL(2)_A\times SO(6,6)$,
\begin{equation}\label{eq:First56Decomp}
\begin{array}{c*{4}{@{\ }c}}
E_{7(7)}    & \supset & SL(2)_A         & \times & SO(6,6), \\
\mathbf{56} & \to     & \mathbf{(2,12)} & +      & \mathbf{(1,32)}.
\end{array}
\end{equation}
Further decomposing $SO(6,6)$ in \eqref{eq:First56Decomp}
\begin{equation}\label{eq:Triality}
\begin{array}{c*{10}{@{\ }c}}
SL(2)_A & \times & SO(6,6) & \supset & SL(2)_A & \times & SL(2)_B & \times & SL(2)_D & \times & SO(4,4), \\
\mathbf{(2,12)} & + & \mathbf{(1,32)} & \to & \mathbf{(2,1,1,8_v)} & + & \mathbf{(1,2,1,8_s)} & + & \mathbf{(1,1,2,8_c)} & + & \mathbf{(2,2,2,1)}.
\end{array}
\end{equation}
Further decomposing $SO(4,4)$,
\begin{equation}
\begin{gathered}
SL(2)_{A} \times SL(2)_{B} \times SL(2)_{D} \times SO(4,4) \\
\supset SL(2)_{A} \times SL(2)_{B} \times SL(2)_{D} \times SO(2,2) \times SO(2,2) \\
\mathbf{(2,2,2,1)+ (2,1,1,8_v) + (1,2,1,8_s) +(1,1,2,8_c)} \\
\to \mathbf{(2,2,2,1,1)+ (2,1,1,4,1) +(2,1,1,1,4)} \\
+~\mathbf{(1,2,1,2,2) + (1,2,1,2,2) + (1,1,2,2,2) +(1,1,2,2,2)}.
\end{gathered}
\end{equation}
Finally, further decomposing each $SO(2,2)$
\begin{equation}
\begin{gathered}
SL(2)_{A} \times SL(2)_{B} \times SL(2)_{D} \times SO(2,2) \times SO(2,2) \\
\supset SL(2)_A \times SL(2)_B \times SL(2)_D \times SL(2)_C \times SL(2)_G \times SL(2)_F \times SL(2)_E \\
\mathbf{(2,2,2,1,1) + (2,1,1,4,1) +(2,1,1,1,4)} \\
+~\mathbf{(1,2,1,2,2) + (1,2,1,2,2) + (1,1,2,2,2) + (1,1,2,2,2)} \\
\to \mathbf{(2,2,2,1,1,1,1) + (2,1,1,2,2,1,1) + (2,1,1,1,1,2,2)} \\
+~\mathbf{(1,2,1,2,1,1,2) +(1,2,1,1,2,2,1) + (1,1,2,2,1,2,1) + (1,1,2,1,2,1,2)}.
\end{gathered}
\end{equation}
In summary,
\begin{equation}
E_{7(7)} \supset SL(2)_A \times SL(2)_B \times SL(2)_C \times SL(2)_D \times SL(2)_E \times SL(2)_F \times SL(2)_G,
\end{equation}
and the $\mathbf{56}$ decomposes as
\begin{equation}\label{eq:56Decomp}
\begin{split}
\mathbf{56} &\to\phantom{+\!}\mathbf{(2,2,1,2,1,1,1)} \\
&\phantom{\to}+\mathbf{(1,2,2,1,2,1,1)} \\
&\phantom{\to}+\mathbf{(1,1,2,2,1,2,1)} \\
&\phantom{\to}+\mathbf{(1,1,1,2,2,1,2)} \\
&\phantom{\to}+\mathbf{(2,1,1,1,2,2,1)} \\
&\phantom{\to}+\mathbf{(1,2,1,1,1,2,2)} \\
&\phantom{\to}+\mathbf{(2,1,2,1,1,1,2)}.
\end{split}
\end{equation}
An analogous decomposition holds for
\begin{equation}
E_7(\mathds{C})\supset [SL(2,\mathds{C})]^7.
\end{equation}
Notice that we find seven copies of the $\mathbf{(2,2,2)}$ appearing in the $STU$ model.  This translates into seven copies of the three qubit Hilbert space:
\begin{equation}\label{eq:7QubitState}
\begin{array}{c@{}c@{\ }c@{\lvert}*{3}{@{}c}@{\rangle}c}
\ket{\Psi}_{56} = &   & a_{ABD} & A       & B       & D \\
                  & + & b_{BCE} & B       & C       & E \\
                  & + & c_{CDF} & C       & D       & F \\
                  & + & d_{DEG} & D       & E       & G \\
                  & + & e_{EFA} & E       & F       & A \\
                  & + & f_{FGB} & F       & G       & B \\
                  & + & g_{GAC} & G       & A       & C &.
\end{array}
\end{equation}
Note that:
\begin{enumerate}
  \item Any pair of states has an individual in common
  \item Each individual is excluded from four out of the seven states
  \item Two given individuals are excluded from two out of the seven states
  \item Three given individuals are never excluded
\end{enumerate}
So we have seven qubits (Alice, Bob, Charlie, Daisy, Emma, Fred and George) but where Alice has tripartite entanglement not only with Bob/Dave but also with Emma/Fred and also Charlie/George, and similarly for the other six individuals. So, in fact, each person has tripartite entanglement with each of the remaining three couples. However, as discussed in \autoref{sec:seven}, this 56-dimensional space is not a subspace of the seven qubit Hilbert space $\mathbf{(2,2,2,2,2,2,2)}$.

\begin{figure}[ht]
  \centering
  \includegraphics[width=\textwidth]{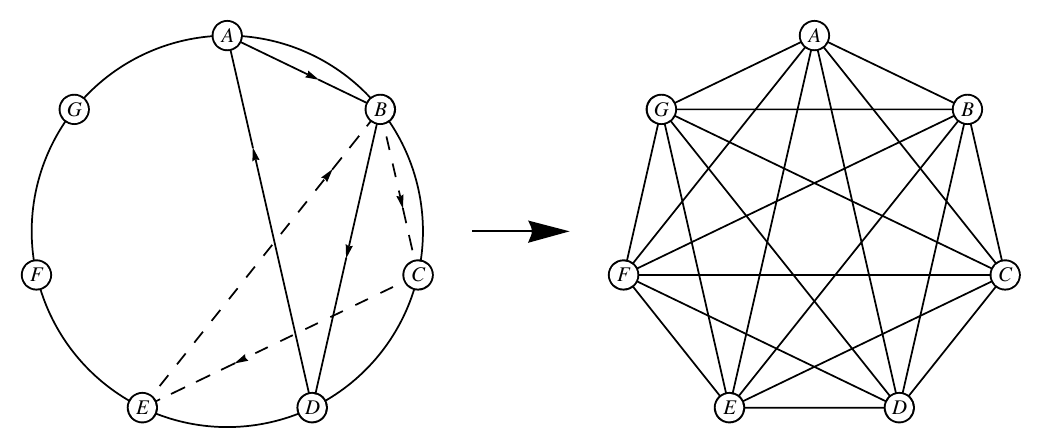}
  \caption[$E_7$ entanglement diagram]{The $E_7$ entanglement diagram corresponding to the decomposition \eqref{eq:56Decomp} and the state \eqref{eq:7QubitState}. Each of the seven vertices $A, B, C, D, E, F, G$ represents a qubit and each of the seven triangles $ABD, BCE, CDF, DEG, EFA, FGB, GAC$ describes a tripartite entanglement. As discussed in \autoref{sec:octandfan} the oriented triangles correspond to quaternionic cycles in the multiplication table of imaginary octonions.}
  \label{fig:E7Entanglement}
\end{figure}

The entanglement may be represented by a heptagon as in  \autoref{fig:E7Entanglement} with seven vertices $A,B,C,D,E,F,G$, and seven triangles
\begin{equation*}
ABD, BCE, CDF, DEG, EFA, FGB,GAC.
\end{equation*}
Each of the seven states transforms as a $\mathbf{(2,2,2)}$ under three of the $SL(2)$'s and are singlets under the remaining four. Note that from \eqref{eq:Triality} we see that the $A$-$B$-$C$ triality of \autoref{sec:Hyperdet} is linked with the $\mathbf{8_v}$-$\mathbf{8_s}$-$\mathbf{8_c}$ triality of the $SO(4,4)$ subgroup. Individually, therefore, the tripartite entanglement of each of the seven states is given by Cayley's hyperdeterminant. Taken together however, we see from \eqref{eq:56Decomp} that they transform as a complex \textbf{56} of $E_7(\mathds{C})$. Their tripartite entanglement must be given by an expression that is quartic in the coefficients $a,b,c,d,e,f,g$ and invariant under $E_7(\mathds{C})$. The unique possibility is the Cartan invariant $I_4$, and so the 3-tangle is given by
\begin{equation}\label{eq:3/7tangle}
\tau(ABCDEFG) = 4|I_4|.
\end{equation}
If the wave-function \eqref{eq:7QubitState} is normalised, then $0 \leq \tau(ABCDEFG) \leq 1$.

\subsection{Octonions and the Fano plane}
\label{sec:octandfan}

\begin{figure}[ht]
  \centering
  \includegraphics[width=7cm]{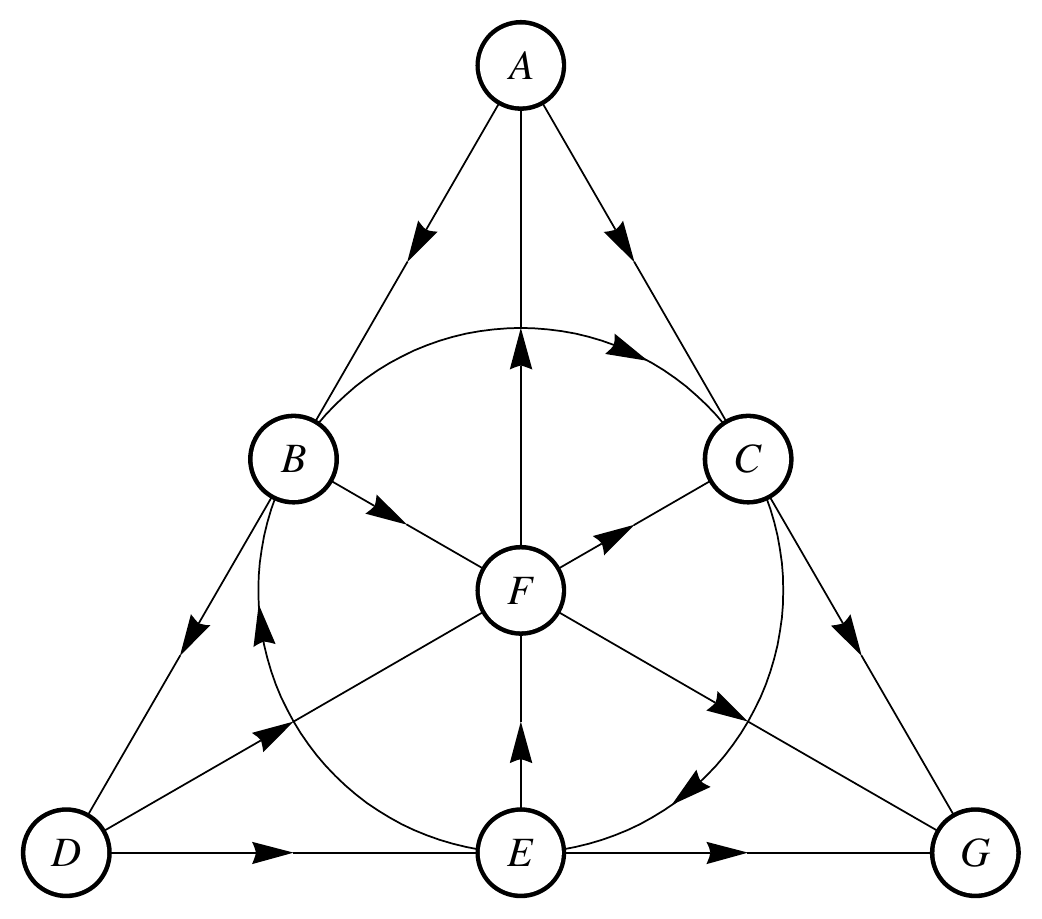}
  \caption[Fano plane]{The Fano plane is a projective plane with seven points and seven lines (the circle counts as a line). We may associate it to the state \eqref{eq:7QubitState} by interpreting the points as the seven qubits $A$-$G$ and the lines as the seven tripartite entanglements. This is consistent as there are three points on every line and three lines through every point. If the plane is oriented like ours, one may use the directed lines to read off the multiplication table of imaginary octonions (\autoref{tab:FanoMult}).}
  \label{fig:FanoPlane}
\end{figure}

An alternative and very useful picture of the state $\ket{\Psi}$ is provided by the Fano plane, as shown in \autoref{fig:FanoPlane}. The Fano plane corresponds to the multiplication table of the imaginary octonions, which we now briefly discuss.

The algebra of octonions $\mathds{O}$ (with product denoted by juxtaposition) possesses numerous interesting properties, some of which are notable absences of familiar properties. The reader is referred to \cite{Baez:2001dm} for extensive details including historical notes. Some interesting physical applications may be found in
\cite{Schray:1994ur,Schafer:1966,Daboul:1999xv,Duff:1997hf,Smolin:2001wc,Toppan:2003yx,Albuquerque:1998,Manogue:1999,Gunaydin:1973,Kugo:1982bn}. We note that the octonions also make their appearance in a QI context in \cite{Mosseri:2001,Mosseri:2003}, but their work seems unrelated to ours.

Typical octonions $a,b,c\in\mathds{O}$ are:
\begin{itemize}
\item 8-tuples of real numbers: $a,b,c\in\mathds{R}^8$ so that they form an 8-dimensional vector space, with basis elements $e_0,\ldots,e_7$.
\item non-real: $a\neq a^*$, like the complexes. The conjugate $\bullet^*:\mathds{O}\to\mathds{O}$ trivially extends the conjugate for $\mathds{R, C}$, and $\mathds{H}$ so that basis element $e_\mu$ is mapped to $\eta_{\mu\nu}e_\nu$ (with $\eta\equiv\diag(1,-\mathds{1}_7)$). Under the familiar partition $\mu=(0,i)$, scalar multiples of $e_0$ are real octonions, and scalar multiples of $e_i$ are imaginary octonions.
\item non-commutative: $ab\neq ba$, like the quaternions.
\item non-associative: $a(bc)\neq (ab)c$, a new property not present in $\mathds{R, C}$, or $\mathds{H}$.
\item alternative - meaning that the subalgebra generated by any two elements is associative, or equivalently, the associator $[\bullet,\bullet,\bullet]:\mathds{O}^3\to\mathds{O}$, $(a,b,c)\mapsto a(bc)-(ab)c$ is alternating: $[a_1,a_2,a_3]=(-)^\pi [a_{\pi(1)},a_{\pi(2)}, a_{\pi(3)}]$ with $\pi\in S_3$.
\item a division algebra, so that when a product of octonions is zero one of the multiplied octonions must have been zero: $ab=0\Rightarrow a=0\wedge b=0$. They share this property with $\mathds{R}$, $\mathds{C}$, $\mathds{H}$, and no other algebras.
\item normed: $|ab|=|a||b|$, which implies the division algebra property. Like the conjugate, the norm $|\bullet|:\mathds{O}\to\mathds{R}$ is also a trivial extension of the norm for the other division algebras: $a\mapsto a^*a$.
\end{itemize}
Clearly the octonions are closely related to the other division algebras, and indeed octonionic multiplication can be defined in terms of multiplication in these algebras, just as $\mathds{H}$ can be defined in terms of $\mathds{C}$, which can in turn be defined through $\mathds{R}$. In this context it is instructive to classify $\mathds{R, C, H}$, and $\mathds{O}$ as $\ast$-algebras. Such algebras are characterised by the possession of a real-linear conjugation map that is involutive ($(a^*)^*\equiv a$) and an anti-automorphism ($(ab)^*\equiv b^*a^*$).

If we denote the imaginary octonions by $e_i$ where $i=1,2,3,4,5,6,7$, the structure constants $C_{ijk}$ are defined by
\begin{equation}
e_i e_j=-\delta_{ij}+C_{ijk}e_k.
\end{equation}
We now establish some useful identities. First
\begin{equation}
C_{pmk}C_{qkn}+C_{qmk}C_{pkn}=\delta_{pm}\delta_{qn}+\delta_{pn}\delta_{qm}-2\delta_{pq}\delta_{mn}.
\end{equation}
Following \cite{Gunaydin:1995as} we define the ``Jacobian'' $C_{ijkl}$
\begin{gather}
\begin{gathered}
[e_{i},[e_{j},e_{k}]]+[e_{k},[e_{i},e_{j}]]+[e_{j},[e_{k},e_{i}]] \\
=4(C_{jkm}C_{imn}+C_{ijm}C_{kmn}+C_{kim}C_{jmn})e_{n} \\
\equiv 3C_{ijkl}e_{l},
\end{gathered}
\shortintertext{where $C_{ijkl}$ satisfies}
C_{ijkl}=\tfrac{1}{6} \varepsilon_{ijklmnp}C_{mnp}\label{eq:associator},
\shortintertext{and}
C_{ijkl}=-C_{mij}C_{mkl}-\delta_{il}\delta_{jk}+\delta_{ik}\delta_{jl}.
\end{gather}

That the Fano plane corresponds to the multiplication table of the imaginary octonions may be seen in \autoref{tab:FanoMult}.
\begin{table}[ht]
\begin{tabular*}{\textwidth}{@{\extracolsep{\fill}}*{10}{M{r}}}
\toprule
&   &  A &  B &  C &  D &  E &  F &  G & \\
\midrule
& A &    &  D &  G & -B &  F & -E & -C & \\
& B & -D &    &  E &  A & -C &  G & -F & \\
& C & -G & -E &    &  F &  B & -D &  A & \\
& D &  B & -A & -F &    &  G &  C & -E & \\
& E & -F &  C & -B & -G &    &  A &  D & \\
& F &  E & -G &  D & -C & -A &    &  B & \\
& G &  C &  F & -A &  E & -D & -B &    & \\
\bottomrule
\end{tabular*}
\caption[Fano plane multiplication table]{Following the oriented lines of \autoref{fig:FanoPlane} allows one to construct the Fano multiplication table, by identifying the qubits with the imaginary basis octonions. The minus signs arise when the Fano lines are followed counter to their orientation.}\label{tab:FanoMult}
\end{table}
The Fano plane multiplication, may be represented numerically as an $8 \times 8$ array as in \autoref{tab:NumericalFanoMult}. The non-vanishing independent components of the octonionic structure constants $C_{ijk}$ and their duals $C_{lmno}$ are then given in  \autoref{tab:FanoStructConst}.
\begin{table}[ht]
\begin{tabular*}{\textwidth}{@{\extracolsep{\fill}}*{10}{c}}
\toprule
& $i$ & $j$ & $k$ & & $l$ & $m$ & $n$ & $o$ & \\
\midrule
& 1   & 2   & 4   & & 3   & 5   & 6   & 7   & \\
& 2   & 3   & 5   & & 4   & 6   & 7   & 1   & \\
& 3   & 4   & 6   & & 5   & 7   & 1   & 2   & \\
& 4   & 5   & 7   & & 6   & 1   & 2   & 3   & \\
& 5   & 6   & 1   & & 7   & 2   & 3   & 4   & \\
& 6   & 7   & 2   & & 1   & 3   & 4   & 5   & \\
& 7   & 1   & 3   & & 2   & 4   & 5   & 6   & \\
\bottomrule
\end{tabular*}
\caption[Fano structure constants]{From \autoref{tab:FanoMult} one can read off which components of the Fano structure constants $C_{ijk}$ are nonzero. Using \eqref{eq:associator} one then obtains the associator coefficients $C_{lmno}$.}\label{tab:FanoStructConst}
\end{table}

\begin{figure}[ht]
 \centering
 \includegraphics[width=7cm]{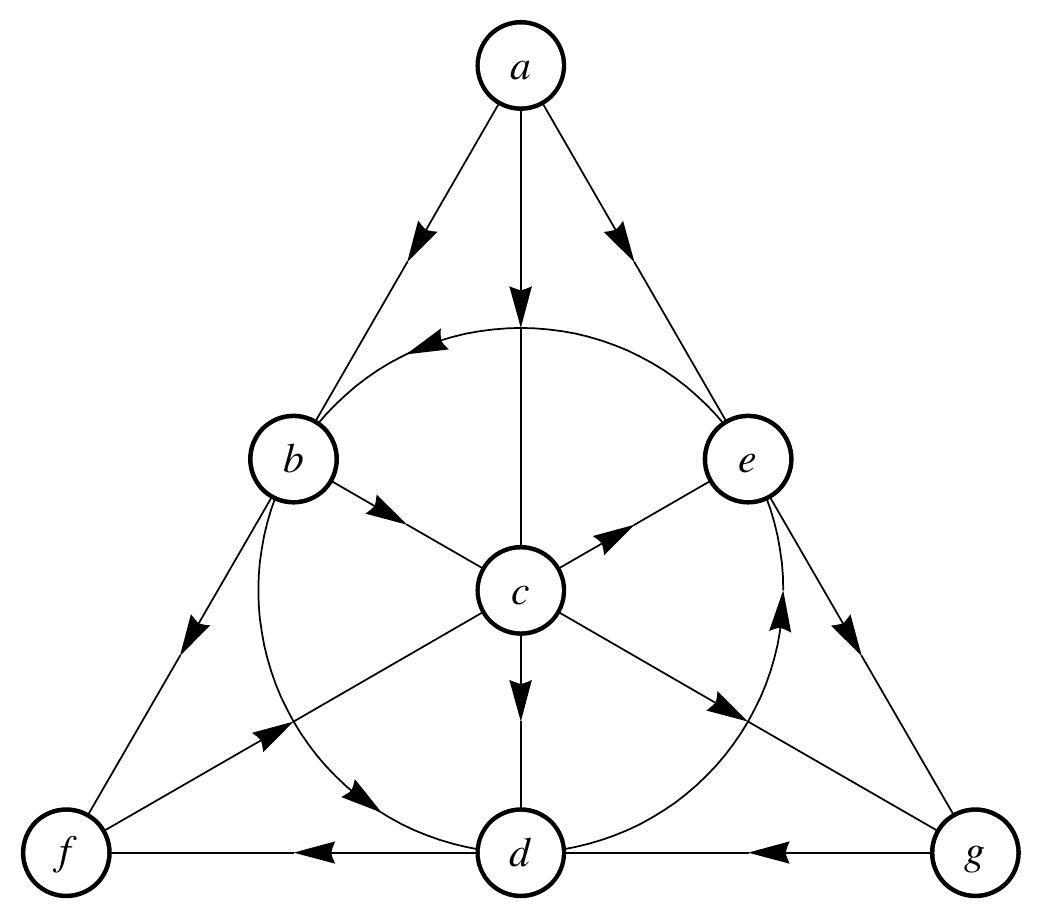}
 \caption[Dual Fano plane]{Like the Fano plane of \autoref{fig:FanoPlane} the dual Fano plane has seven points and seven lines, but this time the plane is associated with the dual state \eqref{eq:Dual7QubitState} interpreting the points as the seven tripartite entanglements and the lines as the seven qubits. The construction of the dual plane is expedited by \autoref{tab:FanoDualFano}. The orientation of our plane provides another multiplication table for imaginary octonions (\autoref{tab:DualFanoMult}).}
 \label{fig:DualFanoPlane}
\end{figure}

Note that in the Fano plane there are seven lines $a,b,c,d,e,f,g$, and seven vertices $A,B,C,D,E,F,G$. Each line passes through three vertices, and three lines meet at a vertex. This means one can define a dual Fano Plane with lines $A,B,C,D,E,F,G$, and vertices $a,b,c,d,e,f,g$. See \autoref{fig:DualFanoPlane}.
Another way to understand the appearance of the dual Fano plane is to recognise the seven rows in \eqref{eq:56Decomp} as the lines of the Fano plane and the seven columns as vertices as in \autoref{tab:FanoDualFano}.
\begin{table}[ht]
\begin{tabular*}{\textwidth}{@{\extracolsep{\fill}}*{10}{c}}
\toprule
&     & $A$ & $B$ & $C$ & $D$ & $E$ & $F$ & $G$ & \\
\midrule
& $a$ & 2   & 2   & 1   & 2   & 1   & 1   & 1   & \\
& $b$ & 1   & 2   & 2   & 1   & 2   & 1   & 1   & \\
& $c$ & 1   & 1   & 2   & 2   & 1   & 2   & 1   & \\
& $d$ & 1   & 1   & 1   & 2   & 2   & 1   & 2   & \\
& $e$ & 2   & 1   & 1   & 1   & 2   & 2   & 1   & \\
& $f$ & 1   & 2   & 1   & 1   & 1   & 2   & 2   & \\
& $g$ & 2   & 1   & 2   & 1   & 1   & 1   & 2   & \\
\bottomrule
\end{tabular*}
\caption[Fano and dual Fano lines and vertices]{The seven terms in decomposition \eqref{eq:56Decomp} may be written in a grid such that Fano lines and vertices are rows and columns. This permits easy identification of the dual lines and vertices, which are simply given by columns and rows.}\label{tab:FanoDualFano}
\end{table}
The dual Fano plane corresponds to the transpose of this matrix and leads to a dual state
\begin{equation}\label{eq:Dual7QubitState}
\begin{array}{c@{}c@{\ }c@{\lvert}*{3}{@{}c}@{\rangle}c}
\ket{\widetilde{\Psi}}_{56} = &   & A_{aeg} & a       & e       & g \\
                              & + & B_{bfa} & b       & f       & a \\
                              & + & C_{cgb} & c       & g       & b \\
                              & + & D_{dac} & d       & a       & c \\
                              & + & E_{ebd} & e       & b       & d \\
                              & + & F_{fce} & f       & c       & e \\
                              & + & G_{gdf} & g       & d       & f & .
\end{array}
\end{equation}
\begin{table}[ht]
\begin{tabular*}{\textwidth}{@{\extracolsep{\fill}}*{10}{M{r}}}
\toprule
&   &  a &  b &  c &  d &  e &  f &  g & \\
\midrule
& a &    &  f &  d & -c &  g & -b & -e & \\
& b & -f &    &  g &  e & -d &  a & -c & \\
& c & -d & -g &    &  a &  f & -e &  b & \\
& d &  c & -e & -a &    &  b &  g & -f & \\
& e & -g &  d & -f & -b &    &  c &  a & \\
& f &  b & -a &  e & -g & -c &    &  d & \\
& g &  e &  c & -b &  f & -a & -d &    & \\
\bottomrule
\end{tabular*}
\caption[Dual Fano plane multiplication table]{In direct analogy with \autoref{fig:FanoPlane} and \autoref{tab:FanoMult}, the oriented dual Fano plane of \autoref{fig:DualFanoPlane} is used to write an octonionic multiplication table.}\label{tab:DualFanoMult}
\end{table}
The dual Fano plane corresponds to the multiplication table of the imaginary octonions given in \autoref{tab:DualFanoMult} and may be represented numerically as an $8 \times 8$ array as in \autoref{tab:NumericalDualFanoMult}. The non-vanishing independent components of the octonionic structure constants $c_{ijk}$ and their duals $c_{lmno}$ are then given by \autoref{tab:DualFanoStructConst}.
\begin{table}[ht]
\begin{tabular*}{\textwidth}{@{\extracolsep{\fill}}*{10}{c}}
\toprule
& $i$ & $j$ & $k$ & & $l$ & $m$ & $n$ & $o$ & \\
\midrule
& 1   & 2   & 6   & & 3   & 4   & 5   & 7   & \\
& 2   & 3   & 7   & & 4   & 5   & 6   & 1   & \\
& 3   & 4   & 1   & & 5   & 6   & 7   & 2   & \\
& 4   & 5   & 2   & & 6   & 7   & 1   & 3   & \\
& 5   & 6   & 3   & & 7   & 1   & 2   & 4   & \\
& 6   & 7   & 4   & & 1   & 2   & 3   & 5   & \\
& 7   & 1   & 5   & & 2   & 3   & 4   & 6   & \\
\bottomrule
\end{tabular*}
\caption[Dual Fano structure constants]{Dual Fano structure constants $c_{ijk}$ and $c_{lmno}$ are obtained from \autoref{tab:DualFanoMult} in the same manner that \autoref{tab:FanoStructConst} is obtained from \autoref{tab:FanoMult}.}\label{tab:DualFanoStructConst}
\end{table}

\subsection{\texorpdfstring{Decomposition of $I_{4}$}{Decomposition of I4}}

To understand better the entanglement we note that, as a result of \eqref{eq:56Decomp}, Cartan's invariant contains not one Cayley hyperdeterminant but seven.  It may be written as the sum of seven terms each of which is invariant under $[SL(2)]^{3}$ plus cross terms. To see this, denote a \textbf{2} in one of the seven entries in \eqref{eq:56Decomp} by $A,B,C,D,E,F,G$. So we may rewrite \eqref{eq:56Decomp} as
\begin{gather}
\textbf{56}=(ABD)+(BCE)+(CDF)+(DEG)+(EFA)+(FGB)+(GAC),\label{eq:Analogue}
\shortintertext{or symbolically}
\textbf{56}=a+b+c+d+e+f+g.
\end{gather}
Then $I_4$ is the singlet in $\textbf{56}\times\textbf{56}\times\textbf{56}\times\textbf{56}$:
\begin{equation}\label{eq:I4}
\begin{gathered}
I_4=a^4+b^4+c^4+d^4+e^4+f^4+g^4 \\
\begin{array}{c*{10}{@{\ }c}}
+~2\Big[a^2b^2 & + & a^2c^2 & + & a^2d^2 & + & a^2e^2 & + & a^2f^2 & + & a^2g^2 \\
               & + & b^2c^2 & + & b^2d^2 & + & b^2e^2 & + & b^2f^2 & + & b^2g^2 \\
               &   &        & + & c^2d^2 & + & c^2e^2 & + & c^2f^2 & + & c^2g^2 \\
               &   &        &   &        & + & d^2e^2 & + & d^2f^2 & + & d^2g^2 \\
               &   &        &   &        &   &        & + & e^2f^2 & + & e^2g^2 \\
               &   &        &   &        &   &        &   &        & + & \phantom{\Big]}f^2g^2\Big]
\end{array} \\
+~8\left[abce+bcdf+cdeg+defa+efgb+fgac+gabd\right],
\end{gathered}
\end{equation}
where products like
\begin{equation}\label{eq:1way}
\begin{split}
a^4&=(ABD)(ABD)(ABD)(ABD) \\
&= \half \varepsilon^{A_1A_2}\varepsilon^{B_1B_2}\varepsilon^{D_1D_4}\varepsilon^{A_3A_4}\varepsilon^{B_3B_4}\varepsilon^{D_2D_3} \\
&\phantom{=}\times a_{A_1B_1D_1}a_{A_2B_2D_2}a_{A_3B_3D_3}a_{A_4B_4D_4},
\end{split}
\end{equation}
exclude four individuals (here Charlie, Emma, Fred, and George), products like
\begin{equation}\label{eq:2way}
\begin{split}
a^2b^2&=(ABD)(ABD)(BCE)(BCE) \\
&= \half \varepsilon^{A_1A_2}\varepsilon^{B_1B_3}\varepsilon^{D_1D_2}\varepsilon^{B_2B_4}\varepsilon^{C_3C_4}\varepsilon^{E_3E_4} \\
&\phantom{=}\times a_{A_1B_1D_1}a_{A_2B_2D_2}b_{B_3C_3E_3}b_{B_4C_4E_4},
\end{split}
\end{equation}
exclude two individuals (here Fred and George), and products like
\begin{equation}\label{eq:4way}
\begin{split}
abce&=(ABD)(BCE)(CDF)(EFA) \\
&= \half \varepsilon^{A_1A_4}\varepsilon^{B_1B_2}\varepsilon^{C_2C_3}\varepsilon^{D_1D_3}\varepsilon^{E_2E_4}\varepsilon^{F_3F_4} \\
&\phantom{=}\times a_{A_1B_1D_1}b_{B_2C_2E_2}c_{C_3D_3F_3}e_{E_4F_4A_4},
\end{split}
\end{equation}
exclude one individual (here George).  These results may be verified using the dictionary between $a,b,c,d,e,f,g$ and the $x$ and $y$ discussed in the next section. Note that $a^4$ is just minus Cayley's hyperdeterminant.

\subsection{Subsectors}

\begin{table}[ht]
\begin{tabular*}{\textwidth}{@{\extracolsep{\fill}}*{6}{M{r}}}
\toprule
&   & a & e &  g & \\
\midrule
& a &   & g & -e & \\
& e &-g & a &    & \\
& g & e &-a &    & \\
\bottomrule
\end{tabular*}
\caption[The $aeg$ multiplication table]{The $aeg$ quaternionic multiplication table is obtained by selecting the $aeg$ quaternionic cycle from \autoref{tab:DualFanoMult}.}\label{tab:aegMult}
\end{table}

Having understood the analogy between $\mathcal{N}=8$ black holes and the tripartite entanglement of seven qubits using $E_{7(7)}$, we may now find the analogy in the $\mathcal{N}=4$ case using $SL(2) \times  SO(6,6)$ and the $\mathcal{N}=2$ case using $SL(2) \times SO(2,2)$.

First we recall the decomposition \eqref{eq:First56Decomp} of the fundamental 56-dimensional representation of $E_{7(7)}$ under its maximal subgroup.  The $\mathcal{N}=4$ subsector consists of just the 24 states belonging to the $\mathbf{(2,12)}$
\begin{equation}
\ket{\Psi} = a_{ABD}\ket{ABD}+e_{EFA}\ket{EFA}+g_{GAC}\ket{GAC}.
\end{equation}
So only Alice talks to all the others. This is described by just those three lines passing through $A$ in the Fano plane or the $aeg$ line in the dual Fano plane. Then the equation analogous to \eqref{eq:Analogue} is
\begin{equation}
\mathbf{(2,12)}= (ABD)+(EFA)+(GAC)= a+e+g,
\end{equation}
and the corresponding quartic invariant, $I_{4}$, reduces to the singlet in $\mathbf{(2,12)\times(2,12)\times(2,12)\times(2,12)}$
\begin{equation}\label{eq:2124}
I_{4} \sim a^{4}+e^{4}+g^{4}+2[e^{2}g^{2}+g^{2}a^{2}+a^{2}e^{2}].
\end{equation}
If we identify the 24 numbers ($a_{ABD},e_{EFA},g_{GAC}$) with $(P^{\mu},Q_{\nu})$ with $\mu,\nu=0,\ldots, 11$ in a way analogous  to \eqref{eq:MiniDict} this becomes the $SL(2) \times SO(6,6)$ invariant \cite{Duff:1995sm,Cvetic:1995kv,Cvetic:1995bj}
\begin{gather}
I_{4}=P^{2}Q^{2}-(P\cdot Q)^{2}.\label{eq:i4n=4}
\shortintertext{So}
I_{4}=I_{aeg}\equiv \det(\gamma^{1}(a)+\gamma^{2}(g)+\gamma^{3}(e)).\label{eq:aeg}
\end{gather}
This reduction from $\mathcal{N}=8$ to $\mathcal{N}=4$ corresponds to a reduction from the imaginary octonions of  \autoref{tab:DualFanoMult} to the imaginary quaternions of \autoref{tab:aegMult}. This suggest that $I_4$ of \eqref{eq:i4n=4} may be written as Cayley's hyperdeterminant over the imaginary quaternions, and this is indeed the case, as shown in \autoref{sec:Cayleyimquat}. From a stringy point of view, this subsector describes just the NS-NS charges.

A different subsector which excludes Alice is obtained by keeping just the $(\textbf{1},\textbf{32})$ in \eqref{eq:First56Decomp}, This is described by just those four lines not passing through $A$ in the Fano plane or the $bcdf$ quadrangle in the dual Fano plane. From a stringy point of view, this subsector describes just the R-R charges.

\begin{equation}\label{eq:563}
\mathbf{(1,32)}=(BCE)+(CDF)+(DEG)+(FGB)= b+c+d+f,
\end{equation}
and the corresponding quartic invariant, $I_{4}$, reduces to the singlet in $\mathbf{(1,32)\times(1,32)\times(1,32)\times(1,32)}$
\begin{equation}\label{eq:567}
I_4\sim b^4+c^4+d^4+f^4+2[b^2c^2+c^2d^2+d^2e^2+d^2f^2+c^2f^2+f^2b^2] +8bcdf.
\end{equation}
This is does not correspond to any $\mathcal{N}=4$ black hole but rather to an $\mathcal{N}=8$ black hole with only the R-R charges switched on.

For $\mathcal{N}=2$, as may be seen from \eqref{eq:Triality}, we only have the $[SL(2)]^{3}$ subgroup of the $STU$ model where there are only 8 states
\begin{equation}
\ket{\Psi} = a_{ABD}\ket{ABD}.
\end{equation}
This is described by just the $ABD$ line in the Fano plane or the $a$ vertex in the dual Fano plane. This is simply the usual tripartite entanglement, for which
\begin{gather}
\mathbf{(2,2,2)}=(ABD)=a,
\shortintertext{and  the corresponding quartic invariant}
I_{4} \sim a^{4},
\shortintertext{is just Cayley's hyperdeterminant}
I_{4}=-\Det a.
\end{gather}

\subsection{\texorpdfstring{$E_7$ as an $\mathds{O}$-graded algebra}{E7 as an O-graded algebra}}
\label{sec:graded}

As we have seen, the seven qubit interpretation relies on the decomposition of the $\mathbf{56}$ under $E_{7(7)} \supset [SL(2)]^7$  and the Fano plane \cite{Duff:2006ue,Levay:2006pt}. Independently, similar observations were being made in pure mathematics, namely  the remarkable fact that $E_7$ has  a natural structure of an $\mathds{O}$-graded algebra, compatible with its action on the minimal 56-dimensional representation \cite{Manivel:2005, Elduque:2005}.

Consider the points
\begin{gather}
i\in\{1,2,3,4,5,6,7\},
\shortintertext{and lines}
l\in\{124,235,346,457,561,672,713\}
\end{gather}
of the Fano plane and  attach to each line a two-dimensional vector space $A_l$. Then we may describe the algebra of $E_7$ and the $\mathbf{56}$ as
\begin{equation}\label{eq:man56}
\begin{gathered}
\textstyle\mathfrak{e}_7=\times_l~\mathfrak{sl}(A_l)e_0\oplus\bigoplus_{1\leq i \leq 7}\left[\bigotimes_{i\notin l} A_l\right]e_i, \\
\textstyle\mathbf{56} = \bigoplus_{1\leq i \leq 7}\left[\bigotimes_{i\in l} A_l\right]e_i.
\end{gathered}
\end{equation}
Alternatively, consider the complementary quadrangles of the Fano plane
\begin{gather}
I\in\{3457,4561,5672,6713,7124,1235,2346\},
\shortintertext{then}
\mathfrak{e}_7=\times_{i=1}^{7}\mathfrak{sl}(A_i)\oplus\bigoplus_{(ijkl)\in I}A_i\otimes A_j\otimes A_k\otimes A_l. \label{eq:quadrangles}
\end{gather}
The same formulae hold good if we go to the dual Fano plane by swapping the roles of points and lines.

Note that there is a quaternionic analogue of this construction where, instead of the Fano plane, we consider just three of its lines. This describes the $\mathcal{N}=4$ subsector discussed above.

We shall return to \eqref{eq:quadrangles} in \autoref{sec:cyclic}.

In fact, a similar construction applies to the algebra of $E_8$ \cite{Manivel:2005}:
\begin{gather}
\mathfrak{e}_8=\times_{i=0}^{7}\mathfrak{sl}(A_i)\oplus\bigoplus_{(ijkl)\in I}A_i\otimes A_j\otimes A_k\otimes A_l, \label{eq:8quadrangles}
\shortintertext{where $I$ now runs over  the 14 quadruples}
I\in\{3457,4561,5672,6713,7124,1235,2346,0124,0235,0346,0457,0561,0672,0713\},
\end{gather}
or their duals appearing in \autoref{tab:CartanDictionary}.

There is a  difference, however, since the fundamental $\mathbf{248}$ of $E_8$ is also the adjoint. So the $224$-dimensional Hilbert space describing the 4-way entanglement of 8-qubits does not transform as a linear representation of $E_8$ but rather as a nonlinear realisation, namely as the coset $E_8/[SL(2)]^8$. Lacking at the moment a good application of this coset construction within quantum information theory, we relegate it to Appendix B.

\subsection{\texorpdfstring{Classification of $\mathcal{N}=8$ black holes and seven-qubit states}{Classification of N=8 black holes and seven-qubit states}}
\label{sec:blackholesand7-qubitstates}

In the $\mathcal{N}=8$ theory, ``large'' and ``small'' black holes are classified by the sign of $I_{4}$:
\begin{subequations}
\begin{align}
  I_{4} &> 0, \label{eq:I4Case1}\\
  I_{4} &= 0, \label{eq:I4Case2}\\
  I_{4} &< 0.
\end{align}
\end{subequations}
Once again, non-zero $I_{4}$ corresponds to large black holes, which are BPS for $I_{4} > 0$ and non-BPS for $I_{4}< 0$, and vanishing  $I_{4}$ corresponds to small black holes. However, in contrast to $\mathcal{N}=2$, case \eqref{eq:I4Case1} requires that only 1/8 of the supersymmetry is preserved, while we may have 1/8, 1/4 or 1/2 for case \eqref{eq:I4Case2}.

The large black hole solutions can be found \cite{Ferrara:2006em} by solving the $\mathcal{N}=8$ classical attractor equations \cite{Ferrara:1995ih} when at the attractor value the $Z_{AB}$ matrix, in normal form, becomes
\begin{gather}
Z_{AB}=
\begin{pmatrix}
Z\varepsilon & 0 & 0 & 0  \\
0 & 0 & 0 & 0 \\
0 & 0 & 0 & 0 \\
0 & 0 & 0 & 0
\end{pmatrix},\label{eq:Z2}
\shortintertext{for positive $I_{4}$ and}
Z_{AB}=e^{i\pi/4}|Z|
\begin{pmatrix}
\varepsilon & 0 & 0 & 0 \\
0 & \varepsilon & 0 & 0 \\
0 & 0 & \varepsilon & 0 \\
0 & 0 & 0 & \varepsilon
\end{pmatrix}\label{eq:Z3}
\end{gather}
for negative $I_{4}$. These matrices exhibit the maximal compact symmetries $SU(6) \times SU(2)$ and $USp(8)$ for the positive and negative $I_{4}$, respectively.

If the phase in \eqref{eq:5ParameterGeneratingSoln} vanishes (which is the case if the configuration preserves at least 1/4 supersymmetry \cite{Ferrara:1997ci}), $I_4$ of \eqref{eq:5ParameterGeneratingSoln} becomes
\begin{gather}
I_4 = s_1s_2s_3s_4,
\shortintertext{where  $s_i$ are given by the $\rho_i$ of \eqref{eq:5ParameterGeneratingSoln}}
\begin{split}
s_1 &= \rho_1 + \rho_2 + \rho_3 + \rho_4, \\
s_2 &= \rho_1 + \rho_2 - \rho_3 - \rho_4, \\
s_3 &= \rho_1 - \rho_2 + \rho_3 - \rho_4, \\
s_4 &= \rho_1 - \rho_2 - \rho_3 + \rho_4,
\end{split}
\end{gather}
and we order the $s_i$ so that $s_1 \ge s_2 \ge s_3 \ge |s_4| $. The charge orbits \cite{Ferrara:1997uz,Ferrara:1997ci,Lu:1997bg} for the black holes depend on the number of unbroken supersymmetries  or the number of vanishing eigenvalues as in \autoref{tab:n=8bh}.
\begin{table}[ht]
\begin{tabular*}{\textwidth}{@{\extracolsep{\fill}}llrrrrrccc}
\toprule
& $\phantom{[SL(2}$Orbit                        & $s_1$ & $s_2$ & $s_3$ & $s_4$ & $I_4$ & Black hole & SUSY          & \\
\midrule
& $[SL(2)]^3/([SO(1,1)]^2\ltimes\mathds{R}^3)$  & $>0$  & 0     & 0     & 0     & 0     & small      & 1/2           & \\
& $[SL(2)]^3/(O(2,1)\times\mathds{R})$          & $>0$  & $>0$  & 0     & 0     & 0     & small      & 1/2           & \\
& $[SL(2)]^3/\mathds{R}^2$                      & $>0$  & $>0$  & $>0$  & 0     & 0     & small      & 1/2           & \\
& $[SL(2)]^3/[U(1)]^2$                          & $>0$  & $>0$  & $>0$  & $>0$  & $>0$  & large      & 1/2           & \\
& $[SL(2)]^3/[U(1)]^2$                          & $>0$  & $>0$  & $>0$  & $<0$  & $<0$  & large      & 0 ($Z=0$)     & \\
& $[SL(2)]^3/[SO(1,1)]^2$                       & $>0$  & $>0$  & $>0$  & $<0$  & $<0$  & large      & 0 ($Z \neq0$) & \\
\bottomrule
\end{tabular*}
\caption[Classification of $D=4,\mathcal{N}=2$  $STU$ black holes.]{Classification of $D=4,\mathcal{N}=2$  $STU$ black holes, see \autoref{sec:stu}. The values of $I_4$ and the eigenvalues $s_i$ distinguish the different charge orbits. Here, small black holes have a vanishing horizon.}
\label{tab:n=2bh}
\end{table}
\begin{table}[ht]
\begin{tabular*}{\textwidth}{@{\extracolsep{\fill}}llrrrrrccc}
\toprule
& $\phantom{E_{7(7)}/}$Orbit                                    & $s_1$ & $s_2$ & $s_3$ & $s_4$ & $I_4$ & Black hole & SUSY & \\
\midrule
& $E_{7(7)}/(E_{6(6)} \ltimes \mathds{R}^{27})$                 & $>0$  & 0     & 0     & 0     & 0     & small      & 1/2  & \\
& $E_{7(7)}/(O(5,6) \ltimes \mathds{R}^{32} \times\mathds{R})$  & $>0$  & $>0$  & 0     & 0     & 0     & small      & 1/4  & \\
& $E_{7(7)}/(F_{4(4)} \ltimes \mathds{R}^{26})$                 & $>0$  & $>0$  & $>0$  & 0     & 0     & small      & 1/8  & \\
& $E_{7(7)}/E_{6(2)} $                                          & $>0$  & $>0$  & $>0$  & $>0$  & $>0$  & large      & 1/8  & \\
& $E_{7(7)}/E_{6(2)} $                                          & $>0$  & $>0$  & $>0$  & $<0$  & $<0$  & large      & 0    & \\
& $E_{7(7)}/E_{6(6)}$                                           & $>0$  & $>0$  & $>0$  & $<0$  & $<0$  & large      & 0    & \\
\bottomrule
\end{tabular*}
\caption[Classification of $D=4,\mathcal{N}=8$  black holes.]{Classification of $D=4,N=8$  black holes. The distinct charge orbits are determined by the number of non-vanishing eigenvalues and $I_4$, as well as the number of preserved supersymmetries.}
\label{tab:n=8bh}
\end{table}

For $\mathcal{N}=8$, as for $\mathcal{N}=2$, the large black holes correspond to the two classes of GHZ-type (entangled) states and small black holes to the separable or W class. As we shall describe in \autoref{sec:binarybasis}, one way of obtaining such states is to go the canonical basis \eqref{eq:x+iy} where the Cartan invariant reduces to Cayley's hyperdeterminant. Then the $A$-$B$-$C$, $A$-$BC$, W and GHZ states are just those of the $STU$ model whose black hole interpretation was given already in \autoref{sec:blackholesandthree-qubit states}. The result is shown in \autoref{tab:n=8bh}. Note, however,  that by embedding the $STU$ model in the $\mathcal{N}=8$ theory, we obtain finer supersymmetry and charge orbit correspondences than those of the
$\mathcal{N}=2$ $STU$ theory of \autoref{tab:n=2bh}. The orbits for the large $\mathcal{N}=2$ black holes were previously found in \cite{Ferrara:1997uz,Bellucci:2006xz}; those of the small black holes are new. The corresponding SLOCC orbits may be found in \cite{Borsten:2008fts}.

Alternatively, having succeeded in writing the Cartan invariant in terms of $a,b,c,d,e,f,g$ in \eqref{eq:I4} we can now look for maximally entangled states in the $\mathcal{N}=4$ subsector and the full $\mathcal{N}=8$ theory. Let us first recall  the normalised GHZ state in the $\mathcal{N}=2$ subsector
\begin{equation}
\ket{\Psi}= \tfrac{1}{\sqrt{2}}\left(\ket{111} +\ket{000}\right).
\end{equation}
It is obviously entangled since if Alice measures 0, Bob and Charlie must also measure 0. This is confirmed by calculating the 3-tangle
\begin{gather}
\gamma^1(a)=
\begin{pmatrix}
2(a_0a_6-a_2a_4) & a_0a_7-a_2a_5+a_1a_6-a_3a_4\\
a_0a_7-a_2a_5+a_1a_6-a_3a_4 & 2(a_1a_7-a_3a_5)
\end{pmatrix},
\shortintertext{and}
a_0=a_7=\tfrac{1}{\sqrt{2}},
\shortintertext{so}
\gamma^1(a)=\begin{pmatrix}0 & \half \\\half & 0\end{pmatrix},
\end{gather}
Moreover it is maximally entangled since
\begin{equation}
\tau=4|\det\gamma_a|=1.
\end{equation}
Now consider  the $\mathcal{N}=4$ subsector state with
\begin{gather}
a_0=a_7=e_0=e_7=g_0=g_7=\tfrac{1}{\sqrt{6}}, \\
\gamma^1(a)+\gamma^2(g)+\gamma^3(e)=
\begin{pmatrix}
0 & \tfrac{1}{6}+\tfrac{1}{6}+\tfrac{1}{6}\\
\tfrac{1}{6}+\tfrac{1}{6}+\tfrac{1}{6}  & 0
\end{pmatrix}, \\
\tau=4|\det(\gamma^1(a)+\gamma^2(g)+\gamma^3(e))|=1,
\end{gather}
also.  Going to the full $\mathcal{N}=8$ sector, consider
\begin{equation}
a_0=a_7=b_0=b_7=c_0=c_7=d_0=d_7=e_0=e_7=f_0=f_7=g_0=g_7=\tfrac{1}{\sqrt{14}},
\end{equation}
we find 7 non-vanishing contributions of the $a^4$ type, and 21 of the $a^2b^2$ type while the $abce$ type terms all vanish. Hence $\tau=1$ once more. Another example involving the $abce$ terms is to keep
\begin{equation}
a_7=a_1=a_2=a_4=\tfrac{1}{\sqrt{28}},
\end{equation}
and similarly for $b,c,d,e,f,g$ when we again find $\tau=1$. The choice
\begin{equation}
a_7=g_1=b_1=d_1=\tfrac{1}{2}
\end{equation}
involves only the $agbd$ term and also yields $\tau=1$.

\subsection{Seven qutrit interpretation}
\label{sec:seven}

We note that the 56-dimensional Hilbert space given in \autoref{eq:7QubitState} is not a subspace of the usual $2^{7}$-dimensional seven-qubit Hilbert space given by $(\mathbf{2,2,2,2,2,2,2})$, but rather a direct sum of seven $2^{3}$-dimensional three-qubit Hilbert spaces $(\mathbf{2,2,2})$:
\begin{equation}\label{eq:56new}
 \begin{split}
   &\mathbf{(2,2,1,2,1,1,1)} \\
+\ &\mathbf{(1,2,2,1,2,1,1)} \\
+\ &\mathbf{(1,1,2,2,1,2,1)} \\
+\ &\mathbf{(1,1,1,2,2,1,2)} \\
+\ &\mathbf{(2,1,1,1,2,2,1)} \\
+\ &\mathbf{(1,2,1,1,1,2,2)} \\
+\ &\mathbf{(2,1,2,1,1,1,2)}.
\end{split}
\end{equation}
This raises an ambiguity about the terminology \emph{tripartite entanglement of seven qubits}. The state corresponding to the usual $2^{7}$-dimensional seven-qubit Hilbert space  $(\mathbf{2,2,2,2,2,2,2})$ is
\begin{equation}
\ket{\Psi}=a_{ABCDEFG}\ket{ABCDEFG},
\end{equation}
and one meaning of tripartite entanglement, $ABD$ say, would be that given by the reduced density matrix
\begin{equation}
\rho_{ABD}=\Tr_{CEFG}\ket{\Psi}\bra{\Psi}.
\end{equation}
So it is important to note that this is clearly different from the meaning we have adopted for seven qubits elsewhere in this Report.

The doublets in \eqref{eq:56new} are interpreted as qubits, but what about the singlets? A natural explanation for the origin of the singlets is to embed each two-valued qubit in a three-valued qutrit and note that under
\begin{gather}
SL(3) \rightarrow SL(2)
\shortintertext{we have}
\mathbf{3 \to 2+1}.
\end{gather}
The seven qutrit system (Alice, Bob, Charlie, Daisy, Emma, Fred and George) is described by the state
\begin{equation}
\ket{\Psi} =a_{\hat A\hat B\hat C\hat D\hat E\hat F\hat G}\ket{\hat A\hat B\hat C\hat D\hat E\hat F\hat G},
\end{equation}
where ${\hat A}=0,1,2$, and the Hilbert space has dimension $3^{7}=2,187$. ${\hat a}_{\hat A\hat B\hat C\hat D\hat E\hat F\hat G}$ transforms as a $\mathbf{(3,3,3,3,3,3,3)}$ of $ SL(3)_A \times SL(3)_B \times SL(3)_C \times SL(3)_D \times SL(3)_E \times SL(3)_F \times SL(3)_G$.
Under
\begin{gather}
\begin{split}
& SL(3)_{A} \times SL(3)_{B} \times SL(3)_{C} \times SL(3)_{D} \times SL(3)_{E} \times SL(3)_{F} \times SL(3)_{G} \\
\supset\ & SL(2)_{A} \times SL(2)_{B} \times SL(2)_{C} \times SL(2)_{D} \times SL(2)_{E} \times SL(2)_{F} \times SL(2)_{G},
\end{split}
\shortintertext{we have}
\begin{array}{c*{4}{@{\ }c}c}
\mathbf{(3,3,3,3,3,3,3)} \to &   & 1  & \text{term\phantom{s} like} & \mathbf{(2,2,2,2,2,2,2)} \\
                             & + & 7  & \text{terms like}           & \mathbf{(2,2,2,2,2,2,1)} \\
                             & + & 21 & \text{terms like}           & \mathbf{(2,2,2,2,2,1,1)} \\
                             & + & 35 & \text{terms like}           & \mathbf{(2,2,2,2,1,1,1)} \\
                             & + & 35 & \text{terms like}           & \mathbf{(2,2,2,1,1,1,1)} \\
                             & + & 21 & \text{terms like}           & \mathbf{(2,2,1,1,1,1,1)} \\
                             & + & 7  & \text{terms like}           & \mathbf{(2,1,1,1,1,1,1)} \\
                             & + & 1  & \text{term\phantom{s} like} & \mathbf{(1,1,1,1,1,1,1)} & ,
\end{array}
\end{gather}
which contains \eqref{eq:56new} as a subspace. Denoting doublets by $A=0,1$ and singlets by $\bullet$, we have
\begin{equation}
\begin{array}{c@{}c@{\ }c@{\lvert}*{7}{@{}c}@{\rangle}@{}c}
\ket{\Psi}_{56} =
&   & a_{AB\bullet D\bullet\bullet\bullet}  & A       & B       & \bullet & D       & \bullet & \bullet & \bullet \\
& + & b_{\bullet BC\bullet E\bullet\bullet} & \bullet & B       & C       & \bullet & E       & \bullet & \bullet \\
& + & c_{\bullet\bullet CD\bullet F\bullet} & \bullet & \bullet & C       & D       & \bullet & F       & \bullet \\
& + & d_{\bullet\bullet\bullet DE\bullet G} & \bullet & \bullet & \bullet & D       & E       & \bullet & G       \\
& + & e_{A\bullet\bullet\bullet EF\bullet}  & A       & \bullet & \bullet & \bullet & E       & F       & \bullet \\
& + & f_{\bullet B\bullet\bullet\bullet FG} & \bullet & B       & \bullet & \bullet & \bullet & F       & G       \\
& + & g_{A\bullet C\bullet\bullet\bullet G} & A       & \bullet & C       & \bullet & \bullet & \bullet & G & ,
\end{array}
\end{equation}
which we abbreviate by \autoref{eq:7QubitState}. So the Fano plane entanglement we have described fits within conventional quantum information theory, but we have discovered a hidden $E_7$ symmetry of this special 56-dimensional subspace.

In the notation ${\hat A}=0,1,2$ the 56 states are
\begin{equation}\label{eq:56States}
\begin{array}{*{8}{K}@{}c}
0020222 & 0021222 & 0120222 & 0121222 & 1020222 & 1021222 & 1120222 & 1121222 \\
2002022 & 2002122 & 2012022 & 2012122 & 2102022 & 2102122 & 2112022 & 2112122 \\
2200202 & 2200212 & 2201202 & 2201212 & 2210202 & 2210212 & 2211202 & 2211212 \\
2220020 & 2220021 & 2220120 & 2220121 & 2221020 & 2221021 & 2221120 & 2221121 \\
0222002 & 0222012 & 0222102 & 0222112 & 1222002 & 1222012 & 1222102 & 1222112 \\
2022200 & 2022201 & 2022210 & 2022211 & 2122200 & 2122201 & 2122210 & 2122211 \\
0202220 & 0202221 & 0212220 & 0212221 & 1202220 & 1202221 & 1212220 & 1212221 & .
\end{array}
\end{equation}

To get a flavour of the qutrit entanglement, we restrict to generalisations of  $\ket{000}+\ket{111}$ GHZ states and get
\begin{equation}
\begin{array}{KK@{}c}
0020222 & 1121222 \\
2002022 & 2112122 \\
2200202 & 2211212 \\
2220020 & 2221121 \\
0222002 & 1222112 \\
2022200 & 2122211 \\
0202220 & 1212221 & .
\end{array}
\end{equation}
Suitably normalised, this generalised GHZ state yields Cartan $\tau=1$ and is maximally entangled.

Suppose Alice measures 0:
\begin{equation}
\begin{array}{K@{}c}0020222\\0222002\\0202220&,\end{array}
\end{equation}
(which incidentally reduces to the $a,e,g$ subsector) then Bob can measure 0:
\begin{equation}
\ket{0020222},
\end{equation}
in which case Charlie, Daisy, Emma, Fred and George can only measure 2,0,2,2,2, respectively or Bob can measure 2
\begin{gather}
\begin{array}{K@{}c}0222002\\0202220&,\end{array}
\shortintertext{then Charlie can measure 0:}
\ket{0202220},
\shortintertext{or 2:}
\ket{0222002},
\end{gather}
and so on.  We may tabulate the various possibilities as in \autoref{tab:MeasurementPossibilities}
\begin{table}[ht]
\begin{tabular*}{\textwidth}{@{\extracolsep{\fill}}*{9}{c}}
\toprule
& $A$ & $B$ & $C$ & $D$ & $E$ & $F$ & $G$ & \\
\midrule
& 0   & 0   & 2   & 0   & 2   & 2   & 2   & \\
&     & 2   & 0   & 2   & 2   & 2   & 0   & \\
&     &     & 2   & 2   & 0   & 0   & 2   & \\
& 1   & 1   & 2   & 1   & 2   & 2   & 2   & \\
&     & 2   & 1   & 2   & 2   & 2   & 1   & \\
&     &     & 2   & 2   & 1   & 1   & 2   & \\
& 2   & 0   & 0   & 2   & 0   & 2   & 2   & \\
&     &     & 2   & 2   & 2   & 0   & 0   & \\
&     & 1   & 1   & 2   & 1   & 2   & 2   & \\
&     &     & 1   & 2   & 1   & 2   & 2   & \\
&     & 2   & 0   & 0   & 2   & 0   & 2   & \\
&     &     & 1   & 1   & 2   & 1   & 2   & \\
&     &     & 2   & 0   & 0   & 2   & 0   & \\
&     &     &     & 1   & 1   & 2   & 1   & \\
\bottomrule
\end{tabular*}
\caption[Measurement possibilities]{The possible sequential results of measurement on the seven qubits starting from the 56 states of \eqref{eq:56States}.}
\label{tab:MeasurementPossibilities}
\end{table}

\newpage
\section{\texorpdfstring{BLACK HOLES AND THE FREUDENTHAL TRIPLE SYSTEM}{Black Holes and the Freudenthal Triple System}}
\label{sec:Freudenthal}

For  the $\mathcal{N}=8$ black hole entropy interpretation,  we considered $I_4$ in a form exhibiting manifest $SO(8)$ symmetry, the 56 charges lying in what we refer to as the \emph{Cartan} basis, that is, the pair of $8 \times 8$ antisymmetric matrices, $x^{IJ}$ and $y_{IJ}$.  For the QI interpretation, we also considered $I_4$ as a quartic polynomial in the 56 coefficients specifying the special seven-qubit state \eqref{eq:7QubitState} with manifest $[SL(2)]^7$ symmetry. We shall refer to this representation as the \emph{Fano} basis. There is actually a third possible representation of the 56 charges, the Freudenthal triple system (FTS) \cite{Jordan:1933vh,Freudenthal:1954,Schafer:1966,McCrimmon:1969,Krutelevich:2004}, which we investigate now.

It is well known that the FTS  may be used as a representation of the black hole charge vector space of $\mathcal{N}=8$ supergravity in $D=4$ \cite{Ferrara:1997uz}. Indeed, for $\mathcal{N}=8$ supergravity in $D=5$ there is a one-to-one correspondence between the black hole charge vector space and the cubic Jordan algebra of $3\times 3$ hermitian matrices defined over the split octonions, which forms a representation of the fundamental $\mathbf{27}$ of $E_{6(6)}$, the U-duality group in this case. Reducing down to $D=4$ the black hole charge vector space is given by the FTS defined over the $D=5$ Jordan algebra which forms a representation of the fundamental $\mathbf{56}$ of $E_{7(7)}$.  We shall refer to this representation as the \emph{Freudenthal} basis, which will facilitate a better understanding of the $E_{7(7)}$ invariant entanglement and its relationship with the black hole charges and entropy.

Interestingly, the $\mathcal{N}=2$ $STU$ model may also be considered from this point of view, giving a different picture of the results already presented, on both the black hole and QI sides of the equation \cite{Borsten:2008fts}.

The Jordan algebra approach presents a possible alternative interpretation of the black hole/qutrit correspondence in $D=5$, considered in \autoref{sec:magic}, allowing one to extend the analogy to $D=6$.  This alternative is inspired by the observation that the  $\mathcal{N}=2$ ``Magic'' supergravity models are themselves intimately related to division algebras, Jordan algebras and the FTS \cite{Gunaydin:1983bi, Gunaydin:1983rk, Gunaydin:1984ak, Ferrara:1997ci, Ferrara:1997uz}.
However, let us postpone these considerations for the moment (we will return to them in \autoref{sec:Freudenthal-Fano} and \autoref{sec:AlternativeJordan}) while we introduce the necessary material and make some observations concerning the properties of the black hole entropy and qubit entanglement from this FTS perspective.

\subsection{Composition algebras}

The Jordan algebras of particular importance to supergravity are, for the most part, conveniently described in terms of Hermitian matrices defined over certain \emph{composition} algebras. An algebra $\mathds{A}$, which need not be associative, defined over a ground field $\mathds{F}$, is said to be composition if it comes equipped with a non-degenerate quadratic norm, $\mathbf{n}:\mathds{A}\rightarrow\mathds{F}$, satisfying,
\begin{equation}
\mathbf{n}(yx)=\mathbf{n}(x)\mathbf{n}(y) \qquad \forall x, y \in \mathds{A}.
\end{equation}
All such algebras are alternative, $x^2y=x(xy)$ and $(xy)x=x(yx)$ for all $x, y\in\mathds{A}$ \cite{Jacobson:1958}. Note, the \emph{associator}, defined as $[x,y,z]=(xy)z-x(yz)$ in direct analogy with the commutator, is an alternating function of its arguments precisely when $\mathds{A}$ is alternative.

An algebra is said to be  \emph{division} if it contains no zero divisors, that is,
\begin{equation}
xy=0 \quad \Rightarrow \quad x=0 \quad\text{or}\quad y=0.
\end{equation}
This condition holds for any composition algebra with positive definite norm, in which case it is referred to as a nicely normed division algebra.  The only four possible nicely normed division algebras are $\mathds{R, C, H}$ and $\mathds{O}$, a celebrated result due to Hurwitz \cite{Hurwitz:1898,Schafer:1966}, cf. \autoref{sec:octandfan}. The norm in each of these cases is given by $\mathbf{n}(x)=x\bar{x}$, where the involution $x\rightarrow\bar{x}$ is just given by  complex conjugation. We shall also be concerned with their \emph{split} cousins, $\mathds{C}^s$, $\mathds{H}^s$ and $\mathds{O}^s$, which have a split signature norm. Consequently, the norm is not positive definite and, while still composition, it is no longer division.

\subsection{Jordan algebras}\label{sec:JordanAlgebras}

Jordan algebras were originally introduced in \cite{Jordan:1933a} as a possible generalisation of the orthodox formulation of quantum mechanics in the hope of addressing certain difficulties in fundamental physics, particularly in the relativistic regime. We will come to the Jordan formulation of quantum mechanics in \autoref{sec:JordanQM}. However, they are interesting objects in their own right and an expansive literature on the subject has developed over the years. See \cite{Jacobson:1968,Gunaydin:2005zz} for a comprehensive account. Their intimate relationship with the exceptional Lie groups is of central importance in their applications to string theory and supergravity.

A vector space, defined over a ground field $\mathds{F}$, equipped with a bilinear product satisfying,
\begin{equation}\label{eq:Jid}
x\circ y =y\circ x; \qquad x^2\circ (x\circ y)=x\circ (x^2\circ y) \qquad\forall x, y \in \mathfrak{J},
\end{equation}
is a \emph{Jordan algebra} $\mathfrak{J}$. An obvious example is given by the set of real matrices with Jordan product defined as $x\circ y = \tfrac{1}{2}(xy+yx)$. More generally, this definition of the Jordan product may be used to construct a Jordan algebra starting from any associative algebra.

Any algebra is said to be \emph{formally real} if,
\begin{equation}\label{eq:formallyreal}
x^2+y^2+z^2\ldots=0 \qquad\implies \qquad x=y=z=\ldots=0.
\end{equation}
Assuming that a given Jordan algebra is formally real it can be shown that the Jordan identity \eqref{eq:Jid} is equivalent to power associativity \cite{Jordan:1933vh},
\begin{equation}\label{eq:powerasso}
x^m\circ x^n=x^{(m+n)}.
\end{equation}
This is significant when considering the application of Jordan algebras to quantum mechanics providing some physical motivation for the Jordan identity as will be discussed in \autoref{sec:JordanQM}.

The full classification of all formally real Jordan algebras was completed in \cite{Jordan:1933vh}. There are four infinite sequences of simple Jordan algebras and one exceptional case. (A Jordan algebra is simple if it contains no proper ideals. All Jordan algebras may be decomposed into a direct sum of simple Jordan algebras.)  Three of the infinite sequences are given by the sets $J_{n}^{\mathds{A}}$ of $n\times n$ Hermitian matrices defined over the three associative division algebras, $\mathds{A}=\mathds{R, C}$ or $\mathds{H}$. The Jordan product in these cases is simply given by $x\circ y = \half(xy+yx)$, where $xy$ denotes conventional matrix multiplication. The fourth is given by $\mathds{R}\oplus Q_n$, where $Q_n$ is a $n$-dimensional real vector space.  The one exceptional simple Jordan algebra is given by $J_{3}^{\mathds{O}}$, the set of $3\times 3$ Hermitian matrices defined over the octonions.

However, we will generally be concerned with the larger class of \emph{cubic} Jordan algebras which need not be formally real. For example, $J_{3}^{\mathds{O}^s}$, the set of $3\times 3$ Hermitian matrices defined over the split octonions, is not formally real but none the less underpins $\mathcal{N}=8$ supergravity.

\subsection{The Freudenthal-Springer-Tits construction of cubic Jordan algebras}
\label{sec:Spring}

In \cite{Springer:1962,McCrimmon:1969} it was shown how to construct a cubic Jordan algebra from any vector space equipped with a cubic form satisfying certain conditions described below.  We sketch this construction here, following closely the conventions of \cite{Krutelevich:2004}.

Let $V$ be a vector space,  defined over a ground field $\mathds{F}$, equipped with both a cubic norm, $N:V\rightarrow \mathds{F}$, satisfying $N(\lambda x)=\lambda^3N(x), \quad \forall \lambda \in \mathds{F}, x\in V$, and a base point $c\in V$ such that $N(c)=1$. If $N(x, y, z)$, referred to as the full \emph{linearisation} of $N$, defined by
\begin{equation}
N(x, y, z):=\tfrac{1}{6}\big(N(x+ y+ z)-N(x+y)-N(x+ z)-N(y+ z)+N(x)+N(y)+N(z)\big)
\end{equation}
is trilinear then  one may define the following four maps,
\begin{subequations}\label{eq:cubicdefs}
\begin{enumerate}
\item The trace,
    \begin{equation}
    \Tr:V\to\mathds{F}, \quad\Tr(x)=3N(c, c, x),
    \end{equation}
\item A quadratic map,
    \begin{equation}
    S:V\to\mathds{F}, \quad S(x)=3N(x, x, c),
    \end{equation}
\item A bilinear map,
    \begin{equation}
    S: V\times V\to\mathds{F}, \quad S(x, y)=6N(x, y, c),
    \end{equation}
\item A trace bilinear form,
    \begin{equation}\label{eq:tracebilinearform}
    \Tr:V\times V\to\mathds{F},  \quad \Tr(x, y)=\Tr(x)\Tr(y)-S(x, y).
    \end{equation}
\end{enumerate}
\end{subequations}
$N$ is said to be \emph{Jordan cubic} if,
\begin{enumerate}
\item The trace bilinear form \eqref{eq:tracebilinearform} is non-degenerate.
\item The quadratic adjoint map, $\#\colon\mathfrak{J}\to\mathfrak{J}$, uniquely defined by $\Tr(x^\#, y) = 3N(x, x, y)$, satisfies
\begin{equation}\label{eq:Jcubic}
    (x^{\#})^\#=N(x)x, \qquad \forall x\in \mathfrak{J}.
    \end{equation}
\end{enumerate}

A cubic Jordan algebra with multiplicative identity $\mathds{1}=c$ may be derived from any such vector space with a Jordan cubic form by defining the Jordan product,
\begin{equation}
x\circ y = \half\big(x\times y+\Tr(x)y+\Tr(y)x-S(x, y)\mathds{1}\big),
\end{equation}
where, $x\times y$ is the linearisation of the quadratic adjoint,
\begin{equation}
x\times y = (x+y)^\#-x^\#-y^\#.
\end{equation}

There are three groups, of particular importance, related to cubic Jordan algebras. The set of automorphisms, $\Aut(\mathfrak{J})$,  is composed of all linear transformations on $\mathfrak{J}$ that preserve the Jordan product,
\begin{equation}\label{eq:JAut}
x\circ y=z\quad\implies\quad g(x)\circ g(y)= g(z), \quad \forall g\in \Aut(\mathfrak{J}).
\end{equation}
The Lie algebra of $\Aut(\mathfrak{J})$ is given by the set of derivations, $\operatorname{Der}(\mathfrak{J})$, that is, all linear maps $D:\mathfrak{J}\rightarrow\mathfrak{J}$ satisfying the Leibniz rule,
\begin{equation}\label{eq:JDer}
D(x\circ y)=D(x)\circ y+x\circ D(y).
\end{equation}
The \emph{structure} group, $\Str(\mathfrak{J})$, is composed of all linear bijections on $\mathfrak{J}$ that leave the cubic norm $N$ invariant up to a fixed scalar factor,
\begin{equation}
N(g(x))=\lambda N(x), \quad \forall g\in \Str(\mathfrak{J}).
\end{equation}
Finally, the \emph{reduced structure} group $\Str_0(\mathfrak{J})$ leaves the cubic norm invariant and therefore consists of those elements in $\Str(\mathfrak{J})$ for which $\lambda =1$ \cite{Brown:1969}. The U-duality group of any $D=5$ supergravity with charge representation $\mathfrak{J}$ is given by $\Str_0(\mathfrak{J})$ \cite{Gunaydin:1983bi}.  Further, the scalar fields of these theories parameterise the symmetric coset spaces $\Str_0(\mathfrak{J})/\Aut(\mathfrak{J})$ and the U-duality charge vector orbits are given by  $\Str_0(\mathfrak{J})/H$, where $H$ is some subgroup of $\Aut(\mathfrak{J})$ \cite{Ferrara:1997uz}.

The usual concept of matrix rank may be generalised to cubic Jordan algebras and is invariant under both $\Str(\mathfrak{J})$ and $\Str_0(\mathfrak{J})$ \cite{Jacobson:1961, Krutelevich:2004}. Explicitly, for any element $x\in\mathfrak{J}$ we have,
\begin{equation}\label{eq:J3rank}
\begin{split}
&\operatorname{Rank} x =3\quad \text{iff}\quad N(x) \not=0,\\
&\operatorname{Rank} x =2\quad \text{iff}\quad  N(x) =0\quad\text{and}\quad x^{\#}\not=0,\\
&\operatorname{Rank} x =1\quad \text{iff}\quad x^{\#} =0\quad\text{and}\quad x\not=0,\\
&\operatorname{Rank} x =0\quad \text{iff}\quad x=0.
\end{split}
\end{equation}
Note, when using the alternative interpretation described in \autoref{sec:AlternativeJordan}, these classifications correspond directly to the different classes of entanglement of two qutrits under SLOCC, as shown in \autoref{tab:2QutritEntangClassif}, since one may consider the reduced density matrices, $\rho_A$ and $\rho_B$, as element of $J_{3}^{\mathds{C}}$.

\subsection{\texorpdfstring{$D=5$ black strings/holes and Jordan algebras}{D=5 black holes and Jordan algebras}}
\label{sec:D=5jordan}

The elements of the cubic Jordan algebras $J_3^{\mathds{A}}$ of degree three are $3 \times 3$ hermitian real $\mathds{A}$ matrices:
\begin{equation}\label{eq:Jordan}
J_3(P)=
\begin{pmatrix}
p^1            & P_v            & \overline{P_s} \\
\overline{P_v} & p^2            & P_c            \\
P_s            & \overline{P_c} & p^3
\end{pmatrix},
\quad \mathrm{where}\quad p^i \in \mathds{R}\quad \mathrm{and}\quad P_{s,c,v} \in \mathds{A},
\end{equation}
where $\mathds{A}$ is one of the normed division algebras, $\mathds{R, C, H, O}$ or the split composition algebras, $\mathds{C}^s, \mathds{H}^s, \mathds{O}^s$. For two elements $X$ and $Y$ in $J_3^{\mathds{A}}$ the Jordan product is given by $X\circ Y= \frac{1}{2}(XY+YX)$, where $XY$ is just the conventional matrix product. The bilinear trace form and quadratic adjoint are given respectively by,
\begin{gather}
\Tr(X,Y)=\Tr(X\circ Y),
\shortintertext{and}
X^{\#}=X^2-\tr (X)X+\half[(\tr X)^2-\tr (X^2)]\mathds{1}.
\end{gather}
The cubic norm $N(J_3(P))$ is given by the appropriate generalisation of the standard matrix determinant,

\begin{equation}\label{eq:Ncubic}
N=p^1p^2p^3 - (p^1P_s\overline{P}_s+p^2P_c\overline{P}_c + p^3P_v\overline{P}_v) + P_sP_cP_v+\overline{P}_s\overline{P}_c\overline{P}_v \equiv I_3(P),
\end{equation}
which just reduces to the conventional determinant definition when $\mathds{A=R, C}$.

The elements $J_3(P)\in J_3^{\mathds{A}}$ transform as the $(3\dim\mathds{A}+3)$ dimensional representation of the norm preserving group, $\Str_0({J_3^{\mathds{A}}})=SL(3,\mathds{A})$. For $\mathds{A=R, C, H, O}$   ($\dim\mathds{A}=1,2,4,8$, respectively) $J_3^{\mathds{A}}$ transforms as the \textbf{6, 9, 15, 27} of $SL(3,\mathds{R})$, $SL(3,\mathds{C})$, $SU^{*}(6)$, $E_{6(-26)}$, respectively. These are the symmetries of the magic $\mathcal{N}=2,D=5$ supergravities \cite{Gunaydin:1984ak,Gunaydin:1983bi,Gunaydin:1983rk} and the magnetic black string charges fall into the corresponding representations. The octonionic case corresponds to the decomposition
\begin{gather}
E_{6(-26)} \supset SO(8),
\shortintertext{under which}
\mathbf{27 \to 1+1+1+8_s+8_c+8_v},\label{eq:27E6underSO8}
\end{gather}
where  $p^1$, $p^2$, $p^3$ are the singlets and $P_s$, $P_c$, $P_v$ are the $\mathbf{8_s, 8_c, 8_v}$. The octonions may be written
\begin{gather}
P^x=P^0_x+P^{1}_xe_1+P^2_xe_2+P^3_xe_3+P^4_xe_4+P^5_xe_5+P^6_xe_6+P^7_xe_7,
\shortintertext{with norm}
P^x\overline{P}^x = {P^0_x}^2+{P^1_x}^2+{P^2_x}^2+{P^3_x}^2+{P^4_x}^2+{P^5_x}^2+{P^6_x}^2+{P^7_x}^2.
\end{gather}
The $\mathcal{N}=8$ case then also follows \cite{Ferrara:1997uz} by using the split octonions in $J_3^{\mathds{A}}$, in which case $\Str_0(J_3^{\mathds{O}^s})=E_{6(6)}$, the $\mathcal{N}=8$ U-duality group. The split octonions have norm
\begin{gather}
P^x\overline{P}^x = {P^0_x}^2+{P^1_x}^2+{P^2_x}^2+{P^3_x}^2-{P^4_x}^2-{P^5_x}^2-{P^6_x}^2-{P^7_x}^2.
\shortintertext{There we have}
E_{6(6)} \supset SO(4,4),
\shortintertext{under which}
\mathbf{27 \to 1+1+1+8_s+8_c+8_v}.
\end{gather}
In all cases the black string entropy is
\begin{equation}
S=\pi \sqrt{|I_3(P)|},
\end{equation}
where $I_3(P)$ is the cubic invariant $N(J_3(P))$ as given in \eqref{eq:Ncubic}.

For the electric black holes, we have the conjugate Jordan matrix
\begin{gather}
J_3(Q)=
\begin{pmatrix}
q_1            & Q_v            & \overline{Q}_s \\
\overline{Q}_v & q_2            & Q_c            \\
Q_s            & \overline{Q}_c & q_3
\end{pmatrix},
\shortintertext{and the entropy is}
S=\pi \sqrt{|I_3(Q)|}.
\end{gather}

\subsubsection{U-duality orbits and Jordan algebras}

Any element $J_3(P)\in J_{3}^{\mathds{A}}$, for $\mathds{A=R,C,H,O}$, may be diagonalised using a $\Aut(J_{3}^{\mathds{A}})$ transformation. In the exceptional octonionic case this corresponds to a $F_4$ transformation, as was shown explicitly in \cite{Gunaydin:1978}. When $\mathds{A=C}$ this operation is related to the Schmidt decomposition of a two-qutrit system. For a detailed discussion of the $D=5$ magic supergravity charge orbits see \cite{Ferrara:1997uz}. For the $\mathcal{N}=8$ case $E_{6(6)}$ acts transitively on the classes of elements of rank 1 or 2 \eqref{eq:J3rank}, the small black holes. For large black holes, those with non-vanishing entropy corresponding to rank 4 elements of $J_{3}^{\mathds{O}^s}$,  $E_{6(6)}$ acts transitively on elements of a given entropy (cubic norm) $I_3$ \cite{Jacobson:1961, Krutelevich:2004}. Any element $J_3(P)\in J_{3}^{\mathds{O}^s}$ may be diagonalised using a $F_{4(4)}$ transformation \cite{Gunaydin:1978} and the representative elements of each of the orbits \cite{Ferrara:1997uz} may be chosen as in \autoref{tab:OrbitRepresentatives} (where $k=I_3\not= 0$).
\begin{table}[ht]
\begin{tabular*}{\textwidth}{@{\extracolsep{\fill}}*{5}{c}}
\toprule
& Rank & Rep            & Orbit                                        & \\
\midrule
& 0    & $\diag(0,0,0)$ & $\{0\}$                                      & \\
& 1    & $\diag(1,0,0)$ & $E_{6(6)}/(O(5,5) \ltimes \mathds{R}^{16})$  & \\
& 2    & $\diag(1,1,0)$ & $E_{6(6)}/(O(5,4) \ltimes \mathds{R}^{16} )$ & \\
& 3    & $\diag(1,1,k)$ & $E_{6(6)}/F_{4(4)}$                          & \\
\bottomrule
\end{tabular*}
\caption[Orbit representatives]{Orbit representatives of $(D=5, \mathcal{N}=8)$, see \autoref{tab:n=8bh5} for details. Each orbit is specified by a Jordan algebra element of a given rank.}
\label{tab:OrbitRepresentatives}
\end{table}
These will appear as the charge orbits of \autoref{tab:n=8bh5} in  the conventional black hole/qutrit correspondence. When making the alternative two-qutrit comparison of \autoref{sec:AlternativeJordan}, the rank corresponds to the Schmidt number of the reduced density matrix \cite{Dur:2000}.

In the quantum theory the black hole charges become integer valued and, consequently, the relevant space is the set of $3\times 3$ Hermitian matrices defined over the \emph{integral} split octonions  $J_{3}^{\mathds{O}^{s}_{\mathds{Z}}}$. The U-duality is broken to $E_{6(6)}(\mathds{Z})$, the norm preserving group in the integral case. It was suggested in \cite{Maldacena:1999bp} that any charge vector, with non-vanishing cubic norm, in the integral theory could be brought in to the standard form originally used in \cite{Strominger:1996sh}. This amounts to the diagonalisability of $3\times 3$ Hermitian matrices defined over the integral split octonions. It was shown in \cite{Krutelevich:2002} that an arbitrary  $J_3(P)$ in $J_{3}^{\mathds{O}^{s}_{\mathds{Z}}}$ is equivalent, under $E_{6(6)}(\mathds{Z})$, to an element of the form,
\begin{equation}
J_3(P)=\textrm{diag}(p^1, p^2, p^3), \qquad p^i\geq 0 \in\mathds{Z},
\end{equation}
where $p^i|p^{(i+1)}$ and all zeros on the diagonal appear in the lower right corner. Interestingly, unlike the continuous case, valid in the low energy effective field theory approximation,  $E_{6(6)}(\mathds{Z})$ acts transitively on elements of cubic norm $n$ if and only if $n$ is squarefree \cite{Krutelevich:2004}. Otherwise said, there exist distinct black hole configurations with the same entropy that are \emph{not} related by the discrete U-duality group.

\subsection{The Freudenthal triple system}

Dimensionally reducing from five to four dimensions the black hole charge configurations of $\mathcal{N}=8$ supergravity may be represented using the FTS, originally introduced in \cite{Freudenthal:1954}, defined over the corresponding $D=5$ Jordan algebra. Following the conventions of \cite{Brown:1969, Krutelevich:2004}, a FTS may always be constructed from a cubic Jordan algebra $\mathfrak{J}$ as follows. Given a cubic Jordan algebra $\mathfrak{J}$ over $\mathds{R}$ define the vector space $\mathfrak{M}(\mathfrak{J})$,
\begin{equation}
\mathfrak{M}(\mathfrak{J})=\mathds{R}\oplus \mathds{R}\oplus \mathfrak{J}\oplus \mathfrak{J}.
\end{equation}
An arbitrary element $x\in \mathfrak{M}(\mathfrak{J})$ may be written as a ``$2\times 2$ matrix'',
\begin{equation}
x=\begin{pmatrix}\alpha&X\\Y&\beta\end{pmatrix} \quad\text{where} ~\alpha, \beta\in\mathds{R}\quad\textrm{and}\quad X, Y\in \mathfrak{J}.
\end{equation}
The bilinear antisymmetric quadratic form $\{x, y\}$ is defined as,
\begin{gather}
\{x, y\}=\alpha\delta-\beta\gamma +\Tr(X, Z) -\Tr(Y, W),\label{eq:bileniearform}
\shortintertext{where,}
\begin{align}
x&=\begin{pmatrix}\alpha&X\\Y&\beta\end{pmatrix},&
y&=\begin{pmatrix}\gamma&W\\Z&\delta\end{pmatrix},
\end{align}
\end{gather}
and the trace bilinear form \eqref{eq:tracebilinearform} is defined as on $\mathfrak{J}$. The quartic norm is given by,
\begin{equation}\label{eq:quarticnorm}
q(x)=-2[\alpha\beta-\Tr(X, Y)]^2 -8[\alpha N(X)+\beta N(Y)-\Tr(X^\#, Y^\#)],
\end{equation}
where all the necessary definitions are inherited from the underlying Jordan algebra $\mathfrak{J}$. A symmetric four-linear form, $q(x, y, w, z)$, is obtained by linearising \eqref{eq:quarticnorm} so that $q(x, x, x, x)=q(x)$. The non-degeneracy of both $q(x, y, w, z)$ and $\{x, y\}$ allows one to uniquely define a trilinear map, the \emph{triple product}, $T:\mathfrak{M}(\mathfrak{J})\times\mathfrak{M}(\mathfrak{J})\times\mathfrak{M}(\mathfrak{J})\rightarrow\mathfrak{M}(\mathfrak{J})$, by \cite{Brown:1969},
\begin{equation}
\{T(x, y, w), z\}=q(x, y, w, z).
\end{equation}

The \emph{automorphism group} $\Aut(\mathfrak{M}(\mathfrak{J}))$ is given by the set of all transformations which leave both $\{x, y\}$ and $q(x, y, w, z)$ invariant. When $\mathfrak{J}$ is either $\mathds{R}\oplus Q_n$ or $J_{3}^{\mathds{A}}$ the set of transformations $\Aut(\mathfrak{M}(\mathfrak{J}))$ forms a Lie group, and $\mathfrak{M}(\mathfrak{J})$ a corresponding irreducible representation as described in \autoref{tab:freliegroups}.
\begin{table}[ht]
\begin{tabular*}{\textwidth}{@{\extracolsep{\fill}}*{6}{M{c}}}
\toprule
& \mathfrak{J} & \dim\mathfrak{J} & \Aut(\mathfrak{M}(\mathfrak{J})) & \dim\mathfrak{M}(\mathfrak{J}) & \\
\midrule
& \mathds{R}\oplus\mathds{R}\oplus\mathds{R} & 3 & [SL(2,\mathds{R})]^3 & 8 & \\
& \mathds{R}\oplus Q_n & 1+n &  SL(2,\mathds{R})\times SO(2,n) & 2n+4 & \\
& J_{3}^{\mathds{R}^{(s)}} & 6  & C_3 & 14 & \\
& J_{3}^{\mathds{C}^{(s)}} & 9  & A_5 & 20 & \\
& J_{3}^{\mathds{H}^{(s)}} & 15 & D_6 & 32 & \\
& J_{3}^{\mathds{O}^{(s)}} & 27 & E_7 & 56 & \\
\bottomrule
\end{tabular*}
\caption[Cubic Jordan algebra dimensionalities]{The Lie group and the dimension of its representation given by the Freudenthal construction defined over the cubic Jordan algebra $\mathfrak{J}$.}\label{tab:freliegroups}
\end{table}
When $\mathds{A=R,C,H,O}$ and $\mathfrak{J}=J_{3}^{\mathds{A}}$, the  group $\Aut(\mathfrak{M}(\mathfrak{J}))$ is generated by the following three maps \cite{Brown:1969}:
\begin{subequations}
\begin{align}\label{eq:FreudenthalConstructionTransformations}
\phi(Z) &: \begin{pmatrix}\alpha&X\\Y&\beta\end{pmatrix} \mapsto \begin{pmatrix}\alpha+(Y,Z)+(X, Z^{\#})+\beta N(Z)&X+\beta Z\\
                                                                                 Y+X\times Z+\beta Z^\# &\beta
                                                                 \end{pmatrix},\\
\psi(Z) &: \begin{pmatrix}\alpha&X\\Y&\beta\end{pmatrix} \mapsto \begin{pmatrix}\alpha& X+Y\times Z+\alpha Z^\#\\
                                                                                Y+\alpha Z &\beta+(X,Z)+(Y, Z^{\#})+\alpha N(Z)
                                                                 \end{pmatrix},\\
T(s) &: \begin{pmatrix}\alpha&X\\Y&\beta\end{pmatrix}    \mapsto \begin{pmatrix}\lambda^{-1}\alpha & s(X)\\
                                                                                {s^*}^{-1}(Y) &\lambda\beta
                                                                 \end{pmatrix}.
\end{align}
\end{subequations}
where $s\in \Str(\mathfrak{J})$  and $s^*$ is its adjoint defined with respect to the trace bilinear form, $\Tr(X, s(Y))=\Tr(s^*(X), Y)$.

The rank of an element $x\in\mathfrak{M}(\mathfrak{J})$ may be uniquely defined using the relations \cite{Krutelevich:2004},
\begin{equation}\label{eq:FTSrank}
\begin{split}
&\operatorname{Rank} x =4\quad     \text{iff}\quad q(x) \not=0,\\
&\operatorname{Rank} x \leq 3\quad \text{iff}\quad q(x) =0,\\
&\operatorname{Rank} x \leq 2\quad \text{iff}\quad T(x, x, x)=0,\\
&\operatorname{Rank} x \leq 1\quad \text{iff}\quad 3T(x, x, y) + \{x, y\}x=0, \qquad \forall y \in \mathfrak{M}(\mathfrak{J}),\\
&\operatorname{Rank} x = 0\quad \text{iff}\quad x=0.
\end{split}
\end{equation}
The rank, defined in this manner, is invariant under $\Aut(\mathfrak{M}(\mathfrak{J}))$. Further, when $\mathfrak{J}=J_{3}^{\mathds{A}}$ for $\mathds{A}=\mathds{C}^{s},\mathds{H}^{s},\mathds{O}^{s}$, $\Aut(\mathfrak{M}(\mathfrak{J}))$ acts transitively on the sets of elements of rank 1, 2 or 3 and on elements of a given norm $q$ in the rank 4 case. These orbits union $x=0$ partition the whole space $\mathfrak{M}(J_{3}^{\mathds{A}})$ \cite{Krutelevich:2004}.

\subsection{\texorpdfstring{$D=4$ black holes and Freudenthal triples}{D=4 black holes and Freudenthal triples}}
\label{sec:D=4freud}

In $D=4$ the black hole charges may be described by the Freudenthal triple system \cite{Ferrara:1997uz} realised as $2 \times 2$ ``matrices''
\begin{equation}\label{eq:FTS}
F(p,q)=\begin{pmatrix}-q_0    & J_3(P) \\J_3(Q) & p^0\end{pmatrix},
\end{equation}
where $p^{0}$ and $q_{0}$ are real numbers that correspond to the graviphoton charges appearing after dimensional reduction from $D=5$. The magic $\mathcal{N}=2,D=4$, given by $\mathfrak{J}=J_{3}^{\mathds{A}}$,  $\mathds{A=R,C,H,O}$, supergravities have symmetries $Sp(6,\mathds{R})$, $SU(3,3)$, $SO^{*}(12)$, and $E_{7(-25)}$ respectively, while the $\mathcal{N}=8$ case, given by $\mathds{A}=\mathds{O}^s$,  has $E_{7(7)}$.  The charge representations have dimensions $(6\dim\mathds{A}+8)$ and correspond to the threefold antisymmetric traceless tensor $(\textbf{14}^\prime)$ of $Sp(6,\mathds{R})$, the threefold antisymmetric self-dual tensor (\textbf{20}) of $SU(3,3)$, the chiral spinor (\textbf{32}) of $SO^{*}(12)$ and the fundamental (\textbf{56}) of $E_{7(-25)}$ or $E_{7(7)}$ as indicated in \autoref{tab:freliegroups}. The real case corresponds to the decomposition
\begin{gather}
Sp(6,\mathds{R}) \supset SL(3,\mathds{R}),
\shortintertext{under which}
\mathbf{14 \to 1+1+6+6'},
\end{gather}
where $p^0$ $q_0$ are the singlets and $J_3(P)$, $J_3(Q)$ are the \textbf{6}, $\mathbf{6}'$. The complex case corresponds to the decomposition
\begin{gather}
SU(3,3) \supset SL(3,\mathds{C}),
\shortintertext{under which}
\mathbf{20^\prime \to 1+1+9+9'}.
\end{gather}
The quaternionic case corresponds to the decomposition
\begin{gather}
SO^*(12) \supset SU^*(6),
\shortintertext{under which}
\mathbf{32 \to 1+1+15+15'}.
\end{gather}
The octonionic case corresponds to the decomposition
\begin{gather}
E_{7(-25)} \supset E_{6(-26)},
\shortintertext{under which}
\mathbf{56 \to 1+1+27+27'}.\label{eq:56ofE7underE6}
\end{gather}
The $\mathcal{N}=8$ case then also follows \cite{Ferrara:1997uz} by using $E_{7(7)}$ and the split octonions in \eqref{eq:Jordan} and \eqref{eq:FTS}. We have
\begin{gather}
E_{7(7)} \supset E_{6(6)},
\shortintertext{under which}
\mathbf{56 \to 1+1 +27+27'}.
\end{gather}
Finally, the $\mathcal{N}=2$ $STU$ model is given by $\mathfrak{J}=\mathds{R}\oplus\mathds{R}\oplus\mathds{R}$, in which case the black hole charge representation has dimension 8 and transforms as the $\mathbf{(2,2,2)}$ of $[SL(2,\mathds{R})]^3$.

In all cases the black hole entropy is
\begin{equation}
S=\pi\sqrt{|I_4|},
\end{equation}
where $I_4$ is Cartan's quartic invariant given by half the quartic norm \eqref{eq:quarticnorm},
\begin{equation}\label{eq:E7InvariantFreudenthalBasis}
\begin{split}
I_4(p^0,P;q_0,Q)=&-\left[p^0q_0+\tr(J_3(P)\circ J_3(Q))\right]^2 \\
&+4\left[-p^0 J_3(Q)+q_0 J_3(P)+\tr({J_3}^{\#}(P)\circ {J_3}^{\#}(Q))\right].
\end{split}
\end{equation}
Note that \eqref{eq:E7InvariantFreudenthalBasis} implies that in the quantum theory with integral charges $I_4$ is itself an integer, $I_4=4n$ or $4n+1$ where $n$ is an integer\footnote{So the square of the black hole entropy is quantised, at least to this order.}. Explicitly we have
\begin{equation}
\begin{split}
I_4&=-\Big[p.q+2\left((P_v\cdot Q_v)+(P_s\cdot Q_s)+(P_c\cdot Q_c)\right)\Big]^2 \\
&\phantom{=}+4\Big[-p^0 q_1 q_2 q_3 + q_0 p^1 p^2 p^3+(p^1q_1)(p^2q_2)+(p^1q_1)(p^3q_3)+(p^3q_3)(p^2q_2) \\
&\phantom{=+4\Big[}-(p^2p^1-p^0q_3){Q_v}^2 - (q_2q_1+p^3q_0){P_v}^2 \\
&\phantom{=+4\Big[}-(p^2p^3-p^0q_1){Q_c}^2 - (q_2q_3+p^1q_0){P_c}^2 \\
&\phantom{=+4\Big[}-(p^1p^3-p^0q_2){Q_s}^2 - (q_1q_3+p^2q_0){P_s}^2 \\
&\phantom{=+4\Big[}-2p^0\Re(Q_vQ_sQ_c)-2q_0\Re(P_cP_sP_v) \\
&\phantom{=+4\Big[}+{P_v}^2{Q_v}^2+{P_s}^2{Q_s}^2+{P_c}^2{Q_c}^2 \\
&\phantom{=+4\Big[}+(\overline{P}_s\overline{P}_c-p^3P_v)(Q_cQ_s-q_3\overline{Q}_v) + (\overline{Q}_s\overline{Q}_c-q_3Q_v)(P_cP_s-p^3\overline{P}_v) \phantom{\Big]} \\
&\phantom{=+4\Big[}+(\overline{P}_s\overline{P}_c-p^3P_v)(Q_cQ_s-q_3\overline{Q}_v) + (\overline{Q}_s\overline{Q}_c-q_3Q_v)(P_cP_s-p^3\overline{P}_v) \phantom{\Big]} \\
&\phantom{=+4\Big[}+(\overline{P}_s\overline{P}_c-p^3P_v)(Q_cQ_s-q_3\overline{Q}_v) + (\overline{Q}_s\overline{Q}_c-q_3Q_v)(P_cP_s-p^3\overline{P}_v) \Big],
\end{split}
\end{equation}
where $P\cdot Q=\half(P\overline{Q}+Q\overline{P})$ and $P^2=P\overline{P}$.

\subsubsection{\texorpdfstring{$\mathcal{N}=2$ $STU$ subsectors}{N=2 STU subsectors}}

The $STU$ model may, in addition to the realisation using $\mathfrak{J}=\mathds{R}\oplus\mathds{R}\oplus\mathds{R}$, be obtained as a consistent truncation of the full $\mathcal{N}=8$ theory. This corresponds to the simple case where we put $P_{s,v,c},Q_{s,v,c}$ all to zero, then
\begin{gather}
F(P,Q)=\begin{pmatrix}-q_{0}& J_3(p^{i}) \\ J_3(q_{i}) & p^{0}\end{pmatrix},\label{eq:STU/FTS}
\shortintertext{where}
\begin{align}
{J_3}(p^i)&=\diag(p^1, p^2, p^3),& {J_{3}}(q_i)&=\diag(q_1, q_2, q_3).
\end{align}
\shortintertext{In this case,}
\begin{align}
I_3(P)&=p^1p^2p^3, &I_3(Q)&=q_1q_2q_3,
\end{align}
\shortintertext{and}
\begin{align}
{J_3}^{\#}(P)&=\diag(p^2p^3, p^1p^3, p^1p^2), & {J_3}^{\#}(Q)&=\diag(q_2q_3, q_1q_3, q_1q_2),
\end{align}
\shortintertext{and $I_4$ becomes}
I_4 =-(p\cdot q)^2 +4\left((p^1q_1)(p^2q_2)+(p^1q_1)(p^3q_3)+(p^3q_3)(p^2q_2)\right)- 4 p^0 q_1 q_2 q_3 + 4q_0 p^1 p^2 p^3. \label{eq:Cayleyfreud}
\end{gather}
If we make the identifications \eqref{eq:charges7}, we recover Cayley's hyperdeterminant \eqref{eq:CayleyHyperdeterminant}. Combined with \eqref{eq:MiniDict} we obtain the transformation between $P,Q$ and $p,q$:
\begin{equation}
\begin{bmatrix}
p^0 \\ p^1 \\ p^2 \\ p^3 \\ q_0 \\ q_1 \\ q_2 \\ q_3
\end{bmatrix}= \frac{1}{\sqrt{2}}
\begin{bmatrix*}[r]
P^0-P^2 \\ Q_0+Q_2 \\ P^3-P^1 \\ -P^3-P^1 \\ Q_0-Q_2 \\ -P^0-P^2 \\ Q_3-Q_1 \\ -Q_3-Q_1
\end{bmatrix*}.
\end{equation}
This transformation gives us the relations:
\begin{gather}
\begin{array}{c@{\ =\ }c@{\ }c@{\ }c}
P^2      & 2(p^2p^3 & - & p^0q_1),   \\
P\cdot Q & p\cdot q & - & 2 p^1 q_1, \\
Q^2      & 2(p^1q_0 & + & q_2q_3),
\end{array}
\shortintertext{hence we find}
I_4=P^2 Q^2 -(P\cdot Q)^2,
\end{gather}
which is manifestly invariant under $SL(2) \times SO(2,2)$. In the Fano basis this is equivalent to just keeping one point of the dual Fano plane i.e. $a_{ABD}$.

\subsubsection{\texorpdfstring{The $\mathcal{N}=4$ subsectors}{The N=4 subsectors}}

The next simplest case is where we keep just the $s$ octonions. This corresponds to a consistent truncation to the $\mathcal{N}=4$ theory. In this case,
\begin{gather}
\begin{align}
I_3(P)&=p^1p^2p^3-p^1P_s\overline{P}_s, & I_3(Q)&=q_1q_2q_3-q_1Q_s\overline{Q}_s,
\end{align}
\shortintertext{and}
\begin{align}
{J_3}^{\#}(P)&=
\begin{pmatrix}
p^2p^3-P_s\overline{P}_s & 0                  & 0       \\
0                        & p^1p^3             & -p^1P_s \\
0                        & -p^1\overline{P}_s & p^1p^2
\end{pmatrix}, &
{J_3}^{\#}(Q)&=
\begin{pmatrix}
q_2q_3-Q_s\overline{Q}_s & 0                  & 0       \\
0                        & q_1q_3             & -q_1Q_s \\
0                        & -q_1\overline{Q}_s & q_1q_2
\end{pmatrix}.
\end{align}
\end{gather}
Therefore $I_4$ becomes
\begin{gather}
\begin{split}
I_4 &=-(p^0q_0+p^iq_i+2(P_{s}\cdot Q_s))^{2}+4\Big[q_0 (p^1 p^2 p^3-p^1{P_{s}}^2)-p^0 (q_1 q_2 q_3 - q_1{Q_s}^2) \\
&\phantom{=}+(p^2p^3-{P_s}^2)(q_2q_3-{Q_s}^2) + p^1p^3q_1q_3+p^1p^2q_1q_2+2p^1q_1(P_s\cdot Q_s)\Big],
\end{split}
\shortintertext{where}
\begin{split}
{P_s}^2      &= P_s\overline{P}_s, \\
P_s\cdot Q_s &= \half(P_s\overline{Q}_s+\overline{Q}_sP_s), \\
{Q_s}^2      &= Q_s\overline{Q}_s.
\end{split}
\shortintertext{Hence we find}
I_4=P^2Q^2 - (P\cdot Q)^2 - 2(P^2{Q_s}^2+Q^2{P_s}^2)+4({P_s}^2{Q_s}^2 -P\cdot QP^2\cdot Q_s-(P_s\cdot Q_s)^2).
\shortintertext{So if we identify}
P_s=\tfrac{1}{\sqrt{2}}(P_{s}^4, P_{s}^5, P_{s}^6, P_{s}^7, P_{s}^8, P_{s}^9, P_{s}^{10}, P_{s}^{11}),
\shortintertext{and}
Q_s=\tfrac{1}{\sqrt{2}}(Q_{s}^4, Q_{s}^5, Q_{s}^6, Q_{s}^7, Q_{s}^8, Q_{s}^9, Q_{s}^{10}, Q_{s}^{11}),
\shortintertext{then}
I_4=P^2Q^2 -(P\cdot Q)^2,
\end{gather}
where the indices now run over $0,\dotsc,11$, which is manifestly invariant under $SL(2) \times SO(2,10)$ or $SL(2) \times SO(6,6)$ according to whether we use the octonions or split octonions. In the Fano basis this is equivalent to just keeping one line $a_{ABD}$ of the Fano plane. In constructing the dictionary between these bases we shall see that this particular case is given by the line $(a_{ABD}, c_{CDF}, d_{DEG})$ defined by the common qubit $D$.

All-in-all there are three $\mathcal{N}=4$ subsectors, transforming as the $\mathbf{(2,12)}$ of $SL(2)\times SO(6,6)$ related to $a_{ABD}$, one for each line defined by the qubits $A,B$ and $D$:
\begin{subequations}
\begin{enumerate}
  \item Keeping only $(p^i, q_i; P_c, Q_c)$:
    \begin{equation}
    \begin{split}
      I_4&=-(P_c\cdot Q_c)^2 \\
      &\phantom{=}+4\big[(p^2p^3-p^0q_1)(q_2q_3+p^1q_0) - (p\cdot q-2p^1q_1)^2 +{P_c}^2{Q_c}^2 \\
      &\phantom{=}-(p^2p^3-p^0q_1){Q_c}^2 - (q_2q_3+p^1q_0){P_c}^2 -(p\cdot q-2p^1q_1)P_c\cdot Q_c\big].
    \end{split}
    \end{equation}
  \item Keeping only $(p^i, q_i; P_v, Q_v)$:
    \begin{equation}
    \begin{split}
      I_4&=-(P_v\cdot Q_v)^2 \\
      &\phantom{=}+4\big[(p^2p^1-p^0q_3)(q_2q_1+p^3q_0) - (p\cdot q-2p^3q_3)^2 +{P_v}^2{Q_v}^2\\
      &\phantom{=}-(p^2p^1-p^0q_3){Q_v}^2 - (q_2q_1+p^3q_0){P_v}^2 -(p\cdot q-2p^3q_3)P_v\cdot Q_v\big].
    \end{split}
    \end{equation}
  \item Keeping only$(p^i, q_i; P_s, Q_s)$:
    \begin{equation}
    \begin{split}
      I_4&=-(P_s\cdot Q_s)^2 \\
      &\phantom{=}+4\big[(p^1p^3-p^0q_2)(q_1q_3+p^2q_0) - (p\cdot q-2p^2q_2)^2 +{P_s}^2{Q_s}^2 \\
      &\phantom{=}-(p^1p^3-p^0q_2){Q_s}^2 - (q_1q_3+p^2q_0){P_s}^2 -(p\cdot q-2p^2q_2)P_s\cdot Q_s\big].
    \end{split}
    \end{equation}
\end{enumerate}
\end{subequations}
Note, the third line of each these equations is related to one of the three equations \eqref{eq:H_i} derived using the transvectants of \autoref{sec:Transvectants}.

Turning our attention now to the complementary situation in which we \emph{exclude} Alice, that is we keep only $(P_s, Q_s)$ and $(P_c, Q_c)$ in the FTS, we find,
\begin{equation}\label{eq:bcdfN=2Subsector}
\begin{gathered}
I_4=4\big[{P_s}^2{Q_s}^2+{P_c}^2{Q_c}^2-(P_s.Q_s)^2-(P_c.Q_c)^2-2(P_s.Q_s)(P_c.Q_c)\big] \\
+4\big[(\overline{P}_s\overline{P}_c)(Q_cQ_s)+(\overline{Q}_s\overline{Q}_c)(P_cP_s)\big].
\end{gathered}
\end{equation}
This corresponds to the complementary $\mathbf{(1,32)}$ of $SL(2)\times SO(6,6)$. The qubit subsector is given by keeping the 32 numbers $(b_{BCE}, c_{CDF}, d_{DEG}, f_{FBG})$. Note the absence of Alice. This is the quadrangle of the dual Fano plane complementary to the line defined by $A$.

\subsubsection{U-duality charge orbits and the FTS}

The U-duality orbits of $\mathcal{N}=8, D=4$ supergravity were calculated in \cite{Ferrara:1997uz} and are presented in \autoref{tab:n=8bh}.  It is not difficult to show \cite{Krutelevich:2004} that using the $E_{7(7)}$ transformations given in \eqref{eq:FreudenthalConstructionTransformations} any non-zero charge configuration
\begin{gather}
F(p,q)=
\begin{pmatrix}
-q_0   & J_3(P) \\
J_3(Q) & p^0
\end{pmatrix},
\shortintertext{may be put in the form,}\
\tilde{F}(p,q)=
\begin{pmatrix}
1 & \tilde{J}_3(P) \\
0 & \tilde{p}^0
\end{pmatrix}.\label{eq:FTSreduced}
\end{gather}
We then have that for each rank 1 to 4, as defined in \eqref{eq:FTSrank}, the configuration \eqref{eq:FTSreduced} is equivalent under $E_{7(7)}$ to one of the standard configurations of \autoref{tab:StandardConfigurations} \cite{Krutelevich:2004}.
\begin{table}[ht]
\begin{tabular*}{\textwidth}{@{\extracolsep{\fill}}*{5}{c}}
\toprule
& Rank & FTS & State & \\
\midrule
& 1 & $\begin{pmatrix}1 & \diag(0,0,0)\\ \diag(0,0,0)& 0\end{pmatrix}$   & $\ket{111}$                                   & \\[8pt]
& 2 & $\begin{pmatrix}1 & \diag(1,0,0)\\ \diag(0,0,0)& 0\end{pmatrix}$   & $\ket{111}+\ket{001}$                         & \\[8pt]
& 3 & $\begin{pmatrix}1 & \diag(1,1,0)\\ \diag(0,0,0)& 0\end{pmatrix}$   & $\ket{111}+\ket{001}+\ket{010}$               & \\[8pt]
& 4 & $\begin{pmatrix}1 & \diag(1,1,p^3)\\ \diag(0,0,0)& 0\end{pmatrix}$ & $\ket{111}+\ket{001}+\ket{010} +p^3\ket{100}$ & \\
\bottomrule
\end{tabular*}
\caption[Standard configurations]{Standard charge configurations, their corresponding rank \eqref{eq:FTSrank} and the equivalent three-qubit states, which belong to distinct entanglement classes. }
\label{tab:StandardConfigurations}
\end{table}
It was proved in \cite{Krutelevich:2004} that $E_{7(7)}$ acts transitively on the elements of rank 1, 2, 3 and on elements of norm $q$ in the rank 4 case. Further, these orbits are distinct and their union with the zero element gives the full charge space. Note, each rank corresponds to a three-qubit entanglement class \cite{Borsten:2008fts}, described in \autoref{tab:3QubitEntangClassif}. The rank 1, 2 and 3 cases are equivalent to the separable $\ket{111}$, bi-partite entangled $\ket{111}+\ket{001}$ and W $\ket{111}+\ket{001}+\ket{010}$ states respectively. The rank 4 case is equivalent to a GHZ $\ket{111}+\ket{001}+\ket{010} +p^3\ket{100}$ state with  unnormalised 3-tangle \eqref{eq:3-tangle} given by $16|p^3|$.

\subsubsection{\texorpdfstring{$[SL(2)]^3\subset E_7$ transformations}{[SL(2)]3 subset E7 transformations}}

The $\phi$ and $\psi$ transformations of the Freudenthal construction may be restricted to $[SL(2)]^3\subset E_7$ transformations by requiring the $\mathfrak{J}$ transformation parameter to be diagonal. This follows directly from their definitions \eqref{eq:FreudenthalConstructionTransformations} and from the fact that a general three-qubit state is of the form \eqref{eq:General8ParamStateInFTSForm}. The $X+\beta Z$ and $Y+\alpha Z$ elements of \eqref{eq:FreudenthalConstructionTransformations} clearly demonstrate that only diagonal transformation parameters can preserve the form of the $\mathfrak{J}$ slots.
\begin{equation}\label{eq:General8ParamStateInFTSForm}
\begin{pmatrix}-a_7 & -\diag(a_1, a_2, a_4) \\ \diag(a_6, a_5, a_3) & a_0\end{pmatrix} =
\begin{pmatrix}-q_0 & \diag(p^1,p^2,p^3) \\ \diag(q_1,q_2,q_3) & p^0\end{pmatrix}.
\end{equation}
In this restricted form $\phi$ and $\psi$ may be used to convert a general eight-parameter state \eqref{eq:General8ParamStateInFTSForm} to various canonical forms. A simple example is the five-parameter state retaining only the 0, 1, 2, 4, and 7 components of the state vector \eqref{eq:FTSState1}, as considered in \cite{Sudbery:2001},
\begin{equation}\label{eq:FTSState1}
\begin{pmatrix}f_1 & \diag(f_2,f_3,f_4) \\ \diag(0,0,0) & f_5\end{pmatrix}.
\end{equation}
This may be obtained via a single $\psi$ transformation,
\begin{equation}\label{eq:FTSTransf1}
\psi(Z)=\psi(\diag(q_1,q_2,q_3)/q_0).
\end{equation}
A more complicated example is a five-parameter state retaining the 0, 4, 5, 6, 7 components \cite{Acin:2001}. A single $\psi$ or $\phi$ transformation will not suffice and instead the transformation is of the form
\begin{equation}\label{eq:FTSTransf2}
\psi(\diag(0,0,1))\circ\phi(-\diag(d_1,d_2,d_3)).
\end{equation}
The $d_i$ are specified most compactly in the notation of \cite{Behrndt:1996hu}, specifically
\begin{equation}\label{eq:BehrndtNotation}
\begin{gathered}
\begin{aligned}
\omega_i &:= 3 d_{ijk}p^jp^k-p^0q_i, \\
z^i &:= \left[(p\cdot q)-2p^iq_i-i\sqrt{D}\right]/(2\omega_i),
\end{aligned} \\
d_{ijk}:=\tfrac{1}{3!}|\varepsilon_{ijk}|,\quad D:=-\Det a,\quad \{z^1,z^2,z^3\}\equiv\{S,T,U\}.
\end{gathered}
\end{equation}
Furthermore, it is instructive to consider the case of a real hyperdeterminant whose value is successively greater than, less than, and equal to zero. See \autoref{tab:FTSTransf2}. In this manner, the conversion of general to canonical states is related explicitly to the $STU$ model scalars.
\begin{table}[ht]
\begin{tabular*}{\textwidth}{@{\extracolsep{\fill}}cccccc}
\toprule
&                        & $d_1$                   & $d_2$                   & $d_3$                     & \\
\midrule
& \multirow{2}{*}{$D>0$} & $S$                     & $T$                     & $1+U+i\sqrt{D}/\omega_3$  & \\
&                        & $S+i\sqrt{D}/\omega_1$  & $T+i\sqrt{D}/\omega_2$  & $1+U$                     & \\[10pt]
& \multirow{2}{*}{$D<0$} & $S$                     & $T$                     & $1+U-\sqrt{|D|}/\omega_3$ & \\
&                        & $S-\sqrt{|D|}/\omega_1$ & $T-\sqrt{|D|}/\omega_2$ & $1+U$                     & \\[10pt]
& $D=0$                  & $S$                     & $T$                     & $1+U$                     & \\
\bottomrule
\end{tabular*}
\caption[Eight to five-parameter state transformation parameters]{Freudenthal construction transformation parameters for the conversion of a general eight-parameter state to a five-parameter canonical form \cite{Acin:2001}. Here $S, T$, and $U$ are the complex scalars of the $STU$ model, $D$ is the entropy, and the $\omega_i$ are defined in \eqref{eq:BehrndtNotation}.}\label{tab:FTSTransf2}
\end{table}

\newpage
\section{\texorpdfstring{CARTAN-FANO-FREUDENTHAL DICTIONARIES}{Cartan-Fano-Freudenthal Dictionaries}}
\label{sec:Cartan}

\subsection{Three descriptions}

The $\mathcal{N}=8$ black hole charges belong to the fundamental 56-dimensional representation of $E_{7(7)}$ and the black hole entropy is given by
\begin{equation}
S = \pi\sqrt{|I_4|},
\end{equation}
where $I_4$ is Cartan's quartic invariant, the singlet in $\mathbf{56 \times 56 \times 56 \times 56}$. We have seen that this invariant also plays the role of an entanglement measure on the QI side. So we need to study its properties in some detail.

As we have discussed, there are three descriptions of the group $E_7$ and its quartic invariant that we will find useful:
\begin{enumerate}
\item Cartan basis:
    \begin{equation}
    \begin{gathered}
    E_7 \supset SO(8),\\
    \mathbf{56 \to 28+ 28},\\
    I_{4}(x,y)=-\tr(x y)^2 + \tfrac{1}{4} (\tr x y)^2 - 4 (\Pf x + \Pf y).\label{eq:cartan}
    \end{gathered}
    \end{equation}
    where $x^{IJ}$ and $y_{IJ}$ are antisymmetric $8 \times 8$ matrices and $\Pf$ is the Pfaffian.
\item Freudenthal/Jordan basis
    \begin{subequations}
    \begin{gather}
    \begin{gathered}
    E_7 \supset E_6,\\
    \mathbf{56 \to 1+27 +1 + 27'},\\
    \begin{split}\label{eq:freud}
    I_4(p^{0},P;q_{0},Q)&=-\bigg{[}p^0q_0+\tr (J_3(P)\circ J_3(Q))\bigg{]}^2\\
    &\phantom{=}+4\bigg{[}-p^0 J_3(Q)+q_0 J_3(P)+\tr ({J_3}^{\#}(P)\circ {J_3}^{\#}(Q))\bigg{]},
    \end{split}
    \end{gathered}
    \shortintertext{where}
    X\circ Y=\half(XY+YX),
    \shortintertext{and}
    X^{\#}=X^2-\tr(X)X+\half[(\tr X)^2-\tr (X^2)]\mathds{1},
    \end{gather}
    \end{subequations}
    where $X$ is a member of Jordan algebra of degree 3.
\item Fano basis
    \begin{subequations}
    \begin{gather}
    E_7 \supset SL(2)^7,\\
    \begin{split}
    \mathbf{56}\to & \mathbf{(2,2,1,2,1,1,1)} \\
    +\             & \mathbf{(1,2,2,1,2,1,1)} \\
    +\             & \mathbf{(1,1,2,2,1,2,1)} \\
    +\             & \mathbf{(1,1,1,2,2,1,2)} \\
    +\             & \mathbf{(2,1,1,1,2,2,1)} \\
    +\             & \mathbf{(1,2,1,1,1,2,2)} \\
    +\             & \mathbf{(2,1,2,1,1,1,2)},
    \end{split}\\
    \begin{gathered}\label{eq:564}
    I_4=a^4+b^4+c^4+d^4+e^4+f^4+g^4 \\
    \begin{array}{c*{10}{@{\ }c}}
    +~2\Big[a^2b^2 & + & a^2c^2 & + & a^2d^2 & + & a^2e^2 & + & a^2f^2 & + & a^2g^2 \\
                   & + & b^2c^2 & + & b^2d^2 & + & b^2e^2 & + & b^2f^2 & + & b^2g^2 \\
                   &   &        & + & c^2d^2 & + & c^2e^2 & + & c^2f^2 & + & c^2g^2 \\
                   &   &        &   &        & + & d^2e^2 & + & d^2f^2 & + & d^2g^2 \\
                   &   &        &   &        &   &        & + & e^2f^2 & + & e^2g^2 \\
                   &   &        &   &        &   &        &   &        & + & \phantom{\Big]}f^2g^2\Big]
    \end{array} \\
    +~8\left[abce+bcdf+cdeg+defa+efgb+fgac+gabd\right],
    \end{gathered}
    \shortintertext{where}
    \begin{split}
    \phantom{aaa^2}a^{4}&= \half\varepsilon^{A_{1}A_{2}}\varepsilon^{B_{1}B_{2}}\varepsilon^{D_{1}D_{4}}\varepsilon^{A_{3}A_{4}}
    \varepsilon^{B_{3}B_{4}}\varepsilon^{D_{2}D _{3}}\\
    &\phantom{=}\times{a}_{A_{1}B_{1}D_{1}}{a}_{A_{2}B_{2}D_{2}}{a}_{A_{3}B_{3}D_{3}}{a}_{A_{4}B_{4}D_{4}},
    \end{split}
    \shortintertext{etc;}
    \begin{split}
    \phantom{aa}a^{2}b^{2}&= \half\varepsilon^{A_{1}A_{2}}\varepsilon^{B_{1}B_{3}}\varepsilon^{D_{1}D_{2}}\varepsilon^{B_{2}B_{4}}
    \varepsilon^{C_{3}C_{4}}\varepsilon^{E_{3}E_{4}}\\
    &\phantom{=}\times{a}_{A_{1}B_{1}D_{1}}{a}_{A_{2}B_{2}D_{2}}{b}_{B_{3}C_{3}E_{3}}{b}_{B_{4}C_{4}E_{4}},
    \end{split}
    \shortintertext{etc;}
    \begin{split}
    \phantom{{}^2{}^2}abce&= \half\varepsilon^{A_{1}A_{4}}\varepsilon^{B_{1}B_{2}}\varepsilon^{C_{2}C_{3}}\varepsilon^{D_{1}D_{3}}
    \varepsilon^{E_{2}E_{4}}\varepsilon^{F_{3}F_{4}}\\
    &\phantom{=}\times{a}_{A_{1}B_{1}D_{1}}{b}_{B_{2}C_{2}E_{2}}{c}_{C_{3}D_{3}F_{3}}{e}_{E_{4}F_{4}A_{4}},
    \end{split}
    \end{gather}
    \end{subequations}
    etc.
\end{enumerate}
Black holes are more conveniently described in either the Cartan or Freudenthal bases, whereas the Fano basis is tailored to the qubits. Hence it is important to find the three dictionaries that relate these three descriptions.  First we discuss each description in more detail.

\subsection{Cartan-Fano dictionary}

\subsubsection{Cyclic basis}
\label{sec:cyclic}

The dual Fano plane structure constants of \autoref{tab:DualFanoStructConst} define antisymmetric matrices $x^{IJ}$ and $y_{IJ}$ according to the dictionary of \autoref{tab:CartanDictionary}.

The 56 state vector coefficients, $a_{ABD}$ through $g_{GAC}$,  are arranged in $x^{IJ}$ and $y_{IJ}$ according to the octonionic multiplication table of the dual Fano plane, compare \autoref{tab:DualFanoMult} with the matrices \eqref{eq:xfano} and \eqref{eq:yfano} below. This uniquely determines the rows of \autoref{tab:CartanDictionary}.

The positions of the binary indices in $x^{IJ}$ and $y_{IJ}$ are specified by the columns of \autoref{tab:CartanDictionary}. The first column describes the position of 111 in $x^{IJ}$ (and 000 in $y_{IJ}$). Note, the first column consists of all pairs $i0$, i.e. the first row and column of $x^{IJ}$ and $y_{IJ}$. To understand the structure of the remaining three columns let us consider a specific example given by considering Alice's qubit $A$. For each row in \autoref{tab:CartanDictionary} one can form a triple $ijk$ from the pair $i0$, appearing in the first column, and any one of the remaining pairs $jk$ in that row. We note that 715 is the unique triple common to rows $a_{ABD}$, $e_{EFA}$ and $g_{GCA}$, the subsector defined by the common qubit $A$. Then, in each case the non-trivial pair $jk$ sits in the column labelled by the position of the common qubit. In our example this is $A$. Therefore the pair 57 belonging to row $a_{ABD}$ sits in the $100$ column where the position of $A$ in $ABD$ corresponds to the position of 1 in 100 or, equivalently, the position of 0 in 011. Similarly, 71 sits in the column labelled $001$ because $A$ is last in $e_{EFA}$. Finally, $15$ sits in the column labelled $010$ because $A$ is  second in $g_{GAC}$. Repeating this procedure for the remaining six qubits, $B$ through $G$, uniquely determines all the columns of \autoref{tab:CartanDictionary}. This procedure may be followed to construct the dictionary based on any octonionic basis.
\begin{table}[ht]
\begin{tabular*}{\textwidth}{@{\extracolsep{\fill}}*{7}{c}}
\toprule
& $x^{IJ}$  & 111 & 010 & 001 & 100 & \\
& $y_{IJ}$  & 000 & 101 & 110 & 011 & \\
\midrule
& $a_{ABD}$ & 10  & 26  & 34  & 57  & \\
& $b_{BCE}$ & 20  & 37  & 45  & 61  & \\
& $c_{CDF}$ & 30  & 41  & 56  & 72  & \\
& $d_{DEG}$ & 40  & 52  & 67  & 13  & \\
& $e_{EFA}$ & 50  & 63  & 71  & 24  & \\
& $f_{FGB}$ & 60  & 74  & 12  & 35  & \\
& $g_{GAC}$ & 70  & 15  & 23  & 46  & \\
\bottomrule
\end{tabular*}
\caption[Cartan basis dictionary]{The Cartan basis dictionary. The binary triples denote the indices on the 56 state vector coefficients, while the pairs give positions within the $x^{IJ}, y_{IJ}$ matrices. These are the positive elements of $x^{IJ}$ and $y_{IJ}$, the remaining elements being fixed by antisymmetry.}\label{tab:CartanDictionary}
\end{table}
\begin{subequations}
\begin{gather}
x^{IJ}=
\begin{pmatrix*}[r]
0~~     & -a_{111} & -b_{111} & -c_{111} & -d_{111} & -e_{111} & -f_{111} & -g_{111} \\
a_{111} &  0~~     &  f_{001} &  d_{100} & -c_{010} &  g_{010} & -b_{100} & -e_{001} \\
b_{111} & -f_{001} &  0~~     &  g_{001} &  e_{100} & -d_{010} &  a_{010} & -c_{100} \\
c_{111} & -d_{100} & -g_{001} &  0~~     &  a_{001} &  f_{100} & -e_{010} &  b_{010} \\
d_{111} &  c_{010} & -e_{100} & -a_{001} &  0~~     &  b_{001} &  g_{100} & -f_{010} \\
e_{111} & -g_{010} &  d_{010} & -f_{100} & -b_{001} &  0~~     &  c_{001} &  a_{100} \\
f_{111} &  b_{100} & -a_{010} &  e_{010} & -g_{100} & -c_{001} &  0~~     &  d_{001} \\
g_{111} &  e_{001} &  c_{100} & -b_{010} &  f_{010} & -a_{100} & -d_{001} & 0~~
\end{pmatrix*},\label{eq:xfano}\\
y_{IJ}=
\begin{pmatrix*}[r]
0~~     & -a_{000} & -b_{000} & -c_{000} & -d_{000} & -e_{000} & -f_{000} & -g_{000} \\
a_{000} &  0~~     &  f_{110} &  d_{011} & -c_{101} &  g_{101} & -b_{011} & -e_{110} \\
b_{000} & -f_{110} &  0~~     &  g_{110} &  e_{011} & -d_{101} &  a_{101} & -c_{011} \\
c_{000} & -d_{011} & -g_{110} &  0~~     &  a_{110} &  f_{011} & -e_{101} &  b_{101} \\
d_{000} &  c_{101} & -e_{011} & -a_{110} &  0~~     &  b_{110} &  g_{011} & -f_{101} \\
e_{000} & -g_{101} &  d_{101} & -f_{011} & -b_{110} &  0~~     &  c_{110} &  a_{011} \\
f_{000} &  b_{011} & -a_{101} &  e_{101} & -g_{011} & -c_{110} &  0~~     &  d_{110} \\
g_{000} &  e_{110} &  c_{011} & -b_{101} &  f_{101} & -a_{011} & -d_{110} & 0~~
\end{pmatrix*}.\label{eq:yfano}
\end{gather}
\end{subequations}
We can summarise the dictionary by writing $I=(0,i), i\in\{1, \dotsc, 7\}, (a^0, a^1, \dotsc, a^7)=(0, a, \ldots, g)$ and
\begin{subequations}
\begin{align}\label{eq:CartanDictionary}
x^{IJ} &=
\begin{cases}
a^I_{111}        & J=0, \\
c_{IJK}a^K_{010} & |I-J|=\text{3 or 4}, \\
c_{IJK}a^K_{001} & |I-J|=\text{1 or 6}, \\
c_{IJK}a^K_{100} & |I-J|=\text{2 or 5},
\end{cases} &
y_{IJ} &=
\begin{cases}
a^I_{000}        & J=0, \\
c_{IJK}a^K_{101} & |I-J|=\text{3 or 4}, \\
c_{IJK}a^K_{110} & |I-J|=\text{1 or 6}, \\
c_{IJK}a^K_{011} & |I-J|=\text{2 or 5}.
\end{cases}
\end{align}
Here we have  extended $c^i_{jk}$ to $c^i_{JK}$ by setting $c^i_{JK}=c^i_{jk}$ whenever $J(=j)$ and $K(=k)$ are not equal to $0$, while defining $c^i_{JK}$ to be zero whenever $J$ or $K$ is equal to $0$.
Alternatively, we may use the more compact formulation:
\begin{equation}
\begin{split}
x^{IJ}&=\eta_{IL}c_{LJK}a^K_{\phi(I,J)},\\
y_{IJ}&=\eta_{IL}c_{LJK}a^K_{\tilde{\phi}(I,J)},
\end{split}
\end{equation}
where $\eta$ is the 8-dimensional Minkowski matrix (negative signature), and $\phi$ and $\tilde{\phi}$ are given by
\begin{equation}
\begin{split}
\phi(I,J)&:=\begin{cases}7 & I=0 \text{ or }J=0 \\ {|(I-J)^2|}_7 & \text{else}\end{cases}\\
\tilde{\phi}(I,J)&:=7-\phi(I,J).
\end{split}
\end{equation}
\end{subequations}
It may be useful to regard the eight components of each $a^i$ as a pair of quaternions.

An alternative way of arriving at \autoref{tab:CartanDictionary} is to note that each entry $(ij)$ represents the row $i$ and column $j$ in which the letters $a,b,c,d,e,f,g$ (or numbers 1, 2, 3, 4, 5, 6, 7) appear with a positive sign in the matrix of \autoref{tab:NumericalDualFanoMult}.
\begin{table}[ht]
\begin{tabular*}{\textwidth}{@{\extracolsep{\fill}}*{10}{c}}
\toprule
& 0 & -1 & -2 & -3 & -4 & -5 & -6 & -7 & \\
& 1 &  0 &  4 &  7 & -2 &  6 & -5 & -3 & \\
& 2 & -4 &  0 &  5 &  1 & -3 &  7 & -6 & \\
& 3 & -7 & -5 &  0 &  6 &  2 & -4 &  1 & \\
& 4 &  2 & -1 & -6 &  0 &  7 &  3 & -5 & \\
& 5 & -6 &  3 & -2 & -7 &  0 &  1 &  4 & \\
& 6 &  5 & -7 &  4 & -3 & -1 &  0 &  2 & \\
& 7 &  3 &  6 & -1 &  5 & -4 & -2 &  0 & \\
\bottomrule
\end{tabular*}
\caption[Numerical Fano plane multiplication table]{The numerical Fano plane multiplication table.}\label{tab:NumericalFanoMult}
\end{table}
\begin{table}[ht]
\begin{tabular*}{\textwidth}{@{\extracolsep{\fill}}*{10}{c}}
\toprule
& 0 & -1 & -2 & -3 & -4 & -5 & -6 & -7 & \\
& 1 &  0 &  6 &  4 & -3 &  7 & -2 & -5 & \\
& 2 & -6 &  0 &  7 &  5 & -4 &  1 & -3 & \\
& 3 & -4 & -7 &  0 &  1 &  6 & -5 &  2 & \\
& 4 &  3 & -5 & -1 &  0 &  2 &  7 & -6 & \\
& 5 & -7 &  4 & -6 & -2 &  0 &  3 &  1 & \\
& 6 &  2 & -1 &  5 & -7 & -3 &  0 &  4 & \\
& 7 &  5 &  3 & -2 &  6 & -1 & -4 &  0 & \\
\bottomrule
\end{tabular*}
\caption[Numerical dual Fano plane multiplication table]{The numerical dual Fano plane multiplication table.}\label{tab:NumericalDualFanoMult}
\end{table}

The  $8 \times 8$ gamma matrices  $\gamma^i_{IJ}$  in seven dimensions, which satisfy the Clifford algebra
\begin{equation}
\{\gamma^i,\gamma^j\}=2\delta^{ij}\mathds{1},
\end{equation}
can be written in terms of the octonionic structure constants. The hermitian (purely imaginary and antisymmetric) gamma matrices in seven dimensions can then be chosen as
\begin{equation}
\gamma^i_{IJ} = i\left(c\indices{^i_{IJ}} \pm \delta_{iI}\delta_{J0} \mp \delta_{iJ}\delta_{I0}\right),
\end{equation}
where the signs are correlated.

The antisymmetric products of gamma matrices are defined as usual, with unit weight, \textit{viz}.
\begin{equation}
\gamma^{ij\cdots k}=\gamma^{[i}\gamma^{j}\cdots \gamma^{k]}.
\end{equation}
The  antisymmetric self-dual and anti-self-dual tensors $c^{\pm}_{IJKL}$,  ($I,J,\ldots = 0,1,2,\ldots, 7 $) in eight dimensions will be defined as:
\begin{equation}
c^{\pm}_{ijkl}=c_{ijkl},\quad\text{and}\quad c^{\pm}_{ijk0}=\pm c_{ijk}.
\end{equation}
With the above choices of gamma matrices one finds
\begin{equation}
\begin{split}
\gamma^{ij}_{IJ}&= c_{ijIJ} + \delta^i_I\delta^j_J-\delta^i_J\delta^j_I \pm c^{~ij}_I\delta_{J0} \mp c^{~ij}_J\delta_{I0} \\
&= c^{\pm}_{ijIJ} + \delta^i_I\delta^j_J-\delta^i_J\delta^j_I.
\end{split}
\end{equation}
Note that $(-i\gamma^i)_{IJ}$ do not form an $8 \times 8$ representation of the octonions;
\begin{gather}
\begin{split}
(-i\gamma^i)_{IK}(-i\gamma^j)_{KJ}&=-\delta^{ij}\delta_{IJ} -\gamma^{ij}_{IJ} \\
&=-\delta^{ij}\delta_{IJ}-c_{ijab} -\delta^i_I\delta^j_J+\delta^i_J\delta^j_I \mp c^{~ij}_I\delta_{J0} \pm c^{~ij}_J\delta_{I0},
\end{split}
\shortintertext{which is to be compared with}
e_ie_j=-\delta_{ij}+c_{ijk} e_k.
\shortintertext{Whereas}
c_{ijk}(-i\gamma^k)_{IJ}=c_{ijk}(c\indices{^i_{IJ}}\pm \delta_{iI}\delta_{J0} \mp \delta_{iJ}\delta_{I0}).
\end{gather}

Accordingly one can rewrite \eqref{eq:CartanDictionary} as
\begin{align}
x^{JK} &=-i\gamma^{iJK}\times
\begin{cases}
a^i_{111} & K=0, \\
a^i_{010} & |J-K|=3, \\
a^i_{001} & |J-K|=1, \\
a^i_{100} & |J-K|=2,
\end{cases} &
y_{JK} &=-i\gamma^{iJK}\times
\begin{cases}
a^i_{000} & K=0, \\
a^i_{101} & |J-K|=3, \\
a^i_{110} & |J-K|=1, \\
a^i_{011} & |J-K|=2.
\end{cases}
\end{align}

\subsubsection{Binary basis}
\label{sec:binarybasis}

So far we have used the cyclic basis of \autoref{tab:FanoMult} to describe the octonions.  However, for some purposes, it is more convenient to use the binary basis.  The addition table for the numbers 0 to 7 written in binaries is given by \autoref{tab:BinaryAdditionTable} or equivalently \autoref{tab:DecimalBinaryAdditionTable}.
\begin{table}[ht]
\begin{tabular*}{\textwidth}{@{\extracolsep{\fill}}*{11}{c}}
\toprule
&       & 000 & 001 & 010 & 011 & 100 & 101 & 110 & 111 & \\
\midrule
& 000   & 000 & 001 & 010 & 011 & 100 & 101 & 110 & 111 & \\
& 001   & 001 & 000 & 011 & 010 & 101 & 100 &  111& 110 & \\
& 010   & 010 & 011 & 000 & 001 & 110 & 111 & 100 & 101 & \\
& 011   & 011 & 010 & 001 & 000 & 111 & 110 & 101 & 100 & \\
& 100   & 100 & 101 & 110 & 111 & 000 & 001 & 010 & 011 & \\
& 101   & 101 & 100 & 111 & 110 & 001 & 000 & 011 & 010 & \\
& 110   & 110 & 111 & 100 & 101 & 010 & 011 & 000 & 001 & \\
& 111   & 111 & 110 & 101 & 100 & 011 & 010 & 001 & 000 & \\
\bottomrule
\end{tabular*}
\caption[Binary addition table]{Binary addition table.}\label{tab:BinaryAdditionTable}
\end{table}
\begin{table}[ht]
\begin{tabular*}{\textwidth}{@{\extracolsep{\fill}}*{11}{c}}
\toprule
&   & 0 & 1 & 2 & 3 & 4 & 5 & 6 & 7 & \\
\midrule
& 0 & 0 & 1 & 2 & 3 & 4 & 5 & 6 & 7 & \\
& 1 & 1 & 0 & 3 & 2 & 5 & 4 & 7 & 6 & \\
& 2 & 2 & 3 & 0 & 1 & 6 & 7 & 4 & 5 & \\
& 3 & 3 & 2 & 1 & 0 & 7 & 6 & 5 & 4 & \\
& 4 & 4 & 5 & 6 & 7 & 0 & 1 & 2 & 3 & \\
& 5 & 5 & 4 & 7 & 6 & 1 & 0 & 3 & 2 & \\
& 6 & 6 & 7 & 4 & 5 & 2 & 3 & 0 & 1 & \\
& 7 & 7 & 6 & 5 & 4 & 3 & 2 & 1 & 0 & \\
\bottomrule
\end{tabular*}
\caption[Decimal binary addition table]{The decimal version of \autoref{tab:BinaryAdditionTable}.}\label{tab:DecimalBinaryAdditionTable}
\end{table}
This defines the octonionic multiplication table in the binary basis. The non-vanishing independent components of the octonionic structure constants $c_{ijk}$ and their duals $c_{lmno}$ are then given by \autoref{tab:ManivelStructConst}.
\begin{table}[ht]
\begin{tabular*}{\textwidth}{@{\extracolsep{\fill}}*{10}{c}}
\toprule
& $i$ & $j$ & $k$ & & $l$ & $m$ & $n$ & $o$ & \\
\midrule
& 1   & 2   & 3   & & 5   & 4   & 7   & 6   & \\
& 2   & 5   & 7   & & 3   & 1   & 6   & 4   & \\
& 3   & 7   & 4   & & 6   & 5   & 1   & 2   & \\
& 4   & 1   & 5   & & 2   & 6   & 3   & 7   & \\
& 5   & 3   & 6   & & 7   & 2   & 4   & 1   & \\
& 6   & 4   & 2   & & 1   & 7   & 5   & 3   & \\
& 7   & 6   & 1   & & 4   & 3   & 2   & 5   & \\
\bottomrule
\end{tabular*}
\caption[Binary structure constants $c_{ijk}$ and $c_{lmno}$]{Structure constants $c_{ijk}$ read off from \autoref{tab:DecimalBinaryAdditionTable}, and the associator coefficients $c_{lmno}$ computed using \eqref{eq:associator}.}\label{tab:ManivelStructConst}
\end{table}
In this basis, it follows from \eqref{eq:quadrangles} that there exists a $\theta_{i,j}=\pm 1$ such that the imaginary octonions obey \cite{Manivel:2005}
\begin{subequations}
\begin{gather}
e_ie_j=\theta_{i,j}e_{i+j},
\shortintertext{where}
\theta_{i,j}=-\theta_{j,i},
\shortintertext{and}
\theta_{i,j}\theta_{i+j,k}=\theta_{j,k}\theta_{j+k,i}=\theta_{k,i}\theta_{k+i,j}.
\end{gather}
\end{subequations}
The binary structure constants of \autoref{tab:ManivelStructConst} define antisymmetric matrices $x^{IJ}$ and $y_{IJ}$ according to the dictionary of \autoref{tab:ManivelDictionary}, namely \eqref{eq:xdictionary} and \eqref{eq:ydictionary}. The dictionary specified in \autoref{tab:ManivelDictionary} may be derived using the procedure described in \autoref{sec:cyclic}. Note that, in this case, \autoref{tab:ManivelDictionary} corresponds precisely to the array of 28 pairs appearing on page 28 of \cite{Manivel:2005}.
\begin{table}[ht]
\begin{tabular*}{\textwidth}{@{\extracolsep{\fill}}*{7}{M{c}}}
\toprule
& x^{IJ}  & 111 & 100 & 010 & 001 & \\
& y_{IJ}  & 000 & 011 & 101 & 110 & \\
\midrule
& a_{ABC} & 10  & 23  & 45  & 67  & \\
& b_{ABE} & 20  & 31  & 46  & 57  & \\
& c_{AFG} & 30  & 12  & 47  & 65  & \\
& d_{BDF} & 40  & 51  & 26  & 73  & \\
& e_{BEG} & 50  & 14  & 72  & 36  & \\
& f_{CDG} & 60  & 17  & 24  & 53  & \\
& g_{CEF} & 70  & 61  & 25  & 34  & \\
\bottomrule
\end{tabular*}
\caption[Binary dictionary]{The binary Cartan dictionary. As with the cyclic dictionary of  \autoref{tab:CartanDictionary} the binary triples denote the indices on the 56 state vector coefficients, while the pairs give positions within the $x^{IJ}, y_{IJ}$ matrices. These are the positive elements of $x^{IJ}$ and $y_{IJ}$, the remaining elements being fixed by antisymmetry.}\label{tab:ManivelDictionary}
\end{table}
\begin{subequations}
\begin{gather}
x^{IJ}=
\begin{pmatrix*}[r]
0~~     & -a_{111} & -b_{111} & -c_{111} & -d_{111} & -e_{111} & -f_{111} & -g_{111} \\
a_{111} & 0~~      & c_{100}  & -b_{100} & e_{100}  & -d_{100} & -g_{100} & f_{100}  \\
b_{111} & -c_{100} & 0~~      & a_{100}  & f_{010}  & g_{010}  & -d_{010} & -e_{010} \\
c_{111} & b_{100}  & -a_{100} & 0~~      & g_{001}  & -f_{001} & e_{001}  & -d_{001} \\
d_{111} & -e_{100} & -f_{010} & -g_{001} & 0~~      & a_{010}  & b_{010}  & c_{010}  \\
e_{111} & d_{100}  & -g_{010} & f_{001}  & -a_{010} & 0~~      & -c_{001} & b_{001}  \\
f_{111} & g_{100}  & d_{010}  & -e_{001} & -b_{010} & c_{001}  & 0~~      & -a_{001} \\
g_{111} & -f_{100} & e_{010}  & d_{001}  & -c_{010} & -b_{001} & a_{001}  & 0~~
\end{pmatrix*},\label{eq:xdictionary}\\
y_{IJ}=
\begin{pmatrix*}[r]
0~~     & -a_{000} & -b_{000} & -c_{000} & -d_{000} & -e_{000} & -f_{000} & -g_{000} \\
a_{000} & 0~~      & c_{011}  & -b_{011} & e_{011}  & -d_{011} & -g_{011} & f_{011}  \\
b_{000} & -c_{011} & 0~~      & a_{011}  & f_{101}  & g_{101}  & -d_{101} & -e_{101} \\
c_{000} & b_{011}  & -a_{011} & 0~~      & g_{110}  & -f_{110} & e_{110}  & -d_{110} \\
d_{000} & -e_{011} & -f_{101} & -g_{110} & 0~~      & a_{101}  & b_{101}  & c_{101}  \\
e_{000} & d_{011}  & -g_{101} & f_{110}  & -a_{101} & 0~~      & -c_{110} & b_{110}  \\
f_{000} & g_{011}  & d_{101}  & -e_{110} & -b_{101} & c_{110}  & 0~~      & -a_{110} \\
g_{000} & -f_{011} & e_{101}  & d_{110}  & -c_{101} & -b_{110} & a_{110}  & 0~~
\end{pmatrix*}.\label{eq:ydictionary}
\end{gather}
\end{subequations}

Kallosh and Linde have shown that in a canonical basis $I_{4}$ depends on 4 complex eigenvalues represented as Cayley's hyperdeterminant of a hypermatrix $a_{ABD}$. Looking at  \eqref{eq:E7InvariantCartanBasis} we note that only the $SO(8)$ symmetry is manifest, yet it was proved in \cite{Cartan} and \cite{Cremmer:1979up} that the sum of all terms in  \eqref{eq:E7InvariantCartanBasis} is invariant under an $SU(8)$ symmetry, which acts as follows
\begin{equation}
\delta(x^{IJ} \pm i y_{IJ})= (2\Lambda\indices{^{[I}_{[K}}\delta\indices{^{J]}_{L]}} \pm i\Sigma_{IJKL}) (x^{KL} \mp i y_{KL}).
\end{equation}
The total number of parameters is 63; 28 are from the manifest $SO(8)$ and 35 from the antisymmetric self-dual $\Sigma_{IJKL}= {}^*\Sigma^{IJKL} $. Thus one can use the $SU(8)$ transformation of the complex matrix $x^{IJ} + i y_{IJ}$ and bring it to the canonical form with four eigenvalues $\lambda_\alpha, \alpha=1,2,3,4$. The value of the quartic invariant \eqref{eq:E7InvariantCartanBasis} will not change.
\begin{equation}
(x^{IJ} + i y_{IJ})_\textrm{can}=
\begin{pmatrix}
0          & \lambda_1 & 0          & 0         & 0          & 0         & 0          & 0         \\
-\lambda_1 & 0         & 0          & 0         & 0          & 0         & 0          & 0         \\
0          & 0         & 0          & \lambda_2 & 0          & 0         & 0          & 0         \\
0          & 0         & -\lambda_2 & 0         & 0          & 0         & 0          & 0         \\
0          & 0         & 0          & 0         & 0          & \lambda_3 & 0          & 0         \\
0          & 0         & 0          & 0         & -\lambda_3 & 0         & 0          & 0         \\
0          & 0         & 0          & 0         & 0          & 0         & 0          & \lambda_4 \\
0          & 0         & 0          & 0         & 0          & 0         & -\lambda_4 & 0
\end{pmatrix}.
\label{eq:x+iy}
\end{equation}
The relation between the complex coefficients $\lambda_\alpha$, the parameters $x^{IJ}$ and $y_{IJ}$, the matrix $a_{ABD}$ and the black hole charges $p^{i}$ and $q_{k}$ \cite{Duff:2006uz} is given by the following dictionary:
\begin{equation}
\begin{split}
\lambda_1 &= x^{01}+i y_{01} = a_{111}+i a_{000} = q_0+i p^0, \\
\lambda_2 &= x^{23}+i y_{23} = a_{100}+i a_{011} = -p^3+ q_3, \\
\lambda_3 &= x^{45}+i y_{45} = -a_{010}-i a_{101} = p^2-i q_2, \\
\lambda_4 &= x^{56}+i y_{56} = -a_{001}-i a_{110} = p^1-iq_1.
\end{split}
\end{equation}
If we now write the quartic $E_{7(7)}$ Cartan invariant in the canonical basis
\begin{equation}
\begin{gathered}
I_4= \\
\begin{aligned}
-(x^{01}y_{01} + x^{23}y_{23} \ +&\  x^{45}y_{45} + x^{67}y_{67})^2 \\
+~4(x^{01}x^{23}y_{01}y_{23} + x^{01}x^{45}&y_{01}y_{45} + x^{23}x^{45}y_{23}y_{45} \\
+~x^{01}x^{67}y_{01}y_{67} + x^{23}x^{67}&y_{23}y_{67} + x^{45}x^{67}y_{45}y_{67}) \\
-~4(x^{01}x^{23}x^{45}x^{67} \ +&\ y_{01}y_{23}y_{45}y_{67}),
\end{aligned}
\end{gathered}
\end{equation}
then it may be compared to Cayley's hyperdeterminant \eqref{eq:CayleyHyperdeterminant}. We find
\begin{equation}
I_{4}=-\Det a.
\end{equation}
The above discussion of $E_{7(7)}$ also applies, \textit{mutatis mutandis}, to $E_{7}(\mathds{C})$.

Evidently, this particular representation of the $\mathbf{56}$ emphasises certain aspects, such as the Fano plane structure, which are not clear in the other representations considered (see \autoref{sec:DiscreteSymmFano} for the an example of how the symmetries of the Fano plane are manifested in the black hole entropy using the Fano basis).

\subsection{Freudenthal-Fano dictionary}
\label{sec:Freudenthal-Fano}

Let us now construct the analogous dictionary relating the 56 charges in the Freudenthal basis with the 56 state vector coefficients specifying the tripartite entanglement of seven qubits. See also \cite{Borsten:2008}.  It is instructive to consider the chain of group decompositions $E_{7(7)}\to E_{6(6)}\to SO(4,4)$. Combining \eqref{eq:56ofE7underE6} and \eqref{eq:27E6underSO8} we have that the $\mathbf{56}$ decomposes as
\begin{equation}\label{eq:569}
\mathbf{56 \to 1 +1 +1 +1 +1 +1 +1 +1 +8_{s} +8_{c} +8_{v} +8_{s} +8_{c} +8_{v}},
\end{equation}
under $E_{7(7)} \supset SO(4,4)$.

Combining this with the $STU$ embedding in the FTS \eqref{eq:STU/FTS}, the decompositions \eqref{eq:56ofE7underE6} and \eqref{eq:27E6underSO8}, and the dictionary as given in \eqref{eq:CayleyHyperdeterminant} it is clear that the 8 state vector coefficients, $a_{ABD}$, are associated with the 8 singlets appearing in \eqref{eq:569}. Now, consider the subgroup containing the three copies of $SL(2)$ associated with the tripartite entanglement of qubits $A$, $B$ and $D$,
\begin{gather}
E_{7(7)} \supset SL(2)_{A} \times SL(2)_{B} \times SL(2)_{D} \times SO(4,4),
\shortintertext{under which,}
\mathbf{56 \to (2,2,2,1)+(2,1,1,8_{v}) +(1,2,1,8_{s}) + (1,1,2,8_{c})}.\label{eq:svc}
\end{gather}
We note that qubit $A$ transforms as doublet with the $\mathbf{8}_{v}$. This suggests that we associate the subsector defined by the common qubit $A$, namely $a_{ABD}\ket{ABD}+e_{EFA}\ket{EFA}+g_{GAC}\ket{GAC}$, with the $\mathbf{8_{v}}$. This is one of the consistent $\mathcal{N}=4$ truncations of the full $\mathcal{N}=8$ theory, the 24 black hole charges transforming as a $(\mathbf{2}, \mathbf{12})$ of $SL(2)_A\times SO(6,6)$ \cite{Duff:2006ue}. Repeating this analysis for qubit $B$ leads us to identify the $b_{BCE}$ and $f_{FGB}$ with the  $\mathbf{8_s}$, while considering  $D$ we identify $c_{CDF}$ and $d_{DEG}$ with the  $\mathbf{8_c}$.

To specify more precisely the dictionary between $(e_{EFA}, g_{GAC})$ and $(P_v, Q_v)$, $(b_{BCE}, f_{FGB})$ and $(P_s, Q_s)$, and finally, $(c_{CDF}, d_{DEG})$ and $(P_c, Q_c)$, we begin by noting that the 8 charges of the $STU$ model may be arranged in a cube as depicted in \autoref{fig:Freudenthalbasiscube}. Following \cite{Bhargava:2004}, the cube may be partitioned into a pair of $2\times 2$ matrices, $(M_i, N_i)$ in three independent ways. These are given by the three possible slicings of the cube along its planes of symmetry,
\begin{gather}
\begin{subequations}
\begin{align}
M_1&=\begin{pmatrix*}[r]-p^3& q_2\\ q_1&q_0\end{pmatrix*}, &
N_1&=\begin{pmatrix*}[r] p^0&-p^1\\-p^2&q_3\end{pmatrix*},\\
M_2&=\begin{pmatrix*}[r]-p^2& q_3\\ q_1&q_0\end{pmatrix*}, &
N_2&=\begin{pmatrix*}[r] p^0&-p^1\\-p^3&q_2\end{pmatrix*},\\
M_3&=\begin{pmatrix*}[r]-p^1& q_3\\ q_2&q_0\end{pmatrix*}, &
N_3&=\begin{pmatrix*}[r] p^0&-p^2\\-p^3&q_1\end{pmatrix*}.
\end{align}
\end{subequations}
\shortintertext{For any element,}
\begin{pmatrix}r&s\\t&u\end{pmatrix}\in SL(2)_i,\qquad 1\leq i\leq3,
\shortintertext{the action on the cube is given by}
(M_i, N_i)\mapsto(rM_i + sN_i, tM_i + uN_i).
\end{gather}
The individual actions of the three $SL(2)_i$ all commute and, therefore, this provides a natural representation of $[SL(2)]^3$ \cite{Bhargava:2004}. Define the three binary quadratic forms, one for each slicing,
\begin{gather}
f_i(x,y)= \det(M_ix+N_iy), \qquad 1\leq i\leq3.
\shortintertext{Explicitly}
\begin{split}
f_1=-(q_2q_1+p^3q_0)x^2+(p\cdot q-2p^3q_3)xy-(p^2p^1-p^0q_3)y^2,\\
f_2=-(q_1q_3+p^2q_0)x^2+(p\cdot q-2p^2q_2)xy-(p^1p^3-p^0q_2)y^2,\\
f_3=-(q_3q_2+p^1q_0)x^2+(p\cdot q-2p^1q_1)xy-(p^3p^2-p^0q_1)y^2.
\end{split}
\end{gather}
These quadratic forms may also be systematically derived using transvectants as presented in \autoref{sec:Transvectants}. Each one is invariant under two of the three factors in $[SL(2)]^3$. For example, $f_1$ is invariant under the subgroup $\{\operatorname{id}_1\}\times SL_2(2)\times SL_3(2)\subset[SL(2)]^3$. Taking the determinant of the Hessian, $H(f_i)$, which is actually given by $\gamma^i(a)$, as defined in \eqref{eq:ABCgammas}, yields Cayley's hyperdeterminant
\begin{equation}
\det H(f_i)=\det \begin{pmatrix}(f_i)_{xx} &(f_i)_{xy}\\(f_i)_{yx}& (f_i)_{yy}\end{pmatrix}
= \det \gamma^i(a)=-\Det a_{ABC},\qquad  1\leq i\leq3.
\end{equation}

Now, consider keeping $(p^i, q_i)$ and only one of $(P_s, Q_s)$, $(P_v, Q_v)$ or $(P_c, Q_c)$ and computing $I_4$ from the FTS. Recall, this gives us the entropy of one of the three $\mathcal{N}=4$ subsectors as defined by one of the three qubits, $A,B$ or $D$. Keeping only $(p^i, q_i, P_v, Q_v)$, the $\mathcal{N}=4$ subsector defined by the common qubit $A$, one finds
\begin{align}
I_{4}&=4(p^2p^1-p^0q_3)(q_2q_1+p^3q_0) - 4(p.q-2p^3q_3)^{2} +4P_{v}{}^2Q_{v}{}^2 -(P^v\cdot Q_v)^{2}\\
&\phantom{=}- 4(p^2p^1-p^0q_3)Q_{v}{}^2 - 4(q_2q_1+p^3q_0)P_{v}{}^2 -4(p\cdot q-2p^3q_3)P_{v}\cdot Q_{v},
\end{align}
where $P_{v}{}^2=P_{v}{\bar P_{v}}$ and $2P_v\cdot Q_v=(P_{v}{\bar Q_{v}}+Q_{v}{\bar P_{v}})$. The terms involving $(p^i, q_i)$ correspond to $\gamma^1(a)$, which is correctly associated with  qubit $A$,  via the dictionary \eqref{eq:CayleyHyperdeterminant},
\begin{align}
2(p^1p^2-p^0q_3)&=-\gamma^1(a)_{00},&
2(q_1q_2+p^3q_0)&=-\gamma^1(a)_{11},&
p\cdot q-2p^3q_3&=\gamma^1(a)_{01}.
\end{align}
This agrees with the conclusions drawn from the decomposition given in \eqref{eq:svc}.  Keeping either $(P^s, Q_s)$ or $(P^c, Q_c)$ instead would have resulted in a different associated slicing of the cube \autoref{fig:Freudenthalbasiscube} and, hence, matrix $\gamma^i(a)$. We then identify $(g_{GAC}, e_{EFA})$ with $( P_v, Q_v)$ such that
\begin{align}
P_{v}{}^2&=\gamma^{2}(g)_{00}+\gamma^{3}(e)_{00},&
Q_{v}{}^2&=\gamma^{2}(g)_{11}+\gamma^{3}(e)_{11},&
P_{v}\cdot Q_v&=\gamma^{2}(g)_{01}+\gamma^{3}(e)_{01},
\end{align}
where, for example, the index on $\gamma^{2}(g)$ is determined by the position of the common qubit $A$ in  the corresponding tripartite subsystem, $GAC$. Computing $I_4$ one finds
\begin{gather}
I_{4}=\det(\gamma^1(a)+\gamma^3(e)+\gamma^2(g)) \sim -\Det a-\Det e-\Det g+2(a^2g^2+a^2e^2+e^2g^2),\label{eq:N4}
\shortintertext{where products like}
\begin{split}
a^{2}e^{2}&= \half\varepsilon^{A_{1}A_{4}}\varepsilon^{B_{1}B_{2}}\varepsilon^{D_{1}D_{2}}\varepsilon^{E_{3}E_{4}}
\varepsilon^{F_{3}F_{4}}\varepsilon^{A_{2}A_{3}}\\
&\phantom{=}\times{a}_{A_{1}B_{1}D_{1}}{a}_{A_{2}B_{2}D_{2}}{e}_{E_{3}F_{3}A_{3}}{e}_{E_{4}F_{4}A_{4}},
\end{split}
\end{gather}
describe the entanglement between two tripartite subsystems connected by a common qubit, in this case $A$. This may be repeated for the remaining two cases, keeping $(P_s, Q_s)$ or $(P_c, Q_c)$, associated with the common qubits $B$ and $D$ respectively, to construct the whole dictionary\footnote{Note, determining the precise form of the full dictionary and verifying that it does indeed give the stated results was done explicitly using \textit{Mathematica}.}. For each of the seven possible $\mathcal{N}=4$ subsectors one obtains the appropriate result, analogous to \eqref{eq:N4}, as presented in \eqref{eq:I_3Nequal4}.

We are now able to select any particular subsector of the full $\mathcal{N}=8$ theory by choosing the appropriate components of the FTS and systematically determine the corresponding qubit system and its measure of entanglement.

In summary the dictionary is
\begin{equation}\label{eq:charges16}
\begin{gathered}
\begin{array}{c@{\ \big(}*{7}{c@{,\ }}c@{\big)}c@{}c}
  & p^0   & p^1    & p^2    & p^3    & q_0   & q_1   & q_2   & q_3   & \\[3pt]
= & a_{0} & -a_{1} & -a_{2} & -a_{4} & a_{7} & a_{6} & a_{5} & a_{3} & ,
\end{array}\\
P_{v}=(g_{G0C},e_{EF0}),\quad Q_{v}=(g_{G1C},e_{EF1}), \\
P_{s}=(f_{FG0},b_{0CE}),\quad Q_{s}=(f_{FG1},b_{1CE}), \\
P_{c}=(d_{0EG},c_{C0F}),\quad Q_{c}=(d_{1EG},c_{C1F}),
\end{gathered}
\end{equation}
where explicitly (see \hyperref[tab:FreudenthalToFano]{Tables~\ref*{tab:FreudenthalToFano}} and \hyperref[tab:FanoToFreudenthal]{\ref*{tab:FanoToFreudenthal}})
\begin{equation}
\begin{array}{*{2}{c@{~=~}c}@{}c}
\begin{pmatrix} d_0 & d_1 \\ d_2 & d_3\end{pmatrix} & \begin{pmatrix*}[r] P_c^0 - P_c^4 & \phantom{-}P_c^1 - P_c^5\phantom{-} \\ -P_c^1 - P_c^5 & P_c^0 + P_c^4\phantom{-}\end{pmatrix*} & \begin{pmatrix} d_4 & d_5 \\ d_6 & d_7\end{pmatrix} & \begin{pmatrix*}[r] Q^0_c - Q^4_c & \phantom{-}Q^1_c - Q^5_c\phantom{-} \\ -Q^1_c - Q^5_c & Q^0_c + Q^4_c\phantom{-}\end{pmatrix*} \\[8pt]
\begin{pmatrix}c_0 & c_1 \\ c_4 & c_5\end{pmatrix} & \begin{pmatrix*}[r] -P_c^6 - P_c^2 & - P_c^3 - P_c^7\phantom{-} \\ P_c^3 - P_c^7 & P_c^6 - P_c^2\phantom{-}\end{pmatrix*} & \begin{pmatrix}c_2 & c_3 \\ c_6 & c_7\end{pmatrix} & \begin{pmatrix*}[r]-Q^6_c - Q^2_c & - Q^3_c - Q^7_c \phantom{-}\\ Q^3_c - Q^7_c & Q^6_c - Q^2_c\phantom{-}\end{pmatrix*} \\[8pt]
\begin{pmatrix}f_ 0 & f_ 2 \\f_ 4 & f_ 6\end{pmatrix} & \begin{pmatrix*}[r]\phantom{-}P_s^3 - P_s^7 & \phantom{-}P_s^6 + P_s^2 \phantom{-}\\ P_s^6 - P_s^2 & P_s^3 + P_s^7\phantom{-}\end{pmatrix*} & \begin{pmatrix}f_ 1 & f_ 3 \\f_ 5 & f_ 7\end{pmatrix} & \begin{pmatrix*}[r]\phantom{-}Q^3_s - Q^7_s & \phantom{-}Q^6_s + Q^2_s \phantom{-}\\ Q^6_s - Q^2_s & Q^3_s + Q^7_s\phantom{-}\end{pmatrix*} \\[8pt]
\begin{pmatrix}b_ 0 & b_ 1 \\b_ 2 & b_ 3\end{pmatrix} & \begin{pmatrix*}[r]\phantom{-}P_s^1 - P_s^5 & \phantom{-}P_s^4 - P_s^0 \phantom{-}\\ P_s^4 + P_s^0 & P_s^1 + P_s^5\phantom{-}\end{pmatrix*} & \begin{pmatrix}b_ 4 & b_ 5 \\b_ 6 & b_ 7\end{pmatrix} & \begin{pmatrix*}[r]\phantom{-} Q^1_s - Q^5_s & \phantom{-}Q^4_s - Q^0_s\phantom{-} \\ Q^4_s + Q^0_s & Q^1_s + Q^5_s\phantom{-}\end{pmatrix*} \\[8pt]
\begin{pmatrix}g_ 0 & g_ 1 \\g_ 4 & g_ 5\end{pmatrix} & \begin{pmatrix*}[r]\phantom{-}P_v^0 + P_v^4 & \phantom{-}P_v^5 + P_v^1 \phantom{-}\\ P_v^5 - P_v^1 & P_v^0 - P_v^4\phantom{-}\end{pmatrix*} & \begin{pmatrix}g_ 2 & g_ 3 \\g_ 6 & g_ 7\end{pmatrix} & \begin{pmatrix*}[r]\phantom{-} Q^0_v + Q^4_v & \phantom{-}Q^5_v + Q^1_v \phantom{-}\\ Q^5_v - Q^1_v & Q^0_v - Q^4_v\phantom{-}\end{pmatrix*} \\[8pt]
\begin{pmatrix}e_ 0 & e_ 2 \\e_ 4 & e_ 6\end{pmatrix} & \begin{pmatrix*}[r]\phantom{-}P_v^2 + P_v^6 & \phantom{-}P_v^7 + P_v^3 \phantom{-}\\P_v^7 - P_v^3 & P_v^2 - P_v^6\phantom{-}\end{pmatrix*} & \begin{pmatrix}e_ 1 & e_ 3 \\e_ 5 & e_ 7\end{pmatrix} & \begin{pmatrix*}[r]\phantom{-}Q^2_v + Q^6_v & \phantom{-}Q^7_v + Q^3_v \phantom{-}\\Q^7_v - Q^3_v & Q^2_v - Q^6_v\phantom{-}\end{pmatrix*} & .
\end{array}
\end{equation}
The $\mathcal{N}=4$ subsector invariant under $SL(2)_{X} \times SO(6,6)$ is given by
\begin{equation}\label{eq:I_3Nequal4}
\begin{split}
I_{dac}&=\det(\gamma^{1}(d)+\gamma^{3}(a)+\gamma^{2}(c)), \quad X=D \\
I_{ebd}&=\det(\gamma^{1}(e)+\gamma^{3}(b)+\gamma^{2}(d)), \quad X=E \\
I_{fce}&=\det(\gamma^{1}(f)+\gamma^{3}(c)+\gamma^{2}(e)), \quad X=F \\
I_{gdf}&=\det(\gamma^{1}(g)+\gamma^{3}(d)+\gamma^{2}(f)), \quad X=G \\
I_{aeg}&=\det(\gamma^{1}(a)+\gamma^{3}(e)+\gamma^{2}(g)), \quad X=A \\
I_{bfa}&=\det(\gamma^{1}(b)+\gamma^{3}(f)+\gamma^{2}(a)), \quad X=B \\
I_{cgb}&=\det(\gamma^{1}(c)+\gamma^{3}(g)+\gamma^{2}(b)), \quad X=C.
\end{split}
\end{equation}
In terms of $a^{i}_{ABD}$, we may write these in two different ways, for example
\begin{gather}
I_{dac}=\det(\gamma^{1}(a^{4})+\gamma^{3}(a^{1})+\gamma^{2}(a^{3})), \quad X=D,
\shortintertext{where}
\begin{split}\label{eq:gooddict}
a^{4}_{DEG}&=d_{DEG}, \\
a^{1}_{ABD}&=a_{ABD}, \\
a^{3}_{CDF}&=c_{CDF},
\end{split}
\shortintertext{or}
I_{dac}=\det(\gamma^{1}(a^{4})+\gamma^{1}(a^{1})+\gamma^{1}(a^{3})), \quad X=D,
\shortintertext{where}
\begin{split}\label{eq:baddict}
a^{4}_{DEG}&=d_{DEG}, \\
a^{1}_{DAB}&=a_{ABD}, \\
a^{3}_{DFC}&=c_{CDF},
\end{split}
\end{gather}
in which case the result is connected to Cayley over the imaginary quaternions
\begin{equation}\label{eq:dac}
\begin{split}
I_{dac}=\Det a&=-\half \varepsilon^{A_{1}A_{2}} \varepsilon^{B_{1}B_{2}} \varepsilon^{A_{3}A_{4}} \varepsilon^{B_{3}B_{4}} \varepsilon^{C_{1}C_{4}} \varepsilon^{C_{2}C_{3}}\\
&\phantom{=}\times{a}^{i}_{A_{1}B_{1}C_{1}}{a}^{j}_{A_{2}B_{2}C_{2}}{a}^{k}_{A_{3}B_{3}C_{3}}{a}^{l}_{A_{4}B_{4}C_{4}} \delta_{ij}\delta_{kl}.
\end{split}
\end{equation}
However, the dictionary \eqref{eq:gooddict} can be applied universally to all the $I_{ijk}$ in \eqref{eq:I_3Nequal4}  whereas the dictionary \eqref{eq:baddict} requires changing for each one. For example, one can write
\begin{equation}
I_{ebd}=\det(\gamma^{1}(a^{5})+\gamma^{1}(a^{2})+\gamma^{1}(a^{4})), \quad X=E,
\end{equation}
but now we require a different $a^{4}$:
\begin{equation}\label{eq:baddict2}
\begin{split}
a^{5}_{EFA}&=e_{EFA}, \\
a^{2}_{EBC}&=b_{BCE}, \\
a^{4}_{EGD}&=d_{DEG}.
\end{split}
\end{equation}
So this dictionary does not lend itself to describing the $\mathcal{N}=8$ invariant.
\newcommand{\mhalf}{\phantom{-}\tfrac{1}{2}}
\begin{sidewaystable}
\begin{tabular*}{\textwidth}{@{\extracolsep{\fill}}*{13}{c}}
\toprule
& Freudenthal & Fano & & Freudenthal & Fano & & Freudenthal & Fano & & Freudenthal & Fano & \\
\midrule
& $p^0$ & $ a_0$ & & $p^1$ & $-a_1$           & & $p^2$ & $-a_2$           & & $p^3$ & $-a_4$           & \\
& $q_0$ & $ a_7$ & & $q_1$ & $\phantom{-}a_6$ & & $q_2$ & $\phantom{-}a_5$ & & $q_3$ & $\phantom{-}a_3$ & \\
\midrule
& $P_c^0$ & $ \half(d_3+d_0)$ & & $P_c^1$ & $ \mhalf(d_1-d_2)$ & & $P_c^2$ & $ -\half(c_5+c_0)$ & & $P_c^3$ & $ \mhalf(c_4-c_1)$ &\\
& $P_c^4$ & $ \half(d_3-d_0)$ & & $P_c^5$ & $ -\half(d_1+d_2)$ & & $P_c^6$ & $ \mhalf(c_5-c_0)$ & & $P_c^7$ & $ -\half(c_4+c_1)$ & \\
\midrule
& $P_s^0$ & $ \half(b_2-b_1)$ & & $P_s^1$ & $ \mhalf(b_3+b_0)$ & & $P_s^2$ & $ \mhalf(f_2-f_4)$ & & $P_s^3$ & $ \mhalf(f_6+f_0)$ & \\
& $P_s^4$ & $ \half(b_2+b_1)$ & & $P_s^5$ & $ \mhalf(b_3-b_0)$ & & $P_s^6$ & $ \mhalf(f_2+f_4)$ & & $P_s^7$ & $ \mhalf(f_6-f_0)$ & \\
\midrule
& $P_v^0$ & $ \half(g_0+g_5)$ & & $P_v^1$ & $ \mhalf(g_1-g_4)$ & & $P_v^2$ & $ \mhalf(e_0+e_6)$ & & $P_v^3$ & $ \mhalf(e_2-e_4)$ & \\
& $P_v^4$ & $ \half(g_0-g_5)$ & & $P_v^5$ & $ \mhalf(g_1+g_4)$ & & $P_v^6$ & $ \mhalf(e_0-e_6)$ & & $P_v^7$ & $ \mhalf(e_2+e_4)$ & \\
\midrule
& $Q^0_c$ & $ \half(d_7+d_4)$ & & $Q^1_c$ & $ \mhalf(d_5-d_6)$ & & $Q^2_c$ & $ -\half(c_7+c_2)$ & & $Q^3_c$ & $ \mhalf(c_6-c_3)$ & \\
& $Q^4_c$ & $ \half(d_7-d_4)$ & & $Q^5_c$ & $ -\half(d_5+d_6)$ & & $Q^6_c$ & $ \mhalf(c_7-c_2)$ & & $Q^7_c$ & $ -\half(c_6+c_3)$ & \\
\midrule
& $Q^0_s$ & $ \half(b_6-b_5)$ & & $Q^1_s$ & $ \mhalf(b_7+b_4)$ & & $Q^2_s$ & $ \mhalf(f_3-f_5)$ & & $Q^3_s$ & $ \mhalf(f_7+f_1)$ & \\
& $Q^4_s$ & $ \half(b_6+b_5)$ & & $Q^5_s$ & $ \mhalf(b_7-b_4)$ & & $Q^6_s$ & $ \mhalf(f_3+f_5)$ & & $Q^7_s$ & $ \mhalf(f_7-f_1)$ & \\
\midrule
& $Q^0_v$ & $ \half(g_2+g_7)$ & & $Q^1_v$ & $ \mhalf(g_3-g_6)$ & & $Q^2_v$ & $ \mhalf(e_1+e_7)$ & & $Q^3_v$ & $ \mhalf(e_3-e_5)$ & \\
& $Q^4_v$ & $ \half(g_2-g_7)$ & & $Q^5_v$ & $ \mhalf(g_3+g_6)$ & & $Q^6_v$ & $ \mhalf(e_1-e_7)$ & & $Q^7_v$ & $ \mhalf(e_3+e_5)$ & \\
\bottomrule
\end{tabular*}
\caption[Freudenthal to Fano dictionary]{The dictionary transforming from the Freudenthal basis to the Fano basis.}
\label{tab:FreudenthalToFano}
\end{sidewaystable}
\begin{sidewaystable}
\begin{tabular*}{\textwidth}{@{\extracolsep{\fill}}*{13}{c}}
\toprule
& Fano & Freudenthal & & Fano & Freudenthal & & Fano & Freudenthal & & Fano & Freudenthal & \\
\midrule
& $a_0$ & $\phantom{-}p^0$ & & $a_1$ & $-p^1$           & & $a_2$ & $-p^2$           & & $a_3$ & $ q_3$ & \\
& $a_4$ & $-p^3$           & & $a_5$ & $\phantom{-}q_2$ & & $a_6$ & $\phantom{-}q_1$ & & $a_7$ & $ q_0$ & \\
\midrule
& $b_0$ & $P_s^1-P_s^5$ & & $b_1$ & $P_s^4-P_s^0$ & & $b_2$ & $P_s^0+P_s^4$ & & $b_3$ & $P_s^1+P_s^5$ & \\
& $b_4$ & $Q^1_s-Q^5_s$ & & $b_5$ & $Q^4_s-Q^0_s$ & & $b_6$ & $Q^0_s+Q^4_s$ & & $b_7$ & $Q^1_s+Q^5_s$ & \\
\midrule
& $c_0$ & $-P_c^2-P_c^6\phantom{-}$ & & $c_1$ & $-P_c^3-P_c^7\phantom{-}$ & & $c_2$ & $-Q^2_c-Q^6_c\phantom{-}$ & & $c_3$ & $-Q^3_c-Q^7_c\phantom{-}$ & \\
& $c_4$ & $P_c^3-P_c^7$ & & $c_5$ & $P_c^6-P_c^2$ & & $c_6$ & $Q^3_c-Q^7_c$ & & $c_7$ & $Q^6_c-Q^2_c$ & \\
\midrule
& $d_0$ & $P_c^0-P_c^4$ & & $d_1$ & $P_c^1-P_c^5$ & & $d_2$ & $-P_c^1-P_c^5\phantom{-}$ & & $d_3$ & $P_c^0+P_c^4$ & \\
& $d_4$ & $Q^0_c-Q^4_c$ & & $d_5$ & $Q^1_c-Q^5_c$ & & $d_6$ & $-Q^1_c-Q^5_c\phantom{-}$ & & $d_7$ & $Q^0_c+Q^4_c$ & \\
\midrule
& $e_0$ & $P_v^2+P_v^6$ & & $e_1$ & $Q^2_v+Q^6_v$ & & $e_2$ & $P_v^3+P_v^7$ & & $e_3$ & $Q^3_v+Q^7_v$ & \\
& $e_4$ & $P_v^7-P_v^3$ & & $e_5$ & $Q^7_v-Q^3_v$ & & $e_6$ & $P_v^2-P_v^6$ & & $e_7$ & $Q^2_v-Q^6_v$ & \\
\midrule
& $f_0$ & $P_s^3-P_s^7$ & & $f_1$ & $Q^3_s-Q^7_s$ & & $f_2$ & $P_s^2+P_s^6$ & & $f_3$ & $Q^2_s+Q^6_s$ & \\
& $f_4$ & $P_s^6-P_s^2$ & & $f_5$ & $Q^6_s-Q^2_s$ & & $f_6$ & $P_s^3+P_s^7$ & & $f_7$ & $Q^3_s+Q^7_s$ & \\
\midrule
& $g_0$ & $P_v^0+P_v^4$ & & $g_1$ & $P_v^1+P_v^5$ & & $g_2$ & $Q^0_v+Q^4_v$ & & $g_3$ & $Q^1_v+Q^5_v$ & \\
& $g_4$ & $P_v^5-P_v^1$ & & $g_5$ & $P_v^0-P_v^4$ & & $g_6$ & $Q^5_v-Q^1_v$ & & $g_7$ & $Q^0_v-Q^4_v$ & \\
\bottomrule
\end{tabular*}
\caption[Fano to Freudenthal dictionary]{The dictionary transforming from the Fano basis to the Freudenthal basis.}
\label{tab:FanoToFreudenthal}
\end{sidewaystable}
\subsection{Cartan-Freudenthal dictionary}

The Cartan-Freudenthal dictionary may be found in \autoref{tab:CartanFreudenthalDictionary} and the reverse (Freudenthal-Cartan) may be found in \autoref{tab:FreudenthalCartanDictionary}.
\begin{sidewaystable}
\begin{tabular*}{\textwidth}{@{\extracolsep{\fill}}*{13}{c}}
\toprule
& Freudenthal & Cartan & & Freudenthal & Cartan & & Freudenthal & Cartan & & Freudenthal & Cartan & \\
\midrule
& $p^0$ & $-y_{01}$ & & $p^1$ & $-x^{34}$           & & $p^2$ & $-x^{26}$           & & $p^3$ & $-x^{57}$            & \\
& $q_0$ & $-x^{01}$ & & $q_1$ & $\phantom{-}y_{34}$ & & $q_2$ & $\phantom{-}y_{26}$ & & $q_3$ & $\phantom{-}y_{57}$  & \\
\midrule
& $P_c^0$ & $\mhalf(y_{13}-y_{04})$ & & $P_c^1$ & $\mhalf(x^{25}+x^{67})$ & & $P_c^2$ & $\mhalf(y_{03}+y_{14})$ & & $P_c^3$ & $-\half(x^{27}+x^{56})$ & \\
& $P_c^4$ & $\mhalf(y_{13}+y_{04})$ & & $P_c^5$ & $\mhalf(x^{25}-x^{67})$ & & $P_c^6$ & $\mhalf(y_{03}-y_{14})$ & & $P_c^7$ & $\mhalf(x^{27}-x^{56})$ & \\
\midrule
& $P_s^0$ & $\mhalf(x^{37}-x^{45})$ & & $P_s^1$ & $-\half(y_{02}+y_{16})$ & & $P_s^2$ & $-\half(x^{35}+x^{47})$ & & $P_s^3$ & $\mhalf(y_{12}-y_{06})$ & \\
& $P_s^4$ & $\mhalf(x^{37}+x^{45})$ & & $P_s^5$ & $\mhalf(y_{02}-y_{16})$ & & $P_s^6$ & $\mhalf(x^{35}-x^{47})$ & & $P_s^7$ & $\mhalf(y_{12}+y_{06})$ & \\
\midrule
& $P_v^0$ & $\mhalf(y_{15}-y_{07})$ & & $P_v^1$ & $\mhalf(x^{23}-x^{46})$ & & $P_v^2$ & $-\half(y_{17}+y_{05})$ & & $P_v^3$ & $-\half(x^{24}+x^{36})$ & \\
& $P_v^4$ & $-\half(y_{15}+y_{07})$ & & $P_v^5$ & $\mhalf(x^{23}+x^{46})$ & & $P_v^6$ & $\mhalf(y_{17}-y_{05})$ & & $P_v^7$ & $\mhalf(x^{24}-x^{36})$ & \\
\midrule
& $Q_0^c$ & $\mhalf(x^{13}-x^{04})$ & & $Q_1^c$ & $-\half(y_{25}+y_{67})$ & & $Q_2^c$ & $\mhalf(x^{14}+x^{03})$ & & $Q_3^c$ & $\mhalf(y_{27}+y_{56})$ & \\
& $Q_4^c$ & $-\half(x^{13}+x^{04})$ & & $Q_5^c$ & $\mhalf(y_{25}-y_{67})$ & & $Q_6^c$ & $\mhalf(x^{14}-x^{03})$ & & $Q_7^c$ & $\mhalf(y_{27}-y_{56})$ & \\
\midrule
& $Q_0^s$ & $\mhalf(y_{45}-y_{37})$ & & $Q_1^s$ & $-\half(x^{16}+x^{02})$ & & $Q_2^s$ & $\mhalf(y_{35}+y_{47})$ & & $Q_3^s$ & $\mhalf(x^{12}-x^{06})$ & \\
& $Q_4^s$ & $\mhalf(y_{45}+y_{37})$ & & $Q_5^s$ & $\mhalf(x^{16}-x^{02})$ & & $Q_6^s$ & $\mhalf(y_{35}-y_{47})$ & & $Q_7^s$ & $-\half(x^{12}+x^{06})$ & \\
\midrule
& $Q_0^v$ & $\mhalf(x^{15}-x^{07})$ & & $Q_1^v$ & $\mhalf(y_{46}-y_{23})$ & & $Q_2^v$ & $-\half(x^{05}+x^{17})$ & & $Q_3^v$ & $\mhalf(y_{24}+y_{36})$ & \\
& $Q_4^v$ & $\mhalf(x^{15}+x^{07})$ & & $Q_5^v$ & $\mhalf(y_{46}+y_{23})$ & & $Q_6^v$ & $\mhalf(x^{05}-x^{17})$ & & $Q_7^v$ & $\mhalf(y_{24}-y_{36})$ & \\
\bottomrule
\end{tabular*}
\caption[Freudenthal to Cartan dictionary]{The dictionary transforming from the Freudenthal basis to the Cartan basis.}
\label{tab:FreudenthalCartanDictionary}
\end{sidewaystable}
\begin{sidewaystable}
\begin{tabular*}{\textwidth}{@{\extracolsep{\fill}}cccc}
\toprule
& Cartan & Freudenthal & \\
\midrule
& $x$ &
$\begin{pmatrix}
0 & -q_0 & -Q_s^1-Q_s^5\phantom{-} & Q_c^2-Q_c^6 & -Q_c^0-Q_c^4\phantom{-} & Q_v^6-Q_v^2 & -Q_s^3-Q_s^7\phantom{-} & Q_v^4-Q_v^0 \\
q_0 & 0 & Q_s^3-Q_s^7 & Q_c^0-Q_c^4 & Q_c^2+Q_c^6 & Q_v^0+Q_v^4 & Q_s^5-Q_s^1 & -Q_v^2-Q_v^6\phantom{-} \\
Q_s^1+Q_s^5 & Q_s^7-Q_s^3 & 0 & P_1^v+P_5^v & P_7^v-P_3^v & P_1^c+P_5^c & -p^2 & P_7^c-P_3^c \\
Q_c^6-Q_c^2 & Q_c^4-Q_c^0 & -P_1^v-P_5^v\phantom{-} & 0 & -p^1 & P_6^s-P_2^s & -P_3^v-P_7^v\phantom{-} & P_0^s+P_4^s \\
Q_c^0+Q_c^4 & -Q_c^2-Q_c^6\phantom{-} & P_3^v-P_7^v & p^1 & 0 & P_4^s-P_0^s & P_5^v-P_1^v & -P_2^s-P_6^s\phantom{-} \\
Q_v^2-Q_v^6 & -Q_v^0-Q_v^4\phantom{-} & -P_1^c-P_5^c\phantom{-} & P_2^s-P_6^s & P_0^s-P_4^s & 0 & -P_3^c-P_7^c\phantom{-} & -p^3 \\
Q_s^3+Q_s^7 & Q_s^1-Q_s^5 & p^2 & P_3^v+P_7^v & P_1^v-P_5^v & P_3^c+P_7^c & 0 & P_1^c-P_5^c \\
Q_v^0-Q_v^4 & Q_v^2+Q_v^6 & P_3^c-P_7^c & -P_0^s-P_4^s\phantom{-} & P_2^s+P_6^s & p^3 & P_5^c-P_1^c & 0
\end{pmatrix}$ & \\
\midrule
& $y$ &
$\begin{pmatrix}
0 & -p^0 & P_5^s-P_1^s & P_2^c+P_6^c & P_4^c-P_0^c & -P_2^v-P_6^v\phantom{-} & P_7^s-P_3^s & -P_0^v-P_4^v\phantom{-} \\
p^0 & 0 & P_3^s+P_7^s & P_0^c+P_4^c & P_2^c-P_6^c & P_0^v-P_4^v & -P_1^s-P_5^s\phantom{-} & P_6^v-P_2^v \\
P_1^s-P_5^s & -P_3^s-P_7^s\phantom{-} & 0 & Q_v^5-Q_v^1 & Q_v^3+Q_v^7 & Q_c^5-Q_c^1 & q_2 & Q_c^3+Q_c^7 \\
-P_2^c-P_6^c\phantom{-} & -P_0^c-P_4^c\phantom{-} & Q_v^1-Q_v^5 & 0 & q_1 & Q_s^2+Q_s^6 & Q_v^3-Q_v^7 & Q_s^4-Q_s^0 \\
P_0^c-P_4^c & P_6^c-P_2^c & -Q_v^3-Q_v^7\phantom{-} & -q_1 & 0 & Q_s^0+Q_s^4 & Q_v^1+Q_v^5 & Q_s^2-Q_s^6 \\
P_2^v+P_6^v & P_4^v-P_0^v & Q_c^1-Q_c^5 & -Q_s^2-Q_s^6\phantom{-} & -Q_s^0-Q_s^4\phantom{-} & 0 & Q_c^3-Q_c^7 & q_3 \\
P_3^s-P_7^s & P_1^s+P_5^s & -q_2 & Q_v^7-Q_v^3 & -Q_v^1-Q_v^5\phantom{-} & Q_c^7-Q_c^3 & 0 & -Q_c^1-Q_c^5\phantom{-} \\
P_0^v+P_4^v & P_2^v-P_6^v & -Q_c^3-Q_c^7\phantom{-} & Q_s^0-Q_s^4 & Q_s^6-Q_s^2 & -q_3 & Q_c^1+Q_c^5 & 0
\end{pmatrix}$ & \\
\bottomrule
\end{tabular*}
\caption[Cartan to Freudenthal]{The dictionary transforming from the Cartan basis to the Freudenthal basis.}
\label{tab:CartanFreudenthalDictionary}
\end{sidewaystable}

\newpage
\section{\texorpdfstring{CAYLEY, CARTAN AND THE OCTONIONS}{Cayley, Cartan and the Octonions}}
\label{sec:Cayley}

\subsection{\texorpdfstring{$\mathcal{N}=4$ Cartan invariant and the quaternions}{N=4 Cartan invariant}}

Define
\begin{gather}
P_{\mu\nu}=P_{\mu}Q_{\nu}-Q_{\mu}P_{\nu},
\shortintertext{where $\mu=0,1,2,3$; then Cayley can be written}
-\Det a=\half P_{\mu\nu}P^{\mu\nu}=P^{2}Q^{2}-(P\cdot Q)^{2},
\end{gather}
where we use the $SO(2,2)$ metric. This is the $\mathcal{N}=2$ expression invariant under $SL(2) \times SO(2,2)$.  Now define
\begin{gather}
P\indices{_{\mu\nu}^{ij}}=P_{\mu}^iQ_{\nu}^j-Q_{\mu}^iP_{\nu}^j,
\shortintertext{where $i=1,2,3$, then}
-\Det a=\half P\indices{_{\mu\nu}^{ij}}P^{\mu\nu ij}=P^{2}Q^{2}-(P\cdot Q)^{2},
\end{gather}
where we now use the $SO(6,6)$ metric. This is the $\mathcal{N}=4$ expression, invariant under $SL(2) \times SO(6,6)$.

So if we write
\begin{gather}
-\Det a=\half P\indices{_{\mu\nu}^{ij}}P^{\mu\nu kl}X_{ijkl},
\shortintertext{we need}
X_{ijkl}=\delta_{ik}\delta_{jl}+\dotsb,
\end{gather}
where the dots refer to terms that vanish when contracted with $P\indices{_{\mu\nu}^{ij}}P^{\mu\nu kl}$. Note that $X_{ijkl}$ is symmetric under
\begin{equation}\label{eq:perm}
\begin{array}{c@{~\to~}c@{~;~}c@{~\to~}c}
i & j & k & l, \\
i & k & j & l, \\
i & l & j & k.
\end{array}
\end{equation}

\subsection{Cayley over the imaginary quaternions}
\label{sec:Cayleyimquat}
If we denote the imaginary quaternions by $e_i$ where $i=1,2,3$, the structure constants are defined by
\begin{gather}
e_i e_j=-\delta_{ij}+\varepsilon_{ijk}e_k.
\shortintertext{and so we obtain the quartic expression}
\begin{split}
e_{i}e_{j}e_{k}e_{l}&=(-\delta_{ij}+\varepsilon_{ijm}e_{m})(-\delta_{kl}+\varepsilon_{kln}e_{n}) \\
&=\delta_{ij}\delta_{kl}-\varepsilon_{ijm}\varepsilon_{klm} -\delta_{ij}\varepsilon_{kln}e_{n} -\delta_{kl}\varepsilon_{ijm}e_{m} +\varepsilon_{ijm}\varepsilon_{kln}\varepsilon_{mnp}e_{p},
\end{split}
\shortintertext{or}
e_{i}e_{j}e_{k}e_{l}=\delta_{ij}\delta_{kl}-\delta_{ik}\delta_{jl}+\delta_{il}\delta_{jk} -\delta_{ij}\varepsilon_{kln}e_{n}-\delta_{kl}\varepsilon_{ijm}e_{m} +\varepsilon_{ijm}\varepsilon_{kln}\varepsilon_{mnp}e_{p}.
\shortintertext{The combination}
\begin{split}
e_{i}e_{j}e_{l}e_{k}+e_{i}e_{k}e_{j}e_{l}=&\ \delta_{ij}\delta_{kl} +\varepsilon_{ijm}\varepsilon_{klm} +\delta_{ij}\varepsilon_{kln}e_{n} -\delta_{kl}\varepsilon_{ijm}e_{m} \\
&-\varepsilon_{ijm}\varepsilon_{kln}\varepsilon_{mnp}e_{p} +\delta_{ik}\delta_{jl} -\varepsilon_{ikm}\varepsilon_{jlm} \\
&-\delta_{ik}\varepsilon_{jln}e_{n} -\delta_{jl}\varepsilon_{ikm}e_{m} -\varepsilon_{ikm}\varepsilon_{jln}\varepsilon_{mnp}e_{p}.
\end{split}
\shortintertext{yields}
\begin{split}
e_ie_je_le_k+e_ie_ke_je_l&=\phantom{+}\delta_{ij}\delta_{kl} -\delta_{il}\delta_{jk} +\delta_{ik}\delta_{jl} \\
&\phantom{\,=}+\delta_{ij}\varepsilon_{kln}e_{n} -\delta_{kl}\varepsilon_{ijm}e_{m} -\varepsilon_{ijm}\varepsilon_{kln}\varepsilon_{mnp}e_{p} \\
&\phantom{\,=}+\delta_{ik}\delta_{jl} -\delta_{ij}\delta_{kl} +\delta_{il}\delta_{jk} \\
&\phantom{\,=}-\delta_{ik}\varepsilon_{jln}e_{n} -\delta_{jl}\varepsilon_{ikm}e_{m} +\varepsilon_{ikm}\varepsilon_{jln}\varepsilon_{mnp}e_{p} \\
&= 2\delta_{ik}\delta_{jl} +\dotsb,
\end{split}
\end{gather}
where the dots refer to terms that vanish when contracted with $P\indices{_{\mu\nu}^{ij}}P^{\mu\nu kl}$. So Cayley over the imaginary quaternions in the form
\begin{equation}
-\Det a=\tfrac{1}{4}P\indices{_{\mu\nu}^{ij}}P^{\mu\nu kl} (e_ie_je_le_k+e_ie_ke_je_l)
\end{equation}
yields the $\mathcal{N}=4, SL(2) \times SO(6,6)$ invariant.

In terms of the $a$'s, on the other hand, the $\mathcal{N}=4$ invariant is given by \eqref{eq:dac}.  So we require the different ordering
\begin{equation}
e_i e_j e_k e_l + e_i e_k e_l e_j=2\delta_{ij}\delta_{kl} +\dotsb,
\label{eq:ijkl}
\end{equation}
where the dots refer to terms that vanish when contracted with
\begin{align}\label{eq:A}
A^{ijkl}=&-\half
\varepsilon^{A_1A_2}\varepsilon^{B_1B_2}\varepsilon^{A_3A_4}\varepsilon^{B_3B_4}\varepsilon^{D_1D_4}\varepsilon^{D_2D_3} \\
&\times a^i_{A_1B_1D_1}a^j_{A_2B_2D_2}a^k_{A_3B_3D_3}a^l_{A_4B_4D_4}.
\end{align}
Let us consider the case where we choose the imaginary quaternions to be
\begin{equation}
q=a e_1+e e_5+g e_7\;.
\end{equation}
In order to define Cayley over imaginary quaternions one needs to be careful with the ordering of the indices in $a_{ABD}$, $e_{EFA}$ and $g_{GAC}$ when contracting them with $\varepsilon^{A_1 A_2}$, etc in \eqref{eq:A}. The final result is given by
\begin{equation}
\begin{gathered}
-\Det a =\\
\begin{array}{c@{\ }c@{}c*{5}{@{\ +\ }c}@{}c}
  &    & a_3^2 a_4^2     & a_2^2 a_5^2     & a_1^2 a_6^2     & a_0^2 a_7^2     & e_3^2 e_4^2     & e_2^2 e_5^2     &   \\
  & +~ & e_1^2 e_6^2     & e_0^2 e_7^2     & g_3^2 g_4^2     & g_2^2 g_5^2     & g_1^2 g_6^2     & g_0^2 g_7^2     &   \\
+ & 2( & a_2 a_5 e_3 e_4 & a_1 a_6 e_3 e_4 & a_2 a_5 e_2 e_5 & a_1 a_6 e_2 e_5 & a_3 a_4 e_1 e_6 & a_0 a_7 e_1 e_6 &   \\
  & +~ & a_3 a_4 e_0 e_7 & a_0 a_7 e_0 e_7 & a_2 a_5 g_3 g_4 & a_1 a_6 g_3 g_4 & e_3 e_4 g_3 g_4 & e_2 e_5 g_3 g_4 &   \\
  & +~ & a_3 a_4 g_2 g_5 & a_0 a_7 g_2 g_5 & e_1 e_6 g_2 g_5 & e_0 e_7 g_2 g_5 & a_2 a_5 g_1 g_6 & a_1 a_6 g_1 g_6 &   \\
  & +~ & e_3 e_4 g_1 g_6 & e_2 e_5 g_1 g_6 & a_3 a_4 g_0 g_7 & a_0 a_7 g_0 g_7 & e_1 e_6 g_0 g_7 & e_0 e_7 g_0 g_7 & ) \\
- & 4( & a_5 a_6 e_2 e_4 & a_1 a_2 e_3 e_5 & a_4 a_7 e_0 e_6 & a_0 a_3 e_1 e_7 & a_5 a_6 g_1 g_4 & e_3 e_5 g_1 g_4 &   \\
  & +~ & a_4 a_7 g_0 g_5 & e_1 e_7 g_0 g_5 & a_1 a_2 g_3 g_6 & e_2 e_4 g_3 g_6 & a_0 a_3 g_2 g_7 & e_0 e_6 g_2 g_7 & ) \\
+ & 4( & a_0 a_3 a_5 a_6 & a_1 a_2 a_4 a_7 & a_4 a_7 e_2 e_4 & a_0 a_3 e_3 e_5 & a_5 a_6 e_0 e_6 & e_0 e_3 e_5 e_6 &   \\
  & +~ & a_1 a_2 e_1 e_7 & e_1 e_2 e_4 e_7 & a_4 a_7 g_1 g_4 & e_1 e_7 g_1 g_4 & a_5 a_6 g_0 g_5 & e_3 e_5 g_0 g_5 &   \\
  & +~ & a_0 a_3 g_3 g_6 & e_0 e_6 g_3 g_6 & g_0 g_3 g_5 g_6 & a_1 a_2 g_2 g_7 & e_2 e_4 g_2 g_7 & g_1 g_2 g_4 g_7 & ) \\
- & 2( & a_2 a_3 a_4 a_5 & a_1 a_3 a_4 a_6 & a_1 a_2 a_5 a_6 & a_0 a_3 a_4 a_7 & a_0 a_2 a_5 a_7 & a_0 a_1 a_6 a_7 &   \\
  & +~ & a_3 a_4 e_3 e_4 & a_0 a_7 e_3 e_4 & a_3 a_4 e_2 e_5 & a_0 a_7 e_2 e_5 & e_2 e_3 e_4 e_5 & a_2 a_5 e_1 e_6 &   \\
  & +~ & a_1 a_6 e_1 e_6 & e_1 e_3 e_4 e_6 & e_1 e_2 e_5 e_6 & a_2 a_5 e_0 e_7 & a_1 a_6 e_0 e_7 & e_0 e_3 e_4 e_7 &   \\
  & +~ & e_0 e_2 e_5 e_7 & e_0 e_1 e_6 e_7 & a_3 a_4 g_3 g_4 & a_0 a_7 g_3 g_4 & e_1 e_6 g_3 g_4 & e_0 e_7 g_3 g_4 &   \\
  & +~ & a_2 a_5 g_2 g_5 & a_1 a_6 g_2 g_5 & e_3 e_4 g_2 g_5 & e_2 e_5 g_2 g_5 & g_2 g_3 g_4 g_5 & a_3 a_4 g_1 g_6 &   \\
  & +~ & a_0 a_7 g_1 g_6 & e_1 e_6 g_1 g_6 & e_0 e_7 g_1 g_6 & g_1 g_3 g_4 g_6 & g_1 g_2 g_5 g_6 & a_2 a_5 g_0 g_7 &   \\
  & +~ & a_1 a_6 g_0 g_7 & e_3 e_4 g_0 g_7 & e_2 e_5 g_0 g_7 & g_0 g_3 g_4 g_7 & g_0 g_2 g_5 g_7 & g_0 g_1 g_6 g_7 & ),
\end{array}
\end{gathered}
\end{equation}
which is twice the $\mathcal{N}=4$ subsector of $I_4$ where we keep the letters $a$, $e$ and $g$ with the common qubit $A$. The factor 2 comes from equation \eqref{eq:ijkl}.

\subsection{\texorpdfstring{$\mathcal{N}=8$ Cartan invariant and the octonions}{N=8 Cartan invariant and the octonions}}

Here we adopt the imaginary octonions, $e_i$ where $i=1,2,3,4,5,6,7$, with the dual Fano structure constants $c_{ijk}$
\begin{equation}
e_i e_j=-\delta_{ij}+c_{ijk}e_k.
\end{equation}
We have the same identities as for the Fano structure constants in \autoref{sec:octandfan}
\begin{gather}
c_{pmk}c_{qkn}+c_{qmk}c_{pkn}=\delta_{pm}\delta_{qn}+\delta_{pn}\delta_{qm}-2\delta_{pq}\delta_{mn}.\\
\begin{gathered}
[ e_{i},[e_{j},e_{k}]]+[ e_{k},[e_{i},e_{j}]]+[ e_{j},[e_{k},e_{i}]] \\
=4(c_{jkm}c_{imn}+c_{ijm}c_{kmn}+c_{kim}c_{jmn})e_{n} \\
=:3c_{ijkl}e_{l},
\end{gathered}
\shortintertext{where $c_{ijkl}$ satisfies}
c_{ijkl}=\tfrac{1}{6} \varepsilon_{ijklmnp}c_{mnp},
\shortintertext{and}
c_{ijkl}=-c_{mij}c_{mkl}-\delta_{il}\delta_{jk}+\delta_{ik}\delta_{jl}.
\shortintertext{We also note}
c_{ijm}c_{klm}=\delta_{ik}\delta_{jl}-\delta_{il}\delta_{jk}-c_{ijkl},
\shortintertext{and}
\begin{split}
c_{ijn}c_{klmn}=&-c_{ikl}\delta_{mj}+ c_{jkl}\delta_{mi}\\
                &-c_{ilm}\delta_{kj}+ c_{jlm}\delta_{ki}\\
                &-c_{imk}\delta_{lj}+ c_{jmk}\delta_{li}.
\end{split}
\end{gather}

With regard to non-associativity, there are two cubic possibilities:
\begin{subequations}
\begin{gather}
\begin{split}
(e_{i}e_{j})e_{k}&=(-\delta_{ij}+c_{ijm}e_{m})e_{k} \\
&=-\delta_{ij}e_{k}+c_{ijm}(-\delta_{mk}+c_{mkn}e_{n}) \\
&=-\delta_{ij}e_{k}-c_{ijk}+c_{ijm}c_{mkn}e_{n} \\
&=-c_{ijk}-\delta_{ij}e_{k}+\delta_{ki}e_{j}-\delta_{jk}e_{i}-c_{ijkl}e_{l},
\end{split}\\
\begin{split}
e_{i}(e_{j}e_{k})&=e_{i}(-\delta_{jk}+c_{jkm}e_{m}) \\
&=-\delta_{jk}e_{i}+c_{jkm}(-\delta_{im}+c_{imn}e_{n}) \\
&=-\delta_{jk}e_{i}-c_{jki}+c_{jkm}c_{imn}e_{n} \\
&=-c_{ijk}-\delta_{ij}e_{k}+\delta_{ki}e_{j}-\delta_{jk}e_{i}+c_{ijkl}e_{l}.
\end{split}
\end{gather}
\end{subequations}
Note that they differ only by the sign of the associator term. Similarly, there are five quartic possibilities:
\begin{subequations}
\begin{gather}
\begin{split}\label{eq:quarticposs1}
(e_{i}e_{j})(e_{k}e_{l})&=(-\delta_{ij}+c_{ijm}e_{m})
                         (-\delta_{kl}+c_{kln}e_{n}) \\
&=\delta_{ij}\delta_{kl}-c_{ijm}c_{klm} -\delta_{ij}c_{kln}e_{n}-\delta_{kl}c_{ijm}e_{m}+c_{ijm}c_{kln}c_{mnp}e_{p} \\
&=\delta_{ij}\delta_{kl}-\delta_{ik}\delta_{jl}+\delta_{il}\delta_{jk}+c_{ijkl} \\
&\phantom{=}-\delta_{ij}c_{kln}e_{n}-\delta_{kl}c_{ijn}e_{n}+c_{kli}e_{j}-c_{klj}e_{i} -c_{klm}c_{ijmn}e_{n},
\end{split}\\
\begin{split}\label{eq:quarticposs2}
[(e_{i}e_{j})e_{k}]e_{l}&=[-\delta_{ij}e_{k}-c_{ijk}+c_{ijm}c_{mkn}e_{n}]e_{l} \\
&=\delta_{ij}\delta_{kl}-c_{ijm}c_{mkl} -\delta_{ij}c_{kln}e_{n}-c_{ijk}e_{l}+c_{ijn}c_{nkm}c_{mlp}e_{p} \\
&=\delta_{ij}\delta_{kl}-\delta_{ik}\delta_{jl}+\delta_{il}\delta_{jk}+c_{ijkl} \\
&\phantom{=}-\delta_{ij}c_{kln}e_{n}+\delta_{ik}c_{jln}e_{n}-\delta_{jk}c_{iln}e_{n}-c_{ijk}e_{l} -c_{ijkm}c_{mln}e_{n},
\end{split}\\
\begin{split}\label{eq:quarticposs3}
e_{i}[e_{j}(e_{k}e_{l})]&=e_{i}[-\delta_{kl}e_{j}-c_{klj}+c_{klm}c_{jmn}e_{n}] \\
&=\delta_{ij}\delta_{kl}-c_{klm}c_{jmi} -\delta_{kl}c_{ijn}e_{n}-c_{klj}e_{i}+c_{klm}c_{jmn}c_{inp}e_{p}\\
&=\delta_{ij}\delta_{kl}-\delta_{ik}\delta_{jl}+\delta_{il}\delta_{jk}+c_{ijkl}\\
&\phantom{=}-\delta_{kl}c_{ijn}e_n+\delta_{jk}c_{lin}e_{n}-\delta_{jl}c_{kin}e_{n}-c_{jkl}e_{i} -c_{jklm}c_{min}e_{n},
\end{split}\\
\begin{split}\label{eq:quarticposs4}
e_{i}[(e_{j}e_{k})e_{l}]&=e_{i}[-\delta_{jk}e_{l}-c_{jkl}+c_{jkn}c_{nlm}e_{m}] \\
&=\delta_{jk}\delta_{il}-c_{jkn}c_{nli}-\delta_{jk}c_{iln}e_{n}-c_{jkl}e_{i}+c_{jkn}c_{nlm}c_{imp}e_{p}\\
&=\delta_{ij}\delta_{kl}-\delta_{ik}\delta_{jl}+\delta_{il}\delta_{jk}-c_{ijkl}\\
&\phantom{=}-\delta_{kl}c_{ijn}e_n+\delta_{lj}c_{ikn}e_{n}-\delta_{jk}c_{iln}e_{n}-c_{jkl}e_{i} +c_{jklm}c_{min}e_{n},
\end{split}\\
\begin{split}\label{eq:quarticposs5}
[e_{i}(e_{j}e_{k})]e_{l}&=[-\delta_{jk}e_{i}-c_{jki}+c_{jkm}c_{imn}e_{n}]e_{l} \\
&=\delta_{jk}\delta_{il}-c_{jkm}c_{iml}-\delta_{jk}c_{iln}e_{n}-c_{jki}e_{l}+c_{jkm}c_{imn}c_{nlp}e_{p}\\
&=\delta_{ij}\delta_{kl}-\delta_{ik}\delta_{jl}+\delta_{il}\delta_{jk}-c_{ijkl}\\
&\phantom{=}-\delta_{ij}c_{kln}e_n+\delta_{ki}c_{jln}e_{n}-\delta_{jk}c_{iln}e_{n}-c_{ijk}e_{l} +c_{ijkm}c_{mln}e_{n}.
\end{split}
\end{gather}
\end{subequations}
Note again that the real parts of \eqref{eq:quarticposs1},\eqref{eq:quarticposs2},\eqref{eq:quarticposs3} differ from the real parts of \eqref{eq:quarticposs4},\eqref{eq:quarticposs5} by the sign of the associator term. The real parts of all the multiplication orderings \eqref{eq:quarticposs1}-\eqref{eq:quarticposs5} are invariant under \eqref{eq:perm}. The imaginary part of \eqref{eq:quarticposs1} is cancelled by itself under this symmetry. The imaginary parts of \eqref{eq:quarticposs2} and \eqref{eq:quarticposs3} cancel each other under this symmetry. The imaginary parts of \eqref{eq:quarticposs4} and \eqref{eq:quarticposs5} also cancel each other under this symmetry.  Note also that for these real parts, there are six distinct combinations since
\begin{equation}
\begin{split}
e_ie_je_ke_l &\sim \phantom{-}\delta_{ij}\delta_{kl}-\delta_{ik}\delta_{jl}+\delta_{il}\delta_{jk}+c_{ijkl}, \\
e_ie_je_le_k &\sim \phantom{-}\delta_{ij}\delta_{kl}+\delta_{ik}\delta_{jl}-\delta_{il}\delta_{jk}-c_{ijkl}, \\
e_ie_ke_je_l &\sim -\delta_{ij}\delta_{kl}+\delta_{ik}\delta_{jl}+\delta_{il}\delta_{jk}-c_{ijkl}, \\
e_ie_ke_le_j &\sim \phantom{-}\delta_{ij}\delta_{kl}+\delta_{ik}\delta_{jl}-\delta_{il}\delta_{jk}+c_{ijkl}, \\
e_ie_le_je_k &\sim -\delta_{ij}\delta_{kl}+\delta_{ik}\delta_{jl}+\delta_{il}\delta_{jk}+c_{ijkl}, \\
e_ie_le_ke_j &\sim \phantom{-}\delta_{ij}\delta_{kl}-\delta_{ik}\delta_{jl}+\delta_{il}\delta_{jk}-c_{ijkl},
\end{split}
\end{equation}
and that the remaining 18 permutations are obtained from these by \eqref{eq:perm}. Note that the combination
\begin{equation}
\begin{split}
e_{i}e_{j}e_{k}e_{l}+e_{i}e_{k}e_{l}e_{j} &\sim\phantom{+}\delta_{ij}\delta_{kl}-\delta_{ik}\delta_{jl}+\delta_{il}\delta_{jk}+c_{ijkl} \\
&\phantom{\,\sim}+\delta_{ik}\delta_{lj}-\delta_{il}\delta_{kj}+\delta_{ij}\delta_{kl}+c_{iklj} \\
&=2(\delta_{ij}\delta_{kl}+c_{ijkl}).
\label{eq:4e}
\end{split}
\end{equation}
Since the $\mathcal{N}=8$ Cartan invariant is just the singlet in ${\bf 56} \times {\bf 56} \times {\bf 56} \times {\bf 56}$, it follows from \eqref{eq:man56} that it may be expressed as a linear combination of the five quartic products given above.  In analogy with \eqref{eq:ijkl}, one might be tempted to construct the $\mathcal{N}=8$ invariant by contracting \eqref{eq:4e} with \eqref{eq:A}. However, although the new associator terms generate new 4-way cross terms of the kind
\begin{equation}
abce+bcdf+cdeg+defa+efgb+fgac+gabd
\end{equation}
they cannot be the correct 4-way cross terms of \eqref{eq:4way}, because the way that the epsilons are contracted is different from the Cayley form \eqref{eq:1way} used in \eqref{eq:A}. So although the $\mathcal{N}=4$ Cartan invariant may be given by Cayley over the quaternions, and although the $\mathcal{N}=8$ Cartan invariant is given by a quartic product over the octonions, it is not simply given by Cayley over the octonions.

\newpage
\section{\texorpdfstring{FIVE-DIMENSIONAL BLACK HOLES}{Five-Dimensional Black Holes}}
\label{sec:five}

\subsection{Five-dimensional supergravity}
\label{sec:fivedimensionalsupergravity}

In five dimensions we might consider:
\begin{enumerate}
\item $\mathcal{N}=2$ supergravity coupled to $l$ vector multiplets where the symmetry is $SO(1,1,\mathds{Z}) \times SO(l-1,1,\mathds{Z})$ and the black holes carry charges belonging to the $(l+1)$ representation (all electric) .
\item $\mathcal{N}=4$ supergravity coupled to $m-1$ vector multiplets where the symmetry is $SO(1,1,\mathds{Z}) \times SO(m-1,5,\mathds{Z})$ where the black holes carry charges belonging to the $(m+4)$ representation (all electric).
\item $\mathcal{N}=8$ supergravity where the symmetry is the non-compact exceptional group $E_{6(6)}(\mathds{Z})$ and the black holes carry charges belonging to the fundamental $27$-dimensional representation (all electric).
\end{enumerate}

The electrically charged objects are point-like and the magnetic duals are one-dimensional, or string-like, transforming according to the contragredient representation. In all three cases above there exist cubic invariants akin to the determinant which yield the corresponding black hole or black string entropy.

In this section we briefly describe the salient properties of maximal $\mathcal{N}=8$ case, following \cite{Ferrara:1997ci}. We have 27 abelian gauge fields which transform in the fundamental representation of $E_{6(6)}$. The first invariant of $E_{6(6)}$ is the cubic invariant \eqref{eq:Ncubic} which may also be written as \cite{Cartan,Ferrara:1996um,Ferrara:1997ci,Ferrara:1997uz,Andrianopoli:1997hb}
\begin{equation}
I_{3}(Q) = q_{ij}\Omega^{jl}q_{lm}\Omega^{mn}q_{np}\Omega^{pi},
\end{equation}
where $q_{ij}$ is the charge vector transforming as a $\mathbf{27}$ which can be represented as traceless $Sp(8)$ matrix. The entropy of a black hole with charges $q_{ij}$ is then given by
\begin{equation}\label{eq:2/3entropy}
S=\pi\sqrt{|I_{3}(Q)|}.
\end{equation}

In five dimensions the compact group $H$ is $USp(8)$. We choose our conventions so that $USp(2)=SU(2)$. In the commutator of the supersymmetry generators we have a central charge matrix $Z_{AB}$ which can be brought to a normal form by a $USp(8)$ transformation. In the normal form the central charge matrix  can be written as
\begin{equation}\label{eq:normal}
Z_{AB} =
\begin{pmatrix}
s_1+s_2-s_3  & 0           & 0           & 0              \\
0            & s_1+s_3-s_2 & 0           & 0              \\
0            & 0           & s_2+s_3-s_1 & 0              \\
0            & 0           & 0           & -(s_1+s_2+s_3) \\
\end{pmatrix}
\times
\begin{pmatrix*}[r]0 & 1  \\-1 & 0\\\end{pmatrix*}.
\end{equation}
We can order the $s_i$ so that $|s_1| \ge |s_2| \ge |s_3|$. The cubic invariant, in this basis, becomes
\begin{equation}\label{eq:cubic}
I_3 = s_1 s_2 s_3.
\end{equation}
Even though the eigenvalues $s_i$ might depend on the moduli, the invariant \eqref{eq:cubic} only depends on the quantised values of the charges. We can write a generic charge configuration as $U e U^t $,  where $e$ is the normal frame as above, and the invariant will then be \eqref{eq:cubic}. There are three distinct possibilities as shown in \autoref{tab:n=8bh5}.
\begin{table}[ht]
\begin{tabular*}{\textwidth}{@{\extracolsep{\fill}}llrrrrccc}
\toprule
& $\phantom{E_{6(6)}/}$Orbit                  & $s_1$ & $s_2$ & $s_3$ & $I_3$      & Black hole & SUSY & \\
\midrule
& $E_{6(6)}/(O(5,5) \ltimes \mathds{R}^{16})$ & $>0$  & 0     & 0     & 0          & small      & 1/2  & \\
& $E_{6(6)}/(O(5,4) \ltimes \mathds{R}^{16})$ & $>0$  & $>0$  & 0     & 0          & small      & 1/4  & \\
& $E_{6(6)}/F_{4(4)}$                         & $>0$  & $>0$  & $>0$  & $>0$       & large      & 1/8  & \\
\bottomrule
\end{tabular*}
\caption[Classification of $D=5,\mathcal{N}=8$  black holes]{Classification of $D=5,\mathcal{N}=8$  black holes. The distinct charge orbits are determined by the number of non-vanishing
eigenvalues and $I_3$, as well as the number of unbroken supersymmetries.}
\label{tab:n=8bh5}
\end{table}

Note that, in contrast to the four-dimensional case where flipping the sign of $I_{4}$ interchanges BPS and non-BPS black holes, the sign of the $I_{3}$ \eqref{eq:cubic} is not important  since it changes under a CPT transformation. There are no non-BPS orbits in five dimensions.

In five dimensions there are also string-like configurations which are the magnetic duals of the  configurations considered here. They transform in the contragredient $27'$ representation and the solutions preserving 1/2, 1/4, 1/8 supersymmetries are characterised in an analogous way. The entropy of a black string with charges $p_{ij}$ is then given by
\begin{equation}
S=\pi\sqrt{|I_{3}(P)|}.
\end{equation}

It is useful to decompose the U-duality group into the T-duality group and the S-duality group. The decomposition reads $E_6 \to SO(5,5) \times SO(1,1)$, leading to
\begin{equation}\label{eq:decompts}
\mathbf{{27} \to {16}_{1} + {10}_{-2}+ {1}_{4}}.
\end{equation}
The last term in \eqref{eq:decompts} corresponds to the NS five-brane charge. The \textbf{16} correspond to the D-brane charges and the
\textbf{10} correspond to the 5 directions of KK momentum and the 5 directions of fundamental string winding, which are the charges that explicitly appear in string perturbation theory. The cubic invariant has the decomposition
\begin{equation}
\mathbf{(27)^3 \to {10}_{-2}\ {10}_{-2}\ {1}_{4} +{16}_{1}\ {16}_1\ {10}_{-2}}.
\end{equation}
This is  saying that in order to have a non-zero area black hole we must have three NS charges (more precisely some ``perturbative'' charges and a solitonic five-brane); or one can have two D-brane charges and one NS charge. In particular, it is not possible to have a black hole with a non-zero horizon area with purely D-brane charges.

Another version of the cubic invariant is  given as follows \cite{Baez:2001dm}: Take two 6-vectors $x_i$ and $y_j$ and a skew-symmetric $6 \times 6$ matrix $z^{ij}$. This corresponds to the decomposition
\begin{gather}
E_{6(6)} \supset  SL(2) \times SL(6),
\shortintertext{under which}
\mathbf{27 \to (2,6) +(1,15)}.
\shortintertext{Now consider the expression}
I_3 = \Pf(z) + z^{ij} x_i y_j,
\end{gather}
where Pf is the Pfaffian. See also \cite{Andrianopoli:1998ah}.

\subsection{\texorpdfstring{$E_{6}$ and the bipartite entanglement of three qutrits}{E6 and the bipartite entanglement of three qutrits}}\label{sec:E6Bipartite3Qutrits}

We have seen that in the case of two qutrits, the bipartite entanglement is classified using by $[SL(3,\mathds{C})]^2$ and that the entanglement measure is given by a determinant.  To give a quantum-theoretic interpretation of the $\mathcal{N}=8$ black hole, we begin by noting that
\begin{gather}
E_{6(6)} \supset [SL(3,\mathds{R})]^3,
\shortintertext{and}
E_{6}(\mathds{C}) \supset [SL(3,\mathds{C})]^3.
\end{gather}
We shall now show that the corresponding system in quantum information theory is that of three qutrits (Alice, Bob and Charlie) \cite{Briand:2003}. However, the larger symmetry requires that they undergo only a bipartite entanglement of a very specific kind. The entanglement measure will be given by the cubic Cartan invariant \eqref{eq:Ncubic}.

A crucial ingredient is that under
\begin{gather}
E_{6(6)} \supset SL(3)_{A} \times SL(3)_{B}  \times SL(3)_{C},
\shortintertext{the \textbf{27} decomposes as}
\mathbf{27 \to(3',3,1)+(1,3',3')+(3,1,3)}.\label{eq:27}
\shortintertext{An analogous decomposition holds for}
E_{6}(\mathds{C}) \supset [SL(3,\mathds{C})]^3.
\end{gather}
Notice that we find three copies of the two qutrit Hilbert space
\begin{equation}\label{eq:3qutrit}
\ket{\Psi} = a_{A'B}\ket{A'B}+b_{B'C'}\ket{B'C'}+c_{CA}\ket{CA},
\end{equation}
where $A,A'=0,1,2$. Note that:
\begin{enumerate}
\item Any pair of states has an individual in common
\item Each individual is excluded from one out of the three states
\end{enumerate}
So we have three qutrits (Alice, Bob, Charlie) but where each person has bipartite entanglement with the other two. However, as discussed in \autoref{sec:three}, this 27-dimensional space is not a subspace of the three qutrit Hilbert space $\mathbf{(3,3,3)}$.

\begin{figure}[ht]
  \centering
  \includegraphics[width=3cm]{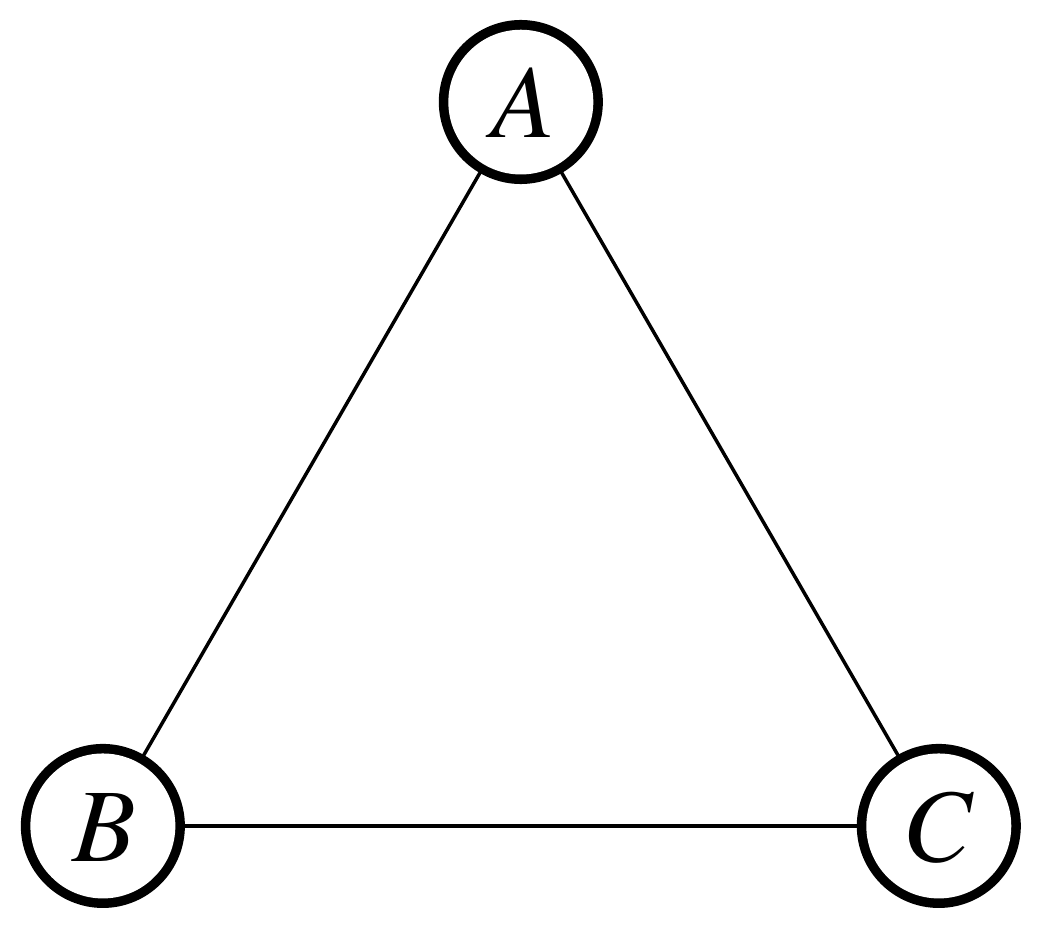}
  \caption[Triangular $E_6$ entanglement diagram]{The $D=5$ analogue of \autoref{fig:E7Entanglement} is the three qutrit entanglement diagram corresponding to the decomposition \eqref{eq:27} and the state \eqref{eq:3qutrit}. It is a triangle with vertices $A, B, C$ representing the qutrits and the lines $AB, BC$ and $CA$ representing the entanglements.}
  \label{fig:Triangle}
\end{figure}

The entanglement may be represented by a triangle with vertices $ABC$ representing the qutrits and the lines $AB,BC$ and $CA$ represent the entanglements. See \autoref{fig:Triangle}.
The three states transforms as a pair of triplets under two of the $SL(3)$'s and singlets under the remaining one. Individually, therefore, the bipartite entanglement of each of the three states is given by the determinant. Taken together however, we see from \eqref{eq:56Decomp} that they transform as a complex \textbf{27} of $E_{6}(\mathds{C})$. Their bipartite entanglement must be given by an expression that is cubic in the coefficients $a,b,c$ and invariant under $E_{6}(\mathds{C})$. The unique possibility is the Cartan invariant $I_3$, and so the 2-tangle is given by generalising \eqref{eq:qutrit2-tangle} to
\begin{equation}\label{eq:2/3tangle}
\tau(ABC) = 27|I_3|^2.
\end{equation}
If the wave-function \eqref{eq:7QubitState} is normalised, then $0 \leq \tau(ABC) \leq 1$.

Note the appearance of both primed and unprimed qutrit representations, a new feature not encountered with qubits. This is because the $\varepsilon$ symbol relates upper and lower indices for qubits
\begin{equation}
a^{A}=\varepsilon^{AB}a_{B},
\end{equation}
while for qutrits $a^{A}$ and $a_{ A}$ are different:
\begin{equation}
a^{A}=\half\varepsilon^{AB_1B_2}a_{[B_1B_2]}.
\end{equation}
The antisymmetry of $a_{[B_1B_2]}$ allows the interpretation of the $3'$ as a pair of indistinguishable ``fermions''.

\subsection{\texorpdfstring{Decomposition of $I_{3}$}{Decomposition of I3}}

To understand better the entanglement we note that, as a result of \eqref{eq:27}, Cartan's invariant contains not one determinant but three.  It may be written as the sum of three terms each of which is invariant under $[SL(3)]^{2}$ plus cross terms. To see this, denote a \textbf{3} in one of the three entries in \eqref{eq:27} by $A$. So we may rewrite \eqref{eq:27} as
\begin{gather}
\mathbf{27}=(AB)+(BC)+(CA),\label{eq:562}
\shortintertext{or symbolically}
\mathbf{27} = a+b+c.\label{eq:568}
\end{gather}
Then $I_{3}$ is the singlet in $\mathbf{27 \times 27 \times 27}$:
\begin{equation}\label{eq:272}
I_{3}= a^{3}+b^{3}+c^{3}+6abc,
\end{equation}
where the products
\begin{subequations}
\begin{gather}
\begin{split}
a^{3}&= (AB)(AB)(AB) \\
&=\tfrac{1}{6}\varepsilon_{A_{1}A_{2}A_{3}} \varepsilon^{B_{1}B_{2}B_{3}}{a}\indices{^{A_{1}}_{B_{1}}}{a}\indices{^{A_{2}}_{B_{2}}}{a}\indices{^{A_{3}}_{B_{3}}},
\end{split} \\
\begin{split}
b^{3}&= (BC)(BC)(BC) \\
&=\tfrac{1}{6}\varepsilon_{B_{1}B_{2}B_{3}} \varepsilon_{C_{1}C_{2}C_{3}}{b}^{B_{1}C_{1}}{b}^{B_{2}C_{2}}{b}^{B_{3}C_{3}},
\end{split} \\
\begin{split}
c^{3}&= (CA)(CA)(CA) \\
&=\tfrac{1}{6}\varepsilon^{C_{1}C_{2}C_{3}} \varepsilon^{A_{1}A_{2}A_{3}}{c}_{C_{1}A_{1}}{c}_{C_{2}A_{2}}{c}_{C_{3}A_{3}},
\end{split}
\end{gather}
\end{subequations}
exclude one individual (Charlie, Alice, and Bob respectively), and the products
\begin{equation}
\begin{split}
abc&= (AB)(BC)(CA) \\
&=\tfrac{1}{6}{a}\indices{^{A}_{B}}{b}^{BC}{c}_{CA},
\end{split}
\end{equation}
exclude none. These results may be verified using the following dictionary between $a,b,c$ and the $P^v,P^s,P^c$ of the Jordan Algebra:
\begin{subequations}
\begin{gather}
p^1=-a^{0}{}_{0}, \quad p^2=-a^{1}{}_{1}, \quad p^3=-a^{2}{}_{2},\\
\begin{split}
P^c=&\half\big(-(a^{1}{}_{2}+a^{2}{}_{1})e_{0}-(b^{00}+c_{00})e_{1}-(b^{01}+c_{10})e_{2}-(b^{02}+c_{20})e_{3}\\
&\phantom{\half\big(}+(a^{1}{}_{2}-a^{2}{}_{1})e_{4}+(b^{00}-c_{00})e_{5}+(b^{01}-c_{10})e_{6}+(b^{02}-c_{20})e_{7}\big),
\end{split}\\
\begin{split}
P^s=&\half\big(-(a^{2}{}_{0}+a^{0}{}_{2})e_{0}-(b^{10}+c_{01})e_{1}-(b^{11}+c_{11})e_{2}-(b^{12}+c_{21})e_{3}\\
&\phantom{\half\big(}+(a^{2}{}_{0}-a^{0}{}_{2})e_{4}+(b^{10}-c_{01})e_{5}+(b^{11}-c_{11})e_{6}+(b^{12}-c_{21})e_{7}\big),
\end{split}\\
\begin{split}
P^v=&\half\big(-(a^{0}{}_{1}+a^{1}{}_{0})e_{0}-(b^{20}+c_{02})e_{1}-(b^{21}+c_{12})e_{2}-(b^{22}+c_{22})e_{3}\\
&\phantom{\half\big(}+(a^{0}{}_{1}-a^{1}{}_{0})e_{4}+(b^{20}-c_{02})e_{5}+(b^{21}-c_{12})e_{6}+(b^{22}-c_{22})e_{7}\big),
\end{split}
\end{gather}
\end{subequations}
which yields,
\begin{equation}\label{eq:I_3}
I_3 = \det J_3(P) = a^3 +b^3+c^3+6abc.
\end{equation}

\subsection{Subsectors}

Just as in the $E_{7}$ case one can truncate to just the $(\textbf{3},\textbf{3})$ in \eqref{eq:27} which excludes Bob
\begin{equation}\label{eq:psi4}
\ket{\Psi} = c_{CA}\ket{CA}.
\end{equation}
This is described by just that line not passing through $B$ in the $ABC$ triangle,
\begin{equation}\label{eq:565}
\mathbf{(3,3)}=(CA)= c,
\end{equation}
and the corresponding cubic invariant, $I_{3}$, reduces to the singlet in $\mathbf{(3,3)\times(3,3)\times(3,3)}$
\begin{equation}\label{eq:273}
I_{3}=\det c_{CA} \sim c^{3}.
\end{equation}

\subsection{\texorpdfstring{Classification of $\mathcal{N}=8$ black holes and three-qutrit states}{Classification of N=8 black holes and three-qutrit states}}
\label{sec:blackholesandthree-qutritstates}

In the $\mathcal{N}=8$ theory, ``large'' and ``small'' black holes are classified by the value of $I_{3}$. Non-zero $I_{3}$ corresponds to large black holes, which are BPS for both signs of $I_{3}$, and vanishing  $I_{3}$ to small black holes. We may have 1/8, 1/4 or 1/2 supersymmetry preserved.

The charge orbits \cite{Ferrara:1997uz,Ferrara:1997ci,Lu:1997bg} for the black holes depend on the number of unbroken supersymmetries  or the number of vanishing eigenvalues as in \autoref{tab:n=8bh5}.  For $\mathcal{N}=8$ the large black holes correspond to the class of rank 3 Bell entangled states and small black holes to the rank 2 Bell or separable class. One way of obtaining such states is to go the canonical basis \eqref{eq:normal} where the Cartan invariant reduces to \eqref{eq:cubic}.  The result is shown in \autoref{tab:n=8bh5}.  For example, a state with non-zero coefficients
\begin{gather}
c_{00}=c_{11}=c_{22}= \tfrac{1}{\sqrt{3}},\\
\ket{\Psi}=\tfrac{1}{\sqrt{3}}(\ket{000}+\ket{111}+\ket{222})
\shortintertext{is maximally entangled with}
\tau=1.
\end{gather}
Alternatively, having succeeded in writing the Cartan invariant in terms of $a,b,c,$ in \eqref{eq:I_3} we can now look for entangled states in  the full $\mathcal{N}=8$ theory. For example, the generalised Bell state with non-zero coefficients
\begin{gather}
a^0{}_0=a^1{}_1=a^2{}_2=b^{00}=b^{11}=b^{22}=c_{00}=c_{11}=c_{22}= \tfrac{1}{{3}}
\shortintertext{is entangled but not maximally since}
\tau=\tfrac{16}{27}.
\end{gather}

\subsection{  Three 7-dit interpretation}
\label{sec:three}

We note that the 27-dimensional Hilbert space given in \eqref{eq:27} is not a subspace of the $3^{3}$-dimensional three qutrit Hilbert space given by $\mathbf{(3,3,3)}$, but rather a direct sum of three $3^{2}$-dimensional Hilbert spaces:
\begin{equation}\label{eq:27new}
\mathbf{(3',3,1)+(3,1,3)+(1,3',3')}.
\end{equation}
This raises an ambiguity about the terminology \emph{bipartite entanglement of three qutrits}. The state corresponding to the usual $3^{3}$-dimensional three qutrit Hilbert space given by $\mathbf{(3,3,3)}$ is
\begin{equation}
\ket{\Psi}=a_{ABC}\ket{ABC}
\end{equation}
and one meaning of bipartite entanglement, $AB$ say, would be that given by the reduced density matrix
\begin{equation}
\rho_{AB}=\Tr_{C}\ket{\Psi}\bra{\Psi}.
\end{equation}
So it is important to note that this is clearly different from the meaning we have adopted in our discussion of three qutrits elsewhere in this section.

The triplets in \eqref{eq:27new} have the interpretation as qutrits but what the singlets? A natural way to interpret the singlets is to embed the qutrits into 7-dits and note that under
\begin{gather}
SL(7) \supset SL(3)
\shortintertext{we have}
\mathbf{7 \to 3+3'+1}.
\end{gather}

The three 7-dit system (Alice, Bob, Charlie) is described by the state
\begin{equation}
\ket{\Psi} = {\hat a}_{{\hat A}{\hat B}{\hat C}}\ket{{\hat A}{\hat B}{\hat C}},
\end{equation}
where ${\hat A}=0,1,2,3,4,5,6$, and the Hilbert space has dimension $7^{3}=343$.  ${\hat a}_{{\hat A}{\hat B}{\hat C}}$ transforms as a $\mathbf{(7,7,7)}$ under $SL(7)_{A} \times SL(7)_{B} \times SL(7)_{C}$. Under
\begin{gather}
SL(7)_{A} \times SL(7)_{B}  \times SL(7)_{C} \supset SL(3)_{A} \times SL(3)_{B}  \times SL(3)_{C}
\shortintertext{we have}
\begin{array}{B{c}@{}c@{\ (}*{2}{B{l}*{2}{@{,\,}B{l}}@{)\ +\ (}}B{l}*{2}{@{,\,}B{l}}@{)}c}
(7,7,7) \to &   & 3'& 3'& 3' &  3 & 3 & 3  &  1 & 1 & 1 \\
            & + & 3'& 3'& 3  &  3'& 3 & 3' &  3 & 3'& 3'\\
            & + & 3'& 3 & 3  &  3 & 3'& 3  &  3 & 3 & 3'\\
            & + & 3'& 3'& 1  &  3'& 1 & 3' &  1 & 3'& 3'\\
            & + & 3 & 3 & 1  &  3 & 1 & 3  &  1 & 3 & 3 \\
            & + & 3'& 1 & 1  &  1 & 3'& 1  &  1 & 1 & 3'\\
            & + & 3 & 1 & 1  &  1 & 3 & 1  &  1 & 1 & 3 \\[5pt]
            & + & 3'& 1 & 3  &  3'& 3 & 1  &  1 & 3 & 3'\\
            & + & 3 & 1 & 3' &  3 & 3'& 1  &  1 & 3'& 3 & ,
\end{array}
\end{gather}
which contains \eqref{eq:3qutrit} as a subspace.
\begin{equation}\label{eq:psi6}
\ket{\Psi} = a_{A'B\bullet}\ket{A'B\bullet}+b_{\bullet BC}\ket{\bullet BC}+c_{A'\bullet C'}\ket{A'\bullet C'},
\end{equation}
which we abbreviate by \eqref{eq:3qutrit}.  So the triangle entanglement we have described fits within conventional quantum information theory, but we have discovered a hidden $E_6$ symmetry of this special 27-dimensional subspace.

\newpage
\section{\texorpdfstring{MAGIC SUPERGRAVITIES}{Magic Supergravities}}
\label{sec:magic}

\subsection{\texorpdfstring{Magic supergravities in $D=4$}{Magic supergravities in D=4}}
\label{sec:magic4}

The black holes described by Cayley's hyperdeterminant are those of $\mathcal{N}=2$ supergravity coupled to three vector multiplets, where the symmetry is $[SL(2,\mathds{Z})]^{3}$. In \autoref{sec:N=8}  the following four-dimensional generalisations were considered: (1) $\mathcal{N}=2$ supergravity coupled to $l+1$ vector multiplets, (2) $\mathcal{N}=4$ supergravity coupled to $m$ vector multiplets, (3) $\mathcal{N}=8$ supergravity. In all three case there exist quartic invariants akin to Cayley's hyperdeterminant whose square root yields the corresponding black hole entropy. We succeeded in giving a quantum theoretic interpretation in the $\mathcal{N}=8$ case together with its truncations to $\mathcal{N}=4$ (with $m=6$) and $\mathcal{N}=2$ (with $l=2$, the case we already knew \cite{Duff:2006uz}).

However, as suggested by Levay \cite{Levay:2006pt}, one might also consider the ``magic'' supergravities \cite{Gunaydin:1984ak,Gunaydin:1983bi,Gunaydin:1983rk,Ferrara:2007pc,Rios:2007qn}. These correspond to the $\mathds{R,C,H,O}$ (real, complex, quaternionic and octonionic) $\mathcal{N}=2,D=4$ supergravity coupled to $6,9,15$ and $27$ vector multiplets with symmetries $Sp(6,\mathds{Z}),SU(3,3),SO^{*}(12)$ and $E_{7(-25)}$, respectively. Once again, as has been shown just recently \cite{Ferrara:2006yb}, in all cases there are quartic invariants whose square root yields the corresponding black hole entropy.

Here we demonstrate that the black-hole/qubit correspondence does indeed continue to hold for magic supergravities. The crucial observation is that, although the black hole charges $a_{ABC}$ are real (integer) numbers and the entropy \eqref{eq:N=8BHEntropy} is invariant under $E_{7(7)}(\mathds{Z})$, the coefficients $a_{ABC}$ that appear in the qubit state \eqref{eq:7QubitState} are complex. So the 3-tangle \eqref{eq:3/7tangle} is invariant under $E_{7}(\mathds{C})$ which contains both $E_{7(7)}(\mathds{Z})$ and $E_{7(-25)}(\mathds{Z})$ as subgroups. To find a supergravity correspondence therefore, we could equally well have chosen the magic octonionic $\mathcal{N}=2$ supergravity rather than the conventional $\mathcal{N}=8$ supergravity. The fact that
\begin{gather}
E_{7(7)}(\mathds{Z}) \supset [SL(2,\mathds{Z})]^{7},
\shortintertext{but}
E_{7(-25)}(\mathds{Z}) \not\supset [SL(2,\mathds{Z})]^{7}
\shortintertext{is irrelevant. All that matters is that}
E_{7}(\mathds{C}) \supset [SL(2,\mathds{C})]^{7}.
\end{gather}
The same argument holds for the magic real, complex and quaternionic $\mathcal{N}=2$ supergravities which are, in any case truncations of $\mathcal{N}=8$ (in contrast to the octonionic).

Having made this observation, one may then revisit the conventional $\mathcal{N}=2$ and $\mathcal{N}=4$ cases (1) and (2) above.  When we looked at the seven qubit subsector $E_7(\mathds{C}) \supset SL(2,\mathds{C})  \times SO(12, \mathds{C})$, we gave an $\mathcal{N}=4$ supergravity interpretation with symmetry $SL(2,\mathds{R}) \times SO(6,6)$ \cite{Duff:2006ue}, but we could equally have given an interpretation in terms of $\mathcal{N}=2$ supergravity coupled to $11$ vector multiplets with symmetry $SL(2,\mathds{R}) \times SO(10,2)$.

Moreover, $SO(l-1,2)$ is contained in $SO(l+1,\mathds{C})$ and $SO(6,m)$ is contained in $SO(12+m,\mathds{C})$ so we can give a qubit interpretation to more vector multiplets for both $\mathcal{N}=2$ and $\mathcal{N}=4$, at least in the case of $SO(4n, \mathds{C})$ which contains $[SL(2,\mathds{C})]^{2n}$.

\subsection{\texorpdfstring{Magic supergravities in $D=5$}{Magic supergravities in D=5}}

One might also consider the ``magic'' supergravities in $D=5$ \cite{Gunaydin:1984ak,Gunaydin:1983bi,Gunaydin:1983rk}. These correspond to the $\mathds{R,C,H,O}$ (real, complex, quaternionic and octonionic) $\mathcal{N}=2$ supergravity coupled to 5, 8, 14 and 26 vector multiplets with symmetries $SL(3,\mathds{R}),SL(3,\mathds{C}),SU^{*}(6)$ and $E_{6(-26)}$ respectively. Once again, in all cases there are cubic invariants whose square root yields the corresponding black hole entropy \cite{Ferrara:2006yb}.

Here we demonstrate that the black-hole/qubit correspondence continue to hold for these $D=5$ magic supergravities, as well as $D=4$ . Once again, the crucial observation is that, although the black hole charges $a_{AB}$ are real (integer) numbers and the entropy \eqref{eq:2/3entropy} is invariant under $E_{6(6)}(\mathds{Z})$,  the coefficients $a_{AB}$ that appear in the wavefunction \eqref{eq:3qutrit} are complex. So the 2-tangle \eqref{eq:2/3tangle} is invariant under $E_{6}(\mathds{C})$ which contains both $E_{6(6)}(\mathds{Z})$ and $E_{6(-26)}(\mathds{Z})$ as subgroups.  To find a supergravity correspondence therefore, we could equally well have chosen the magic octonionic $\mathcal{N}=2$ supergravity rather than the conventional $\mathcal{N}=8$ supergravity. The fact that
\begin{gather}
E_{6(6)}(\mathds{Z}) \supset [SL(3,\mathds{Z})]^{3},
\shortintertext{but}
E_{6(-26)}(\mathds{Z}) \not\supset  [SL(3,\mathds{Z})]^{3}
\shortintertext{is also irrelevant. All that matters is that}
E_{6}(\mathds{C}) \supset [SL(3,\mathds{C})]^{3}.
\end{gather}
Once again, the same argument holds for the magic real, complex and quaternionic $\mathcal{N}=2$ supergravities which are, in any case truncations of $\mathcal{N}=8$ (in contrast to the octonionic).

Having made this observation, one may then revisit the conventional $\mathcal{N}=2$ and $\mathcal{N}=4$ cases (1) and (2) of \autoref{eq:five}. $SO(l,1)$ is contained in $SO(l+1,\mathds{C})$ and $SO(m,5)$ is contained in $SO(5+m,\mathds{C})$, so we can give a qutrit interpretation to more vector multiplets for both $\mathcal{N}=2$ and $\mathcal{N}=4$, at least in the case of $SO(6n, \mathds{C})$ which contains $[SL(3,\mathds{C})]^{n}$.

\subsection{Alternative Jordan algebra interpretation}
\label{sec:AlternativeJordan}

\begin{quotation}
\textit{``Details that could throw doubt on your interpretation must be given, if you know them. You must do the best you can -- if you know anything at all wrong, or possibly wrong -- to explain it. If you make a theory, for example, and advertise it, or put it out, then you must also put down all the facts that disagree with it, as well as those that agree with it. There is also a more subtle problem. When you have put a lot of ideas together to make an elaborate theory, you want to make sure, when explaining what it fits, that those things it fits are not just the things that gave you the idea for the theory; but that the finished theory makes something else come out right, in addition.}

\textit{In summary, the idea is to give all of the information to help others to judge the value of your contribution; not just the information that leads to judgement in one particular direction or another.''}
\newline\newline
\noindent Richard P. Feynman
\newline\newline
\noindent Adapted from a Caltech commencement address given in 1974; from the book ``Surely You're Joking, Mr. Feynman!'' \cite{Feynman:1985}.
\end{quotation}

\subsubsection{New interpretation}

Our analogy between black holes and quantum information remains, for the moment, just that. We know of no physics connecting them. In particular, as far as we can tell our analogy is unconnected with the work of \cite{Maldacena:1997de,Brustein:2005vx,Emparan:2006ni,Hirata:2006jx,Ryu:2006ef,Blasone:2007vw,Ge:2007ig} relating black holes and entanglement entropy in field theory.

Nevertheless, we have seen that the exceptional group $E_{7}$ describes the tripartite entanglement of seven qubits \cite{Duff:2006ue,Levay:2006pt} and  that the exceptional group $E_{6}$ describes the bipartite entanglement of three qutrits. In the $E_{7}$ case, the quartic Cartan invariant provides both the measure of entanglement and the entropy of the four-dimensional $\mathcal{N}=8$ black hole, whereas in the $E_{6}$ case, the cubic Cartan invariant provides both the measure of entanglement and the entropy of the five-dimensional $\mathcal{N}=8$ black hole.

Moreover, we have seen that similar analogies exist not only for the $\mathcal{N}=4$ and $\mathcal{N}=2$ truncations, but also for the magic $\mathcal{N}=2$ supergravities in both four and five dimensions. Murat Gunaydin has suggested (private communication) that the appearance of octonions and split octonions implies a connection to quaternionic and/or octonionic quantum mechanics. This was not apparent (at least to us) in the four-dimensional $\mathcal{N}=8$ case \cite{Duff:2006ue}, but the appearance in the five dimensional magic $\mathcal{N}=2$ case of $SL(3,\mathds{R})$,
$SL(3,\mathds{C})$, $SL(3,\mathds{H})$ and $SL(3,\mathds{O})$ is more suggestive.

With this in mind, we provide in this section an alternative interpretation of these relationships between black hole entropy and entanglement measures in quantum information theory. The black hole charges in the five-dimensional $\mathcal{N}=2$ magic $\mathds{R,C,H,O}$ and $\mathcal{N}=8$ $\mathds{O}^{s}$ (split octonion) supergravities are known to be described by the elements of a Jordan algebra of degree three $J_{3}^{\mathds{A}}$, where $\mathds{A=R,C,H,O}$ and $\mathds{O}^s$, respectively. Here we identify them with the $3 \times 3$ reduced density matrices of two qutrits in quantum mechanics defined over $\mathds{A}$.

Recall  that in the  $D=4$ case, one identifies the black hole charges with two qubit state vector coefficients $a_{AB}$ or three qubit coefficients $a_{ABC}$ etc, where $A=0,1$. Whereas in $D=5$, one identifies the black hole charges with two qutrit state vector coefficients $a_{AB}$ etc, where $A=0,1,2$. In this section, however, we suggest an alternative interpretation for which the black hole charges are identified not the $a$'s themselves but with components of the reduced density matrices, for example
\begin{equation}
\rho_{A_1A_2}=a_{A_1B_1}a^{*}_{A_2B_2}\delta^{B_1B_2},
\end{equation}
where the $a$'s are defined over the reals, complexes, quaternions and octonions in the case of the $\mathcal{N}=2$ magic $\mathds{R,C,H,O}$ supergravities and over the split octonions $\mathds{O}^s$ in the case of $\mathcal{N}=8$.

A similar interpretation is given for the six-dimensional magic $\mathcal{N}=2$ and $\mathcal{N}=8$ black dyonic string charges \cite{Andrianopoli:1998qg}, described by Jordan algebras of degree two $J_{2}^{\mathds{A}}$, in terms of the $2 \times 2$ reduced density matrices of two qubits defined over $\mathds{A}$. This new density matrix interpretation has some advantages over the previous one, while continuing to relate U-duality invariant black entropies to entanglement measures. However, we have yet to establish in this approach the connection between three qubit entanglement and the full set of four-dimensional black hole charges, known to be described by the elements of a Freudenthal triple system $\mathfrak{M(J)}$.

This re-interpretation of the black hole/qubit correspondence makes use of the Jordan algebra formulation of quantum mechanics. Accordingly, in the following section we give a brief overview of Jordan quantum theory and its relationship to the conventional interpretation, following closely the presentation given in \cite{Townsend:1985eq}.

\subsubsection{The Jordan algebra formulation of quantum mechanics}\label{sec:JordanQM}

The advent of the matrix mechanics formulation of quantum theory was shortly followed by a number of investigations into its possible algebraic generalisations. These forays were principally motivated by a certain perception, prevalent at the time, that the irrefutable successes of quantum mechanics as applied to the atom would not be readily extended to the relativistic domain \cite{Jordan:1933vh}. Chief amongst these attempts was the use of Jordan algebras, introduced in \autoref{sec:JordanAlgebras}, as a suitable representation of physical observables \cite{Jordan:1933a,Jordan:1933b}. The intention was to place emphasis on observables, as opposed to states, and in doing so abstract the essential characteristics  of the set of Hermitian operators, displacing the Hilbert space from its central position in the mathematical foundations of the theory \cite{Emch:1997}.

Taken alone, it is not entirely clear that the Jordan identity \eqref{eq:Jid} ought to be of any fundamental importance when representing the algebra of observables. To better understand its physical relevance let us consider what is expected of such a representation starting from the conventional matrix theory. In this case observables are represented by  Hermitian matrices, possibly of infinite dimension. Central to their particular suitability in this role is that their eigenvalues are real, that distinct eigenvalues correspond to distinct orthogonal eigenvectors and that they are formally real in the sense of \eqref{eq:formallyreal}. These properties, coupled with the spectral decomposition theorem, imply that for any polynomial function $F$ and observable $A$,
\begin{equation}
F(A)=\sum_m F(a_m)P_m, \quad \textrm{for}\quad A=\sum_m a_m P_m,
\end{equation}
where $P_m$ and $a_m$ are the projectors and associated eigenvalues, respectively, appearing in the spectral decomposition of $A$. The physical significance of this statement is that, if $a$ is the value of some observable $A$, energy say, for some system in a given state, then one would expect that $F(a)$ be the value of $F(A)$, energy squared say, for the same state. However, this relies crucially on the power associativity \eqref{eq:powerasso} of Hermitian matrices, $A^mA^n=A^{(m+n)}$ ensures that $F(A)$ is defined unambiguously.

Now, given a commutative, formally real algebra, the Jordan identity then \emph{follows} from the physically well motivated assumption of power associativity \cite{Jordan:1933vh}. Equally, given the same initial assumptions, the Jordan identity implies power associativity. However, Hermitian matrices do not in general commute. Consequently, the matrix product of two Hermitian matrices does not necessarily yield a third Hermitian matrix. They do not form a closed algebra under standard matrix multiplication. These considerations, in part, motivate the definition of the Jordan product on Hermitian matrices,
\begin{equation}\label{eq:JPher}
A\circ B=\half(AB+BA),
\end{equation}
which is by definition commutative (but nonassociative) and closed with respect to Hermiticity. The algebra of Hermitian matrices, with multiplicative composition defined by \eqref{eq:JPher},  is closed, formally real, commutative, power associative and, consequently, satisfies the Jordan identity \eqref{eq:Jid}. In this case, Hermitian matrices with the Jordan product \eqref{eq:JPher}, these properties simply amount to a set of identities. However, in seeking generalisations, one could make the paradigm shift and take the Jordan identity as primary, the Jordan product emerging as a secondary consequence. That is, we start from a Jordan algebra and require that, in addition, it be formally real and contain an identity element so as to ensure its suitability as an algebra of observables \cite{Townsend:1985eq}.   In summary, we axiomatise the algebra of observables $\mathfrak{A}$:
\begin{enumerate}
\item $x\circ y=y\circ x,$
\item $x\circ(x^2\circ y)=x^2\circ(x\circ y),$
\item $x^2+y^2+z^2+\dotsb=0 \qquad\implies \qquad x=y=z=\dotsb=0,$
\item $\exists\ \mathds{1}\in\mathfrak{A}\ \textrm{s.t.}\ \mathds{1}\circ x=x\circ \mathds{1}=x\quad\forall x\in\mathfrak{A}.$
\end{enumerate}

Having, to some degree at least, set an axiomatic foundation for the algebra of quantum observables, it is natural to ask what generalisations beyond the orthodox framework this allows for. This question was essentially answered by the classification of all simple formally real Jordan algebras \cite{Jordan:1933vh} (see \autoref{sec:JordanAlgebras}). However, before commenting on the possible alternatives let us address a more immediate issue; how the Jordan formulation captures the statistics of standard quantum mechanics, an obvious minimal prerequisite.

Having articulated the space of quantum observables in terms of Jordan algebras let us now turn our attention to the representation of states and the problem of time evolution. We follow closely the presentation given in \cite{Townsend:1985eq}. States in quantum theory are represented by rays in a Hilbert space. However, given an orthonormal basis, $\{\ket{m}\}$ the projection operators,
\begin{gather}
P_m=\ket{m}\bra{m},
\shortintertext{which satisfy,}
\tr P_m=1,\label{eq:proj1}
\shortintertext{and}
P_m^2=P_m,\label{eq:proj2}
\end{gather}
correspond to an equivalent representation. Any normalised pure state $\ket{\psi}$ may be expressed as a projector, $P_\psi=\ket{\psi}\bra{\psi}$, satisfying \eqref{eq:proj1} and \eqref{eq:proj2}. The expectation value of an observable, represented by a Hermitian matrix $A$, is then given by,
\begin{equation}
\langle A\rangle_\psi=\tr(P_\psi A).
\end{equation}
More generally, any state, pure or mixed, may be represented as a Hermitian, trace one, positive semi-definite density matrix $\rho$ (c.f. \autoref{sec:LOCC}) in which case,
\begin{equation}
\langle A\rangle_\rho=\tr(\rho A).
\end{equation}
To reproduce these results using the Jordan framework it is necessary to introduce a trace form \cite{Townsend:1985eq,Bischoff:1993sr},
\begin{equation}
\tr\mathds{1}=\nu, \quad\tr b_i=0,
\end{equation}
where $\nu\in\mathds{Z}$ and the set $\{b_i\}$ forms a basis for the Jordan algebra such that any element $a$ may be written as $a=a^ib_i$, $a^i\in\mathds{R}$. This defines a positive definite bilinear inner product,
\begin{equation}\label{eq:Jinner}
\langle b_i, b_j \rangle =\tfrac{1}{\nu}\tr (b_i\circ b_j)=\delta_{ij},
\end{equation}
since one can always choose a basis such that $b_i\circ b_j=\delta_{ij}\mathds{1}+f_{ijk}b_k$. Any two idempotents\footnote{Any algebra element $P$ is said to be idempotent if $P^2=P$.}, $P_1$ and $P_2$, are orthogonal, $P_1\circ P_2=0$, if they are orthogonal with respect to the inner product \eqref{eq:Jinner}. A general idempotent $E$ is said to \emph{primitive} if it cannot be decomposed as the sum of two orthogonal idempotents. The highest possible number of orthogonal primitive idempotents is the \emph{degree} of the algebra and is equal to the $\nu$ appearing in the trace form if one normalises $\tr E=1$,  c.f. \eqref{eq:proj1}. Any complete set $\{E_i\}$ of orthogonal primitive idempotents satisfies,
\begin{equation}
\sum_{i=1}^{\nu}E_i=\mathds{1},
\end{equation}
constituting a resolution of the identity. Then, in analogy with the spectral decomposition theorem, any element $a$ in the algebra can be expressed as a linear sum of primitive idempotents,
\begin{equation}
a=\sum_{i=1}^{\nu}a^i E_i(a),
\end{equation}
where the maximal set $\{E_i(a)\}$ depends on particular the element $a$ under question.

It is now possible to represent an arbitrary state $\rho$ in the Jordan formulation,
\begin{equation}
\rho=\sum^{\nu}_{i=1} p_i E_i, \quad\textrm{where}\quad \sum^{\nu}_{i=1} p_i=1\quad\textrm{and}\quad p_i\in [0, 1].
\end{equation}
This clearly relates to the density matrix formalism of conventional quantum mechanics: $\rho$ is a positive semi-definite, trace one element that satisfies $\rho^2\leq\rho$ with equality holding only for pure states i.e. when $\rho$ is a primitive idempotent \cite{Townsend:1985eq}. The expectation value of an observable $a$ in the Jordan algebra with respect to a state $\rho$ is then given by,
\begin{equation}
\langle a\rangle_\rho=\tr (a\circ\rho).
\end{equation}
Hence, this Jordan algebra formulation is essentially equivalent to the density matrix picture of the conventional quantum mechanics \cite{Townsend:1985eq}.

Let us now consider time evolution. Here the Jordan formalism does depart, to some extent, from the standard density matrix picture. If we assume that the affine structure of a general density matrix is preserved by time evolution then,
\begin{equation}\label{eq:DMevo}
\partial_t\rho(t)=-i[H, \rho(t)],
\end{equation}
where $H$ is the Hamiltonian. An important feature of \eqref{eq:DMevo} is that it maps pure states into pure states. Given two Hermitian matrices, $A$ and $B$, we have,
\begin{equation}\label{eq:evoconsistancy}
\begin{split}
\partial_t (AB)&= -i[H, AB]\\
			   &= -i([H, A]B+A[H, B])\\
			   &= (\partial_t A)B+A (\partial_t B).
\end{split}
\end{equation}
The differential evolution operator acts as a derivation as one would expect. Requiring this condition in the Jordan formulation, so that for any two elements $x$ and $y$
\begin{equation}
\partial_t (x\circ y)=D(x\circ y)=x\circ D(y)+D(x)\circ y=x\circ \partial_t(y)+\partial_t(x)\circ y,
\end{equation}
the differential evolution operator $D$ acts as a derivation of the Jordan algebra c.f. \eqref{eq:JDer}. Recall, the set of derivations generates the automorphism group of the algebra \eqref{eq:JAut}. Hence, the corresponding time translation operator, $T_{t_1\to t_2}$, taking an element, $x_{t_1}$, at time $t_1$, to the corresponding element, $x_{t_2}$, at time $t_2$, preserves the Jordan product. If
\begin{gather}
x_{t_1}\circ y_{t_1}=z_{t_1},
\shortintertext{then}
T_{t_1\to t_2}(x_{t_1})\circ T_{t_1\to t_2}(y_{t_1})=T_{t_1\to t_2}(z_{t_1})\quad \text{or, equivalently}\quad x_{t_2}\circ y_{t_2}=z_{t_2}.
\end{gather}
Consequently, $T_{t_1\to t_2}$ takes pure states, represented by primitive idempotents $E_{t_1}$, into pure states, as can be seen from,
\begin{equation}
\begin{split}
E_{t_1}\circ E_{t_1}=E_{t_1} &\implies T_{t_1\to t_2}(E_{t_1})\circ T_{t_1\to t_2}(E_{t_1})=T_{t_1\to t_2}(E_{t_1})\\
                             &\implies E_{t_2}\circ E_{t_2}=E_{t_2}.
\end{split}
\end{equation}
Remarkably, any derivation $D(z)$ can be expressed as,
\begin{equation}
D(z)=D_{x,y}(z)=x\circ(z\circ y)-(x\circ z)\circ y,
\end{equation}
where $x$ and $y$ are any to two traceless elements. Recalling the definition of the associator, this implies that the evolution of any (mixed or pure) state is given by,
\begin{equation}\label{eq:JstateEvo}
\partial_t \rho = [x, \rho, y].
\end{equation}
It would seem that, in the Jordan formalism, the associator plays a role equivalent to that of the commutator in the standard picture. Let us consider the case closest to conventional quantum mechanics, the Jordan algebra of $n\times n$ Hermitian matrices defined over $\mathds{C}$. In this case \eqref{eq:JstateEvo} reduces to,
\begin{equation}\label{eq:JstateEvocomplex}
\partial_t \rho = [[x,y], \rho].
\end{equation}
If $H=i[x,y]+\lambda \mathds{1}$ then $[[x,y], \rho]=-i[H, \rho]$ and \eqref{eq:JstateEvocomplex} reproduces  the conventional unitary evolution equation \eqref{eq:DMevo}. It is as if the Jordan formulation is, in words of \cite{Townsend:1985eq}, the ``square root'' of standard theory.

In light of this relationship between the Jordan formulation and the density matrix formalism, the simple Jordan algebras of $n\times n$ Hermitian matrices over the associative division algebras seem to offer an obvious generalisation. However, it would seem that they simply amount to conventional quantum theory defined over $\mathds{R, C}$ or $\mathds{H}$ \cite{Adler:1995}. This leaves the exceptional octonionic example. In this case a Hilbert space formulation is not possible due to the nonassociative nature of the octonions. It has, however, been shown that the usual axioms of  quantum theory may be satisfied by taking a more abstract ``propositional'' approach \cite{Gunaydin:1978,Townsend:1985eq,Bischoff:1993sr,DeLeo:1996mr,Manogue:1993ja}.

\subsubsection{\texorpdfstring{$D=5$ and Jordan algebras of degree three $J_{3}^{\mathds{A}}$}{D=5 and Jordan algebras of degree three J3(A)}}

Let us compare $J_{3}(P)$ of \autoref{sec:D=5jordan} with the 2 qutrit reduced matrices $\rho_{A}$ and $\rho_{B}$
\begin{equation}
\begin{split}
\rho_{A}&=\Tr_{B}\ket{\Psi}\bra{\Psi},\\
\rho_{B}&=\Tr_{A}\ket{\Psi}\bra{\Psi},
\end{split}
\end{equation}
which are also  $3 \times 3$ hermitian and transform in the same way, at least in the $\mathds{R}$ and $\mathds{C}$ cases:
\begin{equation}
\begin{split}
(\rho_{A})_{A_{1}A_{2}}&=a_{A_{1}B_{1}}a^{*}{}_{A_{2}B_{2}}\delta^{B_{1}B_{2}},\\
(\rho_{B})_{B_{1}B_{2}}&=a_{A_{1}B_{1}}a^{*}{}_{A_{2}B_{2}}\delta^{A_{1}A_{2}}.
\end{split}
\end{equation}
Making contact with the $\mathds{H}$, $\mathds{O}$ and $\mathds{O}^s$ cases, however, would require going to quaternionic or octonionic quantum mechanics.  The two qutrit system (Alice and Bob) is then described by the state
\begin{equation}
\ket{\Psi}=a_{AB}\ket{AB},
\end{equation}
where ${A}=0,1,2$, so the Hilbert space has dimension 9, 18, 36, 72 for $\mathds{A=R,C,H,O}$. The ${a}_{AB}$ transforms as a $(3,3)$ under $SL(3,\mathds{A})_{A} \times SL(3,\mathds{A})_{B}$.

From \autoref{sec:qutrits} the bipartite entanglement is measured by the 2-tangle
\begin{equation}
\tau_{2}(AB)=27|\det\rho|.
\end{equation}
We now propose to identify the black hole charges $J$ with components of the reduced density matrix $\rho$ rather than the state coefficient $a$!

Recall that $SL(3,\mathds{A})$ has dimension 8,16,35,78 for $\mathds{A=R,C,H,O}$. Note that the fundamental reps have real dimension $3\dim\mathds{A}$: 3,6,12,24 but 24 is not a rep of $E_{6}$.

\subsubsection{\texorpdfstring{$D=6$ and Jordan algebras of degree two $J_{2}^{\mathds{A}}$}{D=6 and Jordan algebras of degree two J2(A)}}

The elements of the Jordan algebras $J_{2}^{\mathds{A}}$  of degree two, are $2 \times 2$  hermitian real $\mathds{A}$ matrices:
\begin{equation}
J_2(\mathds{A})=\begin{pmatrix}R_1 & A \\\overline{A} & R_2\end{pmatrix},
\end{equation}
where $R_{i}$ are real numbers and where $A\in\mathds{A=R,C,H,O},\mathds{O}^s$. They transform as the $(\dim\mathds{A}+2)$ representation of $SL(2,\mathds{A})$. In other words as the \textbf{3,4,6,10,10} of $SO(2,1)$, $SO(3,1)$, $SO(5,1)$, $SO(9,1)$, $SO(5,5)$, respectively as in \autoref{tab:QMDivisionAlgebras}. These are the symmetries and dyonic black string charge representations of the magic $\mathcal{N}=2$ and $\mathcal{N}=8$ supergravities. In all cases the black string entropy is
\begin{gather}
S=\pi\sqrt{|\det J_2|},
\shortintertext{where}
\det J_2=R_1R_2-A\overline{A}.
\end{gather}
Let us compare $J_{2}(A)$ with the 2 qubit reduced matrices $\rho_{A}$ and $\rho_{B}$
\begin{equation}
\begin{split}
\rho_{A}&=\Tr_{B}\ket{\Psi}\bra{\Psi},\\
\rho_{B}&=\Tr_{A}\ket{\Psi}\bra{\Psi},
\end{split}
\end{equation}
which are also $2 \times 2$ hermitian and transform in the same way. Generalising to $\mathds{R,C,H,O},\mathds{O}^{s}$, the two qubit system (Alice and Bob) is then described by the state
\begin{equation}
\ket{\Psi} =a_{AB}\ket{AB},
\end{equation}
where ${A}=0,1$, so the Hilbert space has dimension 4, 8, 16, 32, 32 for $\mathds{A=R,C,H,O},\mathds{O}^s$. The ${a}_{AB}$ transforms as a $\mathbf{(2,2)}$ under $SL(2,\mathds{A})_{A} \times SL(2,\mathds{A})_{B}$. The bipartite entanglement is measured by the 2-tangle
\begin{equation}
\tau_{2}(AB)=4|\det\rho|.
\end{equation}

We now propose to identify the black hole charges $J$ with components of the reduced density matrix $\rho$ rather than the state coefficient $a$!

Recall that $SL(2,\mathds{A})$ has dimension $(\dim A+2)(\dim\mathds{A}+1)/2$ namely 3,6,15,45,45 for $\mathds{A=R,C,H,O},\mathds{O}^s$. Note that the fundamental reps have real dimension $2\dim\mathds{A}$ namely 2,4,8,16.

\subsubsection{\texorpdfstring{$D=4$ and Freudenthal triples $\mathfrak{M}(\mathfrak{J})$}{D=4 and Freudenthal triples M(J)}}

Recall from \autoref{sec:D=4freud}  that in $D=4$ the black hole charges are described by the Freudenthal triple system \cite{Ferrara:1997uz} realised as $2 \times 2$ ``matrices''. Unfortunately, we do not know how to proceed with the new interpretation in $D=4$.

\subsubsection{Advantages and drawbacks}

This new density matrix interpretation has some advantages and some drawbacks:
\begin{description}
\item[Advantages:]\ \\[-8pt]
    \begin{enumerate}
    \item It solves the problem of the apparently unconventional Hilbert space appearing in \eqref{eq:3qutrit}. Now it is just that of two qutrits, albeit with unconventional quantum mechanics.  There is no need to describe it as a subspace of 7-dits.
    \item The mismatch between real black hole charges and complex state coefficients now disappears. $\mathds{R,C,H,O}, \mathds{O}^{s}$ charges correspond to $\mathds{R,C,H,O},\mathds{O}^{s}$ density matrices.
    \item It works in $D=6$ as well as $D=5$. The old interpretation does not admit the \textbf{10} of $SO(5,5)$and so cannot explain black strings in $D=6$.
    \item In contrast to the old interpretation, this new interpretation may be extended to $\mathcal{N}=2$ supergravity coupled to an arbitrary number of supermultiplets which also have an interpretation in terms of \emph{reducible} Jordan algebras of degree 3 \cite{Ferrara:1997uz}.
    \end{enumerate}
\item[Drawbacks:]\ \\[-8pt]
    \begin{enumerate}
    \item It seems to contradict our previous interpretation of identifying charges with $a_{AB}$ or $a_{ABC}$ coefficients which seemed to work well, at least for the $\mathcal{N}=2$ $STU$ model and its $\mathcal{N}=8$ generalisations. Moreover the observations that $E_{7} \supset [SL(2)]^{7}$ and $E_{6} \supset [SL(3)]^{3}$ now seem to play no role.
    \item There seems no obvious way to interpret the Freudenthal triples in terms of three qubit density matrices, and so $D=4$ remains a mystery.
    \item Quaternionic \cite{Adler:1995} and octonionic quantum mechanics \cite{DeLeo:1996mr} seem further removed from the real world, as the following example illustrates. Consider a single qudit state defined over $\mathds{A=R,C,H,O}$
    \end{enumerate}
\end{description}
\begin{equation}
\ket{\Psi} = a_A\ket{A},
\end{equation}
where $A=0,1,2,\dotsc,d-1$. The dimension of the Hilbert space is
\begin{equation}
\dim\mathcal{H}=d\dim\mathds{A},
\end{equation}
and the $d \times d$ density matrix
\begin{equation}
\rho_{AB}=a_Aa_B^*,
\end{equation}
belongs to a $d$-dimensional Jordan algebra with
\begin{equation}
\dim J_d=d +\half d(d-1)\dim\mathds{A}
\end{equation}
real parameters, as in \autoref{tab:QMDivisionAlgebras}.
\begin{table}[ht]
\begin{tabular*}{\textwidth}{@{\extracolsep{\fill}}*{5}{M{c}}}
\toprule
& \mathds{A} & \dim\mathds{A} & \dim J_d & \\
\midrule
& \mathds{R} & 1              & d(d+1)/2 & \\
& \mathds{C} & 2              & d^2      & \\
& \mathds{H} & 4              & d(2d-1)  & \\
& \mathds{O} & 8              & d(4d-3)  & \\
\bottomrule
\end{tabular*}
\caption[Jordan algebra dimensions]{Jordan algebra dimensions.}
\label{tab:QMDivisionAlgebras}
\end{table}

Now consider a state of $n$ qudits
\begin{equation}
\ket{\Psi} = a_{A_1A_2\cdots A_n}\ket{A_1A_2\cdots A_n}.
\end{equation}
The dimension of the Hilbert space is
\begin{equation}
\dim\mathcal{H}=d^n \dim\mathds{A},
\end{equation}
and the $d^n \times d^n$ density matrix,
\begin{equation}
\rho_{A_1A_2\cdots A_nB_1B_2\cdots B_n}=a_{A_1A_2\cdots A_n}a_{B_1B_2\cdots B_n}^*,
\end{equation}
belongs to a $d^n$-dimensional Jordan algebra with
\begin{equation}
\dim J_{d^n}=d^n +\half d^n(d^n-1)\dim\mathds{A},
\end{equation}
real parameters.

One might expect that an $n$-qudit state would be described by the same number of parameters as the tensor product of $n$ single qudits \cite{Aaronson:2004}.
\begin{equation}
[\dim J_d]^n=\dim J_{d^n}.
\end{equation}
but this works only in the complex case.

\newpage
\section{\texorpdfstring{WRAPPED BRANES AS QUBITS}{Wrapped Branes as Qubits}}
\label{sec:wrap}

\subsection{\texorpdfstring{$D=4$ black hole origin of qubit two-valuedness}{D=4 black hole origin of qubit two-valuedness}}

In  \autoref{sec:correspondence} we established a correspondence between the tripartite entanglement measure of three qubits and the macroscopic entropy of the four-dimensional 8-charge $STU$ black hole of supergravity. In this section we consider the configurations of intersecting D3-branes, whose wrapping around the six compact dimensions $T^6$ provides the microscopic string-theoretic interpretation of the charges, and associate the three-qubit basis vectors $\ket{ABC}$, ($A,B,C=0$ or $1$) with the corresponding 8 wrapping cycles.  In particular, we relate a well-known fact of quantum information theory, that the most general real three-qubit state can be parameterised by four real numbers and an angle, to a well-known fact of string theory, that the most general $STU$ black hole can be described by four D3-branes intersecting at an angle.

Macroscopically, $S$ is just one quarter the area of the  event horizon of the black hole. To give a microscopic derivation \cite{Strominger:1996sh} we need to invoke ten-dimensional string theory whose associated D$p$-branes wrapping around the six compact  dimensions provide the string-theoretic interpretation of the black holes. A D$p$-brane wrapped around a $p$-dimensional cycle of the compact directions $(x^4,x^5,x^6,x^7,x^8,x^9)$ looks like a D0-brane from the four-dimensional $(x^0,x^1,x^2,x^3)$ perspective \cite{Bergshoeff:1996rn,Gauntlett:1996pb,Gauntlett:1997cv,Tseytlin:1996bh}.

The microscopic analysis is not unique since there are many ways of embedding the $STU$ model in string/M-theory, but a useful one from our point of view is that of four D3-branes of Type IIB wrapping the $(579)$, $(568)$, $(478)$, $(469)$ cycles of $T^6$ with wrapping numbers $N_0$, $N_1$, $N_2$, $N_3$ and intersecting over a string \cite{Klebanov:1996mh}. The wrapped circles are denoted by crosses and the unwrapped circles by noughts as shown in \autoref{tab:3QubitIntersect}. This picture is consistent with the interpretation of the 4-charge black hole as bound state at threshold of four 1-charge black holes \cite{Duff:1994jr,Duff:1996qp,Duff:1995sm}. The fifth parameter $\theta$ is obtained \cite{Balasubramanian:1997ak,Bertolini:2000ei} by allowing the $N_3$ brane to intersect at an angle which induces additional effective charges on the $(579),(569),(479)$ cycles. The microscopic calculation of the entropy consists of taking the logarithm of the number of microstates and yields the same result as the macroscopic one \cite{Bertolini:2000yaa}.

To make the black hole/qubit correspondence we associate the three $T^2$ with the $SL(2)_A \times SL(2)_B \times SL(2)_C$ of the three qubits Alice, Bob, and Charlie. The 8 different cycles then yield 8 different basis vectors $\ket{ABC}$ as in the last column of \autoref{tab:3QubitIntersect}, where $\ket{0}$ corresponds to \textsf{xo} and $\ket{1}$ to \textsf{ox}. To wrap or not to wrap; that is the qubit.
We see immediately that we reproduce the five parameter three-qubit state $\ket{\Psi}$ of \eqref{eq:five}:
\begin{equation}
\begin{split}
\ket{\Psi} &= -N_3\cos^2\theta\ket{001}-N_2\ket{010}+N_3\sin\theta\cos\theta\ket{011}\\
&\phantom{=}-N_1\ket{100}-N_3\sin\theta\cos\theta\ket{101}+(N_0+N_3\sin^2\theta)\ket{111}.
\end{split}
\end{equation}
Note that the GHZ state of \autoref{tab:SLOCCRepresentatives} describes four D3-branes intersecting over a string. Performing a T-duality transformation, one obtains a Type IIA interpretation with zero D6-branes, $N_0$ D0-branes, $N_1$, $N_2$, $N_3$ D4-branes plus effective D2-brane charges, where $\ket{0}$ now corresponds to \textsf{xx} and $\ket{1}$ to \textsf{oo}.
\begin{table}[ht]
\begin{tabular*}{\textwidth}{@{\extracolsep{\fill}}*{11}{c}>{$\lvert}c<{\rangle$}c}
\toprule
& 4    & 5    & & 6    & 7    & & 8    & 9    & macro charges & micro charges              & ABC & \\
\midrule
& \sfx & \sfo & & \sfx & \sfo & & \sfx & \sfo & $p^0$         & 0                          & 000 & \\
& \sfo & \sfx & & \sfo & \sfx & & \sfx & \sfo & $q_1$         & 0                          & 110 & \\
& \sfo & \sfx & & \sfx & \sfo & & \sfo & \sfx & $q_2$         & $-N_3\sin\theta\cos\theta$ & 101 & \\
& \sfx & \sfo & & \sfo & \sfx & & \sfo & \sfx & $q_3$         & $N_3\sin\theta\cos\theta$  & 011 & \\
\midrule
& \sfo & \sfx & & \sfo & \sfx & & \sfo & \sfx & $q_0$         & $N_0+N_3\sin^2\theta$      & 111 & \\
& \sfx & \sfo & & \sfx & \sfo & & \sfo & \sfx & $-p^1$        & $ -N_3\cos^2\theta$        & 001 & \\
& \sfx & \sfo & & \sfo & \sfx & & \sfx & \sfo & $-p^2$        & $-N_2$                     & 010 & \\
& \sfo & \sfx & & \sfx & \sfo & & \sfx & \sfo & $-p^3$        & $-N_1$                     & 100 & \\
\bottomrule
\end{tabular*}
\caption[Wrapped D3-branes]{Three qubit interpretation of the 8-charge $D=4$ black hole from four D3-branes wrapping around the lower four cycles of $T^6$ with wrapping numbers $N_0,N_1,N_2,N_3$ and then allowing $N_3$ to intersect at an angle $\theta$.}\label{tab:3QubitIntersect}
\end{table}

\subsection{\texorpdfstring{$D=5$ black hole origin of qutrit three-valuedness}{D=5 black hole origin of qutrit three-valuedness}}

All this suggests that the analogy \cite{Duff:2007wa} between $D=5$ black holes and qutrits, described in \autoref{sec:E6Bipartite3Qutrits}, should involve the choice of wrapping an M2- brane around one of three circles in $T^3$. This is indeed the case, with the number of qutrits being two.

The 9-charge $\mathcal{N}=2,D=5$ black hole may also be embedded in the $\mathcal{N}=8$ theory in different  ways. The most convenient microscopic description is that of three M2-branes \cite{Papadopoulos:1996uq,Klebanov:1996mh} wrapping the (58), (69), (710) cycles of the $T^6$ compactification of $D=11$ M-theory, with wrapping numbers $N_0$, $N_1$, $N_2$ as in \autoref{tab:2QutritIntersect}. To make the black hole/qutrit  correspondence we associate the two $T^3$ with the $SL(3)_A \times SL(3)_B$ of the two qutrits Alice and Bob. The 9 different cycles then yield the 9 different basis vectors $\ket{AB}$ as in the last column of \autoref{tab:2QutritIntersect},  where $\ket{0}$ corresponds to \textsf{xoo}, $\ket{1}$ to \textsf{oxo} and $\ket{2}$ to \textsf{oox}. We see immediately that we reproduce the three parameter two-qutrit state $\ket{\Psi}$  of \eqref{eq:three}:
\begin{equation}
\ket{\Psi} = N_0\ket{00}+N_1\ket{11}+N_2\ket{22}.
\end{equation}
Note that this rank 3 Bell state describes three M2-branes intersecting over a point.

The black hole entropy, both macroscopic and microscopic, turns out to be given by the 2-tangle
\begin{equation}
S=2\pi\sqrt{|\det a_{AB}|},
\end{equation}
and the classification of the two-qutrit entanglements matches that of the black holes as in \autoref{tab:2QutritIntersect}.
\begin{table}[ht]
\centering
\begin{tabular*}{\textwidth}{@{\extracolsep{\fill}}*{10}{c}>{$\lvert}c<{\rangle$}c}
\toprule
& 5    & 6    & 7    & & 8    & 9    & 10    & macro charges & micro charges & AB & \\
\midrule
& \sfx & \sfo & \sfo & & \sfx & \sfo & \sfo  & $p^0$         & $N_0$         & 00 & \\
& \sfo & \sfx & \sfo & & \sfo & \sfx & \sfo  & $p^1$         & $N_1$         & 11 & \\
& \sfo & \sfo & \sfx & & \sfo & \sfo & \sfx  & $p^2$         & $N_2$         & 22 & \\
\midrule
& \sfx & \sfo & \sfo & & \sfo & \sfx & \sfo  & $p^3$         & $0$           & 01 & \\
& \sfo & \sfx & \sfo & & \sfo & \sfo & \sfx  & $p^4$         & $0$           & 12 & \\
& \sfo & \sfo & \sfx & & \sfx & \sfo & \sfo  & $p^5$         & $0$           & 20 & \\
\midrule
& \sfx & \sfo & \sfo & & \sfo & \sfo & \sfx  & $p^6$         & $0$           & 02 & \\
& \sfo & \sfx & \sfo & & \sfx & \sfo & \sfo  & $p^7$         & $0$           & 10 & \\
& \sfo & \sfo & \sfx & & \sfo & \sfx & \sfo  & $p^8$         & $0$           & 21 & \\
\bottomrule
\end{tabular*}
\caption[Wrapped M2-branes ]{Two qutrit interpretation of the 9-charge $D=5$ black hole from M2-branes in $D=11$ wrapping around the upper three cycles of $T^6$ with wrapping numbers $N_0$, $N_1$, $N_2$. Note that they intersect over a point.}\label{tab:2QutritIntersect}
\end{table}
There is, in fact, a quantum information theoretic interpretation of the 27 charge $\mathcal{N}=8,D=5$ black hole in terms of a Hilbert space consisting of three copies of the two-qutrit Hilbert space \cite{Duff:2007wa}. It relies on the decomposition $E_{6(6)} \supset [SL(3)]^3$ and admits the interpretation of a bipartite entanglement of three qutrits, with the entanglement measure given by Cartan's cubic $E_{6(6)}$ invariant. Once again, however, because the generating solution depends on the same three parameters as the 9-charge model, its classification of states will exactly parallel that of the usual two-qutrit system. Indeed, the Cartan invariant reduces to $\det a_{AB}$ in a canonical basis \cite{Ferrara:1997ci}.

The microscopic interpretation of the $D=5$ black string and its QI correspondence proceeds in a similar way by wrapping three M5-branes around $T^6$. We simply swap the crosses and the noughts in \autoref{tab:2QutritIntersect}.

\newpage
\section{\texorpdfstring{OUTSTANDING PROBLEMS}{Outstanding Problems}}
\label{sec:conclusions}

\subsection{For quantum information theory}

The Fano plane has played a crucial role in our description of the 3-way entanglement of seven qubits. It also finds application in switching networks that can connect any phone to any other phone.  It is the 3-switching network for 7 numbers.  However there also exists a 4-switching network for 13 numbers, a 5-switching network for 21 numbers, and generally an $(n+1)$-switching network for  $(n^{2}+n+1)$ numbers corresponding to the projective planes of order $n$ \cite{Pegg, Penrose:2005}. It would be worthwhile pursuing the corresponding quantum bit entanglements.

Exceptional groups, such as $E_{7(7)}$, have featured in supergravity, string theory, M-theory and other speculative attempts at unification of the fundamental forces. However, it is unusual to find an exceptional group appearing in the context of qubit entanglement. Can this be subject to experimental test?

What is the significance of the coset constructions, such as $E_8/[SL(2)]^8$ that describes the 4-way entanglement of eight qubits?

Cayley's hyperdeterminant provides a good measure of entanglement by virtue of being an entanglement monotone. We intend to return elsewhere to the issue of whether Cartan's $E_7$ invariant is also monotonic.

Jordan algebras were first introduced in the 1930s by Jordan, Wigner and von Neumann to describe the different possible versions of quantum mechanics: defined over the real, complex, quaternionic and octonionic numbers.   As we have seen, they also make their appearance in black hole physics which we have, in turn, related to quantum mechanics. Does this mean that we can come full circle and reformulate the qubit entanglements in terms of quaternionic and/or octonionic quantum mechanics?

The $D=5$ black hole, related to two qutrits, and the $D=4$ black hole, related to three qubits, are themselves related through the 3-dimensional Jordan algebras over the octonions and the Freudenthal triple systems.  Indeed there exists a {\it  5D/4D black hole correspondence} that relates the two but involves black holes carrying angular momentum and Taub-NUT charge \cite{Gaiotto:2005gf,Gaiotto:2005xt}. Is there a QI interpretation of rotating black holes and NUT charge? What does this imply for the relation of qubits to qutrits? Moreover, we have confined out attention to asymptotically flat black holes.  What about those that are asymptotically anti-de Sitter (AdS)?

Although there are no black holes with non-zero entropy in $D \geq 6$ dimensions, there are black strings and other intersecting brane configurations with entropies given by U-duality invariants. Do they have a qubit interpretation?

A pair of entangled qubits may be exploited to perform computational tasks beyond the capability of any classical device. What is more, a three-qubit entangled state can then be used achieve tasks surpassing that of the 2-qubit case. Can we go beyond these examples utilising the special properties of the tripartite entanglement of seven qubits? One possibility is the role of the Fano plane in error-correcting codes.

The tripartite entanglement of three qubits is subject to experimental test and provides
a striking version of Bell's theorem on non-locality versus realism \cite{Mermin:1990}. Can we generalise these experiments to the tripartite entanglement of seven qubits described above by the Fano plane and thereby demonstrate the effects of octonions in the laboratory?

\subsection{For M-theory}

The $STU$ black hole entropy is described by the $2 \times 2 \times 2$ hyperdeterminant, but there is a much richer mathematical structure behind the theory of more general hyperdeterminants \cite{Gelfand:1994} which also enters the classification of entanglement invariants \cite{Miyake:2002,Miyake:2003,Verstraete:2003,Luque:2002}. Do they have a role in M-theory?

We have seen  that the $SL(2) \times SO(6,6)$ Cartan invariant for the 24 NS-NS charges is just the $SL(2)^3$ invariant Cayley's hyperdeterminant when defined over the imaginary quaternions.  Is this just a property of the charges or does this mean that the $\mathcal{N}=4$ supergravity is, in some sense,  just the $\mathcal{N}=2$ $STU$ supergravity defined over the imaginary quaternions? If so, does this mean that the NS-NS sector of $D=10$  string theory compactified on $T^6$ with U-duality $SL(2) \times SO(6,6)$ is somehow equivalent to the NS-NS sector $D=6$ string theory compactified on $T^2$ with U-duality $SL(2)^3$, when defined over the imaginary quaternions?

The $E_7$ Cartan invariant for the full 56 charges, including the 32 R-R,  is not simply given by Cayley's hyperdeterminant over the imaginary octonions but is nevertheless given by a similar expression quartic in the imaginary octonions. So one could again ask whether  this just a property of the charges or does this mean that the $\mathcal{N}=8$ supergravity is, in some sense,  related to the $\mathcal{N}=2$ $STU$ supergravity defined over the imaginary octonions.  If so, does this mean that $D=11$  M-theory compactified on $T^7$ with U-duality $E_7$ is somehow related to the NS-NS sector $D=6$ string theory compactified on $T^2$ with U-duality $SL(2)^3$, when defined over the imaginary octonions?
We are encouraged in these speculations by recent independent advances by both mathematicians \cite{Manivel:2005, Elduque:2005} and physicists \cite{Duff:2006ue,Levay:2006pt} that show, using the Fano heptads,  that $E_7$ has a natural structure of an $\mathds{O}$-graded algebra, compatible with its action on the minimal 56-dimensional representation.

\subsection{For their inter-relation}

A third physical application of Cayley's hyperdeterminant is that of providing the Lagrangian of the Nambu-Goto string in spacetime signature $(2,2)$ \cite{Duff:2006ev,Nishino:2007ke}. (It is then possible to generalise to an $E_7$ invariant string using the Cartan invariant.) Is this related to its other two applications in black holes and  quantum information theory?

The Type IIB microscopic analysis of the black hole has provided an explanation for the appearance of the qubit two-valuedness (0 or 1) that was lacking in the previous treatments: the brane can wrap one circle or the other in each $T^2$.  The number of qubits is three because of the number of string theory extra dimensions is six.  Moreover, the five parameters of the real three-qubit state are seen to correspond to four D3-branes intersecting at an angle. Similar results hold for the two-qutrit system.
Can one now find an underlying physical justification for  \eqref{eq:entropy} relating the 3-tangle to the black hole entropy?

\section{\texorpdfstring{ACKNOWLEDGEMENTS}{Acknowledgements}}

We are grateful to Iosif Bena, Murat Gunaydin, Peter Levay, Christoph Luhn, Martin Plenio and Dan Waldram for useful conversations and correspondence. The sections on $\mathcal{N}=8,D=4$ black holes and seven qubits and $\mathcal{N}=8,D=5$ black holes and three qutrits are based on work by one of us (MJD) in collaboration with Sergio Ferrara, to whom we are grateful for much valuable input. H. E. would like to thank the theory group at Imperial College for their warm hospitality; her work was supported in part by DOE Grant No. DE-FG02-92ER40706, DOE Outstanding Junior Investigator award and by a Marie Curie Fellowship at Imperial College London.

\appendix

\newpage
\section{\texorpdfstring{TRANSVECTANTS AND CAYLEY'S HYPERDETERMINANT}{Transvectants and Cayley's Hyperdeterminant}}
\label{sec:Transvectants}

Cayley's original treatment of the hyperdeterminant \cite{Cayley:1845} made use of a homogeneous polynomial $U$ of third degree in six variables with eight coefficients $a$ to $h$:
\begin{equation}
\begin{split}
U&=\phantom{+}a\,x_1 y_1 z_1+b\,x_1 y_1 z_2+c\,x_1 y_2 z_1+d\,x_1 y_2 z_2 \\
&\,\phantom{=}+e\,x_2 y_1 z_1+f\,x_2 y_1 z_2+g\,x_2 y_2 z_1+h\,x_2 y_2 z_2.
\end{split}
\end{equation}
Successive application of differential operators reduced $U$ to an expression called $u$ - an example of a hyperdeterminant (cf \autoref{sec:Hyperdet}):
\begin{equation}
\begin{split}
u&=\phantom{+}a^2 h^2 + b^2 g^2 + c^2 f^2 + d^2 e^2 \\
&\phantom{=}-2\,(abgh +acfh +aedh +bcfg +bdeg +cdef) \\
&\phantom{=}+4\,(adfg + bceh) \\
&=(ah-bg-cf+de)^2+4(ad-bc)(fg-eh) \\
&=(ah-bg-de+cf)^2+4(af-be)(dg-ch) \\
&=(ah-cf-de+bg)^2+4(ag-ce)(df-bh),
\end{split}
\end{equation}
To begin with the differentials were expressed solely in terms of what we would today recognise as ``momentum space'' duals to the $x, y, z$ variables. This developed into the traditional \emph{symbolic method} or classical \emph{umbral calculus}. Cayley's subsequent treatment made use of determinants of these dual variables, and one of the notations for the determinants was the symbol Omega $\|\Omega\|$. These were combined into a single operator called the hyperdeterminant symbol $\square$ so that $u=\square U$. Cayley eventually moved on from his hyperdeterminant theory and after a sesquicentennial hiatus the work was reestablished in \cite{Gelfand:1994}.

A connection between the hyperdeterminant theory and entanglement measures lies in classical invariant theory \cite{Olver:1999,Luque:2002,Briand:2003a,Briand:2003,Luque:2005,Toumazet:2006,Endrejat:2006}. The objects of interest are forms (homogeneous polynomials), and the functions of these forms that are ``unchanged'' in some sense by transformations of the variables - for this purpose it is instructive to consider the more general notion of a \emph{covariant}\footnote{We rely heavily upon the notation of \cite{Olver:1999} throughout this appendix.}. Covariants of a form are functions $J(\vec{a}, \vec{x})$ in the form's variables $\vec{x}$ and coefficients $\vec{a}$ that are unchanged under general linear transformations modulo the determinant $\Delta$ of the transformation:
\begin{equation}
J(\vec{a}, \vec{x})=\Delta^w J'(\vec{a}', \vec{x}').
\end{equation}
Here, $w$ is called the weight of the covariant. An invariant is simply a covariant in the special case in which there is no $\vec{x}$ dependence. When even the determinant factor drops out ($w=0$) the covariant (or invariant) is called \emph{absolute}. Given a set of forms depending on the same variables one may also construct joint covariants. These depend on the coefficients of all the forms, but satisfy the same transformation formula as ordinary covariants.

One can generate covariants of a form via a process known as \emph{transvection}, which is itself based upon the \emph{Omega process}. The $m$th order Omega process with respect to an $m\times m$ matrix of variables $V$ is a differential operator defined as
\begin{equation}
\begin{split}
\mathbf{\Omega}&:=\begin{vmatrix}\frac{\partial}{\partial V_{11}} & \cdots & \frac{\partial}{\partial V_{1m}} \\
                        \vdots                           & \ddots & \vdots                           \\
                        \frac{\partial}{\partial V_{m1}} & \cdots & \frac{\partial}{\partial V_{mm}}
         \end{vmatrix}, \\
&\ =\textstyle\sum_{\pi\in\mathcal{S}_m}(-)^\pi\prod_{i=1}^m\frac{\partial}{\partial V_{i\pi(i)}}.
\end{split}
\end{equation}
For example, a second order Omega process would be $\frac{\partial}{\partial V_{11}\partial V_{22}}-\frac{\partial}{\partial V_{12}\partial V_{21}}$. We will call the matrix of which  $\mathbf{\Omega}$ is the determinant the Omega matrix $\Omega$. In conjunction with a set $P$ consisting of $p$ polynomials, the Omega process may be used to form transvectants. The essential property of the Omega matrix that permits this construction is the fact that under a simultaneous general linear transformation $\Lambda$ of all the variables, Omega transforms as $\Omega\mapsto(\Lambda^{-1})^{\textsf{T}}\Omega$, so that the Omega process satisfies
\begin{equation}
\mathbf{\Omega}\mapsto(\det\Lambda)^{-1}\mathbf{\Omega},
\end{equation}
making it an example of an invariant process. Other invariant processes are the scaling and polarisation processes
\begin{equation}
\begin{array}{rc@{\ :=\ }c}
\text{Polarisation:} & \pi(\vec{x}, \vec{y}) & x_i\frac{\partial}{\partial y_i}, \\
\text{Scaling:}      & \sigma(\vec{x})       &  \pi(\vec{x}, \vec{x}),
\end{array}
\end{equation}
which obey identities with the Omega process. The covariants resulting from their application find alternative expression as the partial transvectants discussed briefly below.

Defining $P(\overrightarrow{V_i}):=P(V_{i1}, \ldots, V_{ip})$, we use the tensor product notation $\bigotimes_{i=1}^p P_i$ to denote $P_1(\overrightarrow{V_1}), \ldots, P_p(\overrightarrow{V_p})$. A \emph{complete} $n$-transvectant is then defined by
\begin{equation}
(P_1, \ldots, P_p)^{(n)}:=\tr\mathbf{\Omega}^n\textstyle\bigotimes_{i=1}^p P_i,
\end{equation}
where $\tr$ sets all vectors of variables to be equal: $\overrightarrow{V_1}=\overrightarrow{V_2}=\cdots=\overrightarrow{V_p}$. Under the exchange of any of the $P_i$, the transvectant picks up a factor of $(-)^n$. For a complete $n$-transvectant, the dimension $v$ of the $\overrightarrow{V_i}$ vectors satisfies $p=v=m$. However more generally, $v$ can be any integer multiple of $p$: $p=v/d$. In such case, the polynomials are expected to be multiforms; that is, while the polynomials accept $p\times d$ arguments, they are homogeneous in each of the $d$ sets of $v$ variables. Separate Omega processes can then operate on the $d$ sets of variables to form a complete $(n_1, \ldots, n_d)$-transvectant:
\begin{equation}
(P_1, \ldots, P_p)^{(n_1,\dotsc, n_d)}:=\tr\mathbf{\Omega}_1^{n_1}\cdots\mathbf{\Omega}_d^{n_d}\textstyle\bigotimes_{i=1}^p P_i.
\end{equation}
When the $P_i$ have multiweights $(w_{11},\dotsc,w_{1d}),\dotsc,(w_{p1},\dotsc,w_{pd})$ the resulting covariant has multiweight $(w_{11}+\dotsb+w_{p1}+n_1,\dotsc,w_{1d}+\dotsb+w_{pd}+n_d)$. In the multidegree case, transvectants satisfy $p=v/d=m$, but it is possible to generalise yet further to \emph{partial} transvectants in both the single and multidegree cases. The Omega processes of partial transvectants have variables excluded by deleting rows and columns in the full $pd\times pd$ Omega matrix. Partial transvectants subsume the scaling and polarisation processes so that the polynomial covariants of a set of forms can be obtained using only linear aggregates of partial transvectants. In all cases, transvectants obey the inequality
\begin{equation}\label{eq:TransvecIneq}
p \geq v/d= m,
\end{equation}
with saturation for a complete transvectant.

An elementary and eminently recognisable example of a transvectant would be the discriminant of a quadratic form:
\begin{equation}
\begin{gathered}
P(x,y)=ax^2+bxy+cy^2, \\
b^2-4ac\equiv-\half(P,P)^{(2)},
\end{gathered}
\end{equation}
which is an invariant. This is an example of a Hessian covariant $H[P]$:
\begin{equation}
H[P]:=\tfrac{1}{v!}(\underbrace{P,\dotsc,P}_v)^{(2)}\equiv\det\left(\frac{\partial P}{\partial x_i\partial x_j}\right).
\end{equation}
Another example is the Jacobian determinant $J[P_1,\dotsc,P_v]$:
\begin{equation}
J[P_1,\dotsc,P_v]:=(P_1,\dotsc,P_v)^{(1)}\equiv\det\left(\frac{\partial P_i}{\partial x_j}\right),
\end{equation}
which is familiar from the change of variables in integration. As this is a first transvectant it is actually simpler than the example of the Hessian covariant. A yet simpler case is the zeroth transvectant of a set of forms, which is merely the product of the forms.

In similar fashion, transvection can generate the entanglement measures of a three-qubit system using the covariants of a single polynomial $\psi$ in six variables \cite{Toumazet:2006}:
\begin{equation}\label{eq:psi}
\begin{gathered}
\psi:=a_{ijk}x_i y_j z_k \\
=\phantom{+}a_0 x_0 y_0 z_0+a_1 x_0 y_0 z_1+a_2 x_0 y_1 z_0+a_3 x_0 y_1 z_1 \\
\,\phantom{=}+a_4 x_1 y_0 z_0+a_5 x_1 y_0 z_1+a_6 x_1 y_1 z_0+a_7 x_1 y_1 z_1,
\end{gathered}
\end{equation}
where the usual binary to decimal conversion has been performed on the coefficients. The form $\psi$ is in fact a triform of tridegree $(1,1,1)$, which is to say it is a form of degree one in each of the $\vec{x}, \vec{y}, \vec{z}$ variables when the other two are fixed. This immediately suggests that one may form $(n_1,n_2,n_3)$-transvectants of $\psi$, the first of which is trivially $\psi$ itself. The next covariants of interest arise from Hessian-like transvectants whose degree-sum is two:
\begin{equation}
\begin{split}
H_1 &:= \half(\psi, \psi)^{(0,1,1)}, \\
H_2 &:= \half(\psi, \psi)^{(1,0,1)}, \\
H_3 &:= \half(\psi, \psi)^{(1,1,0)}.
\end{split}
\end{equation}
Note that the $H_i$ transvectants are of degree three with two polynomials and six variables, hence they are complete and the Omega processes are second order. Written out more explicitly these are:
\begin{equation}\label{eq:H_i}
\begin{split}
H_1 &= \left(a_0 x_0+a_4 x_1\right) \left(a_3 x_0+a_7 x_1\right)-\left(a_1 x_0+a_5 x_1\right) \left(a_2 x_0+a_6 x_1\right) \\
&= \left(a_0 a_3-a_1 a_2\right) x_0^2+\left(\phantom{-}a_3 a_4-a_2 a_5-a_1 a_6+a_0 a_7\right) x_0 x_1+\left(a_4 a_7-a_5 a_6\right) x_1^2, \\
H_2 &= \left(a_0 y_0+a_2 y_1\right) \left(a_5 y_0+a_7 y_1\right)-\left(a_1 y_0+a_3 y_1\right) \left(a_4 y_0+a_6 y_1\right) \\
&= \left(a_0 a_5-a_1 a_4\right) y_0^2+\left(-a_3 a_4+a_2 a_5-a_1 a_6+a_0 a_7\right) y_0 y_1+\left(a_2 a_7-a_3 a_6\right) y_1^2, \\
H_3 &= \left(a_0 z_0+a_1 z_1\right) \left(a_6 z_0+a_7 z_1\right)-\left(a_2 z_0+a_3 z_1\right) \left(a_4 z_0+a_5 z_1\right) \\
&= \left(a_0 a_6-a_2 a_4\right) z_0^2+\left(-a_3 a_4-a_2 a_5+a_1 a_6+a_0 a_7\right) z_0 z_1+\left(a_1 a_7-a_3 a_5\right) z_1^2.
\end{split}
\end{equation}
The $H_i$ are closely related to the 2-tangles of a three qubit system. Next there is a single Jacobian-like covariant which may be obtained from each of the three $H_i$
\begin{equation}
\begin{split}
T &:= (\psi, H_1)^{(1,0,0)} \\
  &\ \equiv (\psi, H_2)^{(0,1,0)} \\
  &\ \equiv (\psi, H_3)^{(0,0,1)}.
\end{split}
\end{equation}
This is a large expression when written in full:
\begin{equation}
\begin{array}{c@{\ }c@{\ \lparen}c@{}*{8}{c@{\ }}c@{\rparen\,}c}
T = &   &   & 2 a_1 a_2 a_4 & - & a_0 a_3 a_4 & - & a_0 a_2 a_5   & - & a_0 a_1 a_6   & + & a_0^2 a_7   & x_0 y_0 z_0 \\
    & + &   & a_1 a_3 a_4   & + & a_1 a_2 a_5 & - & 2 a_0 a_3 a_5 & - & a_1^2 a_6     & + & a_0 a_1 a_7 & x_0 y_0 z_1 \\
    & + &   & a_2 a_3 a_4   & - & a_2^2 a_5   & + & a_1 a_2 a_6   & - & 2 a_0 a_3 a_6 & + & a_0 a_2 a_7 & x_0 y_1 z_0 \\
    & + &   & a_3^2 a_4     & - & a_2 a_3 a_5 & - & a_1 a_3 a_6   & + & 2 a_1 a_2 a_7 & - & a_0 a_3 a_7 & x_0 y_1 z_1 \\
    & + & - & a_3 a_4^2     & + & a_2 a_4 a_5 & + & a_1 a_4 a_6   & - & 2 a_0 a_5 a_6 & + & a_0 a_4 a_7 & x_1 y_0 z_0 \\
    & + & - & a_3 a_4 a_5   & + & a_2 a_5^2   & - & a_1 a_5 a_6   & + & 2 a_1 a_4 a_7 & - & a_0 a_5 a_7 & x_1 y_0 z_1 \\
    & + & - & a_3 a_4 a_6   & - & a_2 a_5 a_6 & + & a_1 a_6^2     & + & 2 a_2 a_4 a_7 & - & a_0 a_6 a_7 & x_1 y_1 z_0 \\
    & + & - & 2 a_3 a_5 a_6 & + & a_3 a_4 a_7 & + & a_2 a_5 a_7   & + & a_1 a_6 a_7   & - & a_0 a_7^2   & x_1 y_1 z_1,
\end{array}
\end{equation}
which is related to the Kempe invariant \cite{Kempe:1999vk}. Finally there is the hyperdeterminant invariant whose absolute value is proportional to the 3-tangle
\begin{equation}
\Delta := \half(T, \psi)^{(1,1,1)}\equiv\Det a.
\end{equation}
Expressed entirely in terms of $\psi$ this is $\tfrac{1}{4}((\psi, (\psi, \psi)^{(0,1,1)})^{(1,0,0)}, \psi)^{(1,1,1)}$. Interpreting the $x,y,z$ variables as qubits, the LU-invariants can be generated by the self-overlaps of the six polynomials $\psi$, $H_1$, $H_2$, $H_3$, $T$, and $\Delta$. According to \cite{Toumazet:2006}, there is an additional LU-invariant $\braket{\Delta\psi^2}{T^2}$.

Rather than generate the three-qubit invariants through iterated complete transvection, they may alternatively be obtained as partial transvectants. In this case the Omega processes are partitioned into three sets, the first of which act on the $x$ variables, the second on the $y$'s and the third on the $z$'s (remembering that each polynomial's arguments are initially unique and only reduce to $x,y,z$ when the trace is applied). In the case of complete transvectants, there was no ambiguity in the variables involved in the Omega process, but now there is the freedom to choose which polynomials the processes are associated with. Recalling that $v/d=m$, we find that since the Omega processes for our triform \eqref{eq:psi} are second order, the processes need two labels referring to the polynomials involved (and of course there must be at least two polynomials since $p\geq m$). Consequently, we may adopt the notation:
\begin{equation}
\Omega_{\alpha\beta}:=\det\begin{pmatrix}\partial_{x_\alpha} & \partial_{y_\alpha} \\
                                         \partial_{x_\beta}  & \partial_{y_\beta}
                          \end{pmatrix},
\end{equation}
where $x$ and $y$ correspond to $x_0, x_1$ if the process appears in the first partition, $y_0, y_1$ if it appears in the second partition, etc. There is also a bracket notation we may employ:
\begin{equation}
[\alpha\beta]:=\Omega_{\alpha\beta}\equiv-[\beta\alpha],
\end{equation}
where the LHS is called a \emph{bracket factor} of the second kind\footnote{The bracket factors of the first and third kinds are also determinants, but involving the variables themselves. The former are denoted $(\alpha\vec{x})$ and correspond to scaling processes, whereas the latter look like $\llbracket\alpha\beta\rrbracket$ and are ordinary determinants of variables.}. While Cayley introduced the Omega notation, he also employed $\overline{\alpha\beta}$ for bracket factors. The covariants that are written in terms of bracket factors are referred to as \emph{bracket polynomials} and we can rewrite three-qubit covariants as bracket polynomials. To begin with, the $H_i$ are given by
\begin{equation}
\begin{split}
H_1&=\half\tr\,(\phantom{[12]})([12])([12])\,\psi\otimes\psi, \\
H_2&=\half\tr\,([12])(\phantom{[12]})([12])\,\psi\otimes\psi, \\
H_3&=\half\tr\,([12])([12])(\phantom{[12]})\,\psi\otimes\psi,
\end{split}
\end{equation}
where in each case there is a single empty partition (cf \eqref{eq:H_i}). Next, $T$ can be obtained in multiple ways as
\begin{equation}
\begin{split}
T&=\tr\,([12])([23])([23])\,\psi^{\otimes3} = -\tr\,([23])([12])([12])\,\psi^{\otimes3} \\
 &=\tr\,([23])([12])([23])\,\psi^{\otimes3} = -\tr\,([12])([23])([12])\,\psi^{\otimes3} \\
 &=\tr\,([23])([23])([12])\,\psi^{\otimes3} = -\tr\,([12])([12])([23])\,\psi^{\otimes3},
\end{split}
\end{equation}
where the new polynomial is linked to the remaining ones in the previously empty slot of any of the $H_i$. Finally, the hyperdeterminant is realised as
\begin{equation}
\Delta = \half\tr\,([12][34])([14][23])([14][23])\,\psi^{\otimes4}.
\end{equation}
There are of course several bracket polynomials corresponding to the hyperdeterminant, and they are obtained by linking the previously untouched polynomial in each slot with the new polynomial using the appropriate bracket factor.

\newpage
\section{\texorpdfstring{SUPERGRAVITY COMPACTIFICATIONS}{Supergravity Compactifications}}
\label{sec:compact}

\subsection{\texorpdfstring{$\mathcal{N}=1, D=11$}{N=1, D=11}}

In this appendix we look at superalgebras \cite{Townsend:1995gp} for M-theory in $D=11$, Type $IIA$ theory on $D=10$ and Type $IIB$ theory in D=10 in order to see how the central charges give rise to the black hole charges in $D=4$ after compactification on either $T^7$ or $T^6$.  To get the right central charges, we need to know the symmetry properties of Dirac matrices for $SO(1,2n)$ and $SO(1,2n-1)$ \cite{Strathdee:1986jr} which obey the Clifford algebra:
\begin{equation}
\{\Gamma_A, \Gamma_B\}=2\eta_{AB},
\end{equation}
where $A=0,\ldots,2n-1$ and $\eta=\diag(-1,+1,\dotsc,+1)$.
\begin{equation}
(\Gamma^{M_1M_2\cdots M_r}C)_{\alpha\beta}=
(-1)^{(n-r)(n-r+1)/2}(\Gamma^{M_1M_2\cdots M_r}C)_{\beta\alpha},
\end{equation}
where $C$ is the charge conjugation matrix
\begin{equation}
C=\Gamma_{n+1}\Gamma_{n+2}\cdots\Gamma_{2n-1}\Gamma_0.
\end{equation}

Here the symmetric gammas are $\Gamma^M C$, $\Gamma^{MN}C$ and $\Gamma^{MNPQR}C$ so the $D=11$ supersymmetry algebra is
\begin{equation}\label{eq:D=11SUSYAlgebra}
\{Q_\alpha,Q_\beta\} = \big(\Gamma^MC\big)_{\alpha\beta}P_M + \big(\Gamma^{MN}C\big)_{\alpha\beta}\, Z_{MN} + \big(\Gamma^{MNPQR}C\big)_{\alpha\beta}\, Z_{MNPQR}.
\end{equation}
That is, there is a two-form and a five-form charge. The total number of components of all charges on the RHS of \eqref{eq:D=11SUSYAlgebra} is
\begin{equation}
\mathbf{11 + 55 + 462 = 528},
\end{equation}
which is, algebraically, the maximum possible number since the LHS is a symmetric $32 \times 32$ matrix.

Consider an $\mathbf{11=4+7}$ split with
\begin{equation}
SO(1,10) \supset SO(1,3) \times SO(7),
\end{equation}
under which
\begin{subequations}
\begin{gather}
\begin{array}{c@{~\to~}c@{~+~}c}
P_M & P_\mu & P_i, \\
\mathbf{11} & \mathbf{(4,1)} & \mathbf{(1,7)},
\end{array} \\
\begin{array}{c@{~\to~}c@{~+~}c@{~+~}c}
Z_{MN} & Z_{\mu\nu} & Z_{\mu i} & Z_{ij}, \\
\mathbf{55} & \mathbf{(6,1)} & \mathbf{(4,7)} & \mathbf{(1,21)},
\end{array} \\
\begin{array}{c@{~\to~}c*{4}{@{~+~}c}}
Z_{MNPQR} & Z_{\mu\nu\rho\sigma i} & Z_{\mu\nu\rho ij} & Z_{\mu\nu ijk} & Z_{\mu ijkl} & Z_{ijklm}, \\
\mathbf{462} & \mathbf{(1,7)} & \mathbf{(4,21)} & \mathbf{(6,35)} & \mathbf{(4,35)} & \mathbf{(1,21)}.
\end{array}
\end{gather}
\end{subequations}
The 56 0-brane charges are
\begin{equation}
\begin{array}{c}
P_i \sim\, \mathbf{(1,7)}, \\
Z_{ij} \sim\, \mathbf{(1,21)}, \\
Z_{\mu\nu\rho\sigma i}=\varepsilon_{\mu\nu\rho\sigma}{\star Z}_i \sim \mathbf{(1,7)}, \\
Z_{ijklm}=\half\varepsilon_{ijklmno}\widetilde{Z}^{no}\sim \mathbf{(1,21)},
\end{array}
\end{equation}
which combine into a $\mathbf{(1,28+\overline{28})}$ of $SO(1,3) \times SU(8)$.  The 256 1-brane charges are
\begin{equation}
\begin{array}{c}
P_{\mu}   \sim\, \mathbf{(4,1)}, \\
Z_{\mu i} \sim\, \mathbf{(4,7)}, \\
Z_{\mu ijkl}= \tfrac{1}{6}\varepsilon_{ijklmno}\widetilde{Z}\indices{_{\mu}^{mno}} \sim \mathbf{(4,35)}, \\
Z_{\mu\nu\rho ij}= \varepsilon_{\mu\nu\rho\sigma}{\star Z}\indices{^{\sigma}_{ij}} \sim \mathbf{(4,21)},
\end{array}
\end{equation}
which combine into a $\mathbf{(4,1+63)}$ of $SO(1,3) \times SU(8)$.  The 216 2-brane charges are
\begin{equation}
\begin{array}{c}
Z_{\mu\nu} \sim\, \mathbf{(6,1)}, \\
{\star Z}_{\mu\nu ijk}=\half\varepsilon_{\mu\nu\rho\sigma}Z_{\rho\sigma ijk}  \sim \mathbf{(6,35)},
\end{array}
\end{equation}
which combine into a $\mathbf{(6,36)}$ of $SO(1,3) \times SU(8)$. These results are summarised in \autoref{tab:ChargeTable}.
\begin{table}[ht]
\begin{tabular*}{\textwidth}{@{\extracolsep{\fill}}*{7}{c}}
\toprule
& \multirow{2}{*}{$\mathcal{N}=1,D=11$} & \multicolumn{2}{c}{$\mathcal{N}=2A,D=10$} & \multicolumn{2}{c}{$\mathcal{N}=2B,D=10$} & \\
&                             & R-R & NS-NS                     & R-R & NS-NS                     & \\
\midrule
& $P_i\sim\mathbf{(1,7)}$ & $Z\sim\mathbf{(1,1)}$ & $P_i\sim\mathbf{(1,6)}$ & $Z_i\sim\mathbf{(1,6)}$ & $P_i\sim\mathbf{(1,6)}$ & \\
& $Z_{ij}\sim\mathbf{(1,21)}$ & $Z_{ij}\sim\mathbf{(1,15)}$ & $Z_i\sim\mathbf{(1,6)}$ & $Z_{ijk}\sim\mathbf{(1,10+10)}$ & $Z_i\sim\mathbf{(1,6)}$ & \\
& $\star{Z}_i\sim\mathbf{(1,7)}$ & $\star{Z}\sim\mathbf{(1,1)}$ & $\star{Z}_i\sim\mathbf{(1,6)}$ & $\star{Z}^{+}_{\mu i}\sim\mathbf{(1,6)}$ & $\star{Z}^{+}_i\sim\mathbf{(1,6)}$ & \\
& $\widetilde{Z}_{ij}\sim\mathbf{(1,21)}$ & $\widetilde{Z}_{ij}\sim\mathbf{(1,15)}$ & $\widetilde{Z}_i\sim\mathbf{(1,6)}$ & & $\star{\widetilde{Z}}^{+}_i\sim\mathbf{(1,6)}$ & \\
\midrule
& $P_\mu\sim\mathbf{(4,1)}$ & $Z_{\mu i}\sim\mathbf{(4,6)}$ & $P_{\mu}\sim\mathbf{(4,1)}$ & $Z_\mu\sim\mathbf{(4,1)}$ & $P_\mu\sim\mathbf{(4,1)}$ & \\
& $Z_{\mu i}\sim\mathbf{(4,7)}$ & $Z^{+}_{\mu ijk}\sim\mathbf{(4,10)}$ & $Z_{\mu}\sim\mathbf{(4,1)}$ & $Z_\mu\sim\mathbf{(4,1)}$ & $Z_\mu\sim\mathbf{(4,1)}$ & \\
& $\widetilde{Z}_{\mu ijk}\sim\mathbf{(4,35)}$ & $Z^{-}_{\mu ijk}\sim\mathbf{(4,10)}$ & $Z_{\mu ij}\sim\mathbf{(4,15)}$ & $Z_{\mu ij}\sim\mathbf{(4,15)}$ & $Z^{+}_{\mu ij}\sim\mathbf{(4,15)}$ & \\
& $\star{Z}_{\mu ij}\sim\mathbf{(4,21)}$ & $\star{Z}_{\mu i}\sim\mathbf{(4,6)}$ & $\star{Z}_{\mu ij}\sim\mathbf{(4,15)}$ & $Z^{+}_{\mu i}\sim\mathbf{(4,15)}$ & $\widetilde{Z}^{+}_{\mu ij}\sim\mathbf{(4,15)}$ & \\
\midrule
& $Z_{\mu\nu}\sim\mathbf{(6,1)}$ & $Z_{\mu\nu}\sim\mathbf{(6,1)}$ & $\star{Z}^{+}_{\mu\nu ijk}\sim\mathbf{(6,10)}$ & $Z^{+}_{\mu\nu ijk}\sim\mathbf{(6,10)}$ & $\widetilde{Z}^{+}_{\mu\nu ijk}\sim\mathbf{(6,10)}$ & \\
& $\star{Z}_{\mu\nu ijk}\sim\mathbf{(6,35)}$ & $\star{Z}^{+}_{\mu\nu ij}\sim\mathbf{(6,15)}$ & $\star{Z}^{-}_{\mu\nu ijk}\sim\mathbf{(6,10)}$ & $Z_{\mu\nu i}\sim\mathbf{(6,6)}$ & $\widetilde{Z}^{+}_{\mu\nu ijk}\sim\mathbf{(6,10)}$ & \\
\midrule
& \textbf{528} & \textbf{256} & \textbf{272} & \textbf{256} & \textbf{272} & \\
\bottomrule
\end{tabular*}
\caption[Representations appearing in compactifications]{$SO(1,3) \times SO(7)$, and $SO(1,3) \times SO(6)$ representations appearing in compactifications of $(\mathcal{N}=1,D=11)$, and ($\mathcal{N}=2,D=10)$, respectively.}
\label{tab:ChargeTable}
\end{table}

The $D=11$ gamma matrices are related the four- and seven-dimensional gamma matrices by
\begin{gather}
\Gamma_A=(\gamma_\alpha\otimes\mathds{1},\gamma_5\otimes\gamma_i),
\shortintertext{where}
\{\gamma_\alpha,\gamma_\beta\}=2\eta_{\alpha\beta},
\end{gather}
and $\alpha=0,1,2,3$, and where
\begin{equation}
\{\gamma_i,\gamma_j\}=2\delta_{ij},
\end{equation}
and $i=1,\dotsc, 7$. In $D=7, {\gamma}^{i},\gamma^{ij}$ are antisymmetric and $\gamma^{ijk}$ is symmetric, while in $D=4$ $C,\gamma^{\mu}C, \gamma_5C$ are antisymmetric, while $\gamma^{\mu\nu}C,\gamma^{\mu}\gamma_5C$ are symmetric. So the $D=4,\mathcal{N}=8$ algebra is in four-component notation is
\begin{equation}
\begin{array}{ccccc}
\{Q\indices{_{\alpha}^a},Q\indices{_{\beta}^b}\}= & & (\gamma^{\mu})_{\alpha \beta} P_{\mu} \delta^{ab} & + & (\gamma_5 C)_{\alpha \beta} P_i (\gamma^i)^{ab} \\
& + & (\gamma^{\mu\nu}C)_{\alpha\beta}Z_{\mu\nu}\delta^{ab} & + & (\gamma^{\mu} \gamma_5)_{\alpha \beta}Z_{\mu i} (\gamma^i)^{ab} \\
& + & (C)_{\alpha\beta} Z_{ij}(\gamma^{ij})^{ab} & + & (\gamma^{\mu\nu\rho\sigma} \gamma_{5} C)_{\alpha \beta} Z_{\mu\nu\rho\sigma i}(\gamma^i)^{ab} \\
& + & (\gamma^{\mu\nu\rho} )_{\alpha \beta} Z_{\mu\nu\rho ij}(\gamma^{ij})^{ab} & + & (\gamma^{\mu\nu} \gamma_5 )_{\alpha \beta} Z_{\mu\nu ijk}(\gamma^{ijk})^{ab} \\
& + & (\gamma^{\mu}  )_{\alpha \beta} Z_{\mu ijkl}(\gamma^{ijkl})^{ab} & + & (\gamma_5 C)_{\alpha \beta} Z_{ijklm}(\gamma^{ijklm})^{ab},
\end{array}
\end{equation}
where now $\alpha=1,2,3,4$ and $a=0,1,\dotsc,7$. Rewriting in terms of dualised variables:
\begin{gather}
\begin{array}{ccc}
\{Q\indices{_{\alpha}^a},Q\indices{_{\beta}^b}\}= & & [(\gamma^{\mu} )_{\alpha \beta} P_{\mu} + (\gamma^{\mu\nu} C)_{\alpha \beta} Z_{\mu\nu}] \delta^{ab} \\
& + & [(\gamma_5 C)_{\alpha \beta} P_i + (\gamma^{\mu} \gamma_5 )_{\alpha \beta } Z_{\mu i} + (C)_{\alpha \beta} {\star Z}_i] (\gamma^i)^{ab} \\
& + & [(C)_{\alpha\beta} Z_{ij} + (\gamma^{\mu}\gamma_5 )_{\alpha \beta} {\star Z}_{\mu ij} + (\gamma_5 C)_{\alpha \beta} \widetilde{Z}_{ij}](\gamma^{ij})^{ab} \\
& + & [(\gamma^{\mu\nu}C)_{\alpha \beta} {\star Z}_{\mu\nu ijk} + (\gamma^{\mu} )_{\alpha \beta} \widetilde{Z}_{\mu ijk}](\gamma^{ijk})^{ab},
\end{array}
\shortintertext{or}
\begin{array}{ccc}
\{Q\indices{_{\alpha}^a},Q\indices{_{\beta}^b}\}= & & (C)_{\alpha \beta}[{\star Z}_i(\gamma^i)^{ab}+Z_{ij}(\gamma^{ij})^{ab}] \\
& + & (\gamma_{5} C)_{\alpha \beta}[ P_i(\gamma^i)^{ab}+\widetilde{Z}_{ij}(\gamma^{ij})^{ab}] \\
& + & (\gamma^{\mu} )_{\alpha \beta} [P_{\mu}\delta^{ab}+\widetilde{Z}_{\mu ijk}(\gamma^{ijk})^{ab}] \\
& + & (\gamma^{\mu} \gamma_5 )_{\alpha \beta } [Z_{\mu i}(\gamma^{i})^{ab}+{\star Z}_{\mu ij}(\gamma^{ij})^{ab}] \\
& + & (\gamma^{\mu\nu} C)_{\alpha\beta} [Z_{\mu\nu}\delta^{ab}+{\star Z}_{\mu\nu ijk}(\gamma^{ijk})^{ab}].
\end{array}
\end{gather}

Defining
\begin{equation}
\begin{array}{c@{~:=~}c}
Z^{[ab]} & (P_i+i{\star Z}_i) (\gamma^{i})^{ab} + (\widetilde{Z}_{ij}+i Z_{ij})(\gamma^{ij})^{ab}, \\
Z\indices{_{(AB)}^{(ab)}} & Z_{(AB)}\delta^{ab} + {\star Z}_{(AB) ijk}(\gamma^{ijk})^{ab}, \\
Z\indices{_{AA^\prime}^a_b} & Z_{AA^\prime i}(\gamma^i)\indices{^a_b} + {\star Z}_{AA^\prime ij}(\gamma^{ij})\indices{^a_b}+
\widetilde{Z}_{AA^\prime ijk}(\gamma^{ijk})\indices{^a_b},
\end{array}
\end{equation}
the algebra in two-component spinor notation is
\begin{align}
\{Q\indices{_A^a},Q\indices{_B^b}\}&=\varepsilon_{AB}Z^{[ab]}+Z\indices{_{(AB)}^{(ab)}}, \nonumber \\
\{Q\indices{_A^a},Q\indices{^{*}_{A^\prime}_b}\}&=P_{AA^\prime}\delta\indices{^a_b}+Z\indices{_{AA^\prime}^a_b}, \\
\{Q\indices{^{*}_{A^\prime}_a},Q\indices{^{*}_{B^\prime}_b}\}&=\varepsilon_{A'B'}Z^{*}{}_{[ab]}+Z\indices{^{*}_{(A^\prime B^\prime)}_{(ab)}}. \nonumber
\end{align}
In terms of $SO(3,1) \times SU(8)$ representations, we have
\begin{equation}
\begin{array}{c@{\ \times\ }c@{\ =\ }c}
\mathbf{\big(\half,0;8\big)} & \mathbf{\big(\half,0;8\big)} & \mathbf{\big(1,0;28+36\big)}, \\[5pt]
\mathbf{\big(\half,0;8\big)} & \mathbf{\big(0,\half; \overline{8}\big)} & \mathbf{\big(\half,\half;1+63\big)}, \\[5pt]
\mathbf{\big(0,\half; \overline{8}\big)} & \mathbf{\big(0,\half; \overline{8}\big)} & \mathbf{\big(0,1;\overline{28}+\overline{36}\big)}.
\end{array}
\end{equation}

\subsection{\texorpdfstring{$\mathcal{N}=2A, D=10$}{N=2A, D=10}}

Consider next the $\mathcal{N}=2A, D=10$ supersymmetry algebra. Allowing for all algebraically inequivalent p-form charges permitted by symmetry, we have
\begin{equation}\label{eq:N=2AD=10SUSYAlgebra}
\begin{array}{c@{\ }c@{\ }c@{\ +\ }c}
\{Q_\alpha,Q_\beta\}= & & \big(\Gamma^MC\big)_{\alpha\beta} P_M & (\Gamma_{11}C)_{\alpha\beta}Z, \\
& + & \big(\Gamma^M\Gamma_{11}C\big)_{\alpha\beta}Z_M & \big(\Gamma^{MN}C\big)_{\alpha\beta}Z_{MN}, \\
& + & \big(\Gamma^{MNPQ}\Gamma_{11}C\big)_{\alpha\beta}Z_{MNPQ} & \big(\Gamma^{MNPQR}C\big)_{\alpha\beta}Z_{MNPQR}.
\end{array}
\end{equation}
The supersymmetry charges are 32 component non-chiral $D=10$ spinors, so the maximum number of components of charges on the RHS is 528, and this maximum is realised by the above algebra since
\begin{equation}
\mathbf{10 + 1 + 10 + 45 + 210 + 252 = 528}.
\end{equation}
Note that in this case the $\Gamma_{11}$ matrix distinguishes between the term involving the 10-momentum $P$ and that involving the one-form charge carried by the type IIA superstring. This means that on compactification to $D=4$ we obtain an additional six electric central charges from this source, relative to the $\mathcal{N}=1$ case. These are balanced by an additional six magnetic charges due to the fact that the five-form charge is no longer self-dual as it is in the $\mathcal{N}=1$ case. Thus, there is now a total of 24 $D=4$ central charges carried by particles of KK, string, or fivebrane origin. These are the charged particles in the NS-NS sector of the superstring theory.

Consider $\mathbf{10=4+6}$ split with
\begin{equation}
SO(1,9) \supset SO(1,3) \times SO(6),
\end{equation}
under which
\begin{subequations}
\begin{gather}
\begin{array}{c@{\ \to\ }c@{\ +\ }c}
P_M         & P_\mu          & P_i, \\
\mathbf{10} & \mathbf{(4,1)} & \mathbf{(1,6)},
\end{array} \\
\begin{array}{c@{\ \to\ }c@{\ +\ }c}
Z_{M}       & Z_{\mu}        & Z_{i}, \\
\mathbf{10} & \mathbf{(4,1)} & \mathbf{(1,6)},
\end{array} \\
\begin{array}{c@{\ \to\ }c*{4}{@{\ +\ }c}}\label{eq:2A}
Z_{MNPQR}    & Z_{\mu\nu\rho\sigma i} & Z_{\mu\nu\rho ij} & Z_{\mu\nu ijk}     & Z_{\mu ijkl}    & Z_{ijklmn}, \\
\mathbf{252} & \mathbf{(1,6)}         & \mathbf{(4,15)}   & \mathbf{(6,10+10)} & \mathbf{(4,15)} & \mathbf{(1,6)}.
\end{array}
\end{gather}
\end{subequations}
The 24 NS-NS 0-brane charges are
\begin{equation}
\begin{array}{c}
P_i \sim\, \mathbf{(1,6)}, \\
Z_i \sim\, \mathbf{(1,6)}, \\
Z_{\mu\nu\rho\sigma i}=\varepsilon_{\mu\nu\rho\sigma}{\star Z}_i \sim \mathbf{(1,6)}, \\
Z_{ijklm}=\varepsilon_{ijklmn}\widetilde{Z}^n\sim \mathbf{(1,6)},
\end{array}
\end{equation}
which remain a $\mathbf{(1,6+6+6+6)}$ of $SO(3,1) \times SU(4)$. The 128 NS-NS 1-brane charges are
\begin{equation}
\begin{array}{c}
P_{\mu} \sim\, \mathbf{(4,1)}, \\
Z_{\mu} \sim\, \mathbf{(4,1)}, \\
Z_{\mu\nu\rho ij}=\varepsilon_{\mu\nu\rho\sigma}{\star Z}^{\sigma}_{ij} \sim \mathbf{(4,15)}, \\
Z_{\mu ijk} \sim\, \mathbf{(4,15)},
\end{array}
\end{equation}
which remain a $\mathbf{(4,1+1+15+15)}$ of $SO(3,1) \times SU(4)$. The 120 NS-NS 2-brane charges are
\begin{equation}
\begin{split}
Z^{+}_{\mu\nu ijk}&=
\half\varepsilon_{\mu\nu\rho\sigma}{\star Z}\indices{^{+\rho\sigma}_{ijk}} \sim \mathbf{(6,10)}, \\
Z^{-}_{\mu\nu ijk}&=
\half\varepsilon_{\mu\nu\rho\sigma}{\star Z}\indices{^{-\rho\sigma}_{ijk}} \sim \mathbf{(6,10)},
\end{split}
\end{equation}
which remain a $\mathbf{(6,10+10)}$ of $SO(3,1) \times SU(4)$.

The remaining 32 $D=4$ 0-brane central charges of the $D=4,\mathcal{N}=8$ supersymmetry algebra have their $D=10$ origin in the zero-form, two-form and four-form charges of the $D=10$ algebra \eqref{eq:N=2AD=10SUSYAlgebra}. The IIA $p$-branes with $p=(0,6),(1,5),(2,4)$ are the (electric, magnetic) sources for the one-form, two-form and three-form gauge fields, respectively, of type IIA supergravity.
\begin{subequations}
\begin{gather}
\begin{array}{c@{\ \to\ }c}
Z & Z, \\
\mathbf{1} & \mathbf{(1,1)},
\end{array} \\
\begin{array}{c@{\ \to\ }*{2}{c@{\ +\ }}c}
Z_{MN} & Z_{\mu\nu} & Z_{\mu n} & Z_{mn}, \\
\mathbf{45} & \mathbf{(6,1)} & \mathbf{(4,6)} & \mathbf{(1,15)},
\end{array} \\
\begin{array}{c@{\ \to\ }c*{4}{@{\ +\ }c}}
Z_{MNPQ} & Z_{\mu\nu\rho\sigma} & Z_{\mu\nu\rho q} & Z_{\mu\nu pq} & Z_{\mu npq} & Z_{mnpq}, \\
\mathbf{210} & \mathbf{(1,1)} & \mathbf{(4,6)} & \mathbf{(6,15)} & \mathbf{(4,20)} & \mathbf{(1,15)}.
\end{array}
\end{gather}
\end{subequations}

The 32 R-R 0-brane charges are
\begin{equation}
\begin{array}{c}
Z \sim\, \mathbf{1}, \\
Z_{mn} \sim\, \mathbf{(1,15)}, \\
Z_{\mu\nu\rho\sigma}=\varepsilon_{\mu\nu\rho\sigma}\star Z \sim \mathbf{(1,1)}, \\
Z_{mnpq}=\half\varepsilon_{mnpqrs}\widetilde{Z}^{rs}\sim \mathbf{(1,15)},
\end{array}
\end{equation}
which remain a $\mathbf{(1,1+15+1+15)}$ of $SO(3,1) \times SU(4)$.  The 128 R-R 1-brane charges are
\begin{equation}
\begin{array}{c}
Z_{\mu n} \sim (\textbf{4},\textbf{6}), \\
Z_{\mu\nu\rho q}=\varepsilon_{\mu\nu\rho\sigma}Z^{\sigma}_q \sim (\textbf{4},\textbf{6}), \\
Z^{+}_{\mu npq} \sim (\textbf{4},\textbf{10}), \\
Z^{-}_{\mu npq} \sim (\textbf{4},\textbf{10}),
\end{array}
\end{equation}
which remain a $(\textbf{4},\textbf{6}+\textbf{6}+\textbf{10}+\textbf{10})$ of $SO(3,1) \times SU(4)$. The 96 R-R 2-brane charges are
\begin{equation}
\begin{array}{c}
Z_{\mu\nu} \sim \mathbf{(6,1)}, \\
Z_{\mu\nu pq} \sim \mathbf{(6,15)},
\end{array}
\end{equation}
which remain a $\mathbf{(6,15)}$ of $SO(3,1)  \times SU(4)$.  These results are summarised in \autoref{tab:ChargeTable}.

\subsection{\texorpdfstring{$\mathcal{N}=2B,D=10$}{N=2B,D=10}}

Consider next the $\mathcal{N}=2B, D=10$ supersymmetry algebra. Allowing for all algebraically inequivalent p-form charges permitted by symmetry, we have
\begin{equation}\label{eq:N=2BD=10SUSYAlgebra}
\begin{array}{c@{\ }c@{\ }c}
\{Q_\alpha^i,Q_\beta^j\}= & & \delta^{ij}\big(\mathcal{P}\Gamma^MC\big)_{\alpha\beta} P_M + \big(\mathcal{P}\Gamma^MC\big)_{\alpha\beta}\widetilde{Z}_M^{ij} \\
& + & \varepsilon^{ij}\big(\mathcal{P}\Gamma^{MNP}C\big)_{\alpha\beta} Z_{MNP} \\
& + & \delta^{ij} \big(\mathcal{P}\Gamma^{MNPQR}C\big)_{\alpha\beta}(Z^+)_{MNPQR} \\
& + & \big(\mathcal{P}\Gamma^{MNPQR}C\big)_{\alpha\beta}(\tilde Z^+)_{MNPQR}^{ij},
\end{array}
\end{equation}
where the tilde indicates the tracefree symmetric tensor of $SO(2)$, equivalently a $U(1)$ doublet. One component is NS-NS and the other R-R.  The total number of components of all charges on the RHS of \eqref{eq:N=2BD=10SUSYAlgebra} is
\begin{equation}
\mathbf{10 + 2\times 10 + 120 + 126 + 2\times 126 = 528}.
\end{equation}

Under the same $\mathbf{10=4+6}$ split, we have
\begin{subequations}
\begin{gather}
\begin{array}{c@{\ \to\ }c@{\ +\ }c}
P_M & P_\mu & P_i, \\
\mathbf{10} & \mathbf{(4,1)} & \mathbf{(1,6)},
\end{array} \\
\begin{array}{c@{\ \to\ }c@{\ +\ }c}
Z^{\text{NS}}_{M} & Z^{\text{NS}}_{\mu} & Z^{\text{NS}}_{i}, \\
\mathbf{10} & \mathbf{(4,1)} & \mathbf{(1,6)},
\end{array} \\
\begin{array}{c@{\ \to\ }*{2}{c@{\ +\ }}c}
(Z^{+}_{MNPQR})^{\text{NS}} & (Z^{+}_{\mu\nu\rho\sigma i})^{\text{NS}} & (Z^{+}_{\mu\nu\rho i j})^{\text{NS}} & (Z^{+}_{\mu\nu i j k})^{\text{NS}}, \\
\mathbf{126} & \mathbf{(1,6)} & \mathbf{(4,15)} & \mathbf{(6,10)},
\end{array} \\
\begin{array}{c@{\ \to\ }*{2}{c@{\ +\ }}c}\label{eq:2B}
(\widetilde{Z}^{+}_{MNPQR})^{\text{NS}} & (\widetilde{Z}^{+}_{\mu\nu\rho\sigma i})^{\text{NS}} & (\widetilde{Z}^{+}_{\mu\nu\rho i j})^{\text{NS}} & (\widetilde{Z}^{+}_{\mu\nu i j k})^{\text{NS}}, \\
\mathbf{126} & \mathbf{(1,6)} & \mathbf{(4,15)} & \mathbf{(6,10)}.
\end{array}
\end{gather}
\end{subequations}
The 24 NS-NS 0-brane charges are
\begin{equation}
\begin{array}{c}
P_i \sim\, \mathbf{(1,6)}, \\
Z^{\text{NS}}_i \sim\, \mathbf{(1,6)}, \\
(Z^{+}_{\mu\nu\rho\sigma i})^{\text{NS}}=\varepsilon_{\mu\nu\rho\sigma}{{\star Z}^{+}_i}^{\text{NS}} \sim \mathbf{(1,6)}, \\
(\widetilde{Z}^{+}_{\mu\nu\rho\sigma i})^{\text{NS}}=\varepsilon_{\mu\nu\rho\sigma}{{\star \widetilde{Z}}^{+}_i}^{\text{NS}} \sim \mathbf{(1,6)},
\end{array}
\end{equation}
which remain a $\mathbf{(1,6+6+6+6)}$ of $SO(3,1) \times SU(4)$. The 128 NS-NS 1-brane charges are
\begin{equation}
\begin{array}{c}
P_{\mu} \sim\, \mathbf{(4,1)}, \\
Z_{\mu}^{\text{NS}} \sim\, \mathbf{(4,1)}, \\
(Z^{+}_{\mu\nu\rho ij})^{\text{NS}}=\varepsilon_{\mu\nu\rho\sigma}(Z^{+\sigma})^{\text{NS}} \sim \mathbf{(4,15)}, \\
(\widetilde{Z}^{+}_{\mu\nu\rho ij})^{\text{NS}}=\varepsilon_{\mu\nu\rho\sigma}(\widetilde{Z}^{+\sigma})^{\text{NS}} \sim \mathbf{(4,15)},
\end{array}
\end{equation}
which remain a $\mathbf{(4,1+1+15+15)}$ of $SO(3,1) \times SU(4)$. The 120 NS-NS 2-brane charges are
\begin{equation}
\begin{split}
(\widetilde{Z}^{+}_{\mu\nu ijk})^{\text{NS}} & \sim \mathbf{(6,10)}, \\
(\widetilde{Z}^{+}_{\mu\nu ijk})^{\text{NS}} & \sim \mathbf{(6,10)},
\end{split}
\end{equation}

The remaining 32 $D=4$ 0-brane central charges of the $D=4,\mathcal{N}=8$ supersymmetry algebra have their $D=10$ origin in the one-form, three-form and five-form charges of the $D=10$ algebra \eqref{eq:N=2BD=10SUSYAlgebra}. The IIB $p$-branes with $p=(1,5)$ are the (electric) sources for the one-form, and five-form gauge fields and the $p=3$ is the dyonic source of the three-form charge, respectively, of type IIB supergravity.
\begin{gather}
\begin{array}{c@{~\to~}c@{~+~}c}
Z^{\text{RR}}_M & Z^{\text{RR}}_{\mu} & Z^{\text{RR}}_i, \\
\mathbf{10} & \mathbf{(4,1)} & \mathbf{(1,6)},
\end{array} \\
\begin{array}{c@{~\to~}*{3}{c@{~+~}}c}
Z_{MNP} & Z_{\mu\nu\rho} & Z_{\mu\nu i} & Z_{\mu ij} & Z_{ijk}, \\
\mathbf{120} & \mathbf{(4,1)} & \mathbf{(6,6)} & \mathbf{(4,15)} & \mathbf{(1,10+10)},
\end{array} \\
\begin{array}{c@{~\to~}*{2}{c@{~+~}}c}
(Z^{+}_{MNPQR})^{\text{RR}} & (Z^{+}_{\mu\nu\rho\sigma i})^{\text{RR}} & (Z^{+}_{\mu\nu\rho i})^{\text{RR}} & (Z^{+}_{\mu\nu ijk})^{\text{RR}}, \\
\mathbf{126} & \mathbf{(1,6)} & \mathbf{(4,15)} & \mathbf{(6,10)}.
\end{array}
\end{gather}

The 32 R-R 0-brane charges are
\begin{equation}
\begin{array}{c}
Z^{\text{RR}}_i \sim\, \mathbf{(1,6)}, \\
Z_{ijk} \sim\, \mathbf{(1,10+10)}, \\
(Z^{+}_{\mu\nu\rho\sigma i})^{\text{RR}}=\varepsilon_{\mu\nu\rho\sigma}{\star Z}^{+\sigma}_i \sim \mathbf{(1,6)}.
\end{array}
\end{equation}
which remain a $\mathbf{(1,6+6+10+10)}$ of $SO(3,1) \times SU(4)$.  The 128 R-R 1-brane charges are
\begin{equation}
\begin{array}{c}
Z_{\mu}^{\text{RR}} \sim \mathbf{(4,1)}, \\
Z_{\mu\nu\rho}=\varepsilon_{\mu\nu\rho\sigma}Z^{\sigma} \sim \mathbf{(4,1)}, \\
(Z_{\mu ij}) \sim \mathbf{(4,15)}, \\
(Z^{+}_{\mu\nu\rho i})^{\text{RR}}=\varepsilon_{\mu\nu\rho\sigma}(Z^{+\sigma}_i)^{\text{RR}} \sim \mathbf{(4,15)},
\end{array}
\end{equation}
which remain a $\mathbf{(4,1+1+15+15)}$ of $SO(3,1) \times SU(4)$.  The 96 R-R 2-brane charges are
\begin{equation}
\begin{array}{c}
(Z^{+}_{\mu\nu ijk})^{\text{RR}} \sim \mathbf{(6,10)}, \\
Z_{\mu\nu i} \sim \mathbf{(6,6)},
\end{array}
\end{equation}
which remain a $\mathbf{(6,6+10)}$ of $SO(3,1)  \times SU(4)$.  These results are summarised in \autoref{tab:ChargeTable}.

\newpage
\section{\texorpdfstring{MORE HIDDEN SYMMETRIES}{More Hidden Symmetries}}
\label{sec:hidden}

The SLOCC group for an $n$-qudit state is given by $G=[SL(d,\mathds{C})]^n$, but subspaces of the $d^n$-dimensional Hilbert space may display hidden symmetries not contained in $G$. We have already seen that these include an $E_7(\mathds{C})$ symmetry of a 56 dimensional subspace of seven qutrits and $E_6(\mathds{C})$ symmetry of a 27-dimensional subspace of three 7-dits. Here we explore some more possibilities. We thank Peter Levay for his input on early stages of these results.

\subsection{\texorpdfstring{$SL(2)$ decompositions}{SL(2) decompositions}}
\label{sec:sl2}

Consider for example the groups $E_8(\mathds{C})$, $E_7(\mathds{C})$, $SO(12,\mathds{C})$, $SO(8,\mathds{C})$, $Sp(6,\mathds{C})$ and $Sp(4,\mathds{C}) $ and their $SL(2,\mathds{C})^n$ subgroups as in \autoref{tab:CosetPropertiesSL(2)}.
\begin{table}[ht]
\begin{tabular*}{\textwidth}{@{\extracolsep{\fill}}*{8}{M{c}}}
\toprule
& G                 & fund & \dim G & H                  & \dim H & \dim G/H & \\
\midrule
& E_8(\mathds{C})   & 248  & 248    & SL(2,\mathds{C})^8 & 24     & 224      & \\
& E_7(\mathds{C})   & 56   & 133    & SL(2,\mathds{C})^7 & 21     & 112      & \\
& SO(12,\mathds{C}) & 32   & 66     & SL(2,\mathds{C})^6 & 18     & 48       & \\
& SO(8,\mathds{C})  & 8    & 28     & SL(2,\mathds{C})^4 & 12     & 16       & \\
& Sp(6,\mathds{C})  & 6    & 21     & SL(2,\mathds{C})^3 & 9      & 12       & \\
& Sp(4,\mathds{C})  & 4    & 10     & SL(2,\mathds{C})^2 & 6      & 4        & \\
\bottomrule
\end{tabular*}
\caption{Coset properties for $SL(2)$.}\label{tab:CosetPropertiesSL(2)}
\end{table}
The decomposition of their adjoint and fundamental representations are given as follows
\begin{gather}
E_8 \supset SL(2)_A \times SL(2)_B \times SL(2)_C \times SL(2)_D \times SL(2)_E \times SL(2)_F \times SL(2)_G  \times SL(2)_H,
\shortintertext{under which}
\begin{split}
\mathbf{248} &\to\phantom{+\!}\mathbf{(3,1,1,1,1,1,1,1)+(2,2,2,1,2,1,1,1)+(1,1,1,2,1,2,2,2)} \\
&\phantom{\to}+\mathbf{(1,3,1,1,1,1,1,1)+(2,1,2,2,1,2,1,1)+(1,2,1,1,2,1,2,2)} \\
&\phantom{\to}+\mathbf{(1,1,3,1,1,1,1,1)+(2,1,1,2,2,1,2,1)+(1,2,2,1,1,2,1,2)} \\
&\phantom{\to}+\mathbf{(1,1,1,3,1,1,1,1)+(2,1,1,1,2,2,1,2)+(1,2,2,2,1,1,2,1)} \\
&\phantom{\to}+\mathbf{(1,1,1,1,3,1,1,1)+(2,2,1,1,1,2,2,1)+(1,1,2,2,2,1,1,2)} \\
&\phantom{\to}+\mathbf{(1,1,1,1,1,3,1,1)+(2,1,2,1,1,1,2,2)+(1,2,1,2,2,2,1,1)} \\
&\phantom{\to}+\mathbf{(1,1,1,1,1,1,3,1)+(2,2,1,2,1,1,1,2)+(1,1,2,1,2,2,2,1)} \\
&\phantom{\to}+\mathbf{(1,1,1,1,1,1,1,3)},
\end{split}
\shortintertext{and}
E_7 \supset SL(2)_A \times SL(2)_B \times SL(2)_C \times SL(2)_D \times SL(2)_E \times SL(2)_F \times SL(2)_G,
\shortintertext{under which}
\begin{split}
\mathbf{56} &\to\phantom{+\!}\mathbf{(2,2,1,2,1,1,1)} \\
&\phantom{\to}+\mathbf{(1,2,2,1,2,1,1)} \\
&\phantom{\to}+\mathbf{(1,1,2,2,1,2,1)} \\
&\phantom{\to}+\mathbf{(1,1,1,2,2,1,2)} \\
&\phantom{\to}+\mathbf{(2,1,1,1,2,2,1)} \\
&\phantom{\to}+\mathbf{(1,2,1,1,1,2,2)} \\
&\phantom{\to}+\mathbf{(2,1,2,1,1,1,2)},
\end{split}\\
\begin{split}
\mathbf{133} &\to\phantom{+\!}\mathbf{(3,1,1,1,1,1,1)+(1,1,2,1,2,2,2)} \\
&\phantom{\to}+\mathbf{(1,3,1,1,1,1,1)+(2,1,1,2,1,2,2)} \\
&\phantom{\to}+\mathbf{(1,1,3,1,1,1,1)+(2,2,1,1,2,1,2)} \\
&\phantom{\to}+\mathbf{(1,1,1,3,1,1,1)+(2,2,2,1,1,2,1)} \\
&\phantom{\to}+\mathbf{(1,1,1,1,3,1,1)+(1,2,2,2,1,1,2)} \\
&\phantom{\to}+\mathbf{(1,1,1,1,1,3,1)+(2,1,2,2,2,1,1)} \\
&\phantom{\to}+\mathbf{(1,1,1,1,1,1,3)+(1,2,1,2,2,2,1)},
\end{split}
\shortintertext{and}
SO(12) \supset SL(2)_A \times SL(2)_B \times SL(2)_C \times SL(2)_D \times SL(2)_E \times SL(2)_F,
\shortintertext{under which}
\begin{split}
\mathbf{32} &\to\phantom{+\!}\mathbf{(2,2,1,2,1,1)} \\
&\phantom{\to}+\mathbf{(1,2,2,1,2,1)} \\
&\phantom{\to}+\mathbf{(1,1,2,2,1,2)} \\
&\phantom{\to}+\mathbf{(2,1,1,1,2,2)},
\end{split}\\
\begin{split}
\mathbf{66} &\to\phantom{+\!}\mathbf{(3,1,1,1,1,1)+(2,2,2,1,1,2)} \\
&\phantom{\to}+\mathbf{(1,3,1,1,1,1)+(2,1,2,2,2,1)} \\
&\phantom{\to}+\mathbf{(1,1,3,1,1,1)+(1,2,1,2,2,2)} \\
&\phantom{\to}+\mathbf{(1,1,1,3,1,1)} \\
&\phantom{\to}+\mathbf{(1,1,1,1,3,1)} \\
&\phantom{\to}+\mathbf{(1,1,1,1,1,3)},
\end{split}
\shortintertext{and}
SO(8) \supset SL(2)_A \times SL(2)_C \times SL(2)_D \times SL(2)_E,
\shortintertext{under which}
\begin{split}
\mathbf{8} &\to\phantom{+\!}\mathbf{(2,1,2,1)+(1,2,1,2)}, \\
\mathbf{28} &\to\phantom{+\!}\mathbf{(3,1,1,1)+(1,3,1,1)+(1,1,3,1)+(1,1,1,3)} \\
&\phantom{\to}+\mathbf{(2,2,2,2)},
\end{split}
\shortintertext{and}
Sp(6) \supset SL(2)_B \times SL(2)_C \times SL(2)_D,
\shortintertext{under which}
\begin{split}
\mathbf{6}&\to\phantom{+\!}\mathbf{(1,1,2)+(2,1,1)+(1,2,1)}, \\
\mathbf{21} &\to\phantom{+\!}\mathbf{(3,1,1)+(1,3,1)+(1,1,3)} \\
&\phantom{\to}+\mathbf{(2,2,1)+(1,2,2)+(2,1,2)},
\end{split}
\shortintertext{and}
Sp(4) \supset SL(2)_B \times SL(2)_C,
\shortintertext{under which}
\begin{split}
\mathbf{4} &\to\mathbf{(2,1)+(1,2)}, \\
\mathbf{10} &\to\mathbf{(3,1)+(1,3)+(2,2)}.
\end{split}
\end{gather}

\subsection{Qutrit interpretation}
\label{sec:qutrit}

\subsubsection{\texorpdfstring{$Sp(4)$ symmetry from two qutrits}{Sp(4) symmetry from two qutrits}}

Under
\begin{gather}
SL(3)_{A} \times SL(3)_B \supset SL(2)_{A} \times SL(2)_{B},
\shortintertext{we have}
\mathbf{(3,3) \to (2,2)+(2,1)+(1,2)+(1,1)}.
\end{gather}
In particular we find the 4-dimensional subspace
\begin{gather}
\mathbf{(2,1)+(1,2)},
\shortintertext{or}
\ket{\Psi}_{4} =a_{A}\ket{A\bullet}+b_{B}\ket{\bullet B},
\end{gather}
and the 4-dimensional subspace describing the bipartite entanglement of two qubits
\begin{gather}
\mathbf{(2,2)},
\shortintertext{or}
\ket{\Psi'}_{4} =a_{AB}\ket{AB}.
\end{gather}
From \autoref{sec:sl2}, we may assign the $\ket{\Psi}_{4}$ to the fundamental of $Sp(4,\mathds{C})$ and the $\ket{\Psi'}_{4}$ to the coset $Sp(4,\mathds{C})/[SL(2,\mathds{C})]^2$.

\subsubsection{\texorpdfstring{$Sp(6)$ symmetry from three qutrits}{Sp(6) symmetry from three qutrits}}

Under
\begin{gather}
SL(3)_{A} \times SL(3)_{B}  \times SL(3)_{C} \supset SL(2)_{A} \times SL(2)_{B}  \times SL(2)_{C},
\shortintertext{we have}
\begin{split}
\mathbf{(3,3,3)} &\to\phantom{+\!}\mathbf{(2,2,2)} \\
&\phantom{\to}+\mathbf{(2,2,1)+(2,1,2)+(1,2,2)} \\
&\phantom{\to}+\mathbf{(1,1,2)+(1,2,1)+(2,1,1)} \\
&\phantom{\to}+\mathbf{(1,1,1)}.
\end{split}
\end{gather}
In particular we find the 6-dimensional subspace
\begin{gather}
\mathbf{(2,1,1)+(1,2,1)+(1,1,2)},
\shortintertext{or}
\ket{\Psi}_{6} =a_{A}\ket{A\bullet\bullet}+b_{B}\ket{\bullet B\bullet}+c_{C}\ket{\bullet\bullet C},
\end{gather}
and the 12-dimensional subspace describing the bipartite entanglement of three qubits
\begin{gather}
\mathbf{(2,2,1)+(1,2,2)+(2,1,2)},
\shortintertext{or}
\ket{\Psi}_{12} =a_{AB}\ket{AB\bullet}+b_{BC}\ket{\bullet BC}+c_{CA}\ket{A\bullet C}.
\end{gather}
From \autoref{sec:sl2}, we may assign $\ket{\Psi}_{6}$  the  to the fundamental of $Sp(6,\mathds{C})$ and the $\ket{\Psi}_{12}$ to the coset $Sp(6,\mathds{C})/[SL(2,\mathds{C})]^2$.

\subsubsection{\texorpdfstring{$SO(8)$ symmetry from four qutrits}{SO(8) symmetry from four qutrits}}

Under
\begin{gather}
\begin{split}
&SL(3)_{A} \times SL(3)_{B} \times SL(3)_{C} \times SL(3)_{D} \\
\supset\ &SL(2)_{A} \times SL(2)_{B} \times SL(2)_{C} \times SL(2)_{D},
\end{split}
\shortintertext{we have}
\begin{split}
\mathbf{(3,3,3,3)} &\to\phantom{+\!}\mathbf{(2,2,2,2)} \\
&\phantom{\to}+\mathbf{(2,2,2,1)+(2,2,1,2)+(2,1,2,2)+(1,2,2,2)} \\
&\phantom{\to}+\mathbf{(2,2,1,1)+(2,1,2,1)+(1,2,1,2)+(1,1,2,2)} \\
&\phantom{\to}+\mathbf{(2,1,1,1)+(1,2,1,1)+(1,1,2,1)+(1,1,1,2)} \\
&\phantom{\to}+\mathbf{(1,1,1,1)}.
\end{split}
\end{gather}
In particular we find 8-dimensional subspace describing the bipartite entanglement of four qubits
\begin{gather}
\mathbf{(2,1,2,1)+(1,2,1,2)},
\shortintertext{or}
\ket{\Psi}_{8} =a_{AC}\ket{A\bullet C\bullet}+b_{BD}\ket{\bullet B\bullet D},
\end{gather}
and the 16-dimensional subspace describing the four-way entanglement of four qubits.
\begin{gather}
\mathbf{(2,2,2,2)},
\shortintertext{or}
\ket{\Psi}_{16} =a_{ABCD}\ket{ABCD}.
\end{gather}
From \autoref{sec:sl2}, we may assign the $\ket{\Psi}_{8}$ to the fundamental of $SO(8,\mathds{C})$ and the $\ket{\Psi}_{16}$ to the coset $SO(8,\mathds{C})/[SL(2,\mathds{C})]^4$.

\subsubsection{\texorpdfstring{$SO(12)$ symmetry from six qutrits}{SO(12) symmetry from six qutrits}}

Under
\begin{gather}
\begin{split}
& SL(3)_{A} \times SL(3)_{B} \times SL(3)_{C} \times SL(3)_{D} \times SL(3)_{E} \times SL(3)_{F} \\
\supset\ & SL(2)_{A} \times SL(2)_{B} \times SL(2)_{C} \times SL(2)_{D} \times SL(2)_{E} \times SL(2)_{F},
\end{split}
\shortintertext{we have}
\begin{array}{c*{4}{@{\ }c}@{}c}
\mathbf{(3,3,3,3,3,3)} \to &   & 1  & \text{term\phantom{s} like} & \mathbf{(2,2,2,2,2,2)} \\
                           & + & 6  & \text{terms like}           & \mathbf{(2,2,2,2,2,1)} \\
                           & + & 15 & \text{terms like}           & \mathbf{(2,2,2,2,1,1)} \\
                           & + & 20 & \text{terms like}           & \mathbf{(2,2,2,1,1,1)} \\
                           & + & 15 & \text{terms like}           & \mathbf{(2,2,1,1,1,1)} \\
                           & + & 6  & \text{terms like}           & \mathbf{(2,1,1,1,1,1)} \\
                           & + & 1  & \text{term\phantom{s} like} & \mathbf{(1,1,1,1,1,1)} & .
\end{array}
\end{gather}
In particular, we find the 32-dimensional subspace describing the tripartite entanglement of six qubits
\begin{gather}
\begin{split}\label{eq:decomp}
   &\mathbf{(2,2,1,2,1,1)} \\
+\ &\mathbf{(1,2,2,1,2,1)} \\
+\ &\mathbf{(1,1,2,2,1,2)} \\
+\ &\mathbf{(2,1,1,1,2,2)},
\end{split}
\shortintertext{or}
\begin{array}{c@{}c@{\ }c@{\lvert}*{6}{@{}c}@{\rangle}c}\label{eq:ourstate?}
\ket{\Psi}_{32} = &   & a_{ABD} & A       & B       & \bullet & D       & \bullet & \bullet \\
                  & + & b_{BCE} & \bullet & B       & C       & \bullet & E       & \bullet \\
                  & + & c_{CDF} & \bullet & \bullet & C       & D       & \bullet & F       \\
                  & + & e_{EFA} & A       & \bullet & \bullet & \bullet & E       & F & .
\end{array}
\end{gather}
and  the complementary 48-dimensional subspace describing the 4-way entanglement of six qubits
\begin{gather}
\begin{split}
   &\mathbf{(2,2,2,1,1,2)} \\
+\ &\mathbf{(2,1,2,2,2,1)} \\
+\ &\mathbf{(1,2,1,2,2,2)},
\end{split}
\shortintertext{or}
\begin{array}{c@{}c@{\ }c@{\lvert}*{6}{@{}c}@{}c@{}c}
\ket{\Psi}_{48} = &   & d_{FABC} & A       & B       & C       & \bullet & \bullet & F       & \rangle \\
                  & + & f_{ACDE} & A       & \bullet & C       & D       & E       & \bullet & \rangle \\
                  & + & g_{BDEF} & \bullet & B       & \bullet & D       & E       & F       & \rangle & .
\end{array}
\end{gather}
From \autoref{sec:sl2}, we may assign $\ket{\Psi}_{32}$ to the fundamental of $SO(12,\mathds{C})$ and  the complementary $\ket{\Psi}_{48}$ states to the coset $SO(12,\mathds{C})/[SL(2,\mathds{C})]^6$.

\subsubsection{\texorpdfstring{$E_7$ symmetry from seven qutrits}{E7 symmetry from seven qutrits}}

Under
\begin{gather}
\begin{split}
& SL(3)_{A} \times SL(3)_{B} \times SL(3)_{C} \times SL(3)_{D} \times SL(3)_{E} \times SL(3)_{F} \times SL(3)_{G} \\
\supset\ & SL(2)_{A} \times SL(2)_{B} \times SL(2)_{C} \times SL(2)_{D} \times SL(2)_{E} \times SL(2)_{F} \times SL(2)_{G},
\end{split}
\shortintertext{we have}
\begin{array}{c*{4}{@{\ }c}@{}c}
\mathbf{(3,3,3,3,3,3,3)} \to &   & 1  & \text{term\phantom{s} like} & \mathbf{(2,2,2,2,2,2,2)} \\
                             & + & 7  & \text{terms like}           & \mathbf{(2,2,2,2,2,2,1)} \\
                             & + & 21 & \text{terms like}           & \mathbf{(2,2,2,2,2,1,1)} \\
                             & + & 35 & \text{terms like}           & \mathbf{(2,2,2,2,1,1,1)} \\
                             & + & 35 & \text{terms like}           & \mathbf{(2,2,2,1,1,1,1)} \\
                             & + & 21 & \text{terms like}           & \mathbf{(2,2,1,1,1,1,1)} \\
                             & + & 7  & \text{terms like}           & \mathbf{(2,1,1,1,1,1,1)} \\
                             & + & 1  & \text{term\phantom{s} like} & \mathbf{(1,1,1,1,1,1,1)} & .
\end{array}
\end{gather}
In particular, we find the 56-dimensional subspace describing the tripartite entanglement of seven qubits
\begin{gather}
\begin{split}
   &\mathbf{(2,2,1,2,1,1,1)} \\
+\ &\mathbf{(1,2,2,1,2,1,1)} \\
+\ &\mathbf{(1,1,2,2,1,2,1)} \\
+\ &\mathbf{(1,1,1,2,2,1,2)} \\
+\ &\mathbf{(2,1,1,1,2,2,1)} \\
+\ &\mathbf{(1,2,1,1,1,2,2)} \\
+\ &\mathbf{(2,1,2,1,1,1,2)},
\end{split}
\shortintertext{or}
\begin{array}{c@{}c@{\ }c@{\lvert}*{7}{@{}c}@{\rangle}@{}c}
\ket{\Psi}_{56} = &   & a_{ABD} & A       & B       & \bullet & D       & \bullet & \bullet & \bullet \\
                  & + & b_{BCE} & \bullet & B       & C       & \bullet & E       & \bullet & \bullet \\
                  & + & c_{CDF} & \bullet & \bullet & C       & D       & \bullet & F       & \bullet \\
                  & + & d_{DEG} & \bullet & \bullet & \bullet & D       & E       & \bullet & G       \\
                  & + & e_{EFA} & A       & \bullet & \bullet & \bullet & E       & F       & \bullet \\
                  & + & f_{FGB} & \bullet & B       & \bullet & \bullet & \bullet & F       & G       \\
                  & + & g_{GAC} & A       & \bullet & C       & \bullet & \bullet & \bullet & G & ,
\end{array}
\end{gather}
and the complementary 112-dimensional subspace describing the 4-way entanglement of seven qubits, obtained by exchanging $\mathbf{2}$'s and $\mathbf{1}$'s:
\begin{gather}
\begin{split}
   &\mathbf{(1,1,2,1,2,2,2)} \\
+\ &\mathbf{(2,1,1,2,1,2,2)} \\
+\ &\mathbf{(2,2,1,1,2,1,2)} \\
+\ &\mathbf{(2,2,2,1,1,2,1)} \\
+\ &\mathbf{(1,2,2,2,1,1,2)} \\
+\ &\mathbf{(2,1,2,2,2,1,1)} \\
+\ &\mathbf{(1,2,1,2,2,2,1)},
\end{split}
\shortintertext{or}
\begin{array}{c@{}c@{\ }c@{\lvert}*{7}{@{}c}@{\rangle}@{}c}
\ket{\Psi}_{112} = &   & a_{CEFG} & \bullet & \bullet & C       & \bullet & E       & F       & G       \\
                   & + & b_{DFGA} & A       & \bullet & \bullet & D       & \bullet & F       & G       \\
                   & + & c_{EGAB} & A       & B       & \bullet & \bullet & E       & \bullet & G       \\
                   & + & d_{FABC} & A       & B       & C       & \bullet & \bullet & F       & \bullet \\
                   & + & e_{GBCD} & \bullet & B       & C       & D       & \bullet & \bullet & G       \\
                   & + & f_{ACDE} & A       & \bullet & C       & D       & E       & \bullet & \bullet \\
                   & + & g_{BDEF} & \bullet & B       & \bullet & D       & E       & F       & \bullet & .
\end{array}
\end{gather}
From \autoref{sec:sl2}, we may assign $\ket{\Psi}_{56}$ to the fundamental of $E_7(\mathds{C})$ and the complementary $\ket{\Psi}_{112}$ states to the coset $E_7(\mathds{C})/[SL(2,\mathds{C})]^7$.

\subsubsection{\texorpdfstring{$E_8$ symmetry from eight qutrits}{E8 symmetry from eight qutrits}}

Under
\begin{gather}
\begin{split}
& SL(3)_{A} \times SL(3)_{B} \times SL(3)_{C} \times SL(3)_{D} \times SL(3)_{E} \times SL(3)_{F} \times SL(3)_{G} \times SL(3)_{H} \\
\supset\ & SL(2)_{A} \times SL(2)_{B} \times SL(2)_{C} \times SL(2)_{D} \times SL(2)_{E} \times SL(2)_{F} \times SL(2)_{G} \times, SL(2)_{H}
\end{split}
\shortintertext{we have}
\begin{array}{c*{4}{@{\ }c}@{}c}
\mathbf{(3,3,3,3,3,3,3,3)} \to &   & 1  & \text{term\phantom{s} like} & \mathbf{(2,2,2,2,2,2,2,2)} \\
                               & + & 8  & \text{terms like}           & \mathbf{(2,2,2,2,2,2,2,1)} \\
                               & + & 28 & \text{terms like}           & \mathbf{(2,2,2,2,2,2,1,1)} \\
                               & + & 56 & \text{terms like}           & \mathbf{(2,2,2,2,2,1,1,1)} \\
                               & + & 70 & \text{terms like}           & \mathbf{(2,2,2,2,1,1,1,1)} \\
                               & + & 56 & \text{terms like}           & \mathbf{(2,2,2,1,1,1,1,1)} \\
                               & + & 28 & \text{terms like}           & \mathbf{(2,2,1,1,1,1,1,1)} \\
                               & + & 8  & \text{terms like}           & \mathbf{(2,1,1,1,1,1,1,1)} \\
                               & + & 1  & \text{term\phantom{s} like} & \mathbf{(1,1,1,1,1,1,1,1)} & .
\end{array}
\end{gather}
In particular, we find the 224-dimensional subspace describing the 4-way entanglement of eight qubits
\begin{gather}
\begin{split}
   &\mathbf{(2,2,2,1,2,1,1,1)+(1,1,1,2,1,2,2,2)} \\
+\ &\mathbf{(2,1,2,2,1,2,1,1)+(1,2,1,1,2,1,2,2)} \\
+\ &\mathbf{(2,1,1,2,2,1,2,1)+(1,2,2,1,1,2,1,2)} \\
+\ &\mathbf{(2,1,1,1,2,2,1,2)+(1,2,2,2,1,1,2,1)} \\
+\ &\mathbf{(2,2,1,1,1,2,2,1)+(1,1,2,2,2,1,1,2)} \\
+\ &\mathbf{(2,1,2,1,1,1,2,2)+(1,2,1,2,2,2,1,1)} \\
+\ &\mathbf{(2,2,1,2,1,1,1,2)+(1,1,2,1,2,2,2,1)},
\end{split}
\shortintertext{or}
\begin{array}{c@{}c@{\ }c@{\lvert}*{8}{@{}c}@{\rangle\ +\ }c@{\lvert}*{8}{@{}c}@{\rangle}@{}c}\label{eq:ourstste?}
\ket{\Psi}_{224} = &   & a_{HABD}         & H       & A       & B       & \bullet & D       & \bullet & \bullet & \bullet
                       & \tilde{a}_{CEFG} & \bullet & \bullet & \bullet & C       & \bullet & E       &       F & G       \\
                   & + & b_{HBCE}         & H       & \bullet & B       & C       & \bullet & E       & \bullet & \bullet
                       & \tilde{b}_{DFGA} & \bullet & A       & \bullet & \bullet & D       & \bullet &       F & G       \\
                   & + & c_{HCDF}         & H       & \bullet & \bullet & C       & D       & \bullet &       F & \bullet
                       & \tilde{c}_{EGAB} & \bullet & A       & B       & \bullet & \bullet & E       & \bullet & G       \\
                   & + & d_{HDEG}         & H       & \bullet & \bullet & \bullet & D       & E       & \bullet & G
                       & \tilde{d}_{FABC} & \bullet & A       & B       & C       & \bullet & \bullet &       F & \bullet \\
                   & + & e_{HEFA}         & H       & A       & \bullet & \bullet & \bullet & E       &       F & \bullet
                       & \tilde{e}_{GBCD} & \bullet & \bullet & B       & C       & D       & \bullet & \bullet & \bullet \\
                   & + & f_{HFGB}         & H       & \bullet & B       & \bullet & \bullet & \bullet &       F & G
                       & \tilde{f}_{ACDE} & \bullet & A       & \bullet & C       & D       & E       & \bullet & \bullet \\
                   & + & g_{HGAC}         & H       & A       & \bullet & C       & \bullet & \bullet & \bullet & G
                       & \tilde{g}_{BDEF} & \bullet & \bullet & B       & \bullet & D       & E       &       F & \bullet & .
\end{array}
\end{gather}
The 112 are associated with quadrangles of the Fano plane and the other 112 with the quadrangles of the dual Fano plane.  From \autoref{sec:sl2}, we may assign $\ket{\Psi}_{224}$ to the coset $E_8(\mathds{C})/[SL(2,\mathds{C})]^8$.

We have confined these qubit $G/H$ coset constructions, with $H=[SL(2,\mathds{C})]^n$ to the Appendix because we have as yet no good application of them within quantum information theory.

\subsection{\texorpdfstring{$SL(3)$ decompositions}{SL(3) decompositions}}
\label{sec:sl3}

Consider the groups $E_8$ and $E_6$ and their $SL(3,\mathds{C})^n$ subgroups as in \autoref{tab:CosetPropertiesSL(3)}
\begin{table}[ht]
\begin{tabular*}{\textwidth}{@{\extracolsep{\fill}}*{8}{M{c}}}
\toprule
& G   & fund & \dim G & H                  & \dim H & \dim G/H & \\
\midrule
& E_8 & 248  & 248    & SL(3,\mathds{C})^4 & 32     & 216      & \\
& E_6 & 27   & 78     & SL(3,\mathds{C})^3 & 24     & 54       & \\
\bottomrule
\end{tabular*}
\caption{Coset properties for $SL(3)$.}\label{tab:CosetPropertiesSL(3)}
\end{table}

The decomposition of their adjoint and fundamental representations are given as follows.
\begin{gather}
E_8 \supset SL(3)_A \times SL(3)_B \times SL(3)_C  \times SL(3)_D,
\shortintertext{under which}
\begin{array}{B{c}@{}c@{\ (}*{3}{B{l}*{3}{@{,\,}B{l}}@{)\ +\ (}}B{l}*{3}{@{,\,}B{l}}@{)}@{}c}
248 \to &   & 8 & 1 & 1 & 1   &  1 & 8 & 1 & 1  &  1 & 1 & 8 & 1   &  1 & 1 & 1 & 8 \\
        & + & 1 & 3 & 3 & 3'  &  1 & 3'& 3'& 3  &  3 & 3'& 3 & 1   &  3 & 3 & 1 & 3 \\
        & + & 3 & 1 & 3'& 3'  &  3'& 3 & 3'& 1  &  3'& 3'& 1 & 3'  &  3'& 1 & 3 & 3 & ,
\end{array}
\shortintertext{and}
E_6 \supset SL(3)_A \times SL(3)_B \times SL(3)_C,
\shortintertext{under which}
\begin{array}{B{c}@{}c@{\ (}B{l}*{2}{@{,\,}B{l}}@{)}@{\ +\ (}B{l}*{2}{@{,\,}B{l}}@{)}@{}c@{}c@{}@{\ +\ (}B{l}*{2}{@{,\,}B{l}}@{)}@{}c}
27 \to &   & 3'& 3 & 1  &  3 & 1 & 3 &&  &  1 & 3'& 3' \\
78 \to &   & 8 & 1 & 1  &  1 & 8 & 1 &&  &  1 & 1 & 8  \\
       & + & 3 & 3 & 3' &  3'& 3'& 3 &.
\end{array}
\end{gather}

\subsection{7-dit interpretation}

\subsubsection{\texorpdfstring{$E_6$ symmetry from three 7-dits}{E6 symmetry from three 7-dits}}

Under
\begin{gather}
SL(7)_{A} \times SL(7)_{B}  \times SL(7)_{C} \to SL(3)_{A} \times SL(3)_{B}  \times SL(3)_{C},
\shortintertext{we have}
\begin{array}{B{c}@{}c@{\ (}*{2}{B{l}*{2}{@{,\,}B{l}}@{)\ +\ (}}B{l}*{2}{@{,\,}B{l}}@{)}@{}c}
(7,7,7) \to &   & 3'& 3'& 3' &  3 & 3 & 3  &  1 & 1 & 1 \\
            & + & 3'& 3'& 3  &  3'& 3 & 3' &  3 & 3'& 3'\\
            & + & 3'& 3 & 3  &  3 & 3'& 3  &  3 & 3 & 3'\\
            & + & 3'& 3'& 1  &  3'& 1 & 3' &  1 & 3'& 3'\\
            & + & 3 & 3 & 1  &  3 & 1 & 3  &  1 & 3 & 3 \\
            & + & 3'& 1 & 1  &  1 & 3'& 1  &  1 & 1 & 3'\\
            & + & 3 & 1 & 1  &  1 & 3 & 1  &  1 & 1 & 3 \\[5pt]
            & + & 3'& 1 & 3  &  3'& 3 & 1  &  1 & 3 & 3'\\
            & + & 3 & 1 & 3' &  3 & 3'& 1  &  1 & 3'& 3 & .
\end{array}
\end{gather}
In particular we find the 27-dimensional subspace  describing the bipartite entanglement of three qutrits, namely
\begin{gather}
\mathbf{(3',3,1)+(3,1,3)+(1,3',3')},
\shortintertext{or}
\ket{\Psi}_{27} = a_{A'B}\ket{A'B} +b_{BC}\ket{BC} +c_{C'A'}\ket{C'A'},
\end{gather}
and the 54-dimensional subspace describing the tripartite entanglement of three qutrits
\begin{gather}
\mathbf{(3,3,3') + (3',3',3)},
\shortintertext{or}
\ket{\Psi}_{54} = a_{ABC'}\ket{ABC'} +b_{A'B'C}\ket{A'B'C}.
\end{gather}
From \autoref{sec:sl3}, we may assign the $\ket{\Psi}_{27}$ to the fundamental of $E_6(\mathds{C})$ and the $\ket{\Psi}_{54}$ to the coset $E_6(\mathds{C})/[SL(3,\mathds{C})]^3$

\subsubsection{\texorpdfstring{$E_8$ symmetry from four 7-dits}{E8 symmetry from four 7-dits}}
\label{sec:7dits}

Under
\begin{gather}
SL(7)_{A} \times SL(7)_{B}  \times SL(7)_{C}  \times SL(7)_{D}\to SL(3)_{A} \times SL(3)_{B}  \times SL(3)_{C} \times SL(3)_{D},
\shortintertext{we have}
\begin{gathered}
\mathbf{(7,7,7,7)} \to \\
\begin{array}{c@{\ (}B{l}*{3}{@{,\,}B{l}}@{)} *{2}{@{\ +\ (}B{l}*{3}{@{,\,}B{l}}@{)}} @{}c@{} @{\ +\ (}B{l}*{3}{@{,\,}B{l}}@{)} @{}c@{} *{2}{@{\ +\ (}B{l}*{3}{@{,\,}B{l}}@{)}}@{}c}
  & 3'& 3'& 3'& 3' &  3 & 3 & 3 & 3  &  1 & 1 & 1 & 1 \\
+ & 3 & 3'& 3'& 3' &  3'& 3 & 3'& 3' &  3'& 3'& 3 & 3' & &  3'& 3'& 3'& 3 \\
+ & 3'& 3 & 3 & 3  &  3 & 3'& 3 & 3  &  3 & 3 & 3'& 3  & &  3 & 3 & 3 & 3'\\
+ & 1 & 3'& 3'& 3' &  3'& 1 & 3'& 3' &  3'& 3'& 1 & 3' & &  3'& 3'& 3'& 1 \\
+ & 1 & 3 & 3 & 3  &  3 & 1 & 3 & 3  &  3 & 3 & 1 & 3  & &  3 & 3 & 3 & 1 \\
+ & 3'& 1 & 1 & 1  &  1 & 3'& 1 & 1  &  1 & 1 & 3'& 1  & &  1 & 1 & 1 & 3'\\
+ & 3 & 1 & 1 & 1  &  1 & 3 & 1 & 1  &  1 & 1 & 3 & 1  & &  1 & 1 & 1 & 3 \\
+ & 3'& 3'& 3 & 3  &  3 & 3'& 3'& 3  &  3 & 3 & 3'& 3' & &  3'& 3 & 3 & 3' & &  3'& 3 & 3'& 3  &  3 & 3'& 3 & 3'\\
+ & 1 & 1 & 3'& 3' &  3'& 1 & 1 & 3' &  3'& 3'& 1 & 1  & &  1 & 3'& 3'& 1  & &  1 & 3'& 1 & 3' &  3'& 1 & 3'& 1 \\
+ & 1 & 1 & 3 & 3  &  3 & 1 & 1 & 3  &  3 & 3 & 1 & 1  & &  1 & 3 & 3 & 1  & &  1 & 3 & 1 & 3  &  3 & 1 & 3 & 1 \\[5pt]
+ & 1 & 3'& 3 & 3  &  3 & 1 & 3'& 3  &  3 & 3 & 1 & 3' & &  3'& 3 & 3 & 1  & &  1 & 3 & 3'& 3  &  3 & 1 & 3 & 3'\\
+ & 3'& 3 & 1 & 3  &  3 & 3'& 3 & 1  &  1 & 3 & 3 & 3' & &  3'& 1 & 3 & 3  & &  3 & 3'& 1 & 3  &  3 & 3 & 3'& 1 \\[5pt]
+ & 3 & 3'& 1 & 1  &  1 & 3 & 3'& 1  &  1 & 1 & 3 & 3' & &  3'& 1 & 1 & 3  & &  3 & 1 & 3'& 1  &  1 & 3 & 1 & 3'\\
+ & 3'& 1 & 3 & 1  &  1 & 3'& 1 & 3  &  3 & 1 & 1 & 3' & &  3'& 3 & 1 & 1  & &  1 & 3'& 3 & 1  &  1 & 1 & 3'& 3 \\[5pt]
+ & 1 & 3 & 3'& 3' &  3'& 1 & 3 & 3' &  3'& 3'& 1 & 3  & &  3 & 3'& 3'& 1  & &  1 & 3'& 3 & 3' &  3'& 1 & 3'& 3 \\
+ & 3 & 3'& 1 & 3' &  3'& 3 & 3'& 1  &  1 & 3'& 3'& 3  & &  3 & 1 & 3'& 3' & &  3'& 3 & 1 & 3' &  3'& 3'& 3 & 1 & .
\end{array}
\end{gathered}
\end{gather}
In particular we find the 216-dimensional subspace describing the tripartite entanglement of four qutrits, namely
\begin{gather}
\begin{array}{c@{\ (}*{3}{B{l}*{3}{@{,\,}B{l}}@{)\ +\ (}}B{l}*{3}{@{,\,}B{l}}@{)}@{}c}
  & 1 & 3 & 3 & 3' &  1 & 3'& 3'& 3  &  3 & 3'& 3 & 1  &  3 & 3 & 1 & 3\\
+ & 3 & 1 & 3'& 3' &  3'& 3 & 3'& 1  &  3'& 3'& 1 & 3' &  3'& 1 & 3 & 3 & ,
\end{array}
\shortintertext{or}
\begin{array}{c@{}c@{\ }c@{\lvert}*{4}{@{}c}@{\rangle}*{3}{@{\ +\ }c@{\lvert}*{4}{@{}c}@{\rangle}}@{}c}
\ket{\Psi}_{216} = &   & a_{\bullet BCD'}  & \bullet & B & C & D'  & b_{\bullet B'C'D'} & \bullet & B' & C' & D' & c_{AB'C\bullet}   & A & B' & C & \bullet   & d_{AB*D}  & A & B & \bullet & D \\
                   & + & e_{A\bullet C'D'} & A & \bullet & C' & D' & f_{A'BC'\bullet}  & A' & B & C' & \bullet  & g_{A'B'\bullet D'} & A' & B' & \bullet & D' & h_{A'\bullet CD} & A' & \bullet & C & D & .
\end{array}
\end{gather}
From \autoref{sec:sl3}, we may assign $\ket{\Psi}_{216} $ to the coset $E_8(\mathds{C})/[SL(3,\mathds{C})]^4$.

Again we have confined these qutrit $G/H$ coset constructions, with $H=[SL(3,\mathds{C})] ^n$ to the Appendix because we have as yet no good application of them within quantum information theory.

\section{\texorpdfstring{DISCRETE SYMMETRY OF THE FANO PLANE}{Discrete Symmetry of the Fano Plane}}
\label{sec:DiscreteSymmFano}

It ought to be clear by now that the Fano plane plays a central role in the seven qubit interpretation of $\mathcal{N}=8$ black holes. It is therefore natural ask how the symmetries of the Fano plane are manifested in the 56 dimensional seven qubit state.

\subsection{\texorpdfstring{Projective geometry, the Fano plane and $PSL(2,\mathds{F}_7)$}{Projective geometry, the Fano plane and PSL(2, F7)}}

Let $V_0=V(n+1,\mathds{F})/\{0\}$ be a $n+1$-dimensional vector space, defined over a field $\mathds{F}$, with the additive identity 0 removed. Note, $\mathds{F}$ may be a finite field with characteristic $q$,  in which case we denote it by  $\mathds{F}_q$. The $n$-dimensional projective space over $\mathds{F}$, which we write as $PG(n,\mathds{F})$, is the space of equivalence classes defined by the relation, $x\sim y$ iff $x=\alpha y$, where $\alpha \in \mathds{F}/\{0\}$ and $x, y \in V_0$. The set of projectivities\footnote{A projectivity is a linear bijection $PG(n,\mathds{F})\to PG(n,\mathds{F})$ that preserves incidence.} of $PG(n,\mathds{F}_q)$ is the \emph{projective general linear group} $PGL(n+1, \mathds{F}_q)$, i.e. the group of non-singular linear transformations on $V_0$ up to an overall multiplicative factor (see for example, \cite{Hirschfeld:1998}).

The Fano plane is the projective plane over the finite field of order two, $PG(2,\mathds{F}_2)$. It is the smallest example of a projective plane. In this case the projective general linear group, $PGL(3, \mathds{F}_2)$, is isomorphic to the projective \emph{special} linear group $PSL(3, \mathds{F}_2)$\footnote{More generally, $PGL(n, \mathds{F}_q)\simeq PSL(n, \mathds{F}_q)$ iff $\gcd(n+1,q)=1$.}, the set of determinant one projectivities. In fact, in this particular instance,  we have a second useful isomorphism, $PSL(3, \mathds{F}_2)\simeq PSL(2,\mathds{F}_7)$. $PSL(2,\mathds{F}_7)$ is second smallest finite non-abelian simple group, after the alternating group $\mathcal{A}_5$, with 168 elements. It has many guises, but, perhaps most significantly, it is the automorphism group of the Klein quartic. Further, it is the only finite simple subgroup of $SU(3)$ and consequently, in light of the recently measured neutrino mixing patterns, it has been receiving increasing attention as a candidate finite non-abelian flavour group \cite{Luhn:2007yr, Luhn:2008sa}.

\subsection{\texorpdfstring{The 56-dimensional representation }{The 56-dimensional representation)}}

$PSL(2, \mathds{F}_7)$ admits a two generator presentation \cite{Conway:1985,Luhn:2008sa},
\begin{equation}\label{eq:psl27pres}
\langle ~s, t ~| ~s^2=t^3=(st)^7=[s,t]^4=e ~\rangle,
\end{equation}
where the commutator, $[s,t]$, is defined as $s^{-1}t^{-1}st$.

It has six conjugacy classes and, therefore six irreps, as summarised in \autoref{tab:psl27class}.
\begin{table}[ht]
\caption[Character table for $PSL(2,\mathds{F}_7)$]{Character table for $PSL(2,\mathds{F}_7)$. The number in square brackets on each conjugacy class $C_\alpha$ corresponds to the order of the elements in that class. A simple representative for each class is given in the parentheses. Note that geometrical interpretation of the $\mathbf{6}$ is given by its action on the Fano plane \cite{Luhn:2007yr}. The final row corresponds to the compound characters of the reducible 56 dimensional representation described here.}\label{tab:psl27class}
\begin{tabular*}{\textwidth}{@{\extracolsep{\fill}}*{9}{M{c}}}
\toprule
& & C^{[1]}_1 & 21C^{[2]}_2(s) & 56C^{[3]}_3(t) & 42C^{[4]}_4([s,t]) & 24C^{[7]}_5(st) & 24C^{[7]}_6(st^2) & \\
\midrule
& \chi^{[\mathbf{1}]}       & 1  & 1  & 1  &  1 & 1                   & 1                   & \\
& \chi^{[\mathbf{3}]}       & 3  & -1 & 0  &  1 & \half(-1+i\sqrt{7}) & \half(-1-i\sqrt{7}) & \\
& \chi^{[\bar{\mathbf{3}}]} & 3  & -1 & 0  &  1 & \half(-1-i\sqrt{7}) & \half(-1+i\sqrt{7}) & \\
& \chi^{[\mathbf{6}]}       & 6  & 2  & 0  &  0 & -1                  & -1                  & \\
& \chi^{[\mathbf{7}]}       & 7  & -1 & 1  & -1 & 0                   & 0                   & \\
& \chi^{[\mathbf{8}]}       & 8  & 0  & -1 &  0 & 1                   & 1                   & \\
& \chi^{[\mathbf{56}]}      & 56 & 8  & 2  &  0 & 0                   & 0                   & \\
\bottomrule
\end{tabular*}
\end{table}
It has a convenient action defined on the Fano plane given by the permutation of its points. For example, we may consider the permutations,
\begin{equation}
s_{\text{fano}} = (AC)(B)(DE)(F)(G), \qquad t_{\text{fano}}=(ADB)(C)(EFG),
\end{equation}
which are automorphisms of the un-oriented Fano plane. This yields a  7 dimensional real representation
for which it is easily verified that \eqref{eq:psl27pres} is satisfied. This representation is reducible,
\begin{equation}
\mathbf{7}^{\text{fano}} \to \mathbf{1 + 6},
\end{equation}
as can be checked from its characters. The Fano planes initial overall labelling is associated with the singlet \cite{Luhn:2007yr}.

These permutations also interchange the lines of the Fano plane and, consequently, may be considered as permutations of the points of the \emph{dual} Fano plane,
\begin{equation}
s_{\text{dualfano}} = (ab)(ce)(d)(f)(g), \qquad t_{\text{dualfano}}=(a)(bcg)(efd).
\end{equation}

This Fano plane representation may be used to build a 56-dimensional representation of the $PSL(2, \mathds{F}_7)$ generators, denoted $s_{56}$ and $t_{56}$, acting on our particular seven-qubit state,
\begin{equation}\label{eq:psi1}
\begin{split}
\ket{\Psi} &=\phantom{+\!}a_{ABD}\ket{ABD}+b_{BCE}\ket{BCE}+c_{CDF}\ket{CDF}+d_{DEG}\ket{DEG}\\
           &\phantom{=}+ e_{EFA}\ket{EFA}+f_{FGB}\ket{FGB}+g_{GAC}\ket{GAC},
\end{split}
\end{equation}
which leaves quartic entanglement measure, $I_4$, invariant.

Begin by considering the action on both the Fano and dual Fano planes together,
\begin{equation}\label{eq:fdfaction}
\begin{array}{c@{\ \xmapsto{s}\ }c@{,\qquad}c@{\ \xmapsto{t}\ }c@{,}}
a_{ABD} & b_{CBE} & a_{ABD} & a_{BDA} \\
b_{BCE} & a_{BAD} & b_{BCE} & c_{DCF} \\
c_{CDF} & e_{AEF} & c_{CDF} & g_{CAG} \\
d_{DEG} & d_{EDG} & d_{DEG} & e_{AFE} \\
e_{EFA} & c_{DFC} & e_{EFA} & f_{FGB} \\
f_{FGB} & f_{FGB} & f_{FGB} & d_{GED} \\
g_{GAC} & g_{GCA} & g_{GAC} & b_{EBC}
\end{array}
\end{equation}
where $s_{\text{fano}}$ and $s_{\text{dualfano}}$ ($t_{\text{fano}}$ and $t_{\text{dualfano}}$) respectively permute the upper and lower case letters (points and lines) in a consistent manner. Note, however, that the points on each line are permuted away from their original ordering, for example, in \eqref{eq:fdfaction} we find $a_{ABD}\mapsto b_{CBE}$ as opposed to $b_{BCE}$. There is, in addition, a corresponding action on the quadrangles of the dual Fano plane. Label each quadrangle by its excluded qubit as in \autoref{tab:quadlab}. Using this labelling and the composition rule,
\begin{equation}\label{eq:quadcomp}
(bcdf)\cdot (cdeg)=(efgb),
\end{equation}
where the common letters are ``contracted'' over, reproduces the Fano plane. Consequently, the quadrangles transform into each under $s_\textrm{fano}$ and $t_\textrm{fano}$ or, equivalently, under $s_\textrm{dualfano}$ and $t_\textrm{dualfano}$. However, the ordering of the four points within each quadrangle is not left invariant. For example, $s_\textrm{dualfano}:cdeb \mapsto edcg$, which is an odd permutation of the original ordering $cdeg$. This is the case whenever the quadrangle contains $g$ due to the fact that the permutation $s_\textrm{dualfano}$ is defined by the quadrangle, $abce$, excluding George. Similarly, for $t_\textrm{dualfano}$ the points of each quadrangle containing $c$ receive an odd permutation since $t_\textrm{dualfano}$ is associated with the  quadrangle, $defa$, excluding Charlie.
\begin{table}[ht]
\begin{tabular*}{\textwidth}{@{\extracolsep{\fill}}*{9}{M{c}}}
\toprule
& A    & B    & C    & D    & E    & F    & G    & \\
\midrule
& bcdf & cdeg & defa & efgb & fgac & gabd & abce & \\
\bottomrule
\end{tabular*}
\caption[Quadrangle labellings]{The seven quadrangles, identified with points as described here, form an equivalent representation of the Fano plane using the composition rule \eqref{eq:quadcomp}.}\label{tab:quadlab}
\end{table}

In defining a 56-dimensional action leaving $I_4$ invariant one must account not only for the (dual) Fano plane permutations but also the ordering of the points on each line and quadrangle. This is achieved by a composite transformation, first performing the (dual) Fano plane permutations as in \eqref{eq:fdfaction}, then reordering the points on each line to their standard form. Finally, to account for the quadrangles, we must bit flip each qubit not appearing on the line given by the quadrangle associated with the dual Fano plane permutation. For example, the permutation $s_\textrm{dualfano}$  is defined by the quadrangle, $abce$, excluding George and, hence, we bit flip the qubits $B,D,E,$ and $F$, that is those qubits not lying on the line $g_{GAC}$. Similarly, for $t_\textrm{dualfano}$ the relevant quadrangle is given by excluding Charlie and so we bit flip the qubits $A,B,E,$ and $G$, i.e. those not included in the line $c_{CDF}$.

\begin{figure}[ht]
 \centering
 \includegraphics[width=\textwidth]{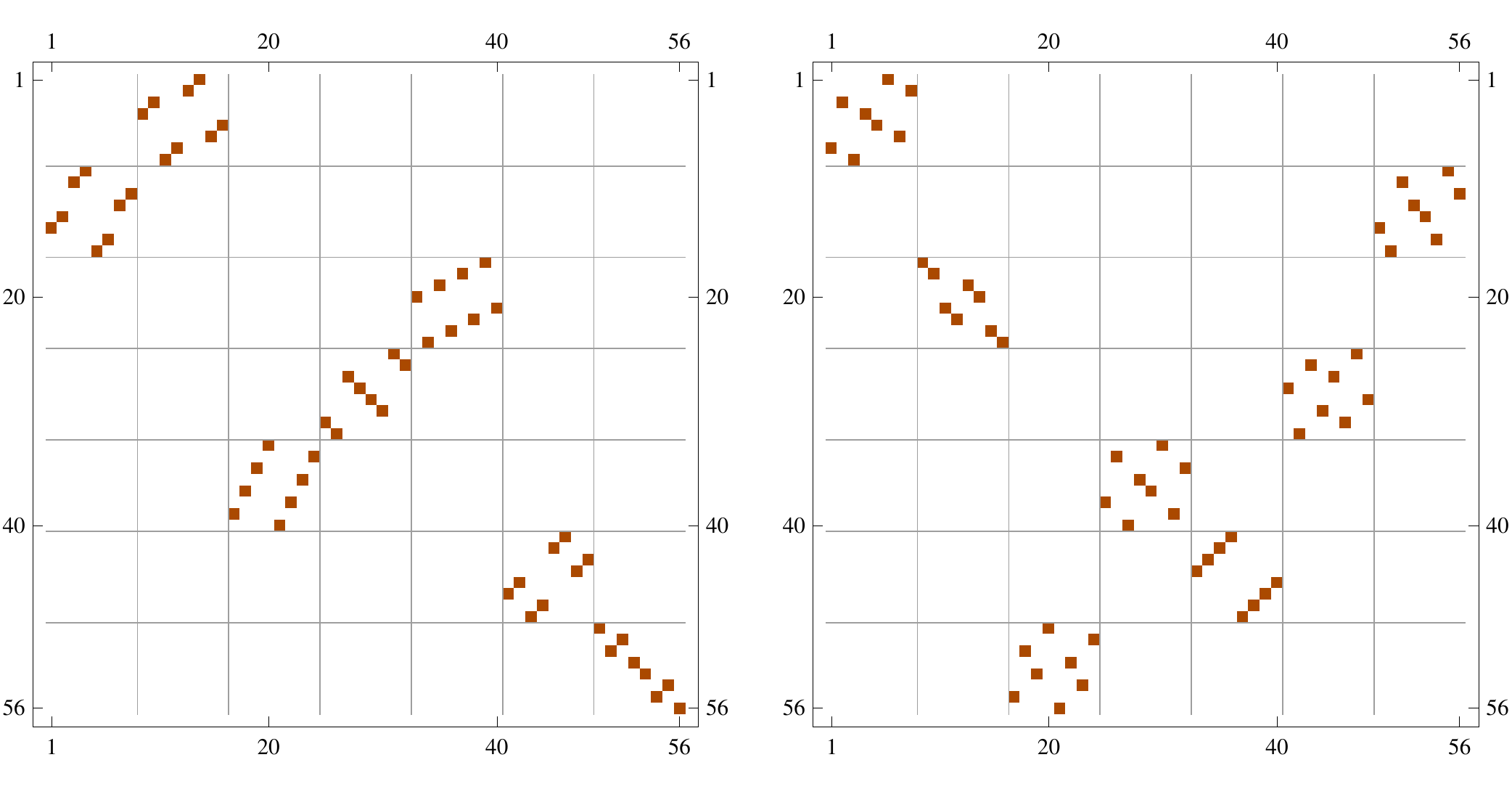}
 \caption[Generators of 56 dimensional rep of $PSL(2,\mathds{F}_7)$]{Graphical representation of the generators of the 56 dimensional representation of $PSL(2,\mathds{F}_7)$. These are the $56\times 56$ matrices $s_{56}$ (left) and $t_{56}$ (right). The gridlines partition the matrices into $8\times 8$ blocks that transform a single letter, while the filled squares correspond to the non-zero unit entries. Clearly each letter octet is transformed into another, without mixing between octets. One can read off, for example, that $s_{56}$ converts $a$'s to $b$'s and \textit{vice versa}, while $t_{56}$ transforms the $a$'s amongst themselves.}
 \label{fig:56psl27}
\end{figure}

As an explicit example let us consider the action of $s_{56}$ on $a_{1_{A}0_{B}1_{D}}$. Note, we have labelled the indices to keep track of orderings. The composite transformation is given by,
\begin{equation}
s_{56}: a_{1_{A}0_{B}1_{D}}\xmapsto{\text{relabel}} b_{1_{C}0_{B}1_{E}}\xmapsto{\text{permute}} b_{0_{B}1_{C}1_{E}}\xmapsto{\text{bit flip} BE} b_{1_{B}1_{C}0_{E}},
\end{equation}
where the relabelling is done according to the $s_{\text{fano}}$ and $s_{\text{dualfano}}$ permutations.
We bit flip $B$ and $E$ (but not $C$) as they do not appear in the line $GAC$. Therefore the complete transformation, for this example, is $a_{101}\mapsto b_{110}$.
All in all this yields a 56 dimensional real representation of $PSL(2,\mathds{F}_7)$, see \autoref{fig:56psl27}.

\subsection{\texorpdfstring{Decomposition of the 56-dimensional representation}{Decomposition of the 56-dimensional representation}}

$PSL(2,\mathds{F}_7)$ has six classes and therefore six irreps, as summarised in \autoref{tab:psl27class} \cite{Luhn:2007yr}. Note, the final row corresponds to the compound characters of the \textbf{56} described here. This 56 dimensional representation is clearly reducible and, to determine how it decomposes, we may use,
\begin{equation}
a_\mu = \frac{1}{g}\sum_\alpha g_\alpha \chi_{\alpha}^{[\mu]*}\chi_{\alpha}^{[\mathbf{56}]},
\end{equation}
where $a_\mu$ counts the number of times irrep $\mu$ appears in the $\mathbf{56}$. Note, $g$ and $g_\alpha$ are the dimension of the group and conjugacy class, $C_\alpha$, respectively. Using \autoref{tab:psl27class} one finds,
\begin{equation}
a_\mu =\{2,0,0,4,2,2\},
\end{equation}
and, hence, we have
\begin{equation}
\mathbf{56=1+1+6+6+6+6+7+7+8+8}.
\end{equation}
This is consistent with the breaking of the fundamental $\mathbf{56}$ of $E_7$ under  $PSL(2,\mathds{F}_7)$, which goes as,
\begin{equation}
\mathbf{56\to1+1+6+6+6+6+7+7+8+8}.
\end{equation}
In \cite{Levay:2008mi} the $PSL(2,\mathds{F}_7)$ symmetry of the $\mathcal{N}=8$ black hole entropy has been related, via a special set of 63 three-qubit operators, to the generalised hexagon of order two using the dictionary constructed in \autoref{sec:cyclic}. They further suggested that the full $G_2(2)$ symmetry of the hexagon may be preserved by the black hole entropy.

\newpage
\phantomsection
\addcontentsline{toc}{section}{References}

\begin{thebibliography}{10%
0}

\bibitem{Bekenstein:1973ur}
J.~D. Bekenstein, ``{Black holes and entropy},''
\href{http://dx.doi.org/10.1103/PhysRevD.7.2333}{{\em Phys. Rev.} {\bf D7}
  (1973) no.~8, 2333--2346}.

\bibitem{Hawking:1974sw}
S.~W. Hawking, ``{Particle creation by black holes},''
\href{http://dx.doi.org/10.1007/BF02345020}{{\em Commun. Math. Phys.} {\bf 43}
  (1975)  199--220}.

\bibitem{Duff:2006uz}
M.~J. Duff, ``{String triality, black hole entropy and Cayley's
  hyperdeterminant},'' \href{http://dx.doi.org/10.1103/PhysRevD.76.025017}{{\em
  Phys. Rev.} {\bf D76} (2007) no.~2, 025017},
\href{http://arxiv.org/abs/hep-th/0601134}{{\tt arXiv:hep-th/0601134}}.

\bibitem{Kallosh:2006zs}
R.~Kallosh and A.~Linde, ``{Strings, black holes, and quantum information},''
  \href{http://dx.doi.org/10.1103/PhysRevD.73.104033}{{\em Phys. Rev.} {\bf
  D73} (2006) no.~10, 104033},
\href{http://arxiv.org/abs/hep-th/0602061}{{\tt arXiv:hep-th/0602061}}.

\bibitem{Levay:2006kf}
P.~L\'evay, ``{Stringy black holes and the geometry of entanglement},''
  \href{http://dx.doi.org/10.1103/PhysRevD.74.024030}{{\em Phys. Rev.} {\bf
  D74} (2006) no.~2, 024030},
\href{http://arxiv.org/abs/hep-th/0603136}{{\tt arXiv:hep-th/0603136}}.

\bibitem{Duff:2006ue}
M.~J. Duff and S.~Ferrara, ``{$E_7$ and the tripartite entanglement of seven
  qubits},'' \href{http://dx.doi.org/10.1103/PhysRevD.76.025018}{{\em Phys.
  Rev.} {\bf D76} (2007) no.~2, 025018},
\href{http://arxiv.org/abs/quant-ph/0609227}{{\tt arXiv:quant-ph/0609227}}.

\bibitem{Levay:2006pt}
P.~L\'evay, ``{Strings, black holes, the tripartite entanglement of seven
  qubits and the Fano plane},''
  \href{http://dx.doi.org/10.1103/PhysRevD.75.024024}{{\em Phys. Rev.} {\bf
  D75} (2007) no.~2, 024024},
\href{http://arxiv.org/abs/hep-th/0610314}{{\tt arXiv:hep-th/0610314}}.

\bibitem{Duff:2007wa}
M.~J. Duff and S.~Ferrara, ``{$E_6$ and the bipartite entanglement of three
  qutrits},'' \href{http://dx.doi.org/10.1103/PhysRevD.76.124023}{{\em Phys.
  Rev.} {\bf D76} (2007) no.~12, 124023},
\href{http://arxiv.org/abs/0704.0507}{{\tt arXiv:0704.0507 [hep-th]}}.

\bibitem{Levay:2007nm}
P.~L\'evay, ``{A three-qubit interpretation of BPS and non-BPS $STU$ black
  holes},'' \href{http://dx.doi.org/10.1103/PhysRevD.76.106011}{{\em Phys.
  Rev.} {\bf D76} (2007) no.~10, 106011},
\href{http://arxiv.org/abs/0708.2799}{{\tt arXiv:0708.2799 [hep-th]}}.

\bibitem{Bellucci:2007zi}
S.~Bellucci, A.~Marrani, E.~Orazi, and A.~Shcherbakov, ``{Attractors with
  vanishing central charge},''
  \href{http://dx.doi.org/10.1016/j.physletb.2007.08.079}{{\em Phys. Lett.}
  {\bf B655} (2007)  185--195},
\href{http://arxiv.org/abs/0707.2730}{{\tt arXiv:0707.2730 [hep-th]}}.

\bibitem{Borsten:2008ur}
L.~Borsten, D.~Dahanayake, M.~J. Duff, W.~Rubens, and H.~Ebrahim, ``{Wrapped
  branes as qubits},''
  \href{http://dx.doi.org/10.1103/PhysRevLett.100.251602}{{\em Phys. Rev.
  Lett.} {\bf 100} (2008) no.~25, 251602},
\href{http://arxiv.org/abs/0802.0840}{{\tt arXiv:0802.0840 [hep-th]}}.

\bibitem{Ferrara:2008hw}
S.~Ferrara, K.~Hayakawa, and A.~Marrani, ``{Lectures on attractors and black
  holes},''
\href{http://dx.doi.org/10.1002/prop.200810569}{{\em Fortschr. Phys.} {\bf 56}
  (2008)  993--1046}.

\bibitem{Borsten:2008}
L.~Borsten, ``{$E_{7(7)}$ invariant measures of entanglement},''
  \href{http://dx.doi.org/10.1002/prop.200810542}{{\em Fortschr. Phys.} {\bf
  56} (2008) no.~7--9, 842--848}.

\bibitem{levay-2008}
P.~L\'{e}vay and P.~Vrana, ``Three fermions with six single-particle states can
  be entangled in two inequivalent ways,''
  \href{http://dx.doi.org/10.1103/PhysRevA.78.022329}{{\em Phys. Rev.} {\bf
  A78} (2008) no.~2, 022329}, \href{http://arxiv.org/abs/0806.4076}{{\tt
  arXiv:0806.4076 [quant-ph]}}.

\bibitem{Bellucci:2008sv}
S.~Bellucci, S.~Ferrara, A.~Marrani, and A.~Yeranyan, ``{$STU$ black holes
  unveiled},'' \href{http://dx.doi.org/10.3390/e10040507}{{\em Entropy} {\bf
  10} (2008) no.~4, 507--555},
\href{http://arxiv.org/abs/0807.3503}{{\tt arXiv:0807.3503 [hep-th]}}.

\bibitem{Levay:2008mi}
P.~L\'{e}vay, M.~Saniga, and P.~Vrana, ``{Three-qubit operators, the split
  Cayley hexagon of order two and black holes},''
  \href{http://dx.doi.org/10.1103/PhysRevD.78.124022}{{\em Phys. Rev.} {\bf
  D78} (2008) no.~12, 124022},
\href{http://arxiv.org/abs/0808.3849}{{\tt arXiv:0808.3849 [quant-ph]}}.

\bibitem{Borsten:2008fts}
L.~Borsten, D.~Dahanayake, M.~J. Duff, H.~Ebrahim, and W.~Rubens,
  ``{Freudenthal triple classificiation of three-qubit entanglement},''
  \href{http://arxiv.org/abs/0812.3322}{{\tt arXiv:0812.3322 [quant-ph]}}.

\bibitem{Coffman:1999jd}
V.~Coffman, J.~Kundu, and W.~K. Wootters, ``{Distributed entanglement},''
  \href{http://dx.doi.org/10.1103/PhysRevA.61.052306}{{\em Phys. Rev.} {\bf
  A61} (2000) no.~5, 052306},
\href{http://arxiv.org/abs/quant-ph/9907047}{{\tt arXiv:quant-ph/9907047}}.

\bibitem{Duff:1995sm}
M.~J. Duff, J.~T. Liu, and J.~Rahmfeld, ``{Four-dimensional
  string-string-string triality},''
  \href{http://dx.doi.org/10.1016/0550-3213(95)00555-2}{{\em Nucl. Phys.} {\bf
  B459} (1996)  125--159},
\href{http://arxiv.org/abs/hep-th/9508094}{{\tt arXiv:hep-th/9508094}}.

\bibitem{Behrndt:1996hu}
K.~Behrndt, R.~Kallosh, J.~Rahmfeld, M.~Shmakova, and W.~K. Wong, ``{$STU$
  black holes and string triality},''
  \href{http://dx.doi.org/10.1103/PhysRevD.54.6293}{{\em Phys. Rev.} {\bf D54}
  (1996) no.~10, 6293--6301},
\href{http://arxiv.org/abs/hep-th/9608059}{{\tt arXiv:hep-th/9608059}}.

\bibitem{Miyake:2002}
A.~Miyake and M.~Wadati, ``{Multipartite entanglement and hyperdeterminants},''
  {\em Quant. Info. Comp.} {\bf 2 (Special)} (2002)  540--555,
  \href{http://arxiv.org/abs/quant-ph/0212146}{{\tt arXiv:quant-ph/0212146}}.

\bibitem{Cayley:1845}
A.~Cayley, ``{On the theory of linear transformations}.''
  \href{http://www.archive.org/download/collmathpapers01caylrich/collmathpaper%
s01caylrich.pdf}{\textit{Camb. Math. J.} \textbf{4} (1845) 193--209}.

\bibitem{Pegg}
{E.~Pegg~Jr}, ``The fano plane.'' MAA online,
  \url{http://www.maa.org/editorial/mathgames/mathgames_05_30_06.html}.

\bibitem{Penrose:2005}
R.~Penrose, {\em The Road to Reality: A Complete Guide to the Laws of the
  Universe}.
\newblock Knopf, 2005.

\bibitem{Streater:2007}
R.~F. Streater, {\em Lost Causes in and beyond Physics}.
\newblock Springer, first~ed., 2007.

\bibitem{Duff:2006rf}
M.~J. Duff and S.~Ferrara, ``{Black hole entropy and quantum information},''
  \href{http://arxiv.org/abs/hep-th/0612036}{{\tt arXiv:hep-th/0612036}}.
Published in \href{http://dx.doi.org/10.1007/978-3-540-79523-0}{\textit{Lect.
  Notes Phys.} \textbf{755} (2008) 93-114} and
  \href{http://dx.doi.org/10.1007/978-3-540-79523-0_2}{\textit{Supersymmetric
  Mechanics - Vol. 3}}, pp. 1-22, Springer Berlin / Heidelberg, 2008.

\bibitem{Gelfand:1994}
I.~M. Gelfand, M.~M. Kapranov, and A.~V. Zelevinsky, {\em {Discriminants,
  Resultants and Multidimensional Determinants}}.
\newblock Birkh\"{a}user, Boston, 1994.

\bibitem{Manivel:2005}
L.~Manivel, ``{Configurations of lines and models of Lie algebras},''
  \href{http://dx.doi.org/10.1016/j.jalgebra.2006.04.029}{{\em J. Algebra} {\bf
  304} (2006) no.~1, 457--486}, \href{http://arxiv.org/abs/math/0507118}{{\tt
  arXiv:math/0507118}}.

\bibitem{Elduque:2005}
{Alberto Elduque}, ``{The magic square and symmetric compositions II},'' {\em
  Rev. Mat. Iberoamericana} {\bf 23} (2007) no.~1, 57--84,
  \href{http://arxiv.org/abs/math/0507282}{{\tt arXiv:math/0507282}}.

\bibitem{Ferrara:1997uz}
S.~Ferrara and M.~G{\"u}naydin, ``{Orbits of exceptional groups, duality and
  BPS states in string theory},''
  \href{http://dx.doi.org/10.1142/S0217751X98000913}{{\em Int. J. Mod. Phys.}
  {\bf A13} (1998)  2075--2088},
\href{http://arxiv.org/abs/hep-th/9708025}{{\tt arXiv:hep-th/9708025}}.

\bibitem{Pioline:2006ni}
B.~Pioline, ``{Lectures on on black holes, topological strings and quantum
  attractors},'' \href{http://dx.doi.org/10.1088/0264-9381/23/21/S05}{{\em
  Class. Quant. Grav.} {\bf 23} (2006)  S981},
\href{http://arxiv.org/abs/hep-th/0607227}{{\tt arXiv:hep-th/0607227}}.

\bibitem{Townsend:1995gp}
P.~K. Townsend, ``{$p$-brane democracy},'' in {\em Proceedings of the
  PASCOS/Johns Hopkins Conference}, J.~Bagger, G.~Domokos, A.~Falk, and
  S.~Kovesi-Domokos, eds.
\newblock World Scientific Singapore, 1995.
\newblock
\href{http://arxiv.org/abs/hep-th/9507048}{{\tt arXiv:hep-th/9507048}}.
\newblock

\bibitem{Horava:1995qa}
P.~Horava and E.~Witten, ``{Heterotic and type I string dynamics from eleven
  dimensions},'' \href{http://dx.doi.org/10.1016/0550-3213(95)00621-4}{{\em
  Nucl. Phys.} {\bf B460} (1996)  506--524},
\href{http://arxiv.org/abs/hep-th/9510209}{{\tt arXiv:hep-th/9510209}}.

\bibitem{Witten:1995em}
E.~Witten, ``{Five-branes and M-theory on an orbifold},''
  \href{http://dx.doi.org/10.1016/0550-3213(96)00032-6}{{\em Nucl. Phys.} {\bf
  B463} (1996)  383--397},
\href{http://arxiv.org/abs/hep-th/9512219}{{\tt arXiv:hep-th/9512219}}.

\bibitem{Duff:1996aw}
M.~J. Duff, ``{M theory (the theory formerly known as strings)},''
  \href{http://dx.doi.org/10.1142/S0217751X96002583}{{\em Int. J. Mod. Phys.}
  {\bf A11} (1996)  5623--5642},
\href{http://arxiv.org/abs/hep-th/9608117}{{\tt arXiv:hep-th/9608117}}.

\bibitem{Salam:1989fm}
A.~Salam and E.~Sezgin, eds., {\em Supergravities in diverse dimensions},
  vol.~1-2.
\newblock Netherlands: North-Holland, Amsterdam, 1989, 1499 p. Singapore: World
  Scientific, Singapore, 1989, 1499 p.

\bibitem{deWit:2002vz}
B.~de~Wit, ``{Supergravity},'' in {\em Les Houches 2001, Gravity, gauge
  theories and strings}, pp.~1--135.
\newblock 2001.
\newblock \href{http://arxiv.org/abs/hep-th/0212245}{{\tt
  arXiv:hep-th/0212245}}.
\newblock
Lecture notes Les Houches Summer School: Session 76: Euro Summer School on
  Unity of Fundamental Physics: Gravity, Gauge Theory and Strings, Les Houches,
  France, 30 Jul - 31 Aug 2001.

\bibitem{Tanii:1998px}
Y.~Tanii, ``{Introduction to supergravities in diverse dimensions},''
  \href{http://arxiv.org/abs/hep-th/9802138}{{\tt arXiv:hep-th/9802138}}.
An Expanded version of a review talk presented at YITP Workshop on
  Supersymmetry, Kyoto, Japan, 27-30 Mar 1996.

\bibitem{Cremmer:1978km}
E.~Cremmer, B.~Julia, and J.~Scherk, ``{Supergravity theory in 11
  dimensions},''
\href{http://dx.doi.org/10.1016/0370-2693(78)90894-8}{{\em Phys. Lett.} {\bf
  B76} (1978)  409--412}.

\bibitem{Duff:1987bx}
M.~J. Duff, P.~S. Howe, T.~Inami, and K.~S. Stelle, ``{Superstrings in $D = 10$
  from supermembranes in $D = 11$},''
\href{http://dx.doi.org/10.1016/0370-2693(87)91323-2}{{\em Phys. Lett.} {\bf
  B191} (1987)  70}.

\bibitem{Witten:1995ex}
E.~Witten, ``{String theory dynamics in various dimensions},''
  \href{http://dx.doi.org/10.1016/0550-3213(95)00158-O}{{\em Nucl. Phys.} {\bf
  B443} (1995)  85--126},
\href{http://arxiv.org/abs/hep-th/9503124}{{\tt arXiv:hep-th/9503124}}.

\bibitem{Polchinski:1995mt}
J.~Polchinski, ``{Dirichlet-branes and Ramond-Ramond charges},''
  \href{http://dx.doi.org/10.1103/PhysRevLett.75.4724}{{\em Phys. Rev. Lett.}
  {\bf 75} (1995) no.~26, 4724--4727},
\href{http://arxiv.org/abs/hep-th/9510017}{{\tt arXiv:hep-th/9510017}}.

\bibitem{Strominger:1996sh}
A.~Strominger and C.~Vafa, ``{Microscopic origin of the Bekenstein-Hawking
  entropy},'' \href{http://dx.doi.org/10.1016/0370-2693(96)00345-0}{{\em Phys.
  Lett.} {\bf B379} (1996)  99--104},
\href{http://arxiv.org/abs/hep-th/9601029}{{\tt arXiv:hep-th/9601029}}.

\bibitem{Green:1987sp}
M.~B. Green, J.~H. Schwarz, and E.~Witten, {\em {Superstring Theory vol. 1:
  Introduction}}.
\newblock Cambridge Monographs on Mathematical Physics. Cambridge University
  Press, Cambridge, UK, 1987.
\newblock 469 p.

\bibitem{Green:1987mn}
M.~B. Green, J.~H. Schwarz, and E.~Witten, {\em {Superstring Theory vol. 2:
  Loop Amplitudes, Anomalies and Phenomenology}}.
\newblock Cambridge Monographs on Mathematical Physics. Cambridge University
  Press, Cambridge, UK, 1987.
\newblock 596 p.

\bibitem{Giveon:1994fu}
A.~Giveon, M.~Porrati, and E.~Rabinovici, ``{Target space duality in string
  theory},'' \href{http://dx.doi.org/10.1016/0370-1573(94)90070-1}{{\em Phys.
  Rept.} {\bf 244} (1994)  77--202},
\href{http://arxiv.org/abs/hep-th/9401139}{{\tt arXiv:hep-th/9401139}}.

\bibitem{Font:1990gx}
A.~Font, L.~E. Ibanez, D.~Lust, and F.~Quevedo, ``{Strong - weak coupling
  duality and nonperturbative effects in string theory},''
\href{http://dx.doi.org/10.1016/0370-2693(90)90523-9}{{\em Phys. Lett.} {\bf
  B249} (1990)  35--43}.

\bibitem{Rey:1989xj}
S.-J. Rey, ``The confining phase of superstrings and axionic strings,''
\href{http://dx.doi.org/10.1103/PhysRevD.43.526}{{\em Phys. Rev.} {\bf D43}
  (1991) no.~2, 526--538}.

\bibitem{Schwarz:1993mg}
J.~H. Schwarz and A.~Sen, ``{Duality symmetries of 4-D heterotic strings},''
  \href{http://dx.doi.org/10.1016/0370-2693(93)90495-4}{{\em Phys. Lett.} {\bf
  B312} (1993)  105--114},
\href{http://arxiv.org/abs/hep-th/9305185}{{\tt arXiv:hep-th/9305185}}.

\bibitem{Duff:1993ij}
M.~J. Duff and R.~R. Khuri, ``{Four-dimensional string / string duality},''
  \href{http://dx.doi.org/10.1016/0550-3213(94)90459-6}{{\em Nucl. Phys.} {\bf
  B411} (1994)  473--486},
\href{http://arxiv.org/abs/hep-th/9305142}{{\tt arXiv:hep-th/9305142}}.

\bibitem{Schwarz:1993vs}
J.~H. Schwarz and A.~Sen, ``{Duality symmetric actions},''
  \href{http://dx.doi.org/10.1016/0550-3213(94)90053-1}{{\em Nucl. Phys.} {\bf
  B411} (1994)  35--63},
\href{http://arxiv.org/abs/hep-th/9304154}{{\tt arXiv:hep-th/9304154}}.

\bibitem{Sen:1992ch}
A.~Sen, ``{Quantization of dyon charge and electric magnetic duality in string
  theory},'' \href{http://dx.doi.org/10.1016/0370-2693(93)90037-I}{{\em Phys.
  Lett.} {\bf B303} (1993)  22--26},
\href{http://arxiv.org/abs/hep-th/9209016}{{\tt arXiv:hep-th/9209016}}.

\bibitem{Sen:1992fr}
A.~Sen, ``{Electric magnetic duality in string theory},''
  \href{http://dx.doi.org/10.1016/0550-3213(93)90475-5}{{\em Nucl. Phys.} {\bf
  B404} (1993)  109--126},
\href{http://arxiv.org/abs/hep-th/9207053}{{\tt arXiv:hep-th/9207053}}.

\bibitem{Montonen:1977sn}
C.~Montonen and D.~I. Olive, ``{Magnetic monopoles as gauge particles?},''
\href{http://dx.doi.org/10.1016/0370-2693(77)90076-4}{{\em Phys. Lett.} {\bf
  B72} (1977)  117}.

\bibitem{Hull:1994ys}
C.~M. Hull and P.~K. Townsend, ``{Unity of superstring dualities},''
  \href{http://dx.doi.org/10.1016/0550-3213(94)00559-W}{{\em Nucl. Phys.} {\bf
  B438} (1995)  109--137},
\href{http://arxiv.org/abs/hep-th/9410167}{{\tt arXiv:hep-th/9410167}}.

\bibitem{Duff:1994zt}
M.~J. Duff, ``{Strong / weak coupling duality from the dual string},''
  \href{http://dx.doi.org/10.1016/S0550-3213(95)00070-4}{{\em Nucl. Phys.} {\bf
  B442} (1995)  47--63},
\href{http://arxiv.org/abs/hep-th/9501030}{{\tt arXiv:hep-th/9501030}}.

\bibitem{Duff:1996rs}
M.~J. Duff, R.~Minasian, and E.~Witten, ``{Evidence for heterotic/heterotic
  duality},'' \href{http://dx.doi.org/10.1016/0550-3213(96)00059-4}{{\em Nucl.
  Phys.} {\bf B465} (1996)  413--438},
\href{http://arxiv.org/abs/hep-th/9601036}{{\tt arXiv:hep-th/9601036}}.

\bibitem{Cremmer:1984hc}
E.~Cremmer and A.~Van~Proeyen, ``{Classification Of K\"ahler manifolds in $N=2$
  vector multiplet supergravity couplings},''
\href{http://dx.doi.org/10.1088/0264-9381/2/4/010}{{\em Class. Quant. Grav.}
  {\bf 2} (1985) no.~4, 445--454}.

\bibitem{Bergshoeff:1985ms}
E.~Bergshoeff, I.~G. Koh, and E.~Sezgin, ``{Coupling of Yang-Mills to $N=4$,
  $D=4$ supergravity},''
\href{http://dx.doi.org/10.1016/0370-2693(85)91034-2}{{\em Phys. Lett.} {\bf
  B155} (1985)  71}.

\bibitem{Duff:1990hn}
M.~J. Duff and J.~X. Lu, ``{Duality rotations in membrane theory},''
\href{http://dx.doi.org/10.1016/0550-3213(90)90565-U}{{\em Nucl. Phys.} {\bf
  B347} (1990)  394--419}.

\bibitem{Polchinski:1998rr}
J.~Polchinski, {\em String Theory vol. 2: Superstring theory and beyond}.
\newblock Cambridge Monographs on Mathematical Physics. Cambridge University
  Press, Cambridge, U.K.; New York, U.S.A., 1998.

\bibitem{Cremmer:1979up}
E.~Cremmer and B.~Julia, ``{The $SO(8)$ supergravity},''
\href{http://dx.doi.org/10.1016/0550-3213(79)90331-6}{{\em Nucl. Phys.} {\bf
  B159} (1979)  141}.

\bibitem{Riccioni:2007au}
F.~Riccioni and P.~C. West, ``{The $E_{11}$ origin of all maximal
  supergravities},''
  \href{http://dx.doi.org/10.1088/1126-6708/2007/07/063}{{\em JHEP} {\bf 07}
  (2007)  063},
\href{http://arxiv.org/abs/0705.0752}{{\tt arXiv:0705.0752 [hep-th]}}.

\bibitem{Cook:2008bi}
P.~P. Cook and P.~C. West, ``{Charge multiplets and masses for $E_{11}$},''
  \href{http://dx.doi.org/10.1088/1126-6708/2008/11/091}{{\em JHEP} {\bf 11}
  (2008)  091},
\href{http://arxiv.org/abs/0805.4451}{{\tt arXiv:0805.4451 [hep-th]}}.

\bibitem{Andrianopoli:1996ve}
L.~Andrianopoli, R.~D'Auria, and S.~Ferrara, ``{U-duality and central charges
  in various dimensions revisited},''
  \href{http://dx.doi.org/10.1142/S0217751X98000196}{{\em Int. J. Mod. Phys.}
  {\bf A13} (1998)  431--490},
\href{http://arxiv.org/abs/hep-th/9612105}{{\tt arXiv:hep-th/9612105}}.

\bibitem{Gunaydin:2005gd}
M.~G{\"u}naydin, ``{Unitary realizations of U-duality groups as conformal and
  quasiconformal groups and extremal black holes of supergravity theories},''
  \href{http://dx.doi.org/10.1063/1.1923339}{{\em AIP Conf. Proc.} {\bf 767}
  (2005)  268--287},
\href{http://arxiv.org/abs/hep-th/0502235}{{\tt arXiv:hep-th/0502235}}.

\bibitem{Hull:1997kt}
C.~M. Hull, ``{Gravitational duality, branes and charges},''
  \href{http://dx.doi.org/10.1016/S0550-3213(97)00501-4}{{\em Nucl. Phys.} {\bf
  B509} (1998)  216--251},
\href{http://arxiv.org/abs/hep-th/9705162}{{\tt arXiv:hep-th/9705162}}.

\bibitem{Sen:2008sp}
A.~Sen, ``{U-duality invariant dyon spectrum in type II on $T^6$},''
  \href{http://dx.doi.org/10.1088/1126-6708/2008/08/037}{{\em JHEP} {\bf 08}
  (2008)  037},
\href{http://arxiv.org/abs/0804.0651}{{\tt arXiv:0804.0651 [hep-th]}}.

\bibitem{Wootters:1997id}
W.~K. Wootters, ``{Entanglement of formation of an arbitrary state of two
  qubits},'' \href{http://dx.doi.org/10.1103/PhysRevLett.80.2245}{{\em Phys.
  Rev. Lett.} {\bf 80} (1998) no.~10, 2245--2248},
\href{http://arxiv.org/abs/quant-ph/9709029}{{\tt arXiv:quant-ph/9709029}}.

\bibitem{Grassl:1997sq}
M.~Grassl, M.~Rotteler, and T.~Beth, ``{Computing local invariants of qubit
  systems},'' \href{http://dx.doi.org/10.1103/PhysRevA.58.1833}{{\em Phys.
  Rev.} {\bf A58} (1998) no.~3, 1833--1839},
\href{http://arxiv.org/abs/quant-ph/9712040}{{\tt arXiv:quant-ph/9712040}}.

\bibitem{Kempe:1999vk}
J.~Kempe, ``{On multi-particle entanglement and its applications to
  cryptography},'' \href{http://dx.doi.org/10.1103/PhysRevA.60.910}{{\em Phys.
  Rev.} {\bf A60} (1999) no.~2, 910--916},
\href{http://arxiv.org/abs/quant-ph/9902036}{{\tt arXiv:quant-ph/9902036}}.

\bibitem{Bennett:1999}
C.~H. Bennett, S.~Popescu, D.~Rohrlich, J.~A. Smolin, and A.~V. Thapliyal,
  ``Exact and asymptotic measures of multipartite pure-state entanglement,''
  \href{http://dx.doi.org/10.1103/PhysRevA.63.012307}{{\em Phys. Rev.} {\bf
  A63} (2000) no.~1, 012307}, \href{http://arxiv.org/abs/quant-ph/9908073}{{\tt
  arXiv:quant-ph/9908073}}.

\bibitem{Brylinski:2000x}
J.-L. Brylinski, ``Algebraic measures of entanglement,''
  \href{http://arxiv.org/abs/quant-ph/0008031}{{\tt arXiv:quant-ph/0008031}}.

\bibitem{Brylinski:2000y}
J.~Brylinski and R.~Brylinski, ``Invariant polynomial functions on $k$
  qudits,'' \href{http://arxiv.org/abs/quant-ph/0010101}{{\tt
  arXiv:quant-ph/0010101}}.

\bibitem{Albeverio:2001ey}
S.~Albeverio and S.-M. Fei, ``{A note on invariants and entanglements},''
  \href{http://dx.doi.org/10.1088/1464-4266/3/4/305}{{\em J. Opt. B Quant.
  Semiclass. Opt.} {\bf B3} (2001) no.~4, 223--227},
\href{http://arxiv.org/abs/quant-ph/0109073}{{\tt arXiv:quant-ph/0109073}}.

\bibitem{Carteret:2000-1}
H.~Carteret and A.~Sudbery, ``Local symmetry properties of pure three-qubit
  states,'' \href{http://dx.doi.org/10.1088/0305-4470/33/28/303}{{\em J. Phys.}
  {\bf A33} (2000) no.~28, 4981--5002},
  \href{http://arxiv.org/abs/quant-ph/0001091}{{\tt arXiv:quant-ph/0001091}}.

\bibitem{Carteret:2000-2}
H.~Carteret, A.~Higuchi, and A.~Sudbery, ``{Multipartite generalisation of the
  Schmidt decomposition},'' \href{http://dx.doi.org/10.1063/1.1319516}{{\em J.
  Math. Phys.} {\bf 41} (2000) no.~12, 7932--7939},
  \href{http://arxiv.org/abs/quant-ph/0006125}{{\tt arXiv:quant-ph/0006125}}.

\bibitem{Verstraete:2003}
F.~Verstraete, J.~Dehaene, and B.~{De Moor}, ``{Normal forms and entanglement
  measures for multipartite quantum states},''
  \href{http://dx.doi.org/10.1103/PhysRevA.68.012103}{{\em Phys. Rev.} {\bf
  A68} (2003) no.~1, 012103}, \href{http://arxiv.org/abs/quant-ph/0105090}{{\tt
  arXiv:quant-ph/0105090}}.

\bibitem{Levay:2003pb}
P.~L\'evay, ``{The geometry of entanglement: metrics, connections and the
  geometric phase},'' \href{http://dx.doi.org/10.1088/0305-4470/37/5/024}{{\em
  J. Phys.} {\bf A37} (2004)  1821--1842},
\href{http://arxiv.org/abs/quant-ph/0306115}{{\tt arXiv:quant-ph/0306115}}.

\bibitem{Toumazet:2006}
F.~Toumazet, J.~Luque, and J.~Thibon, ``Unitary invariants of qubit systems,''
  \href{http://dx.doi.org/10.1017/S0960129507006330}{{\em Mathematical
  Structures in Computer Science} {\bf 17} (2006) no.~6, 1133--1151},
  \href{http://arxiv.org/abs/quant-ph/0604202}{{\tt arXiv:quant-ph/0604202}}.

\bibitem{Abascal:2007}
I.~S. Abascal and G.~Bj\"{o}rk, ``Bipartite entanglement measure based on
  covariances,'' \href{http://dx.doi.org/10.1103/PhysRevA.75.062317}{{\em Phys.
  Rev.} {\bf A75} (2007) no.~6, 062317},
  \href{http://arxiv.org/abs/quant-ph/0703249}{{\tt arXiv:quant-ph/0703249}}.

\bibitem{Plenio:2007}
M.~B. Plenio and S.~Virmani, ``{An introduction to entanglement measures},''
  {\em Quant. Inf. Comp.} {\bf 7} (2007)  1,
  \href{http://arxiv.org/abs/quant-ph/0504163}{{\tt arXiv:quant-ph/0504163}}.

\bibitem{Amico:2007ag}
L.~Amico, R.~Fazio, A.~Osterloh, and V.~Vedral, ``{Entanglement in many-body
  systems},'' \href{http://dx.doi.org/10.1103//RevModPhys.80.517}{{\em Rev.
  Mod. Phys.} {\bf 80} (2008)  517--576},
\href{http://arxiv.org/abs/quant-ph/0703044}{{\tt arXiv:quant-ph/0703044}}.

\bibitem{Horodecki:2007}
R.~Horodecki, P.~Horodecki, M.~Horodecki, and K.~Horodecki, ``Quantum
  entanglement,'' \href{http://arxiv.org/abs/quant-ph/0702225}{{\tt
  arXiv:quant-ph/0702225}}.

\bibitem{Bardeen:1973gs}
J.~M. Bardeen, B.~Carter, and S.~W. Hawking, ``{The four laws of black hole
  mechanics},''
\href{http://dx.doi.org/10.1007/BF01645742}{{\em Commun. Math. Phys.} {\bf 31}
  (1973)  161--170}.

\bibitem{Duff:1991sz}
M.~J. Duff and J.~X. Lu, ``{A Duality between strings and five-branes},''
  \href{http://dx.doi.org/10.1088/0264-9381/9/1/004}{{\em Class. Quant. Grav.}
  {\bf 9} (1992)  1--16}.
Also published in the Proceedings of the International School of Astroparticle
  Physics, Woodlands, TX, Jan 6-12, 1991. Edited by D.V. Nanopoulos. Singapore:
  World Scientific, 1991, pp. 155-182.

\bibitem{Duff:1994an}
M.~J. Duff, R.~R. Khuri, and J.~X. Lu, ``{String solitons},''
  \href{http://dx.doi.org/10.1016/0370-1573(95)00002-X}{{\em Phys. Rept.} {\bf
  259} (1995)  213--326},
\href{http://arxiv.org/abs/hep-th/9412184}{{\tt arXiv:hep-th/9412184}}.

\bibitem{Peet:2000hn}
A.~W. Peet, ``{TASI lectures on black holes in string theory},'' in {\em
  Boulder 1999, Strings, branes and gravity}, pp.~353--433.
\newblock 1999.
\newblock \href{http://arxiv.org/abs/hep-th/0008241}{{\tt
  arXiv:hep-th/0008241}}.
\newblock
Prepared for Theoretical Advanced Study Institute in Elementary Particle
  Physics (TASI 99): Strings, Branes, and Gravity, Boulder, Colorado, 31 May -
  25 Jun 1999.

\bibitem{Ferrara:1997tw}
S.~Ferrara, G.~W. Gibbons, and R.~Kallosh, ``{Black holes and critical points
  in moduli space},''
  \href{http://dx.doi.org/10.1016/S0550-3213(97)00324-6}{{\em Nucl. Phys.} {\bf
  B500} (1997)  75--93},
\href{http://arxiv.org/abs/hep-th/9702103}{{\tt arXiv:hep-th/9702103}}.

\bibitem{Bais:2007pk}
F.~A. Bais and J.~Doyne~Farmer, ``{The physics of information},''
\href{http://arxiv.org/abs/0708.2837}{{\tt arXiv:0708.2837
  [physics.class-ph]}}.

\bibitem{Bell:1964kc}
J.~S. Bell, ``{On the Einstein-Podolsky-Rosen paradox}.''
\href{http://www.physics.princeton.edu/~mcdonald/examples/QM/bell_physics_1_19%
5_64.pdf}{\textit{Physics} \textbf{1} (1964) no. 3, 195}.

\bibitem{Bohm:1951}
D.~Bohm, {\em Quantum Theory}.
\newblock Prentice-Hall, Englewood Cliffs, N.J., 1951.

\bibitem{Einstein:1935rr}
A.~Einstein, B.~Podolsky, and N.~Rosen, ``{Can quantum-mechanical description
  of physical reality be considered complete?},''
\href{http://dx.doi.org/10.1103/PhysRev.47.777}{{\em Phys. Rev.} {\bf 47}
  (1935) no.~10, 777--780}.

\bibitem{Einstein:1970}
A.~Einstein, {\em Autobiographical Notes in Albert Einstein:
  Philosopher-Scientist P. A. Schilpp (ed.), Library of Living Philosophers},
  vol.~III.
\newblock Cambridge Unicersity Press, 1970.

\bibitem{Mermin:1990}
N.~D. {Mermin}, ``{Quantum mysteries revisited},''
  \href{http://dx.doi.org/10.1119/1.16503}{{\em Am. J. Phys.} {\bf 58} (1990)
  731--734}.

\bibitem{Mermin:2006}
N.~D. Mermin, ``In praise of measurement,''
  \href{http://dx.doi.org/10.1007/s11128-006-0017-2}{{\em Quantum Information
  Processing} {\bf 5} (2006) no.~4, 239--260},
  \href{http://arxiv.org/abs/quant-ph/0612216}{{\tt arXiv:quant-ph/0612216}}.

\bibitem{Nielsen:2000}
M.~A. Nielsen and I.~L. Chuang, {\em Quantum Computation and Quantum
  Information}.
\newblock Cambridge University Press, New York, NY, USA, 2000.

\bibitem{Jaeger:2006}
G.~Jaeger, {\em Quantum Information: An Overview}.
\newblock Springer, December, 2006.

\bibitem{Li:2007}
D.~Li, X.~Li, H.~Huang, and X.~Li, ``{Stochastic local operations and classical
  communication invariant and the residual entanglement for $n$ qubits},''
  \href{http://dx.doi.org/10.1103/PhysRevA.76.032304}{{\em Phys. Rev.} {\bf
  A76} (2007) no.~3, 032304}, \href{http://arxiv.org/abs/0704.2087}{{\tt
  arXiv:0704.2087 [quant-ph]}}.

\bibitem{Lamata:2006}
L.~Lamata, J.~Le\'{o}n, D.~Salgado, and E.~Solano, ``Inductive classification
  of multipartite entanglement under stochastic local operations and classical
  communication,'' \href{http://dx.doi.org/10.1103/PhysRevA.74.052336}{{\em
  Phys. Rev.} {\bf A74} (2006) no.~5, 052336},
  \href{http://arxiv.org/abs/quant-ph/0603243}{{\tt arXiv:quant-ph/0603243}}.

\bibitem{Linden:1997qd}
N.~Linden and S.~Popescu, ``{On multi-particle entanglement},'' {\em Fortschr.
  Phys.} {\bf 46} (1998) no.~4--5, 567--578,
\href{http://arxiv.org/abs/quant-ph/9711016}{{\tt arXiv:quant-ph/9711016}}.

\bibitem{Dur:2000}
W.~{D\"ur}, G.~Vidal, and J.~I. Cirac, ``Three qubits can be entangled in two
  inequivalent ways,'' \href{http://dx.doi.org/10.1103/PhysRevA.62.062314}{{\em
  Phys. Rev.} {\bf A62} (2000) no.~6, 062314},
  \href{http://arxiv.org/abs/quant-ph/0005115}{{\tt arXiv:quant-ph/0005115}}.

\bibitem{Sudbery:2001}
A.~Sudbery, ``On local invariants of pure three-qubit states,''
  \href{http://dx.doi.org/10.1088/0305-4470/34/3/323}{{\em J. Phys.} {\bf A34}
  (2001) no.~3, 643--652}, \href{http://arxiv.org/abs/quant-ph/0001116}{{\tt
  arXiv:quant-ph/0001116}}.

\bibitem{Gingrich:2001}
R.~M. Gingrich, ``Properties of entanglement monotones for three-qubit pure
  states,'' \href{http://dx.doi.org/10.1103/PhysRevA.65.052302}{{\em Phys.
  Rev.} {\bf A65} (2002) no.~5, 052302},
  \href{http://arxiv.org/abs/quant-ph/0106042}{{\tt arXiv:quant-ph/0106042}}.

\bibitem{Osterloh:2006}
A.~Osterloh and J.~Siewert, ``Entanglement monotones and maximally entangled
  states in multipartite qubit systems,'' {\em Int. J. Quant. Inf.} {\bf 4}
  (2006)  531, \href{http://arxiv.org/abs/quant-ph/0506073}{{\tt
  arXiv:quant-ph/0506073}}.

\bibitem{Vidal:1998re}
G.~Vidal, ``{On the characterization of entanglement},'' {\em J. Mod. Opt.}
  {\bf 47} (2000)  355,
\href{http://arxiv.org/abs/quant-ph/9807077}{{\tt arXiv:quant-ph/9807077}}.

\bibitem{Levay:2005b}
P.~L\'evay, ``On the geometry of a class of $n$-qubit entanglement monotones,''
  \href{http://dx.doi.org/doi:10.1088/0305-4470/38/41/016}{{\em J. Phys.} {\bf
  A38} (2005) no.~41, 9075--9085},
  \href{http://arxiv.org/abs/quant-ph/0507070}{{\tt arXiv:quant-ph/0507070}}.

\bibitem{Gibbs:2001}
P.~Gibbs, ``{Diophantine quadruples and Cayley's hyperdeterminant},''
  \href{http://arxiv.org/abs/math/0107203}{{\tt arXiv:math/0107203}}.

\bibitem{Fernando:2006}
K.~V. Fernando, ``{Singular $2\times 2\times 2$ arrays},''. Oxford preprint.

\bibitem{Brody:2007}
D.~C. Brody, A.~C.~T. Gustavsson, and L.~P. Hughston, ``Entanglement of
  three-qubit geometry,''
  \href{http://dx.doi.org/doi:10.1088/1742-6596/67/1/012044}{{\em J. Phys.:
  Conf. Ser.} {\bf 67} (2007)  012044},
  \href{http://arxiv.org/abs/quant-ph/0612117}{{\tt arXiv:quant-ph/0612117}}.

\bibitem{Lee:2005}
S.~Lee, J.~Joo, and J.~Kim, ``Entanglement of three-qubit pure states in terms
  of teleportation capability,''
  \href{http://dx.doi.org/10.1103/PhysRevA.72.024302}{{\em Phys. Rev.} {\bf
  A72} (2005) no.~2, 024302}, \href{http://arxiv.org/abs/quant-ph/0502157}{{\tt
  arXiv:quant-ph/0502157}}.

\bibitem{Lee:2007}
S.~Lee, J.~Joo, and J.~Kim, ``Teleportation capability, distillability, and
  nonlocality on three-qubit states,''
  \href{http://dx.doi.org/10.1103/PhysRevA.76.012311}{{\em Phys. Rev.} {\bf
  A76} (2007) no.~1, 012311}, \href{http://arxiv.org/abs/quant-ph/0702247}{{\tt
  arXiv:quant-ph/0702247}}.

\bibitem{Miyake:2003}
A.~Miyake, ``Classification of multipartite entangled states by
  multidimensional determinants,''
  \href{http://dx.doi.org/10.1103/PhysRevA.67.012108}{{\em Phys. Rev.} {\bf
  A67} (2003) no.~1, 012108}, \href{http://arxiv.org/abs/quant-ph/0206111}{{\tt
  arXiv:quant-ph/0206111}}.

\bibitem{Luque:2005}
J.-G. Luque and J.-Y. Thibon, ``{Algebraic invariants of five qubits},''
  \href{http://dx.doi.org/10.1088/0305-4470/39/2/007}{{\em J. Phys.} {\bf A39}
  (2006) no.~2, 371--377}, \href{http://arxiv.org/abs/quant-ph/0506058}{{\tt
  arXiv:quant-ph/0506058}}.

\bibitem{Acin:2001}
A.~Acin, A.~Andrianov, E.~Jane, and R.~Tarrach, ``Three-qubit pure-state
  canonical forms,'' \href{http://dx.doi.org/10.1088/0305-4470/34/35/301}{{\em
  J. Phys.} {\bf A34} (2001) no.~35, 6725--6739},
  \href{http://arxiv.org/abs/quant-ph/0009107}{{\tt arXiv:quant-ph/0009107}}.

\bibitem{Greenberger:1989}
D.~M. Greenberger, M.~Horne, and A.~Zeilinger, {\em {Bell's Theorem, Quantum
  Theory and Conceptions of the Universe}}.
\newblock Kluwer Academic, Dordrecht, 1989.

\bibitem{Greenberger:1990}
D.~M. {Greenberger}, M.~A. {Horne}, A.~{Shimony}, and A.~{Zeilinger}, ``{Bell's
  theorem without inequalities},''
  \href{http://dx.doi.org/10.1119/1.16243}{{\em Am. J. of Phys.} {\bf 58}
  (1990)  1131--1143}.

\bibitem{Acin:2000}
A.~Ac\'in, A.~Andrianov, L.~Costa, E.~Jan\'e, J.~I. Latorre, and R.~Tarrach,
  ``{Generalized Schmidt decomposition and classification of three-quantum-bit
  states},'' \href{http://dx.doi.org/10.1103/PhysRevLett.85.1560}{{\em Phys.
  Rev. Lett.} {\bf 85} (2000) no.~7, 1560--1563},
  \href{http://arxiv.org/abs/quant-ph/0003050}{{\tt arXiv:quant-ph/0003050}}.

\bibitem{Fan:2003}
H.~Fan, K.~Matsumoto, and H.~Imai, ``Quantify entanglement by concurrence
  hierarchy,'' \href{http://dx.doi.org/10.1088/0305-4470/36/14/316}{{\em J.
  Phys.} {\bf A36} (2003) no.~14, 4151--4158},
  \href{http://arxiv.org/abs/quant-ph/0204041}{{\tt arXiv:quant-ph/0204041}}.

\bibitem{Cereceda:2003}
J.~L. Cereceda, ``Degree of entanglement for two qutrits in a pure state,''
  \href{http://arxiv.org/abs/quant-ph/0305043}{{\tt arXiv:quant-ph/0305043}}.

\bibitem{HerrenoFierro:2005}
C.~{Herreno-Fierro} and J.~Luthra, ``Generalized concurrence and limits of
  separability for two qutrits,''
  \href{http://arxiv.org/abs/quant-ph/0507223}{{\tt arXiv:quant-ph/0507223}}.

\bibitem{Pan:2006}
F.~Pan, G.~Lu, and J.~Draayer, ``{Classification and quantification of
  entangled bipartite qutrit pure states},'' {\em Int. J. Mod. Phys.} {\bf B20}
  (2006)  1333--1342, \href{http://arxiv.org/abs/quant-ph/0510178}{{\tt
  arXiv:quant-ph/0510178}}.

\bibitem{Rai:2005}
S.~Rai and J.~R. Luthra, ``Negativity and concurrence for two qutrits,''
  \href{http://arxiv.org/abs/quant-ph/0507263}{{\tt arXiv:quant-ph/0507263}}.

\bibitem{Sen:1995ff}
A.~Sen and C.~Vafa, ``{Dual pairs of type II string compactification},''
  \href{http://dx.doi.org/10.1016/0550-3213(95)00498-H}{{\em Nucl. Phys.} {\bf
  B455} (1995)  165--187},
\href{http://arxiv.org/abs/hep-th/9508064}{{\tt arXiv:hep-th/9508064}}.

\bibitem{Gregori:1999ns}
A.~Gregori, C.~Kounnas, and P.~M. Petropoulos, ``{Non-perturbative triality in
  heterotic and type II $N = 2$ strings},''
  \href{http://dx.doi.org/10.1016/S0550-3213(99)00281-3}{{\em Nucl. Phys.} {\bf
  B553} (1999)  108--132},
\href{http://arxiv.org/abs/hep-th/9901117}{{\tt arXiv:hep-th/9901117}}.

\bibitem{Cvetic:1995uj}
M.~Cvetic and D.~Youm, ``{Dyonic BPS saturated black holes of heterotic string
  on a six torus},'' \href{http://dx.doi.org/10.1103/PhysRevD.53.584}{{\em
  Phys. Rev.} {\bf D53} (1996)  584--588},
\href{http://arxiv.org/abs/hep-th/9507090}{{\tt arXiv:hep-th/9507090}}.

\bibitem{Ceresole:1994cx}
A.~Ceresole, R.~D'Auria, S.~Ferrara, and A.~Van~Proeyen, ``{On electromagnetic
  duality in locally supersymmetric $N=2$ Yang-Mills theory},'' in {\em
  Proceedings of the Workshop on Physics from the Planck Scale to
  Electromagnetic Scale}, P.~Nath, T.~Taylor, and S.~Pokorski, eds.
\newblock World Scientific, 1994.
\newblock
\href{http://arxiv.org/abs/hep-th/9412200}{{\tt arXiv:hep-th/9412200}}.
\newblock

\bibitem{Ferrara:1995ih}
S.~Ferrara, R.~Kallosh, and A.~Strominger, ``{$N=2$ extremal black holes},''
  \href{http://dx.doi.org/10.1103/PhysRevD.52.R5412}{{\em Phys. Rev.} {\bf D52}
  (1995)  5412--5416},
\href{http://arxiv.org/abs/hep-th/9508072}{{\tt arXiv:hep-th/9508072}}.

\bibitem{Bernevig:2003}
B.~A. Bernevig and H.-D. Chen, ``{Geometry of the 3-qubit state, entanglement
  and division algebras},''
  \href{http://dx.doi.org/10.1088/0305-4470/36/30/309}{{\em J. Phys.} {\bf A36}
  (2003) no.~30, 8325--8339}, \href{http://arxiv.org/abs/quant-ph/0302081}{{\tt
  arXiv:quant-ph/0302081}}.

\bibitem{Levay:2004}
P.~L\'evay, ``{Geometry of three-qubit entanglement},''
  \href{http://dx.doi.org/10.1103/PhysRevA.71.012334}{{\em Phys. Rev.} {\bf
  A71} (2005) no.~1, 012334}, \href{http://arxiv.org/abs/quant-ph/0403060}{{\tt
  arXiv:quant-ph/0403060}}.

\bibitem{Hawking:2000da}
S.~Hawking, J.~M. Maldacena, and A.~Strominger, ``{DeSitter entropy, quantum
  entanglement and AdS/CFT},''
  \href{http://dx.doi.org/10.1088/1126-6708/2001/05/001}{{\em JHEP} {\bf 05}
  (2001)  001},
\href{http://arxiv.org/abs/hep-th/0002145}{{\tt arXiv:hep-th/0002145}}.

\bibitem{BrittoPacumio:1999ax}
R.~Britto-Pacumio, J.~Michelson, A.~Strominger, and A.~Volovich, ``{Lectures on
  superconformal quantum mechanics and multi-black hole moduli spaces},'' in
  {\em Cargese 1999, Progress in string theory and M-theory}, pp.~235--264.
\newblock 1999.
\newblock \href{http://arxiv.org/abs/hep-th/9911066}{{\tt
  arXiv:hep-th/9911066}}.
\newblock
Contributed to NATO Advanced Study Institute on Quantum Geometry, Akureyri,
  Iceland, 10-20 Aug 1999.

\bibitem{Batle:2003}
J.~Batle, A.~R. Plastino, M.~Casas, and A.~Plastino, ``{Understanding quantum
  entanglement: Qubits, rebits and the quaternionic approach},''
  \href{http://dx.doi.org/10.1134/1.1576838}{{\em Optics and Spectroscopy} {\bf
  94} (2003) no.~5, 700--705},
  \href{http://arxiv.org/abs/quant-ph/0603060}{{\tt arXiv:quant-ph/0603060}}.

\bibitem{Caves:2000}
C.~M. Caves, C.~A. Fuchs, and C.~A. Rungta, ``{Entanglement of formation of an
  arbitrary state of two rebits},''
  \href{http://dx.doi.org/10.1023/A:1012215309321}{{\em Found. Phys. Lett.}
  {\bf 14} (2001) no.~3, 199--212},
  \href{http://arxiv.org/abs/quant-ph/0009063}{{\tt arXiv:quant-ph/0009063}}.

\bibitem{Duff:1994jr}
M.~J. Duff and J.~Rahmfeld, ``{Massive string states as extreme black holes},''
  \href{http://dx.doi.org/10.1016/0370-2693(94)01638-S}{{\em Phys. Lett.} {\bf
  B345} (1995)  441--447},
\href{http://arxiv.org/abs/hep-th/9406105}{{\tt arXiv:hep-th/9406105}}.

\bibitem{Duff:1996qp}
M.~J. Duff and J.~Rahmfeld, ``{Bound states of black holes and other
  $p$-branes},'' \href{http://dx.doi.org/10.1016/S0550-3213(96)90139-X}{{\em
  Nucl. Phys.} {\bf B481} (1996)  332--352},
\href{http://arxiv.org/abs/hep-th/9605085}{{\tt arXiv:hep-th/9605085}}.

\bibitem{Kallosh:2006bx}
R.~Kallosh, ``{From BPS to non-BPS black holes canonically},''
\href{http://arxiv.org/abs/hep-th/0603003}{{\tt arXiv:hep-th/0603003}}.

\bibitem{Gimon:2007mh}
E.~G. Gimon, F.~Larsen, and J.~Simon, ``{Black holes in supergravity: The
  non-BPS branch},''
  \href{http://dx.doi.org/10.1088/1126-6708/2008/01/040}{{\em JHEP} {\bf 01}
  (2008)  040},
\href{http://arxiv.org/abs/0710.4967}{{\tt arXiv:0710.4967 [hep-th]}}.

\bibitem{Susskind:1993aa}
L.~Susskind, ``{Strings, black holes and Lorentz contraction},''
  \href{http://dx.doi.org/10.1103/PhysRevD.49.6606}{{\em Phys. Rev.} {\bf D49}
  (1994) no.~12, 6606--6611},
\href{http://arxiv.org/abs/hep-th/9308139}{{\tt arXiv:hep-th/9308139}}.

\bibitem{Sen:1994eb}
A.~Sen, ``{Black hole solutions in heterotic string theory on a torus},''
  \href{http://dx.doi.org/10.1016/0550-3213(95)00063-X}{{\em Nucl. Phys.} {\bf
  B440} (1995)  421--440},
\href{http://arxiv.org/abs/hep-th/9411187}{{\tt arXiv:hep-th/9411187}}.

\bibitem{Sen:1995in}
A.~Sen, ``{Extremal black holes and elementary string states},''
  \href{http://dx.doi.org/10.1142/S0217732395002234}{{\em Mod. Phys. Lett.}
  {\bf A10} (1995)  2081--2094},
\href{http://arxiv.org/abs/hep-th/9504147}{{\tt arXiv:hep-th/9504147}}.

\bibitem{Dabholkar:2004dq}
A.~Dabholkar, R.~Kallosh, and A.~Maloney, ``{A stringy cloak for a classical
  singularity},'' \href{http://dx.doi.org/10.1088/1126-6708/2004/12/059}{{\em
  JHEP} {\bf 12} (2004)  059},
\href{http://arxiv.org/abs/hep-th/0410076}{{\tt arXiv:hep-th/0410076}}.

\bibitem{Sinha:2006yy}
A.~Sinha and N.~V. Suryanarayana, ``{Extremal single-charge small black holes:
  Entropy function analysis},''
  \href{http://dx.doi.org/10.1088/0264-9381/23/10/004}{{\em Class. Quant.
  Grav.} {\bf 23} (2006)  3305--3322},
\href{http://arxiv.org/abs/hep-th/0601183}{{\tt arXiv:hep-th/0601183}}.

\bibitem{Ooguri:2004zv}
H.~Ooguri, A.~Strominger, and C.~Vafa, ``{Black hole attractors and the
  topological string},''
  \href{http://dx.doi.org/10.1103/PhysRevD.70.106007}{{\em Phys. Rev.} {\bf
  D70} (2004) no.~10, 106007},
\href{http://arxiv.org/abs/hep-th/0405146}{{\tt arXiv:hep-th/0405146}}.

\bibitem{Dabholkar:2005dt}
A.~Dabholkar, F.~Denef, G.~W. Moore, and B.~Pioline, ``{Precision counting of
  small black holes},''
  \href{http://dx.doi.org/10.1088/1126-6708/2005/10/096}{{\em JHEP} {\bf 10}
  (2005)  096},
\href{http://arxiv.org/abs/hep-th/0507014}{{\tt arXiv:hep-th/0507014}}.

\bibitem{Cardoso:2006bg}
G.~Lopes~Cardoso, B.~de~Wit, J.~Kappeli, and T.~Mohaupt, ``{Black hole
  partition functions and duality},''
  \href{http://dx.doi.org/10.1088/1126-6708/2006/03/074}{{\em JHEP} {\bf 03}
  (2006)  074},
\href{http://arxiv.org/abs/hep-th/0601108}{{\tt arXiv:hep-th/0601108}}.

\bibitem{Alishahiha:2006jd}
M.~Alishahiha and H.~Ebrahim, ``{New attractor, entropy function and black hole
  partition function},''
  \href{http://dx.doi.org/10.1088/1126-6708/2006/11/017}{{\em JHEP} {\bf 11}
  (2006)  017},
\href{http://arxiv.org/abs/hep-th/0605279}{{\tt arXiv:hep-th/0605279}}.

\bibitem{Strominger:1996kf}
A.~Strominger, ``{Macroscopic entropy of $N=2$ extremal black holes},''
  \href{http://dx.doi.org/10.1016/0370-2693(96)00711-3}{{\em Phys. Lett.} {\bf
  B383} (1996)  39--43},
\href{http://arxiv.org/abs/hep-th/9602111}{{\tt arXiv:hep-th/9602111}}.

\bibitem{Ferrara:1996dd}
S.~Ferrara and R.~Kallosh, ``{Supersymmetry and attractors},''
  \href{http://dx.doi.org/10.1103/PhysRevD.54.1514}{{\em Phys. Rev.} {\bf D54}
  (1996) no.~2, 1514--1524},
\href{http://arxiv.org/abs/hep-th/9602136}{{\tt arXiv:hep-th/9602136}}.

\bibitem{Bellucci:2006xz}
S.~Bellucci, S.~Ferrara, M.~G{\"u}naydin, and A.~Marrani, ``{Charge orbits of
  symmetric special geometries and attractors},''
  \href{http://dx.doi.org/10.1142/S0217751X06034355}{{\em Int. J. Mod. Phys.}
  {\bf A21} (2006)  5043--5098},
\href{http://arxiv.org/abs/hep-th/0606209}{{\tt arXiv:hep-th/0606209}}.

\bibitem{Bellucci:2007gb}
S.~Bellucci, S.~Ferrara, and A.~Marrani, ``{Attractor horizon geometries of
  extremal black holes},'' \href{http://arxiv.org/abs/hep-th/0702019}{{\tt
  arXiv:hep-th/0702019}}.
Contribution to the Proceedings of the XVII SIGRAV Conference, Turin, Italy,
  4–7 Sep 2006.

\bibitem{Tripathy:2005qp}
P.~K. Tripathy and S.~P. Trivedi, ``{Non-supersymmetric attractors in string
  theory},'' \href{http://dx.doi.org/10.1088/1126-6708/2006/03/022}{{\em JHEP}
  {\bf 03} (2006)  022},
\href{http://arxiv.org/abs/hep-th/0511117}{{\tt arXiv:hep-th/0511117}}.

\bibitem{Kallosh:2006bt}
R.~Kallosh, N.~Sivanandam, and M.~Soroush, ``{The non-BPS black hole attractor
  equation},'' \href{http://dx.doi.org/10.1088/1126-6708/2006/03/060}{{\em
  JHEP} {\bf 03} (2006)  060},
\href{http://arxiv.org/abs/hep-th/0602005}{{\tt arXiv:hep-th/0602005}}.

\bibitem{Balasubramanian:1997az}
V.~Balasubramanian, F.~Larsen, and R.~G. Leigh, ``{Branes at angles and black
  holes},'' \href{http://dx.doi.org/10.1103/PhysRevD.57.3509}{{\em Phys. Rev.}
  {\bf D57} (1998) no.~6, 3509--3528},
\href{http://arxiv.org/abs/hep-th/9704143}{{\tt arXiv:hep-th/9704143}}.

\bibitem{Gunaydin:2000xr}
M.~G{\"u}naydin, K.~Koepsell, and H.~Nicolai, ``{Conformal and quasiconformal
  realizations of exceptional Lie groups},''
  \href{http://dx.doi.org/10.1007/PL00005574}{{\em Commun. Math. Phys.} {\bf
  221} (2001)  57--76},
\href{http://arxiv.org/abs/hep-th/0008063}{{\tt arXiv:hep-th/0008063}}.

\bibitem{Kallosh:1996uy}
R.~Kallosh and B.~Kol, ``{$E_7$ symmetric area of the black hole horizon},''
  \href{http://dx.doi.org/10.1103/PhysRevD.53.R5344}{{\em Phys. Rev.} {\bf D53}
  (1996)  5344--5348},
\href{http://arxiv.org/abs/hep-th/9602014}{{\tt arXiv:hep-th/9602014}}.

\bibitem{Ferrara:1997ci}
S.~Ferrara and J.~M. Maldacena, ``{Branes, central charges and U-duality
  invariant BPS conditions},''
  \href{http://dx.doi.org/10.1088/0264-9381/15/4/004}{{\em Class. Quant. Grav.}
  {\bf 15} (1998)  749--758},
\href{http://arxiv.org/abs/hep-th/9706097}{{\tt arXiv:hep-th/9706097}}.

\bibitem{Cvetic:1995kv}
M.~Cvetic and D.~Youm, ``{All the static spherically symmetric black holes of
  heterotic string on a six torus},''
  \href{http://dx.doi.org/10.1016/0550-3213(96)00219-2}{{\em Nucl. Phys.} {\bf
  B472} (1996)  249--267},
\href{http://arxiv.org/abs/hep-th/9512127}{{\tt arXiv:hep-th/9512127}}.

\bibitem{Cvetic:1995bj}
M.~Cvetic and A.~A. Tseytlin, ``{Solitonic strings and BPS saturated dyonic
  black holes},'' \href{http://dx.doi.org/10.1103/PhysRevD.53.5619}{{\em Phys.
  Rev.} {\bf D53} (1996) no.~10, 5619--5633},
\href{http://arxiv.org/abs/hep-th/9512031}{{\tt arXiv:hep-th/9512031}}.

\bibitem{Cvetic:1996zq}
M.~Cvetic and C.~M. Hull, ``{Black holes and U-duality},''
  \href{http://dx.doi.org/10.1016/S0550-3213(96)00449-X}{{\em Nucl. Phys.} {\bf
  B480} (1996)  296--316},
\href{http://arxiv.org/abs/hep-th/9606193}{{\tt arXiv:hep-th/9606193}}.

\bibitem{Bertolini:1999je}
M.~Bertolini, P.~Fre, and M.~Trigiante, ``{The generating solution of regular
  $N=8$ BPS black holes},''
  \href{http://dx.doi.org/10.1088/0264-9381/16/9/315}{{\em Class. Quant. Grav.}
  {\bf 16} (1999)  2987--3004},
\href{http://arxiv.org/abs/hep-th/9905143}{{\tt arXiv:hep-th/9905143}}.

\bibitem{Cartan}
E.~Cartan, ``{{\OE}euvres compl\`{e}tes}.'' Editions du centre national de la
  recherche scientifique, 1984.

\bibitem{Baez:2001dm}
J.~C. Baez, ``{The octonions},''
  \href{http://dx.doi.org/10.1090/S0273-0979-01-00934-X}{{\em Bull. Amer. Math.
  Soc.} {\bf 39} (2002)  145--205},
\href{http://arxiv.org/abs/math/0105155}{{\tt arXiv:math/0105155}}.

\bibitem{Schray:1994ur}
J.~Schray and C.~A. Manogue, ``{Octonionic representations of Clifford algebras
  and triality},'' \href{http://dx.doi.org/10.1007/BF02058887}{{\em Found.
  Phys.} {\bf 26} (1996) no.~1, 17--70},
\href{http://arxiv.org/abs/hep-th/9407179}{{\tt arXiv:hep-th/9407179}}.

\bibitem{Schafer:1966}
R.~Schafer, {\em {Introduction to Nonassociative Algebras}}.
\newblock Academic Press Inc., New York, 1966.

\bibitem{Daboul:1999xv}
J.~Daboul and R.~Delbourgo, ``{Matrix representation of octonions and
  generalizations},'' \href{http://dx.doi.org/10.1063/1.532950}{{\em J. Math.
  Phys.} {\bf 40} (1999)  4134--4150},
\href{http://arxiv.org/abs/hep-th/9906065}{{\tt arXiv:hep-th/9906065}}.

\bibitem{Duff:1997hf}
M.~J. Duff, J.~M. Evans, R.~R. Khuri, J.~X. Lu, and R.~Minasian, ``{The
  octonionic membrane},''
  \href{http://dx.doi.org/10.1016/S0370-2693(97)01071-X}{{\em Phys. Lett.} {\bf
  B412} (1997)  281--287},
\href{http://arxiv.org/abs/hep-th/9706124}{{\tt arXiv:hep-th/9706124}}.

\bibitem{Smolin:2001wc}
L.~Smolin, ``{The exceptional Jordan algebra and the matrix string},''
\href{http://arxiv.org/abs/hep-th/0104050}{{\tt arXiv:hep-th/0104050}}.

\bibitem{Toppan:2003yx}
F.~Toppan, ``{On the octonionic M-superalgebra},'' in {\em Sao Paulo 2002,
  Integrable theories, solitons and duality}.
\newblock 2002.
\newblock
\href{http://arxiv.org/abs/hep-th/0301163}{{\tt arXiv:hep-th/0301163}}.
\newblock

\bibitem{Albuquerque:1998}
H.~Albuquerque and S.~Majid, ``{Quasialgebra structure of the octonions},''
  \href{http://dx.doi.org/10.1006/jabr.1998.7850}{{\em J. Algebra} {\bf 220}
  (1999) no.~1, 188--224}, \href{http://arxiv.org/abs/math/9802116}{{\tt
  arXiv:math/9802116}}.

\bibitem{Manogue:1999}
C.~A. Manogue and T.~Dray, ``{Octonionic M\"obius transformations},'' {\em Mod.
  Phys. Lett.} {\bf A14} (1999)  1243--1256,
  \href{http://arxiv.org/abs/math-ph/9905024}{{\tt arXiv:math-ph/9905024}}.

\bibitem{Gunaydin:1973}
M.~G{\"u}naydin and F.~G{\"u}rsey, ``{Quark structure and octonions},''
  \href{http://dx.doi.org/10.1063/1.1666240}{{\em Journal of Mathematical
  Physics} {\bf 14} (1973)  1651--1667}.

\bibitem{Kugo:1982bn}
T.~Kugo and P.~K. Townsend, ``{Supersymmetry and the division algebras},''
\href{http://dx.doi.org/10.1016/0550-3213(83)90584-9}{{\em Nucl. Phys.} {\bf
  B221} (1983)  357}.

\bibitem{Mosseri:2001}
R.~Mosseri and R.~Dandoloff, ``Geometry of entangled states, {Bloch} spheres
  and {Hopf} fibrations,''
  \href{http://dx.doi.org/10.1088/0305-4470/34/47/324}{{\em J. Phys.} {\bf A34}
  (2001) no.~47, 10243--10252},
  \href{http://arxiv.org/abs/quant-ph/0108137}{{\tt arXiv:quant-ph/0108137}}.

\bibitem{Mosseri:2003}
R.~Mosseri, ``{Two-qubits and three qubit geometry and Hopf fibrations},'' in
  {\em {Topology in Condensed Matter}}, M.~I. Monastyrsky, ed.
\newblock Springer Series in Solid-State Sciences, 2006.
\newblock \href{http://arxiv.org/abs/quant-ph/0310053}{{\tt
  arXiv:quant-ph/0310053}}.
\newblock Presented at the Dresde (Germany) ``Topology in Condensed Matter
  Physics'' colloquium in June 2002.

\bibitem{Gunaydin:1995as}
M.~G{\"u}naydin and S.~V. Ketov, ``{Seven-sphere and the exceptional $N=7$ and
  $N=8$ superconformal algebras},''
  \href{http://dx.doi.org/10.1016/0550-3213(96)00088-0}{{\em Nucl. Phys.} {\bf
  B467} (1996)  215--246},
\href{http://arxiv.org/abs/hep-th/9601072}{{\tt arXiv:hep-th/9601072}}.

\bibitem{Ferrara:2006em}
S.~Ferrara and R.~Kallosh, ``{On $N = 8$ attractors},''
  \href{http://dx.doi.org/10.1103/PhysRevD.73.125005}{{\em Phys. Rev.} {\bf
  D73} (2006) no.~12, 125005},
\href{http://arxiv.org/abs/hep-th/0603247}{{\tt arXiv:hep-th/0603247}}.

\bibitem{Lu:1997bg}
H.~Lu, C.~N. Pope, and K.~S. Stelle, ``{Multiplet structures of BPS
  solitons},'' \href{http://dx.doi.org/10.1088/0264-9381/15/3/007}{{\em Class.
  Quant. Grav.} {\bf 15} (1998)  537--561},
\href{http://arxiv.org/abs/hep-th/9708109}{{\tt arXiv:hep-th/9708109}}.

\bibitem{Jordan:1933vh}
P.~Jordan, J.~von Neumann, and E.~P. Wigner, ``{On an algebraic generalization
  of the quantum mechanical formalism}.''
\href{http://www.jstor.org/stable/pdfplus/1968117.pdf}{\textit{Ann. Math.}
  \textbf{35} (1934) no. 1, 29--64}.

\bibitem{Freudenthal:1954}
H.~Freudenthal, ``{Beziehungen der $E_7$ und $E_8$ zur oktavenebene I-II},''
  {\em Nederl. Akad. Wetensch. Proc. Ser.} {\bf 57} (1954)  218--230.

\bibitem{McCrimmon:1969}
K.~McCrimmon, ``{The Freudenthal-Springer-Tits construction of exceptional
  Jordan algebras}.''
  \href{http://www.jstor.org/stable/pdfplus/1995337.pdf}{\textit{Trans. Amer.
  Math. Soc.} \textbf{139} (1969) 495--510}.

\bibitem{Krutelevich:2004}
S.~Krutelevich, ``Jordan algebras, exceptional groups, and {Bhargava}
  composition,'' \href{http://dx.doi.org/10.1016/j.jalgebra.2007.02.060}{{\em
  J. Algebra} {\bf 314} (2007) no.~2, 924–977},
  \href{http://arxiv.org/abs/math/0411104}{{\tt arXiv:math/0411104}}.

\bibitem{Gunaydin:1983bi}
M.~G{\"u}naydin, G.~Sierra, and P.~K. Townsend, ``{The geometry of $N=2$
  Maxwell-Einstein supergravity and Jordan algebras},''
\href{http://dx.doi.org/10.1016/0550-3213(84)90142-1}{{\em Nucl. Phys.} {\bf
  B242} (1984)  244}.

\bibitem{Gunaydin:1983rk}
M.~G{\"u}naydin, G.~Sierra, and P.~K. Townsend, ``{Exceptional supergravity
  theories and the MAGIC square},''
\href{http://dx.doi.org/10.1016/0370-2693(83)90108-9}{{\em Phys. Lett.} {\bf
  B133} (1983)  72}.

\bibitem{Gunaydin:1984ak}
M.~G{\"u}naydin, G.~Sierra, and P.~K. Townsend, ``{Gauging the $d = 5$
  Maxwell-Einstein supergravity theories: More on Jordan algebras},''
\href{http://dx.doi.org/10.1016/0550-3213(85)90547-4}{{\em Nucl. Phys.} {\bf
  B253} (1985)  573}.

\bibitem{Jacobson:1958}
N.~Jacobson, ``{Composition algebras and their automorphisms},'' {\em Rend.
  Circ. Mat. Palermo} {\bf 7} (1958)  58--80.

\bibitem{Hurwitz:1898}
A.~Hurwitz, ``{Uber die komposition der quadratishen formen von beliebig vielen
  variabeln},'' {\em Nachr. Ges. Wiss. Gottingen} (1898)  309--316.

\bibitem{Jordan:1933a}
P.~Jordan, ``{\"Uber die multiplikation quanten-mechanischer grossen},'' {\em
  Zschr. f. Phys.} {\bf 80} (1933)  285.

\bibitem{Jacobson:1968}
N.~Jacobson, {\em Structure and Representations of Jordan Algebras}, vol.~39.
\newblock American Mathematical Society Colloquium Publications, 1968.

\bibitem{Gunaydin:2005zz}
M.~G{\"u}naydin and O.~Pavlyk, ``{Generalized spacetimes defined by cubic forms
  and the minimal unitary realizations of their quasiconformal groups},''
  \href{http://dx.doi.org/10.1088/1126-6708/2005/08/101}{{\em JHEP} {\bf 08}
  (2005)  101},
\href{http://arxiv.org/abs/hep-th/0506010}{{\tt arXiv:hep-th/0506010}}.

\bibitem{Springer:1962}
T.~A. Springer, ``{Characterization of a class of cubic forms},'' {\em Nederl.
  Akad. Wetensch. Proc. Ser. A} {\bf 24} (1962)  259--265.

\bibitem{Brown:1969}
R.~B. Brown, ``{Groups of type $E_7$},'' {\em J. Reine Angew. Math.} {\bf 236}
  (1969)  79--102.

\bibitem{Jacobson:1961}
N.~Jacobson, ``{Some groups of transformations defined by Jordan algebras},''
  {\em J. Reine Angew. Math.} {\bf 207} (1961)  61--85.

\bibitem{Gunaydin:1978}
M.~G{\"u}naydin, C.~Piron, and H.~Ruegg, ``Moufang plane and octonionic quantum
  mechanics,'' \href{http://dx.doi.org/10.1007/BF01609468}{{\em Comm. Math.
  Phys.} {\bf 61} (1978) no.~1, 69--85}.

\bibitem{Maldacena:1999bp}
J.~M. Maldacena, G.~W. Moore, and A.~Strominger, ``{Counting BPS black holes in
  toroidal type II string theory},''
\href{http://arxiv.org/abs/hep-th/9903163}{{\tt arXiv:hep-th/9903163}}.

\bibitem{Krutelevich:2002}
S.~Krutelevich, ``{On a canonical form of a $3\times 3$ Herimitian matrix over
  the ring of integral split octonions},''
  \href{http://dx.doi.org/10.1016/S0021-8693(02)00127-8}{{\em J. Algebra} {\bf
  253} (2002) no.~2, 276--295}.

\bibitem{Bhargava:2004}
M.~Bhargava, ``{Higher composition laws I: A new view on Gauss composition, and
  quadratic generalizations}.''
  {\href{http://annals.math.princeton.edu/issues/2004/BhargavaJan04.pdf}{
  \textit{Ann. Math.} \textbf{159} (2004) no. 1, 217-250}}.

\bibitem{Ferrara:1996um}
S.~Ferrara and R.~Kallosh, ``{Universality of supersymmetric attractors},''
  \href{http://dx.doi.org/10.1103/PhysRevD.54.1525}{{\em Phys. Rev.} {\bf D54}
  (1996) no.~2, 1525--1534},
\href{http://arxiv.org/abs/hep-th/9603090}{{\tt arXiv:hep-th/9603090}}.

\bibitem{Andrianopoli:1997hb}
L.~Andrianopoli, R.~D'Auria, and S.~Ferrara, ``{Five dimensional U-duality,
  black-hole entropy and topological invariants},''
  \href{http://dx.doi.org/10.1016/S0370-2693(97)00949-0}{{\em Phys. Lett.} {\bf
  B411} (1997)  39--45},
\href{http://arxiv.org/abs/hep-th/9705024}{{\tt arXiv:hep-th/9705024}}.

\bibitem{Andrianopoli:1998ah}
L.~Andrianopoli, R.~D'Auria, and S.~Ferrara, ``{U-duality, attractors and
  Bekenstein-Hawking entropy in four and five dimensional supergravities},''
\href{http://dx.doi.org/10.1016/S0920-5632(98)00115-7}{{\em Nucl. Phys. Proc.
  Suppl.} {\bf 67} (1998)  17--24}.

\bibitem{Briand:2003}
E.~Briand, J.-G. Luque, J.-Y. Thibon, and F.~Verstraete, ``The moduli space of
  three-qutrit states,'' \href{http://dx.doi.org/10.1063/1.1809255}{{\em J.
  Math. Phys.} {\bf 45} (2004) no.~12, 4855--4867},
  \href{http://arxiv.org/abs/quant-ph/0306122}{{\tt arXiv:quant-ph/0306122}}.

\bibitem{Ferrara:2007pc}
S.~Ferrara and A.~Marrani, ``{$N=8$ non-BPS attractors, fixed scalars and
  magicvsupergravities},''
  \href{http://dx.doi.org/10.1016/j.nuclphysb.2007.07.028}{{\em Nucl. Phys.}
  {\bf B788} (2008)  63--88},
\href{http://arxiv.org/abs/0705.3866}{{\tt arXiv:0705.3866 [hep-th]}}.

\bibitem{Rios:2007qn}
M.~Rios, ``{Jordan algebras and extremal black holes},''
  \href{http://arxiv.org/abs/hep-th/0703238}{{\tt arXiv:hep-th/0703238}}.
Talk given at 26th International Colloquium on Group Theoretical Methods in
  Physics (ICGTMP26), New York City, New York, 26-30 Jun 2006.

\bibitem{Ferrara:2006yb}
S.~Ferrara, E.~G. Gimon, and R.~Kallosh, ``{Magic supergravities, $N = 8$ and
  black hole composites},''
  \href{http://dx.doi.org/10.1103/PhysRevD.74.125018}{{\em Phys. Rev.} {\bf
  D74} (2006) no.~12, 125018},
\href{http://arxiv.org/abs/hep-th/0606211}{{\tt arXiv:hep-th/0606211}}.

\bibitem{Feynman:1985}
R.~P. Feynman, R.~Leighton~(contributor), and E.~Hutchings~(editor), {\em
  {Surely You're Joking, Mr. Feynman!: Adventures of a Curious Character}}.
\newblock W. W. Norton, 1985.

\bibitem{Maldacena:1997de}
J.~M. Maldacena, A.~Strominger, and E.~Witten, ``{Black hole entropy in
  M-theory},'' \href{http://dx.doi.org/10.1088/1126-6708/1997/12/002}{{\em
  JHEP} {\bf 12} (1997)  002},
\href{http://arxiv.org/abs/hep-th/9711053}{{\tt arXiv:hep-th/9711053}}.

\bibitem{Brustein:2005vx}
R.~Brustein, M.~B. Einhorn, and A.~Yarom, ``{Entanglement interpretation of
  black hole entropy in string theory},''
  \href{http://dx.doi.org/10.1088/1126-6708/2006/01/098}{{\em JHEP} {\bf 01}
  (2006)  098},
\href{http://arxiv.org/abs/hep-th/0508217}{{\tt arXiv:hep-th/0508217}}.

\bibitem{Emparan:2006ni}
R.~Emparan, ``{Black hole entropy as entanglement entropy: A holographic
  derivation},'' \href{http://dx.doi.org/10.1088/1126-6708/2006/06/012}{{\em
  JHEP} {\bf 06} (2006)  012},
\href{http://arxiv.org/abs/hep-th/0603081}{{\tt arXiv:hep-th/0603081}}.

\bibitem{Hirata:2006jx}
T.~Hirata and T.~Takayanagi, ``{AdS/CFT and strong subadditivity of
  entanglement entropy},''
  \href{http://dx.doi.org/10.1088/1126-6708/2007/02/042}{{\em JHEP} {\bf 02}
  (2007)  042},
\href{http://arxiv.org/abs/hep-th/0608213}{{\tt arXiv:hep-th/0608213}}.

\bibitem{Ryu:2006ef}
S.~Ryu and T.~Takayanagi, ``{Aspects of holographic entanglement entropy},''
  \href{http://dx.doi.org/10.1088/1126-6708/2006/08/045}{{\em JHEP} {\bf 08}
  (2006)  045},
\href{http://arxiv.org/abs/hep-th/0605073}{{\tt arXiv:hep-th/0605073}}.

\bibitem{Blasone:2007vw}
M.~Blasone, F.~Dell'Anno, S.~De~Siena, and F.~Illuminati, ``{Entropy,
  entanglement, and transition probabilities in neutrino oscillations},''
\href{http://arxiv.org/abs/0707.4476}{{\tt arXiv:0707.4476 [hep-ph]}}.

\bibitem{Ge:2007ig}
X.-H. Ge and S.~P. Kim, ``{Quantum entanglement and teleportation in higher
  dimensional black hole spacetimes},''
  \href{http://dx.doi.org/10.1088/0264-9381/25/7/075011}{{\em Class. Quant.
  Grav.} {\bf 25} (2008)  075011},
\href{http://arxiv.org/abs/0707.4523}{{\tt arXiv:0707.4523 [quant-ph]}}.

\bibitem{Andrianopoli:1998qg}
L.~Andrianopoli, R.~D'Auria, S.~Ferrara, and M.~A. Lledo, ``{Horizon geometry,
  duality and fixed scalars in six dimensions},''
  \href{http://dx.doi.org/10.1016/S0550-3213(98)00332-0}{{\em Nucl. Phys.} {\bf
  B528} (1998)  218--228},
\href{http://arxiv.org/abs/hep-th/9802147}{{\tt arXiv:hep-th/9802147}}.

\bibitem{Townsend:1985eq}
P.~K. Townsend, ``The {Jordan} formulation of quantum mechanics: A review,''.
  Published in GIFT Seminar (1984) 0346. Print-85-0263 (Cambridge).

\bibitem{Jordan:1933b}
P.~Jordan, ``{\"Uber verallgemeinerungsm\"oglichkeiten des formalismus der
  quantenmechanik},'' {\em Nachr. Ges. Wiss. Gottingen} (1933)  209--214.

\bibitem{Emch:1997}
G.~G. Emch, ``{Foundations of quantum mechanics: Building on von Neumann's
  heritage},''
  \href{http://dx.doi.org/10.1002/(SICI)1097-461X(1997)65:5<379::AID-QUA2>3.0.%
CO;2-T}{{\em Int. J. Quant. Chem.} {\bf 65} (1997) no.~5, 379--387}.

\bibitem{Bischoff:1993sr}
W.~Bischoff, ``{On a Jordan algebraic formulation of quantum mechanics: Hilbert
  space construction},''
\href{http://arxiv.org/abs/hep-th/9304124}{{\tt arXiv:hep-th/9304124}}.

\bibitem{Adler:1995}
S.~L. Adler, {\em {Quaternionic Quantum Mechanics and Quantum Fields}}.
\newblock Oxford University Press, New York, 1995.

\bibitem{DeLeo:1996mr}
S.~De~Leo and K.~Abdel-Khalek, ``{Octonionic quantum mechanics and complex
  geometry},'' \href{http://dx.doi.org/10.1143/PTP.96.823}{{\em Prog. Theor.
  Phys.} {\bf 96} (1996)  823--832},
\href{http://arxiv.org/abs/hep-th/9609032}{{\tt arXiv:hep-th/9609032}}.

\bibitem{Manogue:1993ja}
C.~A. Manogue and J.~Schray, ``{Finite Lorentz transformations, automorphisms,
  and division algebras},'' \href{http://dx.doi.org/10.1063/1.530056}{{\em J.
  Math. Phys.} {\bf 34} (1993)  3746--3767},
\href{http://arxiv.org/abs/hep-th/9302044}{{\tt arXiv:hep-th/9302044}}.

\bibitem{Aaronson:2004}
S.~Aaronson, ``Is quantum mechanics an island in theoryspace?,''
  \href{http://arxiv.org/abs/quant-ph/0401062}{{\tt arXiv:quant-ph/0401062}}.

\bibitem{Bergshoeff:1996rn}
E.~Bergshoeff, M.~de~Roo, E.~Eyras, B.~Janssen, and J.~P. van~der Schaar,
  ``{Multiple intersections of D-branes and M-branes},''
  \href{http://dx.doi.org/10.1016/S0550-3213(97)00151-X}{{\em Nucl. Phys.} {\bf
  B494} (1997)  119--143},
\href{http://arxiv.org/abs/hep-th/9612095}{{\tt arXiv:hep-th/9612095}}.

\bibitem{Gauntlett:1996pb}
J.~P. Gauntlett, D.~A. Kastor, and J.~H. Traschen, ``{Overlapping Branes in
  M-Theory},'' \href{http://dx.doi.org/10.1016/0550-3213(96)00423-3}{{\em Nucl.
  Phys.} {\bf B478} (1996)  544--560},
\href{http://arxiv.org/abs/hep-th/9604179}{{\tt arXiv:hep-th/9604179}}.

\bibitem{Gauntlett:1997cv}
J.~P. Gauntlett, ``{Intersecting branes},'' in {\em Seoul/Sokcho 1997,
  Dualities in gauge and string theories}, pp.~146--193.
\newblock 1997.
\newblock \href{http://arxiv.org/abs/hep-th/9705011}{{\tt
  arXiv:hep-th/9705011}}.
\newblock
Based on the lectures given at APCTP Winter School on Dualities of Gauge and
  String Theories, 17-28 Feb 1997, Seoul and Sokcho, Korea.

\bibitem{Tseytlin:1996bh}
A.~A. Tseytlin, ``{Harmonic superpositions of M-branes},''
  \href{http://dx.doi.org/10.1016/0550-3213(96)00328-8}{{\em Nucl. Phys.} {\bf
  B475} (1996)  149--163},
\href{http://arxiv.org/abs/hep-th/9604035}{{\tt arXiv:hep-th/9604035}}.

\bibitem{Klebanov:1996mh}
I.~R. Klebanov and A.~A. Tseytlin, ``{Intersecting M-branes as four-dimensional
  black holes},'' \href{http://dx.doi.org/10.1016/0550-3213(96)00338-0}{{\em
  Nucl. Phys.} {\bf B475} (1996)  179--192},
\href{http://arxiv.org/abs/hep-th/9604166}{{\tt arXiv:hep-th/9604166}}.

\bibitem{Balasubramanian:1997ak}
V.~Balasubramanian, ``{How to count the states of extremal black holes in $N =
  8$ supergravity},'' in {\em Cargese 1997, Strings, branes and dualities},
  pp.~399--410.
\newblock 1997.
\newblock \href{http://arxiv.org/abs/hep-th/9712215}{{\tt
  arXiv:hep-th/9712215}}.
\newblock
Published in the proceedings of NATO Advanced Study Institute on Strings,
  Branes and Dualities, Cargese, France, 26 May - 14 Jun 1997.

\bibitem{Bertolini:2000ei}
M.~Bertolini and M.~Trigiante, ``{Regular BPS black holes: Macroscopic and
  microscopic description of the generating solution},''
  \href{http://dx.doi.org/10.1016/S0550-3213(00)00216-9}{{\em Nucl. Phys.} {\bf
  B582} (2000)  393--406},
\href{http://arxiv.org/abs/hep-th/0002191}{{\tt arXiv:hep-th/0002191}}.

\bibitem{Bertolini:2000yaa}
M.~Bertolini and M.~Trigiante, ``{Microscopic entropy of the most general
  four-dimensional BPS black hole},''
  \href{http://dx.doi.org/10.1088/1126-6708/2000/10/002}{{\em JHEP} {\bf 10}
  (2000)  002},
\href{http://arxiv.org/abs/hep-th/0008201}{{\tt arXiv:hep-th/0008201}}.

\bibitem{Papadopoulos:1996uq}
G.~Papadopoulos and P.~K. Townsend, ``{Intersecting M-branes},''
  \href{http://dx.doi.org/10.1016/0370-2693(96)00506-0}{{\em Phys. Lett.} {\bf
  B380} (1996)  273--279},
\href{http://arxiv.org/abs/hep-th/9603087}{{\tt arXiv:hep-th/9603087}}.

\bibitem{Gaiotto:2005gf}
D.~Gaiotto, A.~Strominger, and X.~Yin, ``{New connections between 4D and 5D
  black holes},'' \href{http://dx.doi.org/10.1088/1126-6708/2006/02/024}{{\em
  JHEP} {\bf 02} (2006)  024},
\href{http://arxiv.org/abs/hep-th/0503217}{{\tt arXiv:hep-th/0503217}}.

\bibitem{Gaiotto:2005xt}
D.~Gaiotto, A.~Strominger, and X.~Yin, ``{5D black rings and 4D black holes},''
  \href{http://dx.doi.org/10.1088/1126-6708/2006/02/023}{{\em JHEP} {\bf 02}
  (2006)  023},
\href{http://arxiv.org/abs/hep-th/0504126}{{\tt arXiv:hep-th/0504126}}.

\bibitem{Luque:2002}
J.-G. Luque and J.-Y. Thibon, ``{Polynomial invariants of four qubits},''
  \href{http://dx.doi.org/10.1103/PhysRevA.67.042303}{{\em Phys. Rev.} (2003)
  no.~4, 042303}, \href{http://arxiv.org/abs/quant-ph/0212069}{{\tt
  arXiv:quant-ph/0212069}}.

\bibitem{Duff:2006ev}
M.~J. Duff, ``{Hidden symmetries of the Nambu-Goto action},''
  \href{http://dx.doi.org/10.1016/j.physletb.2006.08.050}{{\em Phys. Lett.}
  {\bf B641} (2006)  335--337},
\href{http://arxiv.org/abs/hep-th/0602160}{{\tt arXiv:hep-th/0602160}}.

\bibitem{Nishino:2007ke}
H.~Nishino and S.~Rajpoot, ``{Green-Schwarz, Nambu-Goto actions, and Cayley's
  hyperdeterminant},''
  \href{http://dx.doi.org/10.1016/j.physletb.2007.06.063}{{\em Phys. Lett.}
  {\bf B652} (2007)  135--140},
\href{http://arxiv.org/abs/0709.0973}{{\tt arXiv:0709.0973 [hep-th]}}.

\bibitem{Olver:1999}
P.~J. Olver, {\em {Classical Invariant Theory}}.
\newblock Cambridge University Press, 1999.

\bibitem{Briand:2003a}
E.~Briand, J.-G. Luque, and J.-Y. Thibon, ``A complete set of covariants of the
  four qubit system,''
  \href{http://dx.doi.org/10.1088/0305-4470/36/38/309}{{\em J. Phys.} {\bf A36}
  (2003)  9915--9927}, \href{http://arxiv.org/abs/quant-ph/0304026}{{\tt
  arXiv:quant-ph/0304026}}.

\bibitem{Endrejat:2006}
J.~Endrejat and H.~Buettner, ``{Polynomial invariants and Bell inequalities as
  entanglement measure of 4-qubit states},''
  \href{http://arxiv.org/abs/quant-ph/0606215}{{\tt arXiv:quant-ph/0606215}}.

\bibitem{Strathdee:1986jr}
J.~A. Strathdee, ``Extended {Poincar\'e} supersymmery,''
\href{http://dx.doi.org/10.1142/S0217751X87000120}{{\em Int. J. Mod. Phys.}
  {\bf A2} (1987)  273}.

\bibitem{Hirschfeld:1998}
J.~W.~P. Hirschfeld, {\em Projective Geometries over Finite Fields (Second
  Edition)}.
\newblock Oxford University Press, 1998.

\bibitem{Luhn:2007yr}
C.~Luhn, S.~Nasri, and P.~Ramond, ``{Simple finite non-Abelian flavor
  groups},'' \href{http://dx.doi.org/10.1063/1.2823978}{{\em J. Math. Phys.}
  {\bf 48} (2007)  123519},
\href{http://arxiv.org/abs/0709.1447}{{\tt arXiv:0709.1447 [hep-th]}}.

\bibitem{Luhn:2008sa}
C.~Luhn and P.~Ramond, ``{Anomaly conditions for non-Abelian finite family
  symmetries},'' \href{http://dx.doi.org/10.1088/1126-6708/2008/07/085}{{\em
  JHEP} {\bf 07} (2008)  085},
\href{http://arxiv.org/abs/0805.1736}{{\tt arXiv:0805.1736 [hep-ph]}}.

\bibitem{Conway:1985}
J.~H. Conway, R.~T. Curtis, S.~P. Norton, R.~A. Wilson, and R.~A. Parker, {\em
  {ATLAS of Finite Groups: Maximal Subgroups and Ordinary Characters for Simple
  Groups}}.
\newblock Oxford University Press, 1985.

\end{thebibliography}
\providecommand{\href}[2]{#2}\begingroup\raggedright\endgroup

\end{document}